\documentclass[revtex4,twocolappendix, apj]{emulateapj}
\usepackage[caption = false]{subfig}
\usepackage{color}
\usepackage{lscape}
\usepackage{natbib}
\usepackage{comment}
\usepackage{appendix}
\usepackage{float}
\usepackage{flafter}
\restylefloat{figure}
\shorttitle{Compendium of Benchmark Objects}
\shortauthors{Gonzales et al.}
\begin{document}

\title{The TRENDS High-Contrast Imaging Survey. VIII. \\ Compendium of Benchmark Objects}

\author{Erica J. Gonzales\altaffilmark{1,2,$\dagger$}, Justin R. Crepp\altaffilmark{1}, Eric B. Bechter\altaffilmark{1}, Charlotte M. Wood\altaffilmark{1}, John Asher Johnson\altaffilmark{3}, Benjamin T. Montet\altaffilmark{4,$\ddagger$}, Howard Isaacson \altaffilmark{5}, and Andrew W. Howard \altaffilmark{6}}, 
\altaffiltext{1}{Department of Physics, University of Notre Dame, 225 Nieuwland Science Hall, Notre Dame IN 46556, USA }
\altaffiltext{2}{University of California, Santa Cruz, 1156 High Street, Santa Cruz CA 95065, USA}
\altaffiltext{3}{Harvard Center for Astrophysics, 60 Garden Street, Cambridge MA 02138, USA}
\altaffiltext{4}{University of Chicago, Department of Astronomy, 5640 S Ellis Ave, Chicago IL 60637, USA}
\altaffiltext{5}{University of California, Berkeley, 501 Campbell Hall, Berkeley CA 94720, USA }
\altaffiltext{6}{California Institute of Technology, 1200 E. California Blvd., Pasadena CA 91125, USA}
\altaffiltext{$\dagger$}{National Science Foundation Graduate Research Fellow}
\altaffiltext{$\ddagger$}{Sagan Fellow}
\email{erjagonz@ucsc.edu}

\begin{abstract}
The physical properties of faint stellar and substellar objects often rely on indirect, model-dependent estimates. For example, the masses of brown dwarfs are usually inferred using evolutionary models, which are age dependent and have yet to be properly calibrated. With the goal of identifying new benchmark objects to test low-mass stellar and substellar models, we have carried out a comprehensive adaptive optics survey as part of the TRENDS high-contrast imaging program. Using legacy radial velocity measurements from HIRES at Keck, we have identified several dozen stars that show long-term Doppler accelerations. We present follow-up high-contrast observations from the campaign and report the discovery of 31 co-moving companions, as well as 11 strong candidate companions, to solar-type stars with well-determined parallax and metallicity values. Benchmark objects of this nature lend themselves to orbit determinations, dynamical mass estimates, and independent compositional assessment. This compendium of benchmark objects will serve as a convenient test group to substantiate theoretical evolutionary and atmospheric models near the hydrogen fusing limit. 
\end{abstract}

\section{Introduction}\label{sec:intro}
Very low mass (VLM) stars and brown dwarfs (BD) represent some of the most abundant objects in the universe. Studying their physical properties has implications for our understanding of star formation and evolution, as well as the detection and characterization of extrasolar planets \citep{lada_03,muirhead_18}. Mass is of particular interest as it fundamentally governs the evolution and fate of stellar and substellar objects. Comprising the spectral classes M, L, T, and Y, VLM stars and BDs span $\approx$1.5 dex of dynamic range in mass. Accurate mass estimates however are difficult to obtain due to their cool, dense nature \citep{2002ApJ...571..519L,2012ApJ...753..156K,2017ApJS..231...15D}. Instead, theoretical models are used to infer physical parameters.

Mass, radius, age, and composition often represent degenerate model variables. As such, stellar and substellar ``benchmark" systems with directly measurable physical properties serve as convenient laboratories for testing theoretical models \citep{2000ApJ...542..464C,2001udns.conf...26B,2001ApJ...556..357A,2008ApJ...689.1327S,2015A&A...577A..42B,2008ApJ...689..436L,2012ApJ...757..112B,2013MSAIS..24..128A,2013MmSAI..84.1005K,2015ApJ...805...56D,2015ApJ...813L..23B,2017AJ....153..267M,2016ApJ...831..136C,2017ApJS..231...15D,2018ApJ...853..192C,2018AJ....155..159B}. Dynamical masses of VLM stars and BDs can be determined using laser guide star adaptive optics (AO) to spatially resolve the individual components of tight binary and multi-star systems \citep{2008ApJ...689..436L}. Likewise, faint companions to nearby main-sequence stars can be detected with high-contrast imaging; such objects are particularly valuable as they allow for distance, age, and composition to be inferred from the well-characterized parent star \citep{2008ApJ...689..436L,2014ApJ...781...29C,2016ApJ...831..136C,2018arXiv180302725C,2018AJ....155..159B,2019arXiv190103687W}. 

The TRENDS (\textbf{T}a\textbf{R}getting b\textbf{EN}chmark-objects with \textbf{D}oppler \textbf{S}pectroscopy) high-contrast imaging survey uses AO imaging and coronography to search for companions in the VLM star, BD, and exoplanet regime \citep{2012ApJ...761...39C, 2013ApJ...771...46C,2013ApJ...774....1C}. In this paper, we report the discovery of 31 co-moving companions, as well as 11 strong candidate companions, to nearby solar-type stars. Of these, 6 companions are found to be members of triple star systems. Targets were selected from the California Planet Search (CPS) program \citep{2010ApJ...721.1467H}. In addition to exhibiting a long-term radial velocity (RV) acceleration (``trend"), which was used to identify previously unseen candidate companions, each star also has a well-determined parallax and metallicity (Table 1, Table 2).

\begin{deluxetable*}{lccccc}[h]
\tablecaption{TRENDS System Parameters}
\tablehead{\colhead{Star} & \colhead{RA} & \colhead{Dec} &\colhead{$\pi$ [mas]} & \colhead{$\mu_{\alpha}$[mas/yr]} & \colhead{$\mu_{\delta}$[mas/yr]}}
\startdata
HD 224983 & 00 02 21.54 & +11 00 22.46 & 29.52 $\pm$ 0.06 & -4.95 $\pm$ 0.11 & -42.71 $\pm$ 0.07 \\
HIP 1294 & 00 16 11.59 & +30 31 57.28 & 25.46 $\pm$ 0.12 & 189.89 $\pm$ 0.17 & 47.84 $\pm$ 0.16\\

HD 1205 & 00 16 23.46 & -22 35 16.72 & 15.57 $\pm$ 0.80 & 19.96 $\pm$ 1.43 & -42.68 $\pm$ 1.60\\ 
HD 1384 & 00 18 12.14 & +52 24 45.23 & 6.24 $\pm$ 0.14 & -17.50 $\pm$ 0.20 & -53.50 $\pm$ 0.20 \\ 
HD 6512 & 01 06 12.63& +13 15 09.88&  18.36 $\pm$ 0.06 & 166.03 $\pm$ 0.10 & -10.86 $\pm$ 0.07 \\
HD 31018 & 04 53 22.05 & +24 10 38.47 & 9.93 $\pm$ 0.06 & 21.14 $\pm$ 0.10 & -17.74 $\pm$ 0.06 \\
HD 34721& 05 18 50.47& -18 07 48.19 & 40.47 $\pm$ 0.06 & 386.01 $\pm$ 0.10 & 61.01 $\pm$ 0.10\\
HD 40647 & 06 06 05.77 & +69 28 34.42 & 31.86 $\pm$ 0.16 & -165.11 $\pm$ 0.32 & -82.47 $\pm$ 0.32\\
HD 50639& 06 53 44.83& -09 30 55.67 & 25.68 $\pm$ 0.06 & -69.04 $\pm$ 0.08 & -31.34 $\pm$ 0.07\\
HD 85472 & 09 53 57.96& +57 41 01.33 & 13.16 $\pm$ 0.03 & 36.51 $\pm$ 0.05 & -92.28 $\pm$ 0.05 \\
HD 88986 & 10 16 28.08 & +28 40 56.94 & 30.03 $\pm$ 0.04 & -65.69 $\pm$ 0.09 & -95.51 $\pm$ 0.07 \\
HIP 55507 & 11 22 05.75 & +46 54 30.18 & 39.20 $\pm$ 0.04 & -197.56 $\pm$ 0.04 & -134.88 $\pm$ 0.04\\
HD 110537& 12 42 59.33& -04 02 57.60 &  22.15 $\pm$ 0.05 & -219.37 $\pm$ 0.11 & -179.68 $\pm$ 0.06\\ 
HD 111031& 12 46 30.84 & -11 48 44.79 &  32.02 $\pm$ 0.05 & -279.79 $\pm$ 0.09 & 46.71 $\pm$ 0.07\\
HIP 63762& 13 04 07.28 & +87 06 55.55 & 39.51 $\pm$ 0.04 & -92.16 $\pm$ 0.10 & 181.12 $\pm$ 0.07\\
HD 129191& 14 41 16.18 & -04 56 41.53 & 18.39 $\pm$ 0.05 & 26.42 $\pm$ 0.09 & -152.97 $\pm$ 0.08\\
HD 129814& 14 44 11.69 & +18 27 43.62 & 23.81 $\pm$ 0.06 & -59.34 $\pm$ 0.07 & -165.08 $\pm$ 0.09 \\
HD 136274& 15 18 59.05 & +25 41 30.29 & 29.42 $\pm$ 0.05 & -564.57 $\pm$ 0.06 & -125.14 $\pm$ 0.07 \\
HD 139457& 15 37 59.21 & +10 14 23.56 & 20.82 $\pm$ 0.05 & 129.82 $\pm$ 0.08 & -358.77 $\pm$ 0.08 \\
HD 142229& 15 53 20.01 & +04 15 11.70 & 21.62 $\pm$ 0.04 & -34.24 $\pm$ 0.08 & 9.29 $\pm$ 0.06\\
HD 147231& 16 14 50.25 & +70 55 46.80 & 24.86 $\pm$ 0.02 & -5.28 $\pm$ 0.05 & -287.27 $\pm$ 0.05 \\
HD 155413& 17 11 58.64 & -14 37 13.37 & 11.85 $\pm$ 0.05 & -18.50 $\pm$ 0.09 & -39.82 $\pm$ 0.07 \\
HD 157338& 17 24 08.74 & -34 47 54.44 & 30.18 $\pm$ 0.05 & -0.39 $\pm$ 0.72 & -182.73 $\pm$ 0.37\\
HD 164509& 18 01 31.23 & +00 06 16.40 & 18.80 $\pm$ 0.05 & -7.86 $\pm$ 0.09 & -20.38 $\pm$ 0.09 \\
HD 180684& 19 16 48.59 & +18 58 34.74 & 17.85 $\pm$ 0.04 & 10.68 $\pm$ 0.05 & -65.08 $\pm$ 0.06 \\
HD 183473& 19 27 57.24 & +42 46 35.90 & 11.16 $\pm$ 0.03 & 68.15 $\pm$ 0.05 & 160.24 $\pm$ 0.05\\
HD 196201& 20 35 26.23 & +11 21 25.66 & 24.32 $\pm$ 0.11 & 72.52 $\pm$ 0.15 & 366.03 $\pm$ 0.11\\ 
HD 201924& 21 11 10.85 & +45 27 21.29 & 31.85 $\pm$ 0.04 &  -239.39 $\pm$ 0.06 & -300.78 $\pm$ 0.06\\
HD 213519& 22 31 55.72 & +45 08 42.35 & 24.36 $\pm$ 0.03 & -174.73 $\pm$ 0.05 & 34.55 $\pm$ 0.05 \\ 
\noalign{\vskip 0.1cm}
\hline
\noalign{\vskip 0.1cm}
HD 1293 & 00 17 05.55 & -01 39 10.85 & 5.13 $\pm$ 0.19 & 20.10 $\pm$ 0.44 & 10.41 $\pm$ 0.24 \\ 
HD 1388 & 00 17 58.87 & -13 27 20.31 & 37.11 $\pm$ 0.05 & 401.22 $\pm$ 0.13 & -0.13 $\pm$ 0.05 \\
HD 4406 & 00 46 51.74 & +46 21 48.99 & 8.03 $\pm$ 0.09 & -24.54 $\pm$ 0.10 & -21.47 $\pm$ 0.16 \\
HD 6558 & 01 06 25.72 & -00 44 58.03 & 12.18 $\pm$ 0.06 & 70.80 $\pm$ 0. 10 & -35.26 $\pm$ 0.06 \\
HIP 46199& 09 25 10.78 & +46 05 53.65 &  44.40 $\pm$ 0.82 &  -221.04 $\pm$ 1.20 & 24.98 $\pm$ 1.20 \\ 
HD 103829& 11 57 31.21& +53 33 15.09& 11.25 $\pm$ 1.37& 44.02 $\pm$ 0.98& -10.98 $\pm$ 0.97\\
HD 105618& 12 09 36.62 & +11 12 41.68 & 14.24 $\pm$ 0.05 & -104.61 $\pm$ 0.10& -1.96 $\pm$ 0.05\\
HD 131509& 14 53 11.92 & +28 30 29.78 & 12.91 $\pm$ 0.05 & -94.09 $\pm$ 0.08 & 86.55 $\pm$ 0.10\\
HD 156826& 17 19 59.55 & -05 55 03.02 & 20.94 $\pm$ 0.04 & 45.29 $\pm$ 0.08 & -193.00 $\pm$ 0.06 \\
HD 217165& 22 58 29.88 & +09 49 32.01 & 22.74 $\pm$ 0.05 & 109.66 $\pm$ 0.10 & -1.44 $\pm$ 0.07
\enddata
\tablecomments{Right ascension (RA), declination (DEC), parallax ($\pi$), and proper motions ($\mu_{\alpha}$, $\mu_{\delta}$) are taken from the Gaia Data Release 2 \citep{2018arXiv180409365G}. \\
The horizontal line separates systems with confirmed companions (above) and systems with candidate companions (below).}
\end{deluxetable*}

\begin{deluxetable*}{lccccccc}[h]
\tablecaption{Spectral-type, Metallicity, and Apparent Magnitudes}
\tablehead{\colhead{Star} & \colhead{SpTy} &  \colhead{[Fe/H]}& \colhead{$B$}  & \colhead{$V$}  & \colhead{$J$}   & \colhead{$H$}  & \colhead{$K_{s}$}}
\startdata

HD 224983 & K0V & 0.00 $\pm$ 0.03& 9.33 $\pm$ 0.01 & 8.47 $\pm$ 0.01 & \nodata& \nodata& \nodata\\

HIP 1294 &K4V& 0.17 $\pm$ 0.04 & 9.84 $\pm$ 0.03 & 8.82 $\pm$ 0.01 & 6.92 $\pm$ 0.02 & 6.40 $\pm$ 0.04& 6.31 $\pm$ 0.02 \\

HD 1205 & G3V & 0.30 $\pm$ 0.04 & 8.58 $\pm$ 0.02 & 7.90 $\pm$ 0.01 & 6.73 $\pm$ 0.02 &  6.44 $\pm$ 0.02 &  6.36 $\pm$ 0.02 \\

HD 1384 & G5& 0.26 $\pm$ 0.04 &9.13 $\pm$ 0.01 & 8.10 $\pm$ 0.01 & 6.37 $\pm$ 0.02 & 5.90 $\pm$ 0.02 & 5.78 $\pm$ 0.02  \\ 

HD 6512 & G0& 0.12 $\pm$ 0.04& 8.81 $\pm$ 0.02 & 8.15 $\pm$ 0.01 & 6.89 $\pm$ 0.02 & 6.64 $\pm$ 0.02 & 6.57 $\pm$ 0.02  \\

HD 31018 & F8& 0.18 $\pm$ 0.04 & 8.28 $\pm$ 0.02 & 7.62 $\pm$ 0.01 & 6.45 $\pm$ 0.03 & 6.16 $\pm$ 0.02 & 6.08 $\pm$ 0.02 \\

HD 34721& G0V& -0.07 $\pm$ 0.03 & 6.54 & 5.966 & 5.18 $\pm$ 0.27 & 4.75 $\pm$ 0.27 & 4.55 $\pm$ 0.02 \\

HD 40647 & G5V& -0.06 $\pm$ 0.04 & 9.06 $\pm$ 0.02 & 8.26 $\pm$ 0.01 & 6.81 $\pm$ 0.03 & 6.43 $\pm$ 0.04 & 6.33 $\pm$ 0.02\\

HD 50639& F8.5V& -0.02 $\pm$ 0.03& 7.59 $\pm$ 0.02 & 7.05 $\pm$ 0.01 & 6.04 $\pm$ 0.03 & 5.81 $\pm$ 0.03 & 5.69 $\pm$ 0.02\\

HD 85472 & G8IV& -0.05 $\pm$ 0.04 & 8.25 $\pm$ 0.01 & 7.45 $\pm$ 0.01 & 5.99 $\pm$ 0.02 & 5.59 $\pm$ 0.03 & 5.55 $\pm$ 0.06\\

HD 88986 & G2V& 0.09 $\pm$ 0.03 & 7.10 $\pm$ 0.02 & 6.47 $\pm$ 0.01 & 5.2 $\pm$ 0.02 & 4.95 $\pm$ 0.02 & 4.88 $\pm$ 0.02 \\

HIP 55507 & K6V& -0.05 $\pm$ 0.04 & 11.18 $\pm$ 0.07 & 9.79 $\pm$ 0.03 & 7.36 $\pm$ 0.02 & 6.76 $\pm$ 0.02 & 6.61 $\pm$ 0.02\\

HD 110537&  G6/8V& 0.12 $\pm$ 0.03& 8.51 $\pm$ 0.02 & 7.83 $\pm$ 0.01 & 6.58 $\pm$ 0.02 & 6.33 $\pm$ 0.05 & 6.20 $\pm$ 0.02\\ 

HD 111031& G5V& 0.28 $\pm$ 0.03& 7.57 & 6.87 & 5.74 $\pm$ 0.02 & 5.41 $\pm$ 0.04 & 5.32 $\pm$ 0.03 \\

HIP 63762& K0& 0.09 $\pm$ 0.04& 9.91 $\pm$ 0.03 & 8.81 $\pm$ 0.01 & 6.81 $\pm$ 0.02 & 6.26 $\pm$ 0.02 & 6.16 $\pm$ 0.02 \\

HD 129191& G6V& 0.24 $\pm$ 0.03& 8.87 $\pm$ 0.02 & 8.21 $\pm$ 0.02 & 7.04 $\pm$ 0.04 & 6.74 $\pm$ 0.04 & 6.68 $\pm$ 0.02\\

HD 129814& G5V& 0.00 $\pm$ 0.03& 8.16 $\pm$ 0.02 & 7.52 $\pm$ 0.01 & 6.33 $\pm$ 0.02 & 6.07 $\pm$ 0.02 & 5.99 $\pm$ 0.02  \\

HD 136274& G8V& -0.23 $\pm$ 0.03& 8.70 & 7.96 & 6.53 $\pm$ 0.02 & 6.21  $\pm$ 0.03 & 6.12  $\pm$ 0.02\\

HD 139457& F8V& -0.48 $\pm$ 0.03& 7.62 $\pm$ 0.01 & 7.10 $\pm$ 0.01 & 6.05 $\pm$ 0.03 & 5.77 $\pm$ 0.03 & 5.70 $\pm$ 0.02  \\

HD 142229& G5V& 0.05 $\pm$ 0.03& 8.70 $\pm$ 0.02 & 8.08 $\pm$ 0.01 & 6.97 $\pm$ 0.04 & 6.67 $\pm$ 0.03 & 6.61 $\pm$ 0.02 \\

HD 147231& G5V& -0.00 $\pm$ 0.03& 8.53 & 7.90 & 6.56 $\pm$ 0.03 & 6.23 $\pm$ 0.02 & 6.15 $\pm$ 0.02 \\

HD 155413& G3III/IV& 0.26 $\pm$ 0.04& 7.94$\pm$ 0.02 & 7.26 $\pm$ 0.01 & 6.03 $\pm$ 0.02 & 5.73 $\pm$ 0.03 & 5.63 $\pm$ 0.02  \\

HD 157338& F9.5V& -0.08 $\pm$ 0.03& 7.50 $\pm$ 0.01 & 6.92 $\pm$ 0.01 & 5.87 $\pm$ 0.03 & 5.58 $\pm$ 0.04 & 5.49 $\pm$ 0.03\\

HD 164509& G2V& 0.21 $\pm$ 0.04& 8.72 $\pm$ 0.02 & 8.10 $\pm$ 0.01 & 6.94 $\pm$ 0.03 & 6.67 $\pm$ 0.06 & 6.58 $\pm$ 0.02  \\

HD 180684& F8V& 0.05 $\pm$ 0.03& 7.57 $\pm$ 0.01 & 7.02 $\pm$ 0.01 & 5.98 $\pm$ 0.03 & 5.74 $\pm$ 0.02 & 5.71 $\pm$ 0.02\\

HD 183473& G5& -0.06 $\pm$ 0.04& 8.61 $\pm$ 0.02 & 7.88 $\pm$ 0.01 & 6.49 $\pm$ 0.02 & 6.17 $\pm$ 0.02 & 6.16 $\pm$ 9.99 \\

HD 196201& G5& -0.14 $\pm$ 0.03& 9.25 $\pm$ 0.02 & 8.48 $\pm$ 0.01 & 7.07 $\pm$ 0.03 & 6.67 $\pm$ 0.04 & 6.56 $\pm$ 0.02 \\ 

HD 201924& K0& 0.10 $\pm$ 0.04& 8.59 & 7.81 & 6.41 $\pm$ 0.03 & 6.06 $\pm$ 0.02 & 5.99 $\pm$ 0.02\\

HD 213519& G5& -0.00 $\pm$ 0.03& 8.32 $\pm$ 0.01 & 7.68 $\pm$ 0.01 & 6.47 $\pm$ 0.02 & 6.20 $\pm$ 0.02 & 6.14 $\pm$ 0.02\\ 
\noalign{\vskip 0.1cm}
\hline
\noalign{\vskip 0.1cm}
HD 1293& G6V& -0.38 $\pm$ 0.04& 9.14 $\pm$ 0.02 & 8.37 $\pm$ 0.01 & 6.88 $\pm$ 0.02 & 6.42 $\pm$ 0.03 & 6.35 $\pm$ 0.02 \\
HD 1388 &G0V & -0.00 $\pm$ 0.04& 7.09 & 6.50 & 5.38 $\pm$ 0.03 & 5.07 $\pm$ 0.03 & 4.99 $\pm$ 0.02 \\

HD 4406 & G5& 0.41 $\pm$ 0.04& 8.20 $\pm$ 0.02 & 7.48 $\pm$ 0.01 & 6.20 $\pm$ 0.02 & 5.93 $\pm$ 0.02 & 5.82 $\pm$ 0.02  \\

HD 6558 & G2V& 0.26 $\pm$ 0.03& 8.81 & 8.20 & 7.19 $\pm$ 0.02 & 6.95 $\pm$ 0.03 & 6.86 $\pm$ 0.02  \\

HIP 46199& K4V& \nodata& 10.22 $\pm$ 0.03 & 9.14 $\pm$ 0.02 & 6.91 $\pm$ 0.02 & 6.35 $\pm$ 0.03 & 6.19 $\pm$ 0.02\\ 

HD 103829& F8& 0.12 $\pm$ 0.03& 9.91 $\pm$ 0.01 & 9.24 $\pm$ 0.01 & 7.97 $\pm$ 0.03 & 7.67 $\pm$ 0.02 & 7.57 $\pm$ 0.03\\

HD 105618& G0& \nodata& 9.30 & 8.64 & 7.42 $\pm$ 0.03 & 7.14 $\pm$ 0.03& 7.01 $\pm$ 0.02\\

HD 131509& K0V& -0.11 $\pm$ 0.03& 8.82 & 7.93 & 6.31 $\pm$ 0.03 & 5.84 $\pm$ 0.02 & 5.76 $\pm$ 0.02\\

HD 156826& K0V& -0.13 $\pm$ 0.03& 7.17 & 6.32 & 5.02 $\pm$ 0.02 & 4.41 $\pm$ 0.20 & 4.25 $\pm$ 0.02 \\

HD 217165& G0& 0.01 $\pm$ 0.03& 8.25 $\pm$ 0.01 & 7.67 $\pm$ 0.01 & 6.57 $\pm$ 0.02 & 6.29 $\pm$ 0.03 & 6.19 $\pm$ 0.02\\

\enddata
\tablecomments{Metallicity values are derived using Spectroscopy Made Easy (SME) and taken from \citep{2005yCat..21590141V}. Visual ($B, V$) magnitudes are taken from the optical photometry catalogs of \citep{1996yCat.2182....0O,2008yCat..34850571J,2010MNRAS.403.1949K}. Near-infrared ($J,H,K$) magnitudes are from the 2-Micron All Sky Survey (2MASS) catalog of point sources \citep{2003yCat.2246....0C,2006AJ....131.1163S}. The horizontal line separates systems with confirmed companions (above) versus candidate companions (below).} 
\end{deluxetable*}

The Doppler RV method benefits from the longest time baselines of any exoplanet detection method. Beyond the orbital radii of planets with already well-characterized Keplerian motions, many stars exhibit accelerations for which only partial orbits or systemic trends in their time series measurements can be discerned \citep{2012ApJ...751...97C}. Largely unpublished to date, these data sets lend themselves to early studies of more distant, and necessarily more massive, objects that reside in the outer extremities of other solar systems at projected separations accessible to direct imaging. The combination of methods, precision RVs with high resolution AO imaging, produces a more holistic view of stellar and extrasolar planet orbital architectures by allowing for further inferences of formation, dynamical hierarchies, interactions, and their evolution.    

We present RV time series data, AO imaging observations, and multi-epoch astrometry measurements for the companions detected. from photometry and theoretical evolutionary models are compared to lower limits available from orbital monitoring. A number of the detected companions will lend themselves to precise dynamical mass estimates with continued RV and AO imaging follow-up observations. 

\section{Target Selection}\label{sec:selection}

Targets were selected by analyzing legacy RV time series data obtained from the High Resolution \textbf{E}chelle \textbf{S}pectrometer (HIRES) at Keck \citep{1992PASP..104..270M,1994SPIE.2198..362V,2010ApJ...721.1467H}. Complementary RV data was also obtained from the publicly available Lick-Carnegie Exoplanet Survey \citep{2017AJ....153..208B}. Visual inspection of the RV time series was performed and, if need be, analyzed using a statistical Bayesian Inference Criteria (BIC) to compare marginal accelerations with the null hypothesis of a straight line of zero slope. Stars exhibiting long-term RV accelerations were (re-)observed based in part on the amount of time baseline that could be gained from an additional measurement. In the case where systems showed marginal trends, several measurements were obtained to improve time sampling and the signal-to-noise ratio of the acceleration. Those with fewer than $N\approx10$ measurements over a several year time span were generally included as lower priority targets for planning purposes depending on the logistics of a given observing run. Those confirmed as co-moving using multi-epoch TRENDS imaging observations now have a median time baseline of 10.41 years from precision stellar RVs, and 2.2 years from our direct imaging astrometric follow-up. A table of observations and observational set-up is listed in Appendix $\S$ 10.1.

Observing lists were otherwise based primarily on the strength of the RV trend and distance to the star, so as to gauge whether the companion could potentially be imaged directly. The companion minimum mass $(M_{2}$),
\begin{equation}
M_2 \geq 5.34 \times 10^{-6}M_{\odot}\left(\frac{d}{\mbox{pc}} \frac{\rho}{\mbox{arcsec}}\right)^{2} \left| \frac{\dot{v_{r}}}{\mbox{m}\:\mbox{s}^{-1}\mbox{yr}^{-1}}\right|\sqrt{27}/2,
\end{equation}
was evaluated at hypothetical angular separations, $\rho$, and for an inferred instantaneous RV acceleration, $\dot{v_r}$ \citep{2002ApJ...571..519L}. The constant, $\sqrt{27}$/2, is the minimum value of the orbital function $F(i,e,\omega,\phi)$. The dependent values of the function are $i$ inclination, $e$ eccentricity, $\omega$ longitude of periastron, and $\phi$ orbital phase. A derivation of this function can be found in \citet{1999PASP..111..169T}.

\begin{figure}
\centering
    \includegraphics[width = .5\textwidth]{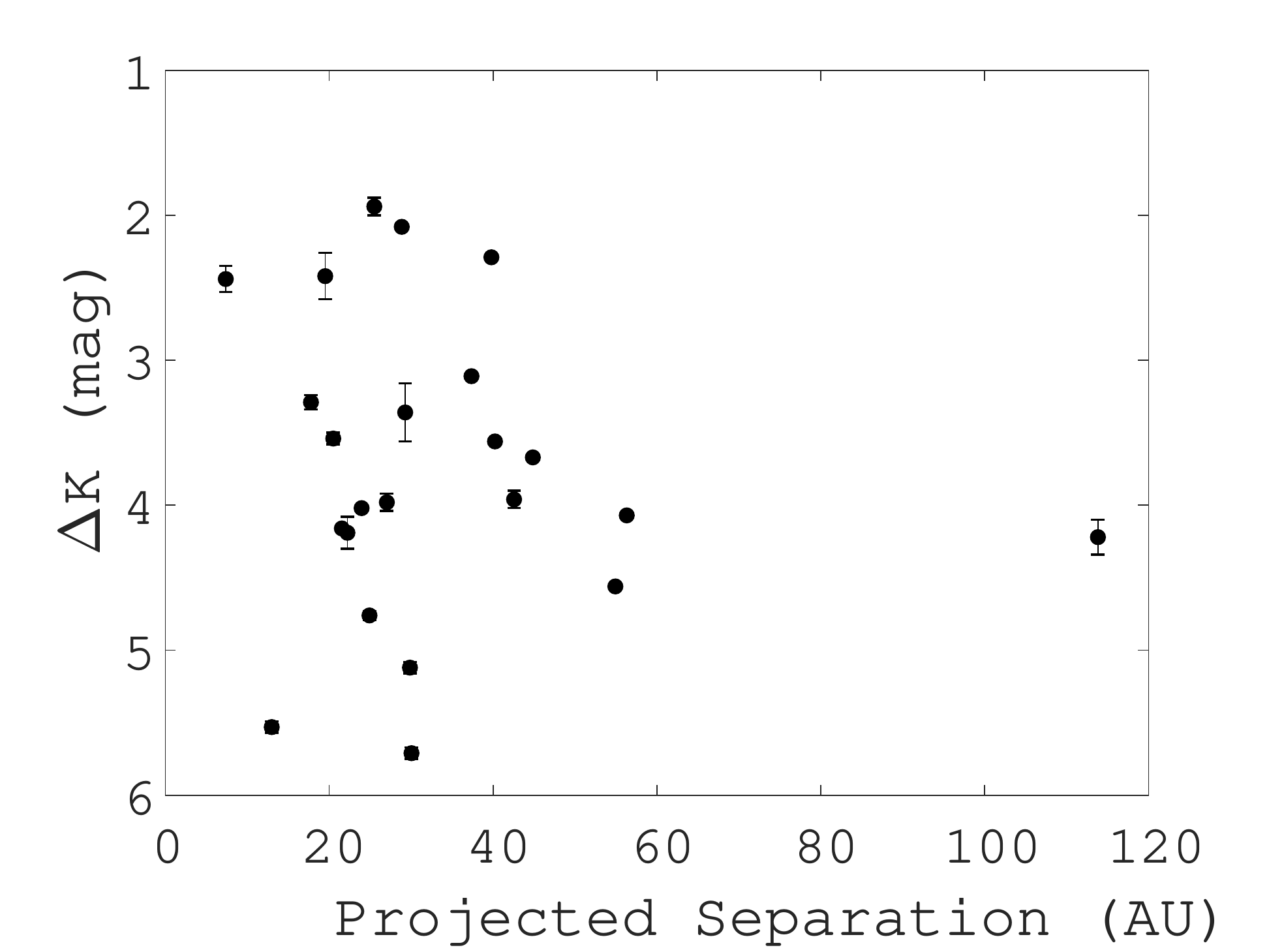}
    \caption{\textbf{Demographics of Compendium Objects:} Projected physical separation of confirmed companions versus $\Delta$ K magnitude illustrating that most objects are between 20 and 100 AU away from the primary star and are on long orbits.}
    \label{fig:1}
\end{figure}

As the original goal of the program was to maximize detection efficiency by discovering as many benchmark objects as possible (for atmospheric model and evolutionary model comparison purposes), no firm cuts were made to target selection criteria to establish a uniform sample, minimize observational bias, or for other statistical reasons, such as determining the occurrence rate of planets, given a non-detection of a low-mass stellar companion or brown dwarf \citep{2014ApJ...781...28M}. Instead, targets were observed mainly depending on the likelihood of being detected from high-resolution imaging observations (within a threshold $\Delta$Kmag $\geq$ 8 and projected separation $\geq$ 10 AU relative to the primary star). Figure \ref{fig:1} shows the demographics of the observed companions which span this range of expected detections from high-resolution imaging. 

To further refine our sample, based on the inferred Doppler acceleration, parallax, and whether the system already showed orbital motion through the exhibition of a jerk, i.e. a change in the RV acceleration ($\ddot{v_r} = \frac{d\dot{v_r}}{dt}$), we organized follow-up imaging observations if any additional constraints on the companion angular separation and relative flux ratio could be determined \textit{a priori}. 

Obvious binaries, where the minimum mass from the RV time series corresponded to $M\approx0.5M_{\odot}$ or greater, were observed only during gaps in the AO schedule or poor seeing conditions. Already built into the time series velocity scatter, basic checks involving the presence of stellar noise were performed, e.g. using $R'_{HK}$ and $v \sin{i}$ values as a proxy for youth, to help distinguish between accelerations caused by dynamics versus activity.

\section{Radial Velocity Observations}

\begin{figure}[htbp!]
\centering
    \includegraphics[width = .5\textwidth]{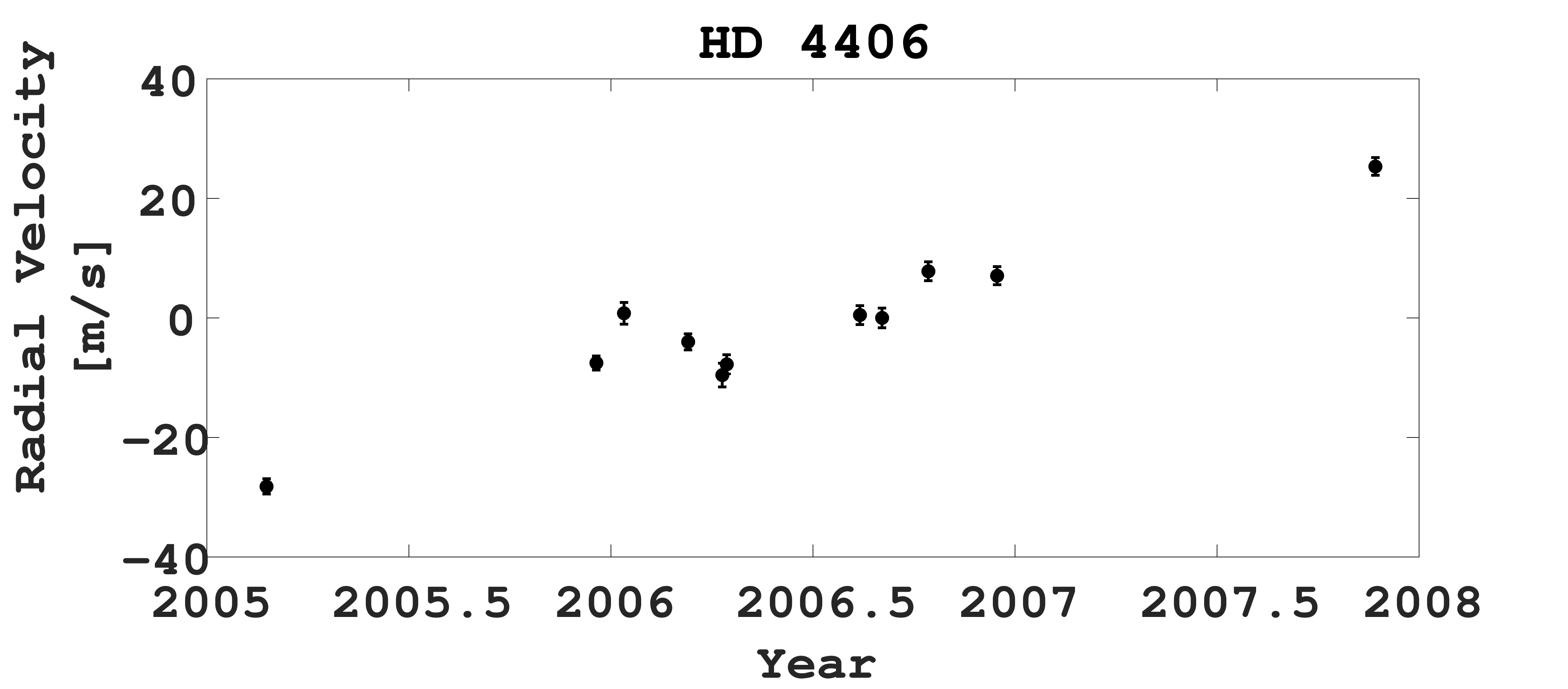}
    \caption{Radial velocity plot of HD 4406. The slope in data over time indicates the existence of a, now confirmed, companion causing the acceleration.}
    \label{fig:2}
\end{figure}

Standard observing methods used by the CPS were implemented for RV observations as has been done with previous TRENDS discoveries \citep{2010ApJ...721.1467H}. An iodine gas cell mounted in front of the spectrometer allowed for calibration of instrumental drifts and other systematics \citep{1999ASPC..185..121M}. Relative RV data are visualized as plots, as exemplified in Figure \ref{fig:2}, and are available in machine-readable form \citep{2017AJ....153..208B}. Tables 3 - 5 are objects where the data used were not available in \citet{2017AJ....153..208B}. All RV plots are found in Appendix $\S$ 10.2. It is noted that for some of the plots, the error bars are smaller than the data points due to the high resolution of HIRES.

\begin{deluxetable}{ccc}[ht] 
\tablecaption{Doppler RV measurements for HD1384} 
\tablehead{ \colhead{BJD-2,440,000} & \colhead{RV [m/s]} & \colhead{Uncertainty [m/s]}} 
\startdata 
 14339.91  &  165.2  &  1.7 \\
 14399.83  &  159.4  &  1.6 \\
 14461.83  &  167.6  &  1.6 \\
 14642.07  &  98.5  &  1.6 \\
 14675.00  &  111.0  &  1.5 \\
 14726.92  &  64.0  &  1.5 \\
 14809.82  &  85.3  &  1.4 \\
 14838.85  &  104.4  &  1.5 \\
 14866.71  &  92.0  &  1.4 \\
 14984.12  &  34.1  &  1.7 \\
 15015.10  &  8.9  &  1.4 \\
 15016.09  &  16.3  &  1.5 \\
 15029.08  &  55.8  &  1.7 \\
 15049.05  &  36.8  &  1.6 \\
 15078.14  &  65.3  &  1.6 \\
 15079.05  &  30.9  &  1.4 \\
 15080.04  &  15.8  &  1.4 \\
 15081.05  &  25.9  &  1.4 \\
 15082.14  &  72.3  &  1.4 \\
 15083.15  &  55.4  &  1.6 \\
 15084.09  &  44.2  &  1.4 \\
 15085.14  &  62.8  &  1.4 \\
 15106.92  &  43.7  &  1.5 \\
 15134.04  &  61.5  &  1.6 \\
 15169.71  &  -24.2  &  1.6 \\
 15190.72  &  14.0  &  1.4 \\
 15196.72  &  13.5  &  1.3 \\
 15231.77  &  30.6  &  1.5 \\
 15256.72  &  40.1  &  1.6 \\
 15373.13  &  -47.7  &  1.4 \\
 15401.10  &  -9.1  &  1.5 \\
 15471.90  &  -81.3  &  1.5 \\
 15584.71  &  -28.0  &  1.4 \\
 16114.11  &  -152.3  &  1.6 \\
 16166.11  &  -130.1  &  1.5 \\
 16584.99  &  -196.1  &  1.7 \\
 16851.13  &  -282.6  &  1.5 \\
 16895.06  &  -283.2  &  1.5 \\
 18393.05  &  -584.5  &  1.4 \\
\enddata 
\tablecomments{HD 1384 was not included in data reduction of \citet{2017AJ....153..208B}. The reported measurements are produced with a distinct yet similar code as described in \citet{2010ApJ...721.1467H}. This table also serves as an example of the RV data used to create the plots in Figures \ref{fig7}-\ref{fig12}. All RV data products can be found in a machine-readable table in \cite{2017AJ....153..208B}.}
\end{deluxetable} 
\clearpage

\begin{deluxetable}{ccc} 
\tablecaption{Doppler RV measurements for HD155413} 
\tablehead{ \colhead{BJD-2,440,000} & \colhead{RV [m/s]} & \colhead{Uncertainty [m/s]}} 
\startdata 
 13842.13  &  30.4  &  1.8 \\
 13926.90  &  27.6  &  1.6 \\
 13960.84  &  30.8  &  1.4 \\
 13961.78  &  29.3  &  1.3 \\
 13962.77  &  19.2  &  1.3 \\
 13963.81  &  19.9  &  1.5 \\
 13981.75  &  22.3  &  1.3 \\
 13982.72  &  17.8  &  1.3 \\
 13983.72  &  20.2  &  1.4 \\
 15471.71  &  -64.5  &  1.6 \\
 16862.75  &  -154.4  &  1.7 \\
\enddata 
\tablecomments{HD 155413 was not included in data reduction of \citet{2017AJ....153..208B}. The reported measurements are produced with a distinct yet similar code as described in \citet{2010ApJ...721.1467H}.}
\end{deluxetable}

\begin{deluxetable}{ccc}
\tablecaption{Doppler RV measurements for HD 4406}
\tablehead{
\colhead{BJD-2,440,000} & \colhead{RV [m/s]} & \colhead{Uncertainty [m/s]} 
}
\startdata
13425.77 & -28.23 & 1.28 \\
13723.90 & -7.54 & 1.20 \\
13748.81 & 0.78 & 1.80 \\
13806.84 & -4.01 & 1.33 \\
13837.73 & -9.57 & 1.96 \\
13841.75 & -7.77 & 1.60 \\
13962.13 & 0.48 & 1.60 \\
13982.13 & 0.00 & 1.64 \\
14023.97 & 7.80 & 1.59 \\
14086.08 & 7.06 & 1.50 \\
14428.00 & 25.34 & 1.49 \\
\enddata
\tablecomments{HD 4406 was not included in data reduction of \citet{2017AJ....153..208B}. The reported measurements are produced with a distinct yet similar code as described in \citet{2010ApJ...721.1467H}.}
\end{deluxetable}

\section{High-Contrast Imaging}\label{sec:imaging}

TRENDS campaign targets are generally nearby ($d \lesssim 100$ pc), bright ($V \lesssim 10$), main sequence stars (Table 1, Table 2). Reconnaissance imaging observations span several observing seasons starting as early as May 2010. Observations were obtained in natural guide star AO mode using NIRC2 (Instrument PI: Keith Matthews) on the Keck II telescope \citep{2000PASP..112..315W}. Targets were first vetted for binarity using unocculted, snapshot images.\footnote{Additional observations for lower-priority targets were obtained using PHARO at Palomar \citep{hayward_01}. Binaries discovered by PHARO will be reported in a separate paper as they require a different reduction pipeline and have not yet been fully processed.} 

Lower-dynamic-range observations, which generally did not incorporate angular differential (ADI) imaging, often used a narrow filter (e.g. $K_{\rm cont}$) to prevent saturation, allowing for many targets to be observed per time such that meaningful upper-limits could be placed on the mass and separation of companions prior to committing to deeper imaging observations. A three-point dither pattern was used for the quick snap-shot program to subtract the near-infrared sky background. Otherwise, the 600 mas diameter coronagraphic spot was used with ADI to search for RV trend companions that evaded detection using shallow observations. The objects reported herein do not utilize coronography or ADI techniques. Such faint objects can be found in previous TRENDS publications. 

As with previous TRENDS publications, the narrow mode camera setting was used (9.952 $\pm$ 0.002 mas pix$^{-1}$) \citep{2010ApJ...725..331Y} with a 1024 x 1024 pixel FOV.\footnote{The plate scale from \citet{2016PASP..128i5004S} was used for observations following the NIRC2 hardware upgrade in April 2015.} Standard background subtraction, flat-fielding, distortion correction, and speckle suppression methods were used to reduce science frames \citep{2012ApJ...761...39C,2010ApJ...725..331Y,2016PASP..128i5004S}. Observations were generally carried out in the $K_s$ and $K'$ filters. When appropriate, multi-epoch observations were recorded in complementary filters to obtain color information, assess proper motions, and begin to track orbits. Table 6 lists the dates of observations, integration times, and filters used. In some cases, small sub-array camera modes were required to minimize integration time and avoid saturation of the bright on-axis star in order to calculate contrast ratios.

\begin{figure}[h]
\centering
    \includegraphics[width = .5\textwidth]{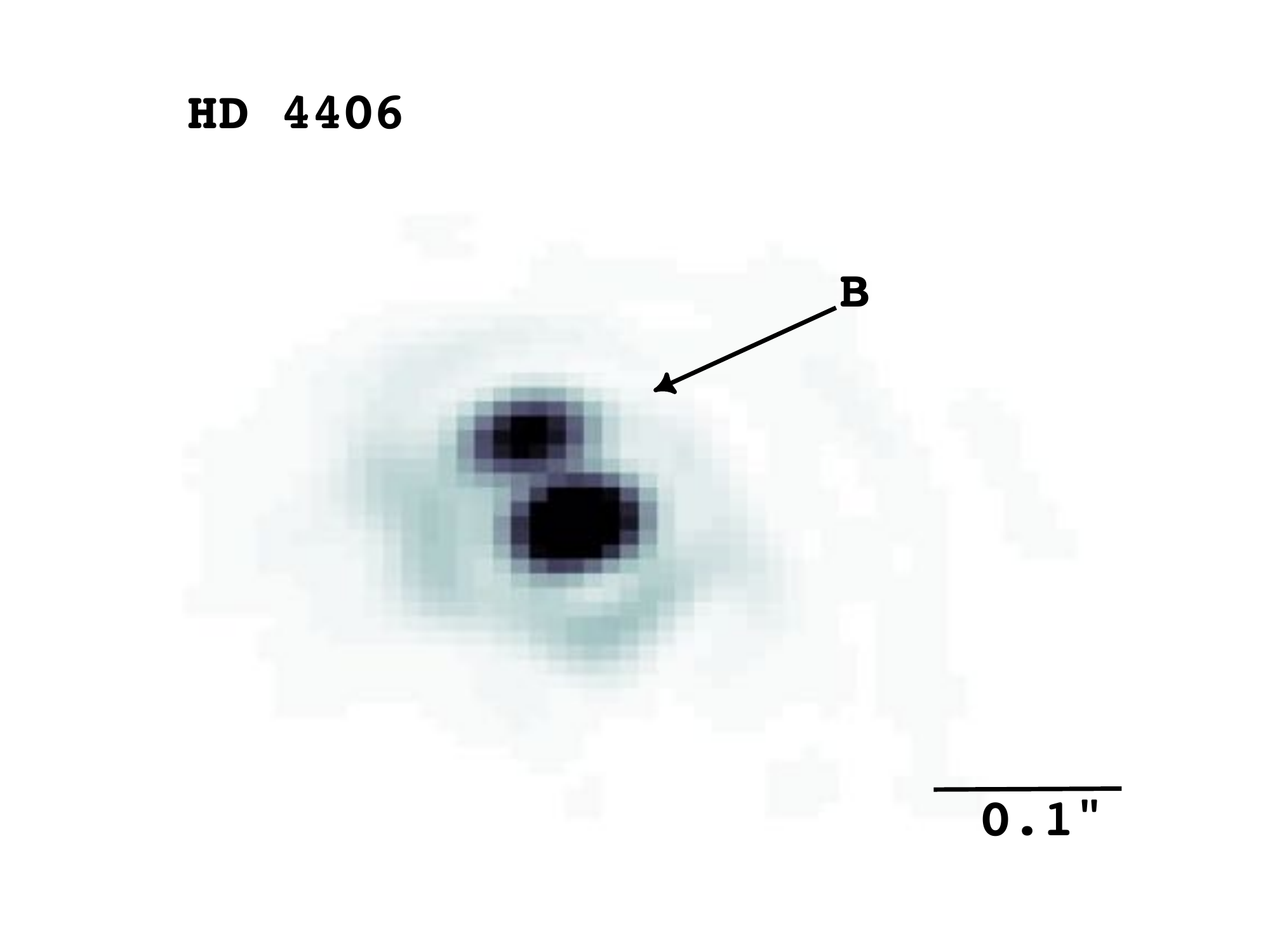}
    \caption{AO image of a confirmed binary system, HD 4406. Angular scale size legend is on the lower right hand corner.}
    \label{fig:3}
\end{figure}

Figure 3 displays an example our AO images. 31 companions were found to be co-moving and an additional 11 candidate companions were detected that we identify as likely to be gravitationally bound based on self-consistency arguments involving mass and projected orbital separation compared to the inferred Doppler trend ($\S$\ref{sec:mass}). Images of all companions can be found in Appendix $\S$ 10.3. Of the 31 companions proven to be associated with their parent star, six were found to be members of hierarchical triple star systems (HD 1205ABC, HD 136274ABC, and HD 196201ABC). 

\section{Astrometry \& Photometry}\label{sec:followup}

Astrometric positioning and photometric flux values were measured relative to the primary star. For companions detected with unocculted (non-coronagraphic) AO data, standard routines were used to determine the angular separation, position angle, and relative flux value  for each reduced frame \citep{2012ApJ...761...39C}. Generally ten to twenty images of each companion were recorded per epoch to estimate frame-to-frame statistical variance.

As with previous TRENDS discoveries, statistical uncertainties and systematic effects were folded into the analysis using Monte Carlo error propagation methods \citep{2014ApJ...781...29C, 2016ApJ...831..136C, 2018arXiv180302725C}. For triple star systems, we used the Bayesian methods described in \cite{2014ApJ...788....2B} to model the NIRC2 PSF, de-blend sources, and determine uncertainties for each companion's astrometric position and relative flux value. 

Tables in Appendix $\S$ 10.4 list photometric and astrometric analysis. \textbf{A machine-readable table is available for the complete analysis of confirmed companions.}  Absolute magnitudes and projected physical separations are calculated using recent parallax measurements from Gaia \citep{2016A&A...595A...4L, 2018arXiv180409365G}. Time baselines between observations span several months to several year with an average of 2.2 years for confirmed companions. The high proper-motion of TRENDS targets and signal-to-noise ratio of AO-assisted NIRC2 astrometry allows candidate companions to be evaluated for sharing common space motion with the on-axis star using only two or three epochs.\\

\begin{figure}[hb!]
    \centering
    \includegraphics[width =.5\textwidth]{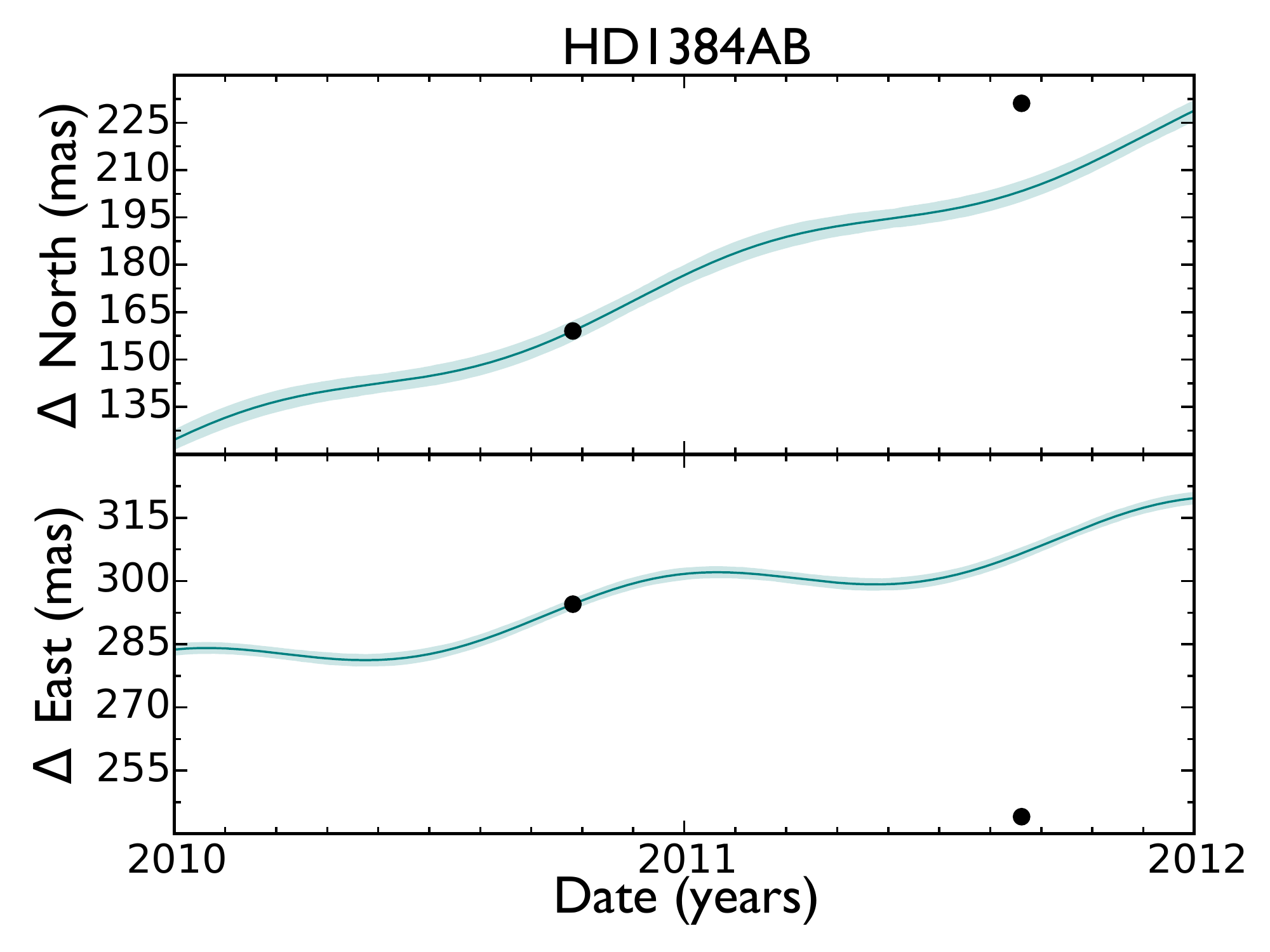}
    \caption{Example common proper motion plot showing space motion of HD1384 B, compared to a background object at infinity as indicated by the track plotted in teal.}
\end{figure}
Figure 4 is an example of the space motion diagrams of each star and detected companion taking into account proper motion and parallactic motion. Each epoch of relative astrometry is plotted as a point with coordinates of DN ($\Delta$North) and DE ($\Delta$East). The track plotted in teal shows the motion of a background object, given the proper motion of the primary star. The companions data points fall off this track and thus show they are co-moving with the primary star. All plots can be found in Appendix $\S$ 10.5.
\clearpage
 
\begin{deluxetable*}{lcccc}[ht]
\tablecaption{Colors, Inferred Spectral Types, and Effective Temperatures of Companions}
\tablewidth{0pt}
\tablehead{\colhead{a) Name} & \colhead{b) J-K}  & \colhead{c) H-K} & \colhead{d) Spectral Type} & \colhead{e) T$_{eff}$ (K)} }
\startdata
HIP 1294 B 	&	0.97	$\pm$	0.04	&	 		 --  	&	M6.5V	&	2710	\\
HD 1384 B 	&	0.22	$\pm$	0.33	&	0.00	$\pm$	0.28	&	F5V	&	6510	\\
HD 6512 B 	&	0.85	$\pm$	0.11	&		 --  	&	M4.5V	&	3100	\\
HD 31018 B  	&	 $\star\star$ 		&	0.96 $\pm$ 0.26	&	 --	&	 --	\\
HD 34721 B 	&	1.00	$\pm$	0.27	&	   	  	      --	&	M7V	&	2650	\\
HD 50639 B 	&	0.48	$\pm$	0.45	&	   	  --	    	&	K0.5V	&	5240	\\
HD 85472 B 	&	0.18    $\pm$	0.41	&	    	  --	    	&	K3V	&	4830	\\
HIP 55507 B 	&	0.93	$\pm$	0.13	&	   	  --	    	&	M5.5V	&	3000	\\
HD 110537 B$^\ddagger$ 	&	0.70	$\pm$	0.03	&	   	  --	    	&	K5V	&	4410	\\
	&	1.59	$\pm$	0.08	&	   	  --	    	&	L3V	&	1830	\\
HD 111031 B 	&	0.69	$\pm$	0.04	&	 	  --	    	&	K5V	&	4410	\\
HIP 63762 B 	&	1.02	$\pm$	0.09	&	 	  --	    	&	M7V	&	2650	\\
HD 129814 B 	&	2.18	$\pm$	0.18	&	   	  --	    	&	 --	&	 --	\\
HD 136274 B 	&	0.81	$\pm$	0.15	&	   	  --	    	&	M0V	&	3870	\\
HD 136274 C 	&	0.83	$\pm$	0.13	&	   	  --	    	&	M4V	&	3200	\\
HD 139457 B 	&	0.68	$\pm$	0.19	&	   	  --	    	&	K4.5V	&	4540	\\
HD 142229 B 	&	0.62	$\pm$	0.08	&	   	  --	    	&	K3.5V	&	4700	\\
HD 147231 B$^\ddagger$ 	&	1.55	$\pm$	0.04	&	0.55	  $\pm$	0.04	&	L2V	&	1960	\\
	&	0.72	$\pm$	0.04	&	  	  --	 	&	K5.5V	&	4330	\\
HD 155413 B 	&	0.79	$\pm$	0.07	&		  --	 	&	K8V	&	4000	\\
HD 164509 B 	&	0.97	$\pm$	0.04	&	 	  --		&	M6.5V	&	2710	\\
HD 180684 B 	&	0.50	$\pm$	0.06	&	 	  --	  	&	K1.5V	&	5140	\\
HD 183473 B$^{\dagger}$ 	&	0.69	$\pm$	 9.99 	&	0.03	  $\pm$	 9.99  	&	M4V	&	3200	\\
HD 196201 B 	&	1.01	$\pm$	0.16	&	0.34	$\pm$	0.17	&	M7V	&	2650	\\
HD 196201 C 	&	1.41	$\pm$	0.13	&	0.82	$\pm$	0.17	&	L2V	&	1960	\\
HD 201924 B 	&	0.80	$\pm$	0.04	&	   	  --	 	&	K8V	&	4000	\\
HD 213519 B 	&	0.53	$\pm$	0.03	&	 	  --	  	&	K3.5V	&	4700	\\
\hline 										
\rule{0pt}{3ex}HIP 46199 B  & 1.19 $\pm$ 0.50   &       --      & M8.5V & 2440    \\
HD 131509 B 	&	0.86	$\pm$	0.06	&	 	  --	   	&	M4.5V	&	3100	\\
HD 156826 B 	&	1.94	$\pm$	0.34	&	 	  --	 	&	 L3V	&	 1830	\\
HD 156826 C 	&	1.09	$\pm$	0.15	&	 	  --	   	&	M7.5V	&	2600 
\enddata
\tablecomments{\rule{0pt}{3ex}J - K colors are used to estimate the spectral type using the \citet{2013ApJS..208....9P} values that were most recently updated in August 2018. Colors are averaged for targets with multiple epochs or multi-filter observations unless their is large uncertainty between the two epochs$^{\ddagger}$. Confirmed and candidate companions are respectively listed above and below the horizontal line. \\
-- Observations in filter were not obtained for this object.\\
$\dagger$ The large uncertainty in HD 183473 B is due to the large uncertainty in K-magnitude of the primary star. J - H = 0.83 $\pm$ 0.38 for this object and are used to estimate the spectral type and effective temperature. \\
$\ddagger$ Large uncertainties in color exists between different epochs and thus we list both possible spectral types.}
\end{deluxetable*}

\section{Companion Mass Estimates}\label{sec:mass}
The majority of TRENDS detections are found to be bound companions ($\S$\ref{sec:followup}). This is due to the selection bias of identifying stars based on the existence of RV accelerations. Of 39 stars observed to have a nearby companion, we confirm 31 companions as co-moving. We further identify 11 as highly probable companions based on having only one epoch in one filter. These targets have an infrared color (Table 6, Figure 5) expected of VLM stars (\citet{2012ApJS..201...19D},\citet{2016ApJS..225...10F}). \\

\begin{figure}[!hb]
\includegraphics[width = .5\textwidth]{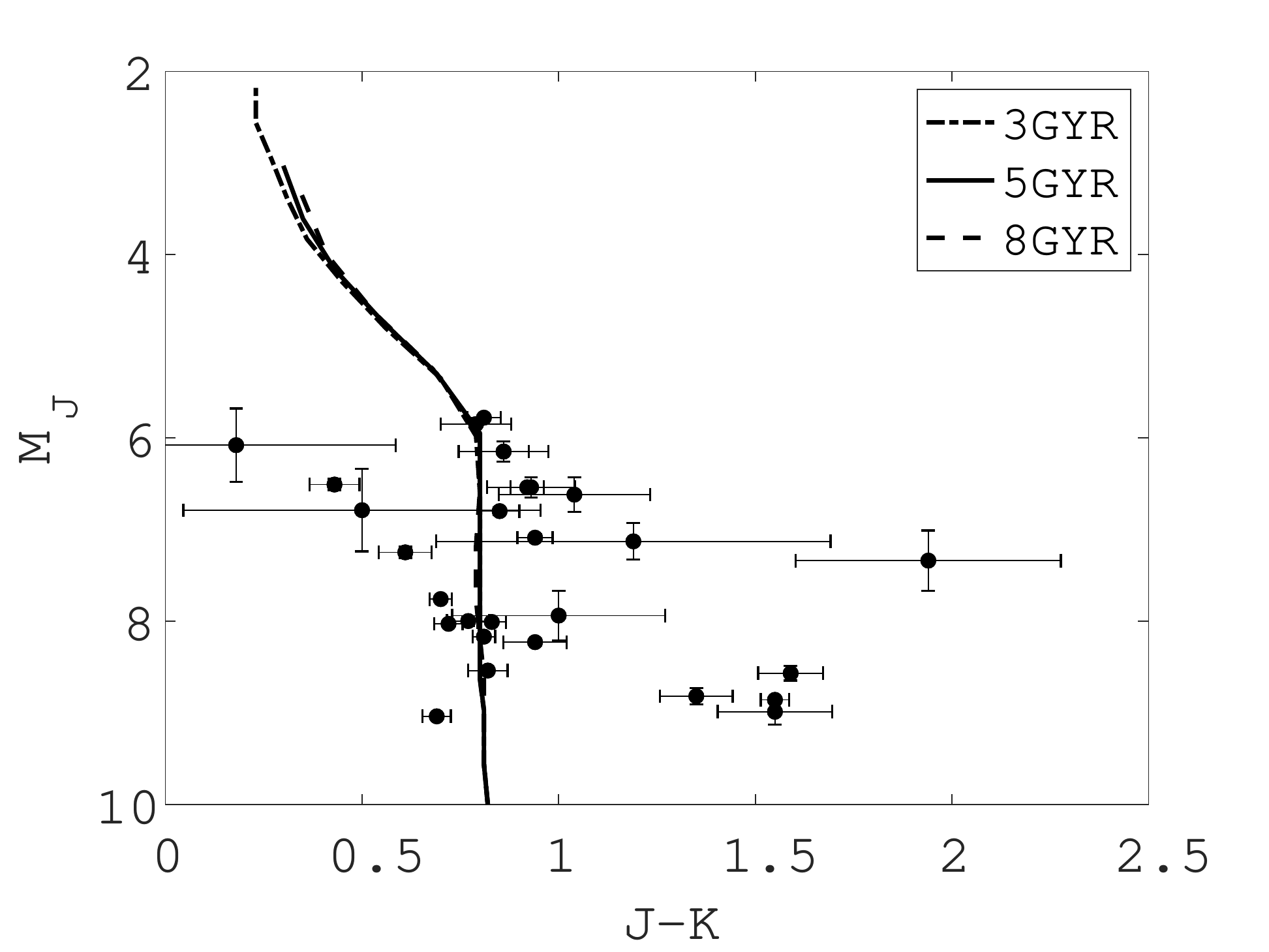}\\
   \caption{\textbf{Color Magnitude Diagram:} Column b) of Table 7 vs. Column d) of Tables 8 and 9. See Table 7 for inferred spectral types and effective temperatures. We have removed outliers with an error deviation greater than 3-$\sigma$ of the average. We have also removed the two objects, HD 1384 and HD 129814, in need of further spectroscopic observations and thus unreliable inferred spectral types. We overlay isochrones of 3, 5, and 8 GYR models from \citet{2015A&A...577A..42B}}
\end{figure}

We estimate companion masses using two methods: (i) evolutionary models that relate near-infrared absolute magnitude to mass \citet{2015A&A...577A..42B}; and (ii) minimum mass limits, $M_2$ from Eq. 1, based on the companion projected separation and inferred RV acceleration \citet{1999PASP..111..169T}. A number of Doppler trend
s are not linear but show low levels of curvature. In either case, we evaluate the instantaneous acceleration to calculate the companion minimum mass based on Newtonian dynamics.  

Using the contrast ratio between the companion and host star, we first de-blend the signal(s) by correcting the apparent magnitude of each source based on the system combined light. In the case of a triple star, we further de-blend the apparent magnitude of the combined secondary and tertiary companions. To estimate the companion masses using evolutionary models, we interpolate model grids by \cite{2015A&A...577A..42B} across both mass and age to give a finer grid of magnitudes for each band. Using a Markov-Chain Monte Carlo (MCMC) simulation, we then explore the three-dimensional (age, mass, magnitude) space including the J-, H-, or K-band magnitudes and an assumed age estimate of 5 +/- 3 Gyr as priors for each companion. We perform the MCMC simulation using {\tt emcee} \citep{2013PASP..125..306F}, a Python package which implements an affine-invariant ensemble sampler \citep{2010CAMCS...5...65G}. 

Figure 6 shows a comparison between the photometric masses based on evolutionary models and the minimum masses based on dynamics. Targets with a ratio of less than 1 are indicative of another companion that exists that is not responsible for the RV trend (See notes on specific objects in Section \ref{sec:notes}).

\begin{figure}
\includegraphics[width = .5\textwidth]{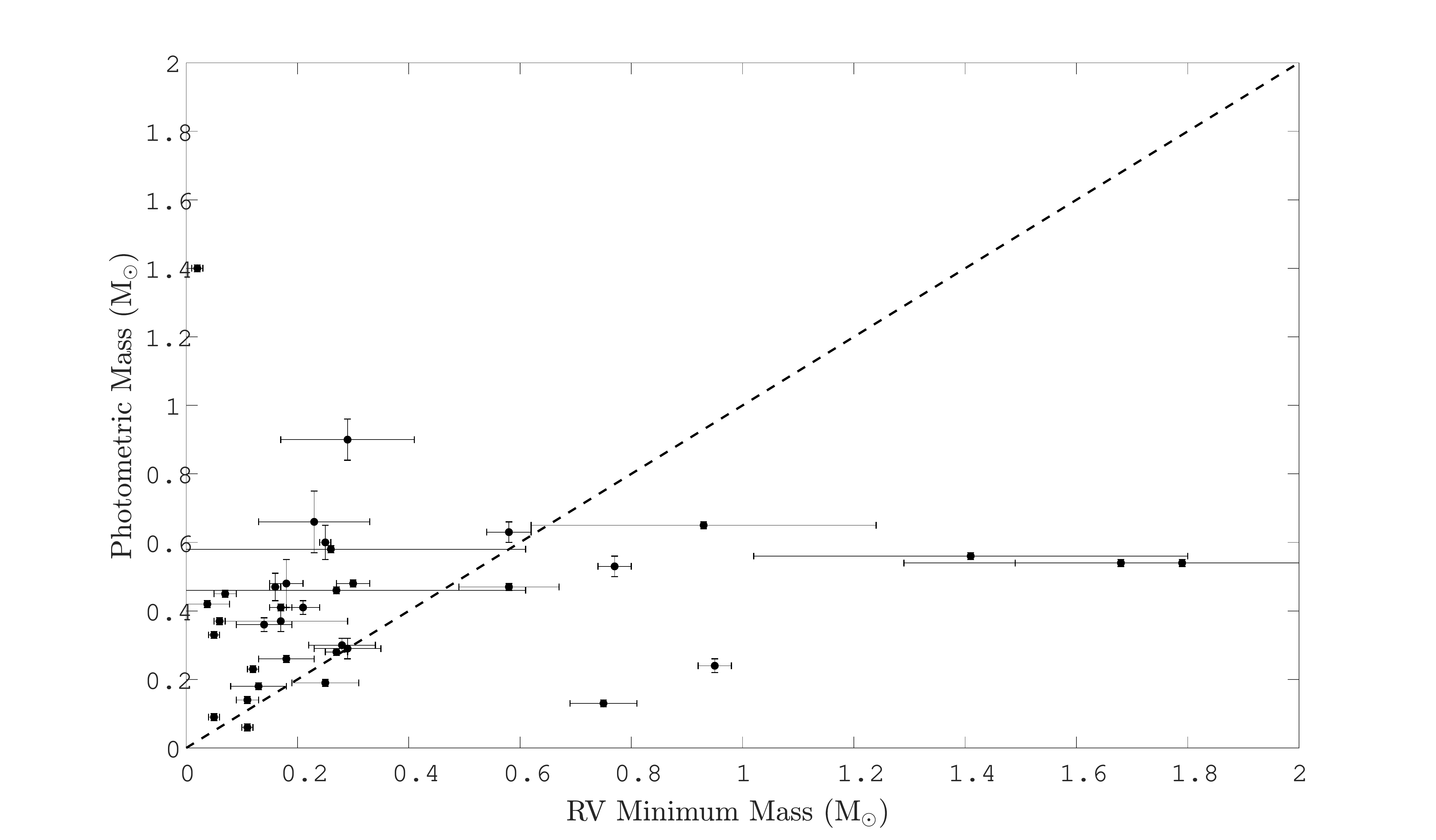}
   \caption{\textbf{Comparison of Inferred Radial Velocity and Photometric Masses:}  Columns i) and j) of Tables 7 and 8 are plotted to demonstrate the distribution of imaged companions in this survey. The dashed line is of the form RV Minimum Mass = Photometric Mass. Objects \textbf{above} the dashed line represent companions responsible for the RV acceleration and expected from this survey. Objects \textbf{below} the dash line represent imaged not responsible for the RV acceleration at the measured separation of the observed companion. See Section 7 for details on these objects.}
   \end{figure}

\section{Notes on Specific Objects}\label{sec:notes}
Specific objects in this survey are of interest for future observations for the following reasons; 
\begin{enumerate}
    \item They have a confirmed companion observed that does not cause the RV acceleration as emphasized in Figure 2.
    \item They have a tertiary companion previously unresolved yet a prior known binary component from previous surveys.
    \item The colors measured with this survey do not correspond to an identifiable spectral type from the extensive work done by \citet{2013ApJS..208....9P} (Table 6, Figure 5) and require more detailed observations with an IFS.
\end{enumerate}

\subsection*{HIP 1294}
HIP 1294 is a triple star system with an unresolved tight binary and a faint tertiary companion. We treat the tight binary as a point source and the measured separation and delta magnitude of the tertiary are with respect to the point source. At a separation of 739.69 $\pm$ 0.92 mas, a companion causing an RV acceleration of 130.22 $\pm$ 30.14 m$\:$s$^{-1}$yr$^{-1}$ would have a minimum mass of 1.68$\pm$ 0.39 M$\_{\odot}$. The imaged tertiary located at the aforementioned separation has a photometric mass of 0.54 $\pm$ 0.01 M$\_{\odot}$ and thus we asses the acceleration must be largely due to the unresolved secondary companion.

\subsection*{HD 1205} 
HD 1205 is a known binary \citep{2000A&A...355L..27H}. We image the known companion at 1093.2 $\pm$ 6.29 mas and reveal a closer companion at 216.5 $\pm$ 4.72 mas. The radial velocity trend is likely due to the bright secondary that was unresolved until this survey. 

\subsection*{HD 1384}
At a separation of 335.95 $\pm$ 4.38 mas, a companion causing an RV acceleration of -65.92 $\pm$ 0.32 m$\:$s$^{-1}$yr${-1}$ would have a minimum mass of 2.67$\pm$ 0.48 M$\_{\odot}$. The imaged companion located at the aforementioned separation has a photometric mass of 1.03 $\pm$ 0.07 M$\_{\odot}$ and thus further observations of HD 1384 are recommended.\\
Of further interest, HD 1384 B has both $J - K$ and $H - K$ colors that indicate it is an F5 star with an effective temperature of 6510 K (Table 6, Figure 5). HD 1384 A is a G5 star (Table 2). Of the two epochs of photometric measurements, the smallest difference in magnitude between HD 1384 AB is $\Delta$K = 3.32 $\pm$ 0.06. HD 1384 B is fainter and thus it is not likely to be an F5 star. Further observations with IFS are needed to determine the spectral type. 

\subsection*{HD 6558}
At a separation of 4777.50 $\pm$ 2.93 mas, a companion causing an RV acceleration of -37.28 $\pm$ 0.88 m$\:$s$^{-1}$yr${-1}$ would have a minimum mass of 79.38 $\pm$ 4.92 M$\_{\odot}$. The imaged companion located at the aforementioned separation has a photometric mass of 0.33 $\pm$ 0.01 M$\_{\odot}$. This indicates there may be a smaller, closer-in companion responsible for the trend.

\subsection*{HD 31018}
HD 31018 B has an  $H - K$ = 0.96 $\pm$ 0.26  color. Using the \citet{2013ApJS..208....9P} table, we were unable to identify a spectral type for this object (Table 6, Figure 5). We suggest another epoch of imaging in J, H, and K filters or follow up with IFS to determine the spectral type.

\subsection*{HIP 63762}
At a separation of 1077.53 $\pm$ 10.34 mas, a companion causing an RV acceleration of 642.27 $\pm$ 2.06 m$\:$s$^{-1}$yr${-1}$ would have a minimum mass of 6.70 $\pm$ 0.07 M$\_{\odot}$. The imaged companion located at the aforementioned separation has a photometric mass of 0.28 $\pm$ 0.01 M$\_{\odot}$. The RV fit is made using 6 data points over a 1.5 yr baseline. Curvature is evident in these data points and may be responsible for the high estimate of the RV acceleration. Further observations of HIP 63762 are recommended.\\

\subsection*{HD 129814}
HD 129814 B has an  $J - K$ = 2.18 $\pm$ 0.18 color. Using \citet{2013ApJS..208....9P}, we were unable to identify a spectral type for this object (Table 6, Figure 5). We suggest another epoch of imaging in J, H, and K filters or follow up with IFS to determine the spectral type.

\subsection*{HD 136274}
HD 136274 is a triple star system. The RV acceleration of -31.09 $\pm$ 0.70 m$\:$s$^{-1}$yr$^{-1}$ suggests a companion minimum mass of 0.16 $\pm$ 0.01 M$_{\odot}$ at 575.15 $\pm$ 1.5 mas separation. Imaging reveals a 0.47 $\pm$ 0.04 M$_{\odot}$ secondary companion at this separation. We also reveal a 0.22 $\pm$ 0.03 M$_{\odot}$ tertiary companion at 3339.14 $\pm$ 4.03 mas. The RV acceleration would require a tertiary companion of 5.58 $\pm$ 0.026 M$_{\odot}$ at the wider separation. We conclude that the RV acceleration is due to the close-in secondary and we serendipitously discover a tertiary companion. 

\subsection*{HD 164509}
\citet{2012ApJ...744....4G} found a .48 M$_{Jup}$ planet, HD 164509b,  with a RV signature of 14.2 $\pm$ 2.7 m$\:$s$^{–1}$. They allude to another companion or stellar jitter being responsible for the residual RV trend of -5.1 $\pm$ 0.7 m$\:$s$^{-1}$yr$^{-1}$. \citet{2017AJ....153..242N} confirm the co-movement of a VLM stellar companion and thus the binarity of the HD 164509 system.  We imaged the 0.45 $\pm$ 0.01 $_{\odot}$ companion on 2013 August 18 UT and confirmed the companion on 2016 April 21.

\subsection*{HD 183473}
HD 183473 exhibits a distinct linear trend with an RV acceleration fit of 40.85 $\pm$ 0.52 m$\:$s$^{-1}$yr$^{-1}$. At a separation of 446.52 $\pm$ 0.76 mas, the suggested minimum mass of a companion is 0.93 $\pm$ 0.31 M$_{\odot}$. High resolution imaging in the H and K$^{'}$ filters reveal a fainter companion with a suggested photometric mass of 0.65 $\pm$ 0.01. Due to the low resolution magnitude of the primary star in the K-band, from 2MASS, we use the H-band magnitude to compute a photometric mass. The imaged companion is less massive than suggested by the RV acceleration and thus further observations are suggested to determine if a fainter closer-in companion is contributing to the linear trend.

\subsection*{HD 196201}
HD 196201 is a triple star system.  Imaging reveals a 0.53 $\pm$ 0.03 M$_{\odot}$ secondary companion at 608.51 $\pm$ 0.28 mas away. We also reveal a 0.24 $\pm$ 0.02 M$_{\odot}$ tertiary companion at 657.83 $\pm$ 0.29 mas. We asses the RV acceleration may be due to the combined effect of the tight binary comprised of the secondary and tertiary stars.

\section{Summary and Concluding Remarks}\label{sec:conclusions}

The TRENDS high-contrast imaging program has been active since May 2010, and has produced several notable discoveries including two benchmark brown dwarfs and two compact objects \citep{2012ApJ...761...39C,2013ApJ...771...46C,2013ApJ...774....1C,2014ApJ...781...29C,2016ApJ...831..136C,2018ApJ...853..192C}. In the course of detecting and studying progressively fainter companions as AO technologies continually improve, TRENDS has uncovered several dozen stellar-mass objects in the process. This paper reports the detection of all candidate and bona-fide companions that we have imaged using Keck/NIRC2 since the program's inception. 

We find that many of the slowly varying Doppler accelerations measured as part of the Lick-Carnegie and California Planet Search campaigns are caused by stellar companions. We have used these data sets to place constraints on the companions orbits and masses from dynamics with initial comparisons to evolutionary models. Several systems have been found to be hierarchical triple stars, often with the binary components (that produce the RV acceleration) separated at the spatial resolution limit offered by Keck NGS AO.

One objective of the survey was to provide a sample of VLM stars and BD objects to follow-up for future dynamical studies. Future observations obtaining astrometry of the objects reported herein can be used to trace the orbit of these benchmark stars and thus ascertain a dynamical mass to subsequently  study the SED of the lower mass companions. \\ 

The next step, as shown in the mass benchmark HD 4747B \citep{2018ApJ...853..192C}, IFS spectroscopy can be used to obtain a high-resolution SED of these VLM stars in an attempt to refine their spectral types and effective temperatures which currently are based on color information (Figure 3). The lack of benchmark M-stars, and knowledge of M-stars and/or ability to fit their spectra, will be enhanced by mass benchmark M-stars that are companions to well-characterized F, G, and K stars. In addition, M-stars are the optimal targets for finding small, Earth-like planets as pursued by the current Transiting Exoplanet Survey Satellite (TESS) \citep{2015JATIS...1a4003R,2018AJ....155..180M}. Thus, better characterization of low mass stars will translate to refined characterization and inferred planet parameters of planets around these small, faint host stars . 
\section{Acknowledgements}
We thank Greg P. Laughlin for providing an early release of the RV data that was later produced in the manuscript by \citet{2017AJ....153..208B}. We thank the entire CPS team for their contribution to the legacy of RV's that inspired this work. E.J. Gonzales thanks Ian J.M. Crossfield and Andy Skemer for their critique and support during the writing process. E.J. Gonzales and this material is based upon work supported by the National Science Foundation Graduate Research Fellowship under Grant No. 1339067. J.R.C. acknowledges support from the NASA Early Career and NSF CAREER fellowship programs.

This work has made use of data from the European Space Agency (ESA)
mission {\it Gaia} (\url{https://www.cosmos.esa.int/gaia}), processed by
the {\it Gaia} Data Processing and Analysis Consortium (DPAC,
\url{https://www.cosmos.esa.int/web/gaia/dpac/consortium}). Funding
for the DPAC has been provided by national institutions, in particular
the institutions participating in the {\it Gaia} Multilateral Agreement. We  are  appreciative of the
vision  and  support  of  the  Potenziani  and  Wolfe
families. \textbf{We are grateful to the island of Hawaii and the use of the W.M. Keck Observatory on the revered mountain of Mauna Kea, mahalo.}
\\
\facility{Keck:II (HIRES, NIRC2)}

\bibliographystyle{apalike2}

\begin{thebibliography}{}

\bibitem[{Allard} et~al., 2001]{2001ApJ...556..357A}
{Allard}, F., {Hauschildt}, P.~H., {Alexander}, D.~R., {Tamanai}, A., \&
  {Schweitzer}, A. (2001).
\newblock {The Limiting Effects of Dust in Brown Dwarf Model Atmospheres}.
\newblock {\em \apj}, 556, 357--372.

\bibitem[{Allard} et~al., 2013]{2013MSAIS..24..128A}
{Allard}, F., {Homeier}, D., {Freytag}, B., {Schaffenberger}, {}, W., \&
  {Rajpurohit}, A.~S. (2013).
\newblock {Progress in modeling very low mass stars, brown dwarfs, and
  planetary mass objects.}
\newblock {\em Memorie della Societa Astronomica Italiana Supplementi}, 24,
  128.

\bibitem[{Baraffe} et~al., 2015]{2015A&A...577A..42B}
{Baraffe}, I., {Homeier}, D., {Allard}, F., \& {Chabrier}, G. (2015).
\newblock {New evolutionary models for pre-main sequence and main sequence
  low-mass stars down to the hydrogen-burning limit}.
\newblock {\em \aap}, 577, A42.

\bibitem[{Bechter} et~al., 2014]{2014ApJ...788....2B}
{Bechter}, E.~B., {Crepp}, J.~R., {Ngo}, H., {Knutson}, H.~A., {Batygin}, K.,
  {Hinkley}, S., {Muirhead}, P.~S., {Johnson}, J.~A., {Howard}, A.~W.,
  {Montet}, B.~T., {Matthews}, C.~T., \& {Morton}, T.~D. (2014).
\newblock {WASP-12b and HAT-P-8b are Members of Triple Star Systems}.
\newblock {\em \apj}, 788, 2.

\bibitem[{Biller} et~al., 2015]{2015ApJ...813L..23B}
{Biller}, B.~A., {Vos}, J., {Bonavita}, M., {Buenzli}, E., {Baxter}, C.,
  {Crossfield}, I.~J.~M., {Allers}, K., {Liu}, M.~C., {Bonnefoy}, M., {Deacon},
  N., {Brandner}, W., {Schlieder}, J.~E., {Dupuy}, T., {Kopytova}, T.,
  {Manjavacas}, E., {Allard}, F., {Homeier}, D., \& {Henning}, T. (2015).
\newblock {Variability in a Young, L/T Transition Planetary-mass Object}.
\newblock {\em \apjl}, 813, L23.

\bibitem[{Bowler} et~al., 2018]{2018AJ....155..159B}
{Bowler}, B.~P., {Dupuy}, T.~J., {Endl}, M., {Cochran}, W.~D., {MacQueen},
  P.~J., {Fulton}, B.~J., {Petigura}, E.~A., {Howard}, A.~W., {Hirsch}, L.,
  {Kratter}, K.~M., {Crepp}, J.~R., {Biller}, B.~A., {Johnson}, M.~C., \&
  {Wittenmyer}, R.~A. (2018).
\newblock {Orbit and Dynamical Mass of the Late-T Dwarf GL 758 B}.
\newblock {\em \aj}, 155, 159.

\bibitem[{Boyajian} et~al., 2012]{2012ApJ...757..112B}
{Boyajian}, T.~S., {von Braun}, K., {van Belle}, G., {McAlister}, H.~A., {ten
  Brummelaar}, T.~A., {Kane}, S.~R., {Muirhead}, P.~S., {Jones}, J., {White},
  R., {Schaefer}, G., {Ciardi}, D., {Henry}, T., {L{\'o}pez-Morales}, M.,
  {Ridgway}, S., {Gies}, D., {Jao}, W.-C., {Rojas-Ayala}, B., {Parks}, J.~R.,
  {Sturmann}, L., {Sturmann}, J., {Turner}, N.~H., {Farrington}, C.,
  {Goldfinger}, P.~J., \& {Berger}, D.~H. (2012).
\newblock {Stellar Diameters and Temperatures. II. Main-sequence K- and
  M-stars}.
\newblock {\em \apj}, 757, 112.

\bibitem[{Burrows}, 2001]{2001udns.conf...26B}
{Burrows}, A. (2001).
\newblock {Alkali Metals and the Colour of Brown Dwarfs}.
\newblock In H.~R.~A. {Jones} \& I.~A. {Steele} (Eds.), {\em Ultracool Dwarfs:
  New Spectral Types L and T}  (pp.\~26).

\bibitem[{Butler} et~al., 2017]{2017AJ....153..208B}
{Butler}, R.~P., {Vogt}, S.~S., {Laughlin}, G., {Burt}, J.~A., {Rivera}, E.~J.,
  {Tuomi}, M., {Teske}, J., {Arriagada}, P., {Diaz}, M., {Holden}, B., \&
  {Keiser}, S. (2017).
\newblock {The LCES HIRES/Keck Precision Radial Velocity Exoplanet Survey}.
\newblock {\em \aj}, 153, 208.

\bibitem[{Chabrier} et~al., 2000]{2000ApJ...542..464C}
{Chabrier}, G., {Baraffe}, I., {Allard}, F., \& {Hauschildt}, P. (2000).
\newblock {Evolutionary Models for Very Low-Mass Stars and Brown Dwarfs with
  Dusty Atmospheres}.
\newblock {\em \apj}, 542, 464--472.

\bibitem[{Cheetham} et~al., 2018]{2018arXiv180302725C}
{Cheetham}, A., {Bonnefoy}, M., {Desidera}, S., {Langlois}, M., {Vigan}, A.,
  {Schmidt}, T., {Olofsson}, J., {Chauvin}, G., {Klahr}, H., {Gratton}, R.,
  {D'Orazi}, V., {Henning}, T., {Janson}, M., {Biller}, B., {Peretti}, S.,
  {Hagelberg}, J., {S{\'e}gransan}, D., {Udry}, S., {Mesa}, D., {Sissa}, E.,
  {Kral}, Q., {Schlieder}, J., {Maire}, A.-L., {Mordasini}, C., {Menard}, F.,
  {Zurlo}, A., {Beuzit}, J.-L., {Feldt}, M., {Mouillet}, D., {Meyer}, M.,
  {Lagrange}, A.-M., {Boccaletti}, A., {Keppler}, M., {Kopytova}, T., {Ligi},
  R., {Rouan}, D., {Le Coroller}, H., {Dominik}, C., {Lagadec}, E., {Turatto},
  M., {Abe}, L., {Antichi}, J., {Baruffolo}, A., {Baudoz}, P., {Blanchard}, P.,
  {Buey}, T., {Carbillet}, M., {Carle}, M., {Cascone}, E., {Claudi}, R.,
  {Costille}, A., {Delboulb{\'e}}, A., {De Caprio}, V., {Dohlen}, K.,
  {Fantinel}, D., {Feautrier}, P., {Fusco}, T., {Giro}, E., {Gluck}, L.,
  {Hubin}, N., {Hugot}, E., {Jaquet}, M., {Kasper}, M., {Llored}, M., {Madec},
  F., {Magnard}, Y., {Martinez}, P., {Maurel}, D., {Le Mignant}, D.,
  {M{\"o}ller-Nilsson}, O., {Moulin}, T., {Orign{\'e}}, A., {Pavlov}, A.,
  {Perret}, D., {Petit}, C., {Pragt}, J., {Puget}, P., {Rabou}, P., {Ramos},
  J., {Rigal}, F., {Rochat}, S., {Roelfsema}, R., {Rousset}, G., {Roux}, A.,
  {Salasnich}, B., {Sauvage}, J.-F., {Sevin}, A., {Soenke}, C., {Stadler}, E.,
  {Suarez}, M., {Weber}, L., \& {Wildi}, F. (2018).
\newblock {Discovery of a brown dwarf companion to the star HIP 64892}.
\newblock {\em ArXiv e-prints}.

\bibitem[{Crepp} et~al., 2016]{2016ApJ...831..136C}
{Crepp}, J.~R., {Gonzales}, E.~J., {Bechter}, E.~B., {Montet}, B.~T.,
  {Johnson}, J.~A., {Piskorz}, D., {Howard}, A.~W., \& {Isaacson}, H. (2016).
\newblock {The TRENDS High-contrast Imaging Survey. VI. Discovery of a Mass,
  Age, and Metallicity Benchmark Brown Dwarf}.
\newblock {\em \apj}, 831, 136.

\bibitem[{Crepp} et~al., 2012a]{2012ApJ...751...97C}
{Crepp}, J.~R., {Johnson}, J.~A., {Fischer}, D.~A., {Howard}, A.~W., {Marcy},
  G.~W., {Wright}, J.~T., {Isaacson}, H., {Boyajian}, T., {von Braun}, K.,
  {Hillenbrand}, L.~A., {Hinkley}, S., {Carpenter}, J.~M., \& {Brewer}, J.~M.
  (2012a).
\newblock {The Dynamical Mass and Three-dimensional Orbit of HR7672B: A
  Benchmark Brown Dwarf with High Eccentricity}.
\newblock {\em \apj}, 751, 97.

\bibitem[{Crepp} et~al., 2014]{2014ApJ...781...29C}
{Crepp}, J.~R., {Johnson}, J.~A., {Howard}, A.~W., {Marcy}, G.~W., {Brewer},
  J., {Fischer}, D.~A., {Wright}, J.~T., \& {Isaacson}, H. (2014).
\newblock {The TRENDS High-contrast Imaging Survey. V. Discovery of an Old and
  Cold Benchmark T-dwarf Orbiting the Nearby G-star HD 19467}.
\newblock {\em \apj}, 781, 29.

\bibitem[{Crepp} et~al., 2012b]{2012ApJ...761...39C}
{Crepp}, J.~R., {Johnson}, J.~A., {Howard}, A.~W., {Marcy}, G.~W., {Fischer},
  D.~A., {Hillenbrand}, L.~A., {Yantek}, S.~M., {Delaney}, C.~R., {Wright},
  J.~T., {Isaacson}, H.~T., \& {Montet}, B.~T. (2012b).
\newblock {The TRENDS High-contrast Imaging Survey. I. Three Benchmark M Dwarfs
  Orbiting Solar-type Stars}.
\newblock {\em \apj}, 761, 39.

\bibitem[{Crepp} et~al., 2013a]{2013ApJ...771...46C}
{Crepp}, J.~R., {Johnson}, J.~A., {Howard}, A.~W., {Marcy}, G.~W., {Fischer},
  D.~A., {Yantek}, S.~M., {Wright}, J.~T., {Isaacson}, H., \& {Feng}, Y.
  (2013a).
\newblock {The TRENDS High-contrast Imaging Survey. II. Direct Detection of the
  HD 8375 Tertiary}.
\newblock {\em \apj}, 771, 46.

\bibitem[{Crepp} et~al., 2013b]{2013ApJ...774....1C}
{Crepp}, J.~R., {Johnson}, J.~A., {Howard}, A.~W., {Marcy}, G.~W., {Gianninas},
  A., {Kilic}, M., \& {Wright}, J.~T. (2013b).
\newblock {The TRENDS High-contrast Imaging Survey. III. A Faint White Dwarf
  Companion Orbiting HD 114174}.
\newblock {\em \apj}, 774, 1.

\bibitem[{Crepp} et~al., 2018]{2018ApJ...853..192C}
{Crepp}, J.~R., {Principe}, D.~A., {Wolff}, S., {Giorla Godfrey}, P.~A.,
  {Rice}, E.~L., {Cieza}, L., {Pueyo}, L., {Bechter}, E.~B., \& {Gonzales},
  E.~J. (2018).
\newblock {GPI Spectroscopy of the Mass, Age, and Metallicity Benchmark Brown
  Dwarf HD 4747 B}.
\newblock {\em \apj}, 853, 192.

\bibitem[{Cutri} et~al., 2003]{2003yCat.2246....0C}
{Cutri}, R.~M., {Skrutskie}, M.~F., {van Dyk}, S., {Beichman}, C.~A.,
  {Carpenter}, J.~M., {Chester}, T., {Cambresy}, L., {Evans}, T., {Fowler}, J.,
  {Gizis}, J., {Howard}, E., {Huchra}, J., {Jarrett}, T., {Kopan}, E.~L.,
  {Kirkpatrick}, J.~D., {Light}, R.~M., {Marsh}, K.~A., {McCallon}, H.,
  {Schneider}, S., {Stiening}, R., {Sykes}, M., {Weinberg}, M., {Wheaton},
  W.~A., {Wheelock}, S., \& {Zacarias}, N. (2003).
\newblock {VizieR Online Data Catalog: 2MASS All-Sky Catalog of Point Sources
  (Cutri+ 2003)}.
\newblock {\em VizieR Online Data Catalog}, 2246.

\bibitem[{Dupuy} \& {Liu}, 2012]{2012ApJS..201...19D}
{Dupuy}, T.~J. \& {Liu}, M.~C. (2012).
\newblock {The Hawaii Infrared Parallax Program. I. Ultracool Binaries and the
  L/T Transition}.
\newblock {\em \apjs}, 201(2), 19.

\bibitem[{Dupuy} \& {Liu}, 2017]{2017ApJS..231...15D}
{Dupuy}, T.~J. \& {Liu}, M.~C. (2017).
\newblock {Individual Dynamical Masses of Ultracool Dwarfs}.
\newblock {\em \apjs}, 231, 15.

\bibitem[{Dupuy} et~al., 2015]{2015ApJ...805...56D}
{Dupuy}, T.~J., {Liu}, M.~C., {Leggett}, S.~K., {Ireland}, M.~J., {Chiu}, K.,
  \& {Golimowski}, D.~A. (2015).
\newblock {The Mass-Luminosity Relation in the L/T Transition: Individual
  Dynamical Masses for the New J-band Flux Reversal Binary
  SDSSJ105213.51+442255.7AB}.
\newblock {\em \apj}, 805, 56.

\bibitem[{Faherty} et~al., 2016]{2016ApJS..225...10F}
{Faherty}, J.~K., {Riedel}, A.~R., {Cruz}, K.~L., {Gagne}, J., {Filippazzo},
  J.~C., {Lambrides}, E., {Fica}, H., {Weinberger}, A., {Thorstensen}, J.~R.,
  {Tinney}, C.~G., {Baldassare}, V., {Lemonier}, E., \& {Rice}, E.~L. (2016).
\newblock {Population Properties of Brown Dwarf Analogs to Exoplanets}.
\newblock {\em \apjs}, 225(1), 10.

\bibitem[{Foreman-Mackey} et~al., 2013]{2013PASP..125..306F}
{Foreman-Mackey}, D., {Hogg}, D.~W., {Lang}, D., \& {Goodman}, J. (2013).
\newblock {emcee: The MCMC Hammer}.
\newblock {\em \pasp}, 125, 306.

\bibitem[{Gaia Collaboration} et~al., 2018]{2018arXiv180409365G}
{Gaia Collaboration}, {Brown}, A.~G.~A., {Vallenari}, A., {Prusti}, T., {de
  Bruijne}, J.~H.~J., {Babusiaux}, C., \& {Bailer-Jones}, C.~A.~L. (2018).
\newblock {Gaia Data Release 2. Summary of the contents and survey properties}.
\newblock {\em ArXiv e-prints}.

\bibitem[{Giguere} et~al., 2012]{2012ApJ...744....4G}
{Giguere}, M.~J., {Fischer}, D.~A., {Howard}, A.~W., {Johnson}, J.~A., {Henry},
  G.~W., {Wright}, J.~T., {Marcy}, G.~W., {Isaacson}, H.~T., {Hou}, F., \&
  {Spronck}, J. (2012).
\newblock {A High-eccentricity Component in the Double-planet System around HD
  163607 and a Planet around HD 164509}.
\newblock {\em \apj}, 744, 4.

\bibitem[{Goodman} \& {Weare}, 2010]{2010CAMCS...5...65G}
{Goodman}, J. \& {Weare}, J. (2010).
\newblock {Ensemble samplers with affine invariance}.
\newblock {\em Communications in Applied Mathematics and Computational Science,
  Vol.~5, No.~1, p.~65-80, 2010}, 5, 65--80.

\bibitem[{Hayward} et~al., 2001]{hayward_01}
{Hayward}, T.~L., {Brandl}, B., {Pirger}, B., {Blacken}, C., {Gull}, G.~E.,
  {Schoenwald}, J., \& {Houck}, J.~R. (2001).
\newblock {PHARO: A Near-Infrared Camera for the Palomar Adaptive Optics
  System}.
\newblock {\em \pasp}, 113, 105--118.

\bibitem[{H{\o}g} et~al., 2000]{2000A&A...355L..27H}
{H{\o}g}, E., {Fabricius}, C., {Makarov}, V.~V., {Urban}, S., {Corbin}, T.,
  {Wycoff}, G., {Bastian}, U., {Schwekendiek}, P., \& {Wicenec}, A. (2000).
\newblock {The Tycho-2 catalogue of the 2.5 million brightest stars}.
\newblock {\em \aap}, 355, L27--L30.

\bibitem[{Howard} et~al., 2010]{2010ApJ...721.1467H}
{Howard}, A.~W., {Johnson}, J.~A., {Marcy}, G.~W., {Fischer}, D.~A., {Wright},
  J.~T., {Bernat}, D., {Henry}, G.~W., {Peek}, K.~M.~G., {Isaacson}, H.,
  {Apps}, K., {Endl}, M., {Cochran}, W.~D., {Valenti}, J.~A., {Anderson}, J.,
  \& {Piskunov}, N.~E. (2010).
\newblock {The California Planet Survey. I. Four New Giant Exoplanets}.
\newblock {\em \apj}, 721, 1467--1481.

\bibitem[{Jenkins} et~al., 2008]{2008yCat..34850571J}
{Jenkins}, J.~S., {Jones}, H.~R.~A., {Pavlenko}, Y., {Pinfield}, D.~J.,
  {Barnes}, J.~R., \& {Lyubchik}, Y. (2008).
\newblock {VizieR Online Data Catalog: Metallicities activities of southern
  stars (Jenkins+, 2008)}.
\newblock {\em VizieR Online Data Catalog}, 348.

\bibitem[{Kirkpatrick} et~al., 2012]{2012ApJ...753..156K}
{Kirkpatrick}, J.~D., {Gelino}, C.~R., {Cushing}, M.~C., {Mace}, G.~N.,
  {Griffith}, R.~L., {Skrutskie}, M.~F., {Marsh}, K.~A., {Wright}, E.~L.,
  {Eisenhardt}, P.~R., {McLean}, I.~S., {Mainzer}, A.~K., {Burgasser}, A.~J.,
  {Tinney}, C.~G., {Parker}, S., \& {Salter}, G. (2012).
\newblock {Further Defining Spectral Type ``Y'' and Exploring the Low-mass End
  of the Field Brown Dwarf Mass Function}.
\newblock {\em \apj}, 753, 156.

\bibitem[{Koen} et~al., 2010]{2010MNRAS.403.1949K}
{Koen}, C., {Kilkenny}, D., {van Wyk}, F., \& {Marang}, F. (2010).
\newblock {UBV(RI)$_{C}$ JHK observations of Hipparcos-selected nearby stars}.
\newblock {\em \mnras}, 403, 1949--1968.

\bibitem[{Konopacky}, 2013]{2013MmSAI..84.1005K}
{Konopacky}, Q.~M. (2013).
\newblock {The fundamental importance of brown dwarf binaries }.
\newblock {\em \memsai}, 84, 1005.

\bibitem[{Lada} \& {Lada}, 2003]{lada_03}
{Lada}, C.~J. \& {Lada}, E.~A. (2003).
\newblock {Embedded Clusters in Molecular Clouds}.
\newblock {\em \araa}, 41, 57--115.

\bibitem[{Lindegren} et~al., 2016]{2016A&A...595A...4L}
{Lindegren}, L., {Lammers}, U., {Bastian}, U., {Hern{\'a}ndez}, J., {Klioner},
  S., {Hobbs}, D., {Bombrun}, A., {Michalik}, D., {Ramos-Lerate}, M.,
  {Butkevich}, A., {Comoretto}, G., {Joliet}, E., {Holl}, B., {Hutton}, A.,
  {Parsons}, P., {Steidelm{\"u}ller}, H., {Abbas}, U., {Altmann}, M., {Andrei},
  A., {Anton}, S., {Bach}, N., {Barache}, C., {Becciani}, U., {Berthier}, J.,
  {Bianchi}, L., {Biermann}, M., {Bouquillon}, S., {Bourda}, G.,
  {Br{\"u}semeister}, T., {Bucciarelli}, B., {Busonero}, D., {Carlucci}, T.,
  {Casta{\~n}eda}, J., {Charlot}, P., {Clotet}, M., {Crosta}, M., {Davidson},
  M., {de Felice}, F., {Drimmel}, R., {Fabricius}, C., {Fienga}, A.,
  {Figueras}, F., {Fraile}, E., {Gai}, M., {Garralda}, N., {Geyer}, R.,
  {Gonz{\'a}lez-Vidal}, J.~J., {Guerra}, R., {Hambly}, N.~C., {Hauser}, M.,
  {Jordan}, S., {Lattanzi}, M.~G., {Lenhardt}, H., {Liao}, S., {L{\"o}ffler},
  W., {McMillan}, P.~J., {Mignard}, F., {Mora}, A., {Morbidelli}, R.,
  {Portell}, J., {Riva}, A., {Sarasso}, M., {Serraller}, I., {Siddiqui}, H.,
  {Smart}, R., {Spagna}, A., {Stampa}, U., {Steele}, I., {Taris}, F., {Torra},
  J., {van Reeven}, W., {Vecchiato}, A., {Zschocke}, S., {de Bruijne}, J.,
  {Gracia}, G., {Raison}, F., {Lister}, T., {Marchant}, J., {Messineo}, R.,
  {Soffel}, M., {Osorio}, J., {de Torres}, A., \& {O'Mullane}, W. (2016).
\newblock {Gaia Data Release 1. Astrometry: one billion positions, two million
  proper motions and parallaxes}.
\newblock {\em \aap}, 595, A4.

\bibitem[{Liu} et~al., 2008]{2008ApJ...689..436L}
{Liu}, M.~C., {Dupuy}, T.~J., \& {Ireland}, M.~J. (2008).
\newblock {Keck Laser Guide Star Adaptive Optics Monitoring of 2MASS
  J15344984-2952274AB: First Dynamical Mass Determination of a Binary T Dwarf}.
\newblock {\em \apj}, 689, 436--460.

\bibitem[{Liu} et~al., 2002]{2002ApJ...571..519L}
{Liu}, M.~C., {Fischer}, D.~A., {Graham}, J.~R., {Lloyd}, J.~P., {Marcy},
  G.~W., \& {Butler}, R.~P. (2002).
\newblock {Crossing the Brown Dwarf Desert Using Adaptive Optics: A Very Close
  L Dwarf Companion to the Nearby Solar Analog HR 7672}.
\newblock {\em \apj}, 571, 519--527.

\bibitem[{Mann} et~al., 2017]{2017AJ....153..267M}
{Mann}, A.~W., {Dupuy}, T., {Muirhead}, P.~S., {Johnson}, M.~C., {Liu}, M.~C.,
  {Ansdell}, M., {Dalba}, P.~A., {Swift}, J.~J., \& {Hadden}, S. (2017).
\newblock {The Gold Standard: Accurate Stellar and Planetary Parameters for
  Eight Kepler M Dwarf Systems Enabled by Parallaxes}.
\newblock {\em \aj}, 153, 267.

\bibitem[{Marcy} \& {Butler}, 1992]{1992PASP..104..270M}
{Marcy}, G.~W. \& {Butler}, R.~P. (1992).
\newblock {Precision radial velocities with an iodine absorption cell}.
\newblock {\em \pasp}, 104, 270--277.

\bibitem[{Marcy} et~al., 1999]{1999ASPC..185..121M}
{Marcy}, G.~W., {Butler}, R.~P., \& {Fischer}, D.~A. (1999).
\newblock {Doppler Detection of Extra-Solar Planets}.
\newblock In J.~B. {Hearnshaw} \& C.~D. {Scarfe} (Eds.), {\em IAU Colloq. 170:
  Precise Stellar Radial Velocities}, volume 185 of {\em Astronomical Society
  of the Pacific Conference Series}  (pp.\ 121).

\bibitem[{Montet} et~al., 2014]{2014ApJ...781...28M}
{Montet}, B.~T., {Crepp}, J.~R., {Johnson}, J.~A., {Howard}, A.~W., \& {Marcy},
  G.~W. (2014).
\newblock {The TRENDS High-contrast Imaging Survey. IV. The Occurrence Rate of
  Giant Planets around M Dwarfs}.
\newblock {\em \apj}, 781, 28.

\bibitem[{Muirhead} et~al., 2018a]{muirhead_18}
{Muirhead}, P.~S., {Dressing}, C.~D., {Mann}, A.~W., {Rojas-Ayala}, B.,
  {L{\'e}pine}, S., {Paegert}, M., {De Lee}, N., \& {Oelkers}, R. (2018a).
\newblock {A Catalog of Cool Dwarf Targets for the Transiting Exoplanet Survey
  Satellite}.
\newblock {\em \aj}, 155, 180.

\bibitem[{Muirhead} et~al., 2018b]{2018AJ....155..180M}
{Muirhead}, P.~S., {Dressing}, C.~D., {Mann}, A.~W., {Rojas-Ayala}, B.,
  {L{\'e}pine}, S., {Paegert}, M., {De Lee}, N., \& {Oelkers}, R. (2018b).
\newblock {A Catalog of Cool Dwarf Targets for the Transiting Exoplanet Survey
  Satellite}.
\newblock {\em \aj}, 155, 180.

\bibitem[{Ngo} et~al., 2017]{2017AJ....153..242N}
{Ngo}, H., {Knutson}, H.~A., {Bryan}, M.~L., {Blunt}, S., {Nielsen}, E.~L.,
  {Batygin}, K., {Bowler}, B.~P., {Crepp}, J.~R., {Hinkley}, S., {Howard},
  A.~W., \& {Mawet}, D. (2017).
\newblock {No Difference in Orbital Parameters of RV-detected Giant Planets
  between 0.1 and 5 au in Single versus Multi-stellar Systems}.
\newblock {\em \aj}, 153, 242.

\bibitem[{Oja}, 1996]{1996yCat.2182....0O}
{Oja}, T. (1996).
\newblock {VizieR Online Data Catalog: UBV Photometry of Stars with Accurate
  Positions (Oja 1984-1993)}.
\newblock {\em VizieR Online Data Catalog}, 2182.

\bibitem[{Pecaut} \& {Mamajek}, 2013]{2013ApJS..208....9P}
{Pecaut}, M.~J. \& {Mamajek}, E.~E. (2013).
\newblock {Intrinsic Colors, Temperatures, and Bolometric Corrections of
  Pre-main-sequence Stars}.
\newblock {\em \apjs}, 208, 9.

\bibitem[{Ricker} et~al., 2015]{2015JATIS...1a4003R}
{Ricker}, G.~R., {Winn}, J.~N., {Vanderspek}, R., {Latham}, D.~W., {Bakos},
  G.~{\'A}., {Bean}, J.~L., {Berta-Thompson}, Z.~K., {Brown}, T.~M.,
  {Buchhave}, L., {Butler}, N.~R., {Butler}, R.~P., {Chaplin}, W.~J.,
  {Charbonneau}, D., {Christensen-Dalsgaard}, J., {Clampin}, M., {Deming}, D.,
  {Doty}, J., {De Lee}, N., {Dressing}, C., {Dunham}, E.~W., {Endl}, M.,
  {Fressin}, F., {Ge}, J., {Henning}, T., {Holman}, M.~J., {Howard}, A.~W.,
  {Ida}, S., {Jenkins}, J.~M., {Jernigan}, G., {Johnson}, J.~A., {Kaltenegger},
  L., {Kawai}, N., {Kjeldsen}, H., {Laughlin}, G., {Levine}, A.~M., {Lin}, D.,
  {Lissauer}, J.~J., {MacQueen}, P., {Marcy}, G., {McCullough}, P.~R.,
  {Morton}, T.~D., {Narita}, N., {Paegert}, M., {Palle}, E., {Pepe}, F.,
  {Pepper}, J., {Quirrenbach}, A., {Rinehart}, S.~A., {Sasselov}, D., {Sato},
  B., {Seager}, S., {Sozzetti}, A., {Stassun}, K.~G., {Sullivan}, P.,
  {Szentgyorgyi}, A., {Torres}, G., {Udry}, S., \& {Villasenor}, J. (2015).
\newblock {Transiting Exoplanet Survey Satellite (TESS)}.
\newblock {\em Journal of Astronomical Telescopes, Instruments, and Systems},
  1(1), 014003.

\bibitem[{Saumon} \& {Marley}, 2008]{2008ApJ...689.1327S}
{Saumon}, D. \& {Marley}, M.~S. (2008).
\newblock {The Evolution of L and T Dwarfs in Color-Magnitude Diagrams}.
\newblock {\em \apj}, 689, 1327--1344.

\bibitem[{Service} et~al., 2016]{2016PASP..128i5004S}
{Service}, M., {Lu}, J.~R., {Campbell}, R., {Sitarski}, B.~N., {Ghez}, A.~M.,
  \& {Anderson}, J. (2016).
\newblock {A New Distortion Solution for NIRC2 on the Keck II Telescope}.
\newblock {\em \pasp}, 128(9), 095004.

\bibitem[{Skrutskie} et~al., 2006]{2006AJ....131.1163S}
{Skrutskie}, M.~F., {Cutri}, R.~M., {Stiening}, R., {Weinberg}, M.~D.,
  {Schneider}, S., {Carpenter}, J.~M., {Beichman}, C., {Capps}, R., {Chester},
  T., {Elias}, J., {Huchra}, J., {Liebert}, J., {Lonsdale}, C., {Monet}, D.~G.,
  {Price}, S., {Seitzer}, P., {Jarrett}, T., {Kirkpatrick}, J.~D., {Gizis},
  J.~E., {Howard}, E., {Evans}, T., {Fowler}, J., {Fullmer}, L., {Hurt}, R.,
  {Light}, R., {Kopan}, E.~L., {Marsh}, K.~A., {McCallon}, H.~L., {Tam}, R.,
  {Van Dyk}, S., \& {Wheelock}, S. (2006).
\newblock {The Two Micron All Sky Survey (2MASS)}.
\newblock {\em \aj}, 131, 1163--1183.

\bibitem[{Torres}, 1999]{1999PASP..111..169T}
{Torres}, G. (1999).
\newblock {Substellar Companion Masses from Minimal Radial Velocity or
  Astrometric Information: a Monte Carlo Approach}.
\newblock {\em \pasp}, 111, 169--176.

\bibitem[{Valenti} \& {Fischer}, 2005]{2005yCat..21590141V}
{Valenti}, J.~A. \& {Fischer}, D.~A. (2005).
\newblock {VizieR Online Data Catalog: Spectroscopic properties of cool stars.
  I. (Valenti+, 2005)}.
\newblock {\em VizieR Online Data Catalog}, 215.

\bibitem[{Vogt} et~al., 1994]{1994SPIE.2198..362V}
{Vogt}, S.~S., {Allen}, S.~L., {Bigelow}, B.~C., {Bresee}, L., {Brown}, B.,
  {Cantrall}, T., {Conrad}, A., {Couture}, M., {Delaney}, C., {Epps}, H.~W.,
  {Hilyard}, D., {Hilyard}, D.~F., {Horn}, E., {Jern}, N., {Kanto}, D.,
  {Keane}, M.~J., {Kibrick}, R.~I., {Lewis}, J.~W., {Osborne}, J.,
  {Pardeilhan}, G.~H., {Pfister}, T., {Ricketts}, T., {Robinson}, L.~B.,
  {Stover}, R.~J., {Tucker}, D., {Ward}, J., \& {Wei}, M.~Z. (1994).
\newblock {HIRES: the high-resolution echelle spectrometer on the Keck 10-m
  Telescope}.
\newblock In D.~L. {Crawford} \& E.~R. {Craine} (Eds.), {\em Instrumentation in
  Astronomy VIII}, volume 2198 of {\em \procspie}  (pp.\ 362).

\bibitem[{Wizinowich} et~al., 2000]{2000PASP..112..315W}
{Wizinowich}, P., {Acton}, D.~S., {Shelton}, C., {Stomski}, P., {Gathright},
  J., {Ho}, K., {Lupton}, W., {Tsubota}, K., {Lai}, O., {Max}, C., {Brase}, J.,
  {An}, J., {Avicola}, K., {Olivier}, S., {Gavel}, D., {Macintosh}, B., {Ghez},
  A., \& {Larkin}, J. (2000).
\newblock {First Light Adaptive Optics Images from the Keck II Telescope: A New
  Era of High Angular Resolution Imagery}.
\newblock {\em \pasp}, 112, 315--319.

\bibitem[{Wood} et~al., 2019]{2019arXiv190103687W}
{Wood}, C.~M., {Boyajian}, T., {von Braun}, K., {Brewer}, J.~M., {Crepp},
  J.~R., {Schaefer}, G., {Adams}, A., \& {White}, T.~R. (2019).
\newblock {Benchmarking Substellar Evolutionary Models Using New Age Estimates
  for HD 4747 B and HD 19467 B}.
\newblock {\em arXiv e-prints}, (pp.\ arXiv:1901.03687).

\bibitem[{Yelda} et~al., 2010]{2010ApJ...725..331Y}
{Yelda}, S., {Lu}, J.~R., {Ghez}, A.~M., {Clarkson}, W., {Anderson}, J., {Do},
  T., \& {Matthews}, K. (2010).
\newblock {Improving Galactic Center Astrometry by Reducing the Effects of
  Geometric Distortion}.
\newblock {\em \apj}, 725, 331--352.

\end{thebibliography}

\section{Appendix}
\subsection{Table of Observations and Observational Set-up}
\LongTables
\begin{deluxetable}{lcccc}
\tabletypesize{\scriptsize}

\tablecaption{Adaptive Optics Imaging Observations}
\tablewidth{0pt}
\tablehead{\colhead{Name} & \colhead{JD-2,440,000} & \colhead{Filter} & \colhead{$\Delta t$ [s]} & \colhead{Array}}
\startdata
HD224983 AB & 15803.8 & K$^{'}$ & 33.75 & 128 \\
& 16522.9 & J$_{cont}$ & 126 & 1024 \\
& 16522.9 & K$_{cont}$ & 240 & 1024 \\
& 17213.1 & K$_{cont}$ &  36 & 1024 \\

HIP1294 AB & 15803.8 & J & 37.5 & 128\\
& 15803.8 & K$^{'}$ & 37.5 & 128\\
& 16943.8 & K$_{s}$ & 45.12 & 128\\

HD1205 ABC & 16514.1& K$_{cont}$ & 300 & 1024\\
& 16943.9 & K$_{s}$ & 280 & 256\\
& 17213.1 & K$_{cont}$ & 198 & 1024\\

HD1384 AB & 15482.8 & J & 1.2 & 128\\
& 15482.8 & H & 1.2 & 128\\
& 15482.8 & K$^{'}$& 1.2 &128\\
& 15803.8 & K$^{'}$&101.25 &1024\\

HD6512 AB &16523.1 &J$_{cont}$ &180 &1024\\
&16523.1& K$_{cont}$ & 150 & 1024\\
&16943.8 & K$_{cont}$ & 54.3 &1024\\

HD31018 AB & 15483.5 & J$^{\star}$ & 2.25 & 128 \\
& 15483.5 & K$^{'}$ &1.125 & 128\\
& 15483.5 & L$^{'\star}$ & 1.125 & 128 \\
& 15933.9 & H & 20.25 & 128\\
& 15933.9 & K$^{'\star}$ & 15 & 1024 \\
& 15933.9& K$^{'}$ & 20.25 & 128\\

HD34721 AB & 16943.9& J$_{cont}^{\star}$&150 &1024  \\
& 16943.9& K$_{cont}$&150 &1024  \\
&17290.1& J$_{cont}$ &106 & 512\\
&17290.1& K$_{cont}$& 106& 512\\

HD40647 AB & 15482.1& K$^{'}$&2.25 & 1024   \\
& 15933.9& K$^{'}$ & 18 & 256 \\

HD50639 AB &16222.1& J & 34.3 & 256 \\
&16222.1& K$^{'}$ & 34.3 & 256\\
&17291.2 & J$^{\star}$ &200 &256 \\
&17291.2 & K$_{s}$&200 &256 \\
  
HD85472 AB & 16054.8 & J& 18.75 &128 \\
& 16054.8 & K$_{s}$  &18.75 & 128\\
& 16403.8 & J$^{\star}$& 22.5 &128 \\
& 16403.8 & K$_{s}$& 22.5 &128 \\

HIP55507 AB & 15934.1 & K$^{'}$&7.5 &256\\
&17171.7 &J$_{cont}$ & 150 & 1024 \\
&17171.7 &K$_{cont}$& 100 & 1024 \\

HD110537 AB & 15960.1 &J& 50 &256\\
& 15960.1 &K$^{'}$ & 50 &256\\
&17499.9 & J$_{cont}$ & 54 & 1024 \\
&17499.9 & K$_{cont}$ & 54 & 1024 \\

HD111031 AB &15934.1 &K$^{'}$ &7.5 &128\\
&17499.9 & J$_{cont}$ & 54 & 1024 \\
&17499.9 & K$_{cont}$ & 54 & 1024 \\

HIP63762 AB &16076.8 &J &5.3&512\\
&16076.8 &K$^{'}$ &5.3&512\\
&17171.8&K$_{cont}$  &60 & 512\\

HD129191 AB &16054.9 & J$^{\star}$ & 17.6&256\\
&16054.9 &K$^{'}$ & 17.6&256\\
&16493.8&K$_{cont}$&270&1024\\
 
HD129814 AB & 15615.9 & K$^{'}$ &175 & 512\ \\
& 16054.9 & J &18 &256\\
& 16054.9 & K$^{'}$& 18&256\\
& 17171.8 & K$_{cont}$ & 180 &1024\\

HD136274 ABC &16503.8 &J$_{cont}$&180 &1024 \\
&16503.8 &K$_{cont}$&135 &1024 \\
& 17171.8 & K$_{cont}$ & 45 &1024  \\

HD139457 AB&15615.1 &K$^{'}$ & 5 & 512\\
&16054.9 &J & 15 & 224\\
&16054.9 &K$^{'}$ & 15 & 224\\

HD142229 AB & 15615.1 & K$^{'}$ & 10&512  \\
& 16076.8 & J &  52.7 & 512 \\
& 16076.8 & K$^{'}$& 52.7 &  512\\
& 17171.8 & K$_{cont}$ & 45 & 1024  \\

HD147231 AB& 15803.8 & H & 37.5& 128 \\
& 15803.8 & J &37.5& 128 \\
& 15803.8 & K$^{'}$ & 33.75&128 \\
& 16054.9 & J &11.43 & 256\\
& 16054.9  &K$^{'}$ & 11.43& 256\\

HD155413 AB& 16076.9 & J$^{\star}$ & 108 & 300 \\
& 16076.9 & K$^{'}$  & 108 & 300 \\
& 16522.7& J$_{cont}$ & 180 & 1024 \\
& 16522.7& K$_{cont}$ & 162 & 1024 \\
& 17171.8 & K$_{cont}$ & 27 & 1024  \\

HD157338 AB& 15803.8 & K$^{'}$ & 18.75 & 128  \\
& 16076.9 & J$^{\star}$ & 25 & 300\\
& 16076.9 & K$^{'}$ & 25 & 300\\

HD164509 AB &16522.8 &K$_{cont}$ & 180 &1024\\
&17500.0 &  J$_{cont}$ & 100 & 1024 \\
&17500.0 &  K$_{cont}$ & 100 & 1024 \\

HD180684 AB &16055.1 &J&13 & 140 \\
&16055.1 &K$^{'}$&13 & 140 \\
&17500.5&  J$_{cont}$ &54 &1024\\
&17500.5& K$_{cont}$ &54 &1024\\

HD183473 AB &15342.1 &H &.5 & 512\\
&16102.9&J& 22.58&128\\
&16102.9&H& 22.58&128\\
&16102.9&K$^{'}$& 35.25&256\\

HD196201 ABC &15803.8& J &37.5 &128 \\
&15803.8& H & 37.5& 128 \\
& 15803.8 &  K$^{'}$ & 37.5 & 128 \\
&16055.1 & J & 45&256 \\
&16055.1 & K$^{'}$ & 45 &256 \\
&16942.8 & K$_{cont}$ & 200 & 1024\\

HD201924 AB &16113.0&J& 38.125&192\\
&16113.0& K$^{'}$ &38.125 &256  \\
&16165.9 & J&14.5 & 256\\
&16165.9 &  K$^{'}$ & 14.5& 256\\

HD213519 AB & 15803.9 & J$^{\star}$ & 10 & 1024 \\ 
& 15803.9 & K$^{'}$ & 67.5 & 128\\
& 15803.9 & K$^{'}$ & 10 & 1024 \\
& 16103.0& J & 18 & 256 \\
& 16103.0 & K$^{'}$ & 18 & 256  \\

\hline

\rule{0pt}{3ex}HD1293 AB &15483.8 & K$^{'\star}$&1.125 &128\\
&15483.8 &H&1.125 &128\\
&15483.8 & J$^{\star}$ &1.125 &128\\

HD1388 AB & 15803.9 & H$^{\star}$ & 90 & 512\\
& 15803.9 & K$^{'\star}$ & 100 & 512\\
& 16514.1 & J$^{\star}_{cont}$ & 100 & 1024 \\
& 16514.1 & K$_{cont}$ & 300 & 1024 \\

HD4406 ABC& 15481.9 & H & 1.2 &128\\
& 15481.9 & H$^{\star}$ & 18.1 & 1024\\
& 15803.9 & J$^{\star}$ & 56.25 & 128 \\
& 15803.9 & K$^{'}$ & 56.25 & 128 \\

HD6558 AB&15788.9 & K$^{'\star}$ & 33.75 & 128 \\
&15788.9&K$^{'}$ &45 & 512\\
&15788.9 & K$^{'\star}$ & 2490 & 1024\\
&15933.7& K$^{'\star}$ & 12 & 256\\
&15933.7& K$^{'\star}$&1470 & 1024\\

HD88986 AB & 15934.1 & K$^{'}$ & 13& 156 \\
& 16403.8 &K$_{s}^{\star}$ &11.25 &1024\\

HIP46199 AB & 16403.8&J &15&128 \\
& 16403.8&K$^{'}$ &15&128 \\

HD103829 AB &17171.8&K$_{cont}$ &270 & 1024\\

HD105618 AB &17171.9&K$_{cont}$ &300 & 1024\\

HD131509 AB &17171.8 & J$_{cont}$ & 187.5 & 1024\\
&17171.8&K$_{cont}$ &150 & 1024\\

HD156826 ABC & 17171.9&K$_{cont}$&75&512 \\
& 17171.9&J$_{cont}$&75 &512 \\

HD217165 AB&17213.1  &K$_{cont}$&150 &1024  \\
\enddata
\tablecomments{$\star$ Companion not resolved in filter, bad data, too small of collecting area, or coronograph used and occulted companion. Confirmed and candidate companion observations are listed respectively above and below the horizontal line.}
\end{deluxetable}
\clearpage

\subsection{Astrometry, Photometry, and Inferred Physical Measurements}
\begin{landscape}
\begin{deluxetable}{lccccccccc}
\tabletypesize{\scriptsize}
\tablecaption{Astrometry and Photometry: Confirmed Companions}
\tablehead{\colhead{a) Name} & \colhead{b) Filter} &\colhead{c) $\Delta$ mag}  & \colhead{d) M$_{comp}$}& \colhead{e)$\rho$[mas]} & \colhead{f)$\ddagger$P.A.[$^\circ]$}  & \colhead{g) Proj. Sep.[AU]}& \colhead{h) $\dot{v}$ [ms$^{-1}$yr$^{-1}$]} & \colhead{i) $\ast$M$_{min}$[M$_{\odot}$]} & \colhead{j) M$_{phot}$[M$_{\odot}$]}}
\startdata
HD224983 AB & K$^{'}$ & -- -- & -- --& 4640.90 $\pm$ 6.91  & 346.30 $\pm$ 0.09 & 157.19 $\pm$ 0.40 & -1.75 $\pm$ 0.35 & 0.56 $\pm$ 0.34 & -- -- \\
& J$_{cont}$ & 3.20 $\pm$ 0.50 & -- -- & 4660.99 $\pm$ 1.27 & 346.18 $\pm$ 0.02 & 157.87 $\pm$ 0.33 & -- & -- & -- --   \\
& K$_{cont}$ & 2.64 $\pm$ 0.19 & -- -- & 4658.88 $\pm$ 1.08 & 346.23 $\pm$ 0.02 & 157.80 $\pm$ 0.33 & -- & -- & -- -- \\
& K$_{cont}$ & 2.78 $\pm$ 0.08 & -- -- & 4603.12 $\pm$ 2.33 & 350.38 $\pm$ 0.10 & 155.91 $\pm$ 0.33 & -- & -- & -- --  \\
HIP1294 AB & J & 2.48 $\pm$ 0.02 & 6.54 $\pm$ 0.03 &  750.73 $\pm$ 0.86 & 98.18 $\pm$ 0.28 & 29.37 $\pm$ 0.62 & 141.32 $\pm$ 16.08 & 1.68 $\pm$ 0.39 &  0.51 $\pm$ 0.01 \\
& K$^{'}$ & 2.14 $\pm$ 0.02 & 5.62 $\pm$ 0.03 & 747.34 $\pm$ 0.21 & 98.14 $\pm$ 0.01 & 29.36 $\pm$ 0.14 & -- & -- & 0.53 $\pm$ 0.01 \\
& K$_{s}$ & 2.02 $\pm$ 0.03 & 5.52 $\pm$ 0.03 & 720.97 $\pm$ 0.25 & 104.16 $\pm$ 0.01 & 28.33 $\pm$ 0.14 & -- & -- & 0.54 $\pm$ 0.01 \\
HD1205 AB & K$_{cont}$ &1.38 $\pm$ 0.01 & 3.97 $\pm$ 0.11 & 215.7 $\pm$ 1.2& 213.15 $\pm$ 0.68 & 16.52 $\pm$ 0.09 & 108.65 $\pm$ 0.82 & 0.29 $\pm$ 0.12 & 0.85 $\pm$ 0.03 \\
& K$_{s}$ & 0.96 $\pm$ 0.03 & 3.65 $\pm$ 0.11 & 219.4 $\pm$ 4.2 & 242.22 $\pm$ 0.50 & 16.78 $\pm$ 0.10 & -- & -- & 0.92 $\pm$ 0.04 \\
& K$_{cont}$ & 0.94 $\pm$ 0.03 & 3.64 $\pm$ 0.11 & 214.4 $\pm$ 1.8 & 240.19 $\pm$ 0.50 & 16.42 $\pm$ 0.09 & -- & -- & 0.92 $\pm$ 0.04 \\
HD1205 AC & K$_{cont}$ & 4.01 $\pm$ 0.02 & 6.35 $\pm$ 0.11 &1100.8 $\pm$ 2.2 & 165.53 $\pm$ 0.68 & 84.29 $\pm$ 0.09 & -- & 7.74 $\pm$ 3.17 & 0.41 $\pm$ 0.02 \\ 
& K$_{s}$ & 3.47 $\pm$ 0.03 & 5.83 $\pm$ 0.12 & 1091.7 $\pm$ 5.3 & 197.09 $\pm$ 0.50 & 83.59 $\pm$ 0.09 & -- & -- & 0.49 $\pm$ 0.02 \\
& K$_{cont}$ & 3.39 $\pm$ 0.02 & 5.75 $\pm$ 0.11 & 1087.1 $\pm$ 2.5 & 195.17 $\pm$ 0.50 & 83.24 $\pm$ 0.09 & -- & -- & 0.51 $\pm$ 0.02 \\
HD1384 AB & J & 2.99 $\pm$ 0.34  & 3.41 $\pm$ 0.32   & 336.1 $\pm$ 0.7 & 134.87 $\pm$ 0.68 & 60.01 $\pm$ 0.05 & -65.92 $\pm$ 0.32 & 2.67 $\pm$ 0.48 & 1.05 $\pm$ 0.08  \\
& H & 3.26 $\pm$ 0.28  & 3.19 $\pm$ 0.27   & 336.9 $\pm$ 1.1 & 133.93 $\pm$ 0.68 & 59.89 $\pm$ 0.06 & -- & -- & 1.03 $\pm$ 0.08 \\
& K${'}$& 3.32 $\pm$ 0.06  & 3.13 $\pm$ 0.08 & 334.7 $\pm$ 4.0 & 134.63 $\pm$ 0.68 & 60.09 $\pm$ 0.07 & -- & --& 1.04 $\pm$ 0.05 \\
& K${'}$& 3.45 $\pm$ 0.01 & 3.25 $\pm$ 0.05 & 336.1 $\pm$ 1.4 & 133.44 $\pm$ 0.09 & 60.34 $\pm$ 0.07 & -- & -- &  1.01 $\pm$ 0.05 \\
HD6512 AB & J$_{cont}$ & 2.86 $\pm$ 0.12 & 6.15 $\pm$ 0.11 &  738.34 $\pm$ 3.12 & 86.37 $\pm$ 0.18 & 39.18 $\pm$ 0.55 & -10.41 $\pm$ 0.41 & 0.21 $\pm$ 0.03 & 0.56 $\pm$ 0.02  \\
& K$_{cont}$ & 2.27 $\pm$ 0.02 & 5.29 $\pm$ 0.03 & 739.11 $\pm$ 1.48 & 86.97 $\pm$ 0.12 & 40.26 $\pm$ 0.16 & -- & -- & 0.58 $\pm$ 0.01  \\
& K$_{cont}$ & 2.30 $\pm$ 0.01 & 5.31 $\pm$ 0.02 & 721.91 $\pm$ 0.29 &  88.04 $\pm$ 0.04 & 39.32 $\pm$ 0.13 & -- & -- & 0.58 $\pm$ 0.01 \\
HD31018 AB & K$^{'}$ & 3.36 $\pm$ 0.35 & 4.47 $\pm$ 0.34 & 287.32 $\pm$ 1.33 & 65.03 $\pm$ 0.27 & 28.94 $\pm$ 0.22 & 18.58 $\pm$ 1.39 & 0.23 $\pm$ 0.10 & 0.74 $\pm$ 0.08 \\
& H & 4.26 $\pm$ 0.18 & 5.43 $\pm$ 0.18 & 293.11 $\pm$ 1.59 & 77.61 $\pm$ 0.08 & 29.76 $\pm$ 0.27 & -- & -- & 0.58 $\pm$ 0.03 \\
&K$^{'}$ & 3.35 $\pm$ 0.04 & 4.46 $\pm$ 0.04 & 293.87 $\pm$ 1.27 & 77.48 $\pm$ 0.06 & 29.60 $\pm$ 0.22 & -- & -- & 0.73 $\pm$ 0.01 \\
HD34721 AB & K$_{cont}$&4.33 $\pm$ 0.01 & 6.94 $\pm$ 0.02 & 2176.0 $\pm$ 5.7 & 132.26 $\pm$ 0.05 & 54.45 $\pm$ 0.01 & 1.26 $\pm$ 0.10 & 0.05 $\pm$ 0.01 &0.31 $\pm$ 0.01 \\
& J$_{cont}$ & 4.71 $\pm$ 0.02 & 7.94 $\pm$ 0.27 & 2163.6 $\pm$ 4.3 & 131.60 $\pm$ 0.68 & 46.65 $\pm$ 0.33 & -- & -- & 0.29 $\pm$ 0.04 \\
& K$_{cont}$ & 4.10 $\pm$ 0.02 & 6.71 $\pm$ 0.03 & 2159.4 $\pm$ 4.1 & 131.57 $\pm$ 0.68 & 46.56 $\pm$ 0.33 & -- & -- &0.35 $\pm$ 0.01 \\
HD40647 AB & K$^{'}$ & 2.40 $\pm$ 0.09 & 6.36 $\pm$ 0.08 & 254.86 $\pm$ 1.76 & 223.45 $\pm$ 0.38  & 8.00 $\pm$ 0.07 & 169.90 $\pm$ 0.90 & 0.17 $\pm$ 0.02 & 0.41 $\pm$ 0.01  \\
& K$^{'}$ & 2.48 $\pm$ 0.09 & 6.43 $\pm$ 0.08 & 214.80 $\pm$ 0.54 & 270.11 $\pm$ 0.24 & 6.74 $\pm$ 0.04 & -- & -- & 0.40 $\pm$ 0.01 \\
HD50639 AB & J & 3.66 $\pm$ 0.47 & 6.79 $\pm$ 0.45 & 489.79 $\pm$ 1.19 & 10.61 $\pm$ 0.68 & 18.65 $\pm$ 0.17 & -86.98 $\pm$ 0.77 & 0.38 $\pm$ 0.04 & 0.47 $\pm$ 0.07 \\
& K$^{'}$ &3.51 $\pm$ 0.06 & 6.29 $\pm$ 0.06 & 490.99 $\pm$ 1.06 & 9.56 $\pm$ 0.55 & 19.12 $\pm$ 0.06 & -- & -- & 0.42 $\pm$ 0.01 \\
& K$_{s}$ & 3.56 $\pm$ 0.01 & 6.34 $\pm$ 0.02 & 560.9 $\pm$ 4.2 & 21.96 $\pm$ 0.45 & 20.29 $\pm$ 0.14 & -- & -- & 0.41 $\pm$ 0.01 \\
HD85472 AB & J & 4.48 $\pm$ 0.41 & 6.08 $\pm$ 0.40 & 323.83 $\pm$ 0.43 & 101.18 $\pm$ 0.68 & 21.45 $\pm$ 0.07 & -37.17 $\pm$ 0.30 & 0.30 $\pm$ 0.03 & 0.58 $\pm$ 0.06\\
& K$_{s}$ & 4.74 $\pm$ 0.03 & 5.90 $\pm$ 0.07 & 325.44 $\pm$ 0.35 & 100.93 $\pm$ 0.08 & 24.73 $\pm$ 0.06 & -- & -- &0.48 $\pm$ 0.01 \\
& K$^{'}$ & 4.78 $\pm$ 0.02 & 5.94 $\pm$ 0.06 & 329.99 $\pm$ 0.29 & 109.88 $\pm$ 0.04 & 25.08 $\pm$ 0.06 & -- & -- &0.48 $\pm$ 0.01 \\
HIP55507 AB & K$^{'}$  & 5.78 $\pm$ 0.05 & 10.36 $\pm$ 0.05 & 473.04 $\pm$ 1.44 & 178.48 $\pm$ 0.07 & 12.07 $\pm$ 0.04 & 24.14 $\pm$ 0.74 & 0.05 $\pm$ 0.01 &  0.08 $\pm$ 0.01 \\
& J$_{cont}$ & 5.45 $\pm$ 0.12 & 10.78 $\pm$ 0.12 & 547.62 $\pm$ 2.38 & 175.31 $\pm$ 0.31 & 12.26 $\pm$ 0.06 & -- & -- & 0.09 $\pm$ 0.01 \\
& K$_{cont}$ & 5.27 $\pm$ 0.02 & 9.85 $\pm$ 0.03 & 544.19 $\pm$ 2.24 & 174.90 $\pm$ 0.31 & 13.88 $\pm$ 0.06 & -- & -- & 0.09 $\pm$ 0.01  \\
HD110537 AB & J & 4.44 $\pm$ 0.01 & 7.76 $\pm$ 0.02 & 1246.54 $\pm$ 1.43 & 69.76 $\pm$ 0.15 & 56.28 $\pm$ 0.14 & 7.15 $\pm$ 0.12 & 0.28 $\pm$ 0.06 & 0.31 $\pm$ 0.01 \\
& K$^{'}$  & 4.11 $\pm$ 0.01 & 7.06 $\pm$ 0.02 & 1247.80 $\pm$ 1.22 & 67.75 $\pm$ 0.11 & 56.34 $\pm$ 0.14 & -- & -- & 0.29 $\pm$ 0.02   \\
& J$_{cont}$ & 5.25 $\pm$ 0.08 & 8.57 $\pm$ 0.08 & 1283.8 $\pm$ 3.5 & 87.60 $\pm$ 0.68 & 50.57 $\pm$ 0.53 & -- & -- & 0.21 $\pm$ 0.01 \\
& K$_{cont}$ & 4.03 $\pm$ 0.01 & 6.98 $\pm$ 0.02 & 1288.6 $\pm$ 7.1 & 87.72 $\pm$ 0.68 & 50.76 $\pm$ 0.72 & -- & -- & 0.30 $\pm$ 0.01  \\
HD111031 AB & K$^{'}$ & 5.91 $\pm$ 0.06 & 8.76 $\pm$ 0.07 & 961.77 $\pm$ 2.11 & 282.84 $\pm$ 0.05 & 30.04 $\pm$ 0.08 & 7.8 $\pm$ 0.48 & 0.11 $\pm$ 0.02 & 0.13 $\pm$ 0.01 \\
& J$_{cont}$ & 5.77 $\pm$ 0.01 & 9.04 $\pm$ 0.02 & 1036.0 $\pm$ 4.5 & 303.99 $\pm$ 0.68 & 28.23 $\pm$ 0.30 &  -- & -- & 0.16 $\pm$ 0.01\\
& K$_{cont}$ & 5.50 $\pm$ 0.01 & 8.35 $\pm$ 0.03 & 1040.9 $\pm$ 2.8 & 304.42 $\pm$ 0.68 &  28.37 $\pm$ 0.23 & -- & -- & 0.15 $\pm$ 0.01\\
\enddata
\tablecomments{This is an excerpt of Table 8. The entire table is available in a machine readable format.\\
--  -- Observation set-up (Table 7) did not allow for analysis of this value due to any of the following: Companion not resolved in filter, bad data, too small of collecting area, or coronograph used and occulted companion.\\
$\ddagger$ Position Angle (P.A.) is measured as East of North. In the event the images were taken in vertical angle mode, de-rotation of the images was performed to place each frame back into position angle mode.\\
$\ast$SPOCS catalog used to obtain stellar mass measurement of the primary star and infer the dynamical minimum mass of the companion.}
\end{deluxetable}
\clearpage
\end{landscape}

\begin{landscape}
\begin{deluxetable}{lccccccccc}
\tabletypesize{\scriptsize}
\tablecaption{Astrometry and Photometry: Candidate Companions}
\tablehead{\colhead{a) Name} & \colhead{b) Filter} &\colhead{c) $\Delta$mag} & \colhead{d) M$_{comp}$}& \colhead{e) $\rho$ [mas]} & \colhead{f)$\ddagger$ P.A. [$^\circ$]} & \colhead{g) Proj. Sep.[AU]}& \colhead{h) $\dot{v}$ [ms$^{-1}$yr$^{-1}$]} & \colhead{i) $\ast$M$_{min}$ [M$_{\odot}$]} & \colhead{j) M$_{phot}$ [M$_{\odot}$]}}
\startdata
HD1293 AB & H & 3.05 $\pm$ 0.12 & 3.08 $\pm$ 0.14 & 126.77 $\pm$ 1.84 & 185.20 $\pm$ 0.61 & 24.74 $\pm$ 0.99 & 68.56 $\pm$ 1.13 & 0.57 $\pm$ 0.43 & 1.05 $\pm$ 0.06 \\

HD1388 AB & K$_{cont}$ & 5.87 $\pm$ 0.22 & 8.71 $\pm$ 0.22 & 1677.18 $\pm$ 2.40  & 319.70 $\pm$ 0.09 & 45.20 $\pm$ 0.09 & 26.10 $\pm$ 0.25 & 0.75 $\pm$ 0.06 & 0.13 $\pm$ 0.01\\ 

HD4406 AB & K$^{'}$ &0.43 $\pm$ 0.01 & 1.33 $\pm$ 0.03 & 75.96 $\pm$ 4.98 & 246.79 $\pm$ 3.77  & 9.46 $\pm$ 0.63 & 18.20 $\pm$ 1.68 & 0.02 $\pm$ 0.01 & 1.40 $\pm$ 0.01 \\

HD6558 AB & K$^{'}$ & 4.54 $\pm$ 0.01 & 6.84 $\pm$ 0.02  & 4777.50 $\pm$ 2.93 & 286.20 $\pm$ 0.06 & 392.23 $\pm$ 1.93 & -37.28 $\pm$ 0.88 & 79.38 $\pm$ 4.92 & 0.33 $\pm$ 0.01 \\

HIP46199 AB & J & 1.79 $\pm$ 0.23 & 7.13 $\pm$ 0.20 & 411.4 $\pm$ 1.2 & 288.69 $\pm$ 0.68 & 8.08 $\pm$ 0.60 & -152.10 $\pm$ 2.21 & 0.18 $\pm$ 0.03 & 0.41 $\pm$ 0.03 \\ 
& K$^{'}$ & 1.16 $\pm$ 0.62 & 5.94 $\pm$ 0.46 & 412.3 $\pm$ 1.0 & 288.56 $\pm$ 0.68 & 8.10 $\pm$ 0.60 & -- & -- & 0.48 $\pm$ 0.07 \\

HD88986 AB &  K$^{'}$ & 4.28 $\pm$ 0.01 & 6.57 $\pm$ 0.02 & 1521.11 $\pm$ 1.66 & 59.74 $\pm$ 0.09 & 50.65 $\pm$ 0.08 & -1.65 $\pm$ 0.07 & 0.06 $\pm$ 0.01 & 0.37 $\pm$ 0.01 \\

HD105618 AB & K$_{cont}$ & 5.20 $\pm$ 0.10 & 7.99 $\pm$ 0.10 & 1040.08 $\pm$ 3.99 & 316.75 $\pm$ 0.22 & 73.05 $\pm$ 0.38 & 3.36 $\pm$ 0.40 & 0.25 $\pm$ 0.06 & 0.19 $\pm$ 0.01 \\

HD131509 AB & J$_{cont}$ & 4.26 $\pm$ 0.03 & 6.15 $\pm$ 0.04 & 589.07 $\pm$ 1.08 & 35.78 $\pm$ 0.14 & 45.63 $\pm$ 0.20 & -7.73 $\pm$ 0.10 & 0.26 $\pm$ 0.35 & 0.57 $\pm$ 0.01  \\
& K$_{cont}$ & 3.95 $\pm$ 0.05 & 5.29 $\pm$ 0.05 & 598.75 $\pm$ 0.46 & 35.99 $\pm$ 0.06 & 46.38 $\pm$ 0.18 & -- & -- & 0.58 $\pm$ 0.01  \\

HD156826 AB & J$_{cont}$ & 5.71 $\pm$ 0.33 & 7.34 $\pm$ 0.33 & 1720.28 $\pm$ 2.92 & 338.78 $\pm$ 0.04 & 82.16 $\pm$ 0.21 & -9.80 $\pm$ 0.16 &1.41 $\pm$ 0.39 & 0.38 $\pm$ 0.05  \\
& K$_{cont}$ & 4.53 $\pm$ 0.07 & 5.40 $\pm$ 0.07 & 1721.39 $\pm$ 0.47 & 338.77 $\pm$ 0.01 & 82.21 $\pm$ 0.16 & -- & -- & 0.56 $\pm$ 0.01  \\

HD156828 AC & J$_{cont}$ & 4.97 $\pm$ 0.13 & 6.61 $\pm$ 0.13 & 1939.54 $\pm$ 2.10 & 341.91 $\pm$ 0.04 & 92.63 $\pm$ 0.20 & -- & 1.79 $\pm$ 0.40  & 0.50 $\pm$ 0.02 \\
& K$_{cont}$ & 4.65 $\pm$ 0.08 & 5.52 $\pm$ 0.08 & 1940.01 $\pm$ 0.66 & 341.88 $\pm$ 0.01 & 92.66 $\pm$ 0.18 & -- & -- & 0.54 $\pm$ 0.01 \\

HD217165 AB & K$_{cont}$ & 2.98 $\pm$ 0.01 & 6.02 $\pm$ 0.02 & 318.76 $\pm$ 0.37 & 73.88 $\pm$ 0.17 & 14.02 $\pm$ 0.04 & -365.60 $\pm$ 0.54 & 0.27 $\pm$ 0.34 & 0.46 $\pm$ 0.01 \\
\enddata
\tablecomments{\noindent Astrometry and photometry for all candidate companions. The entire table can be found in machine readable format with \textbf{Table 7}.\\$\ddagger$Position Angle (P.A) is measured as East of North. In the event the images were taken in vertical angle mode, de-rotation of the images was performed to place each frame back into position angle mode.\\
$\ast$SPOCS catalog used to obtain stellar mass measurement of the primary star and infer the dynamical minimum mass of the companion.}
\end{deluxetable}
\clearpage
\end{landscape}

\newpage
\subsection{Relative Radial Velocity Plots of Compendium Ohjects}
\begin{figure}[t]
\subfloat{\subfloat{\includegraphics[width=.5\textwidth]{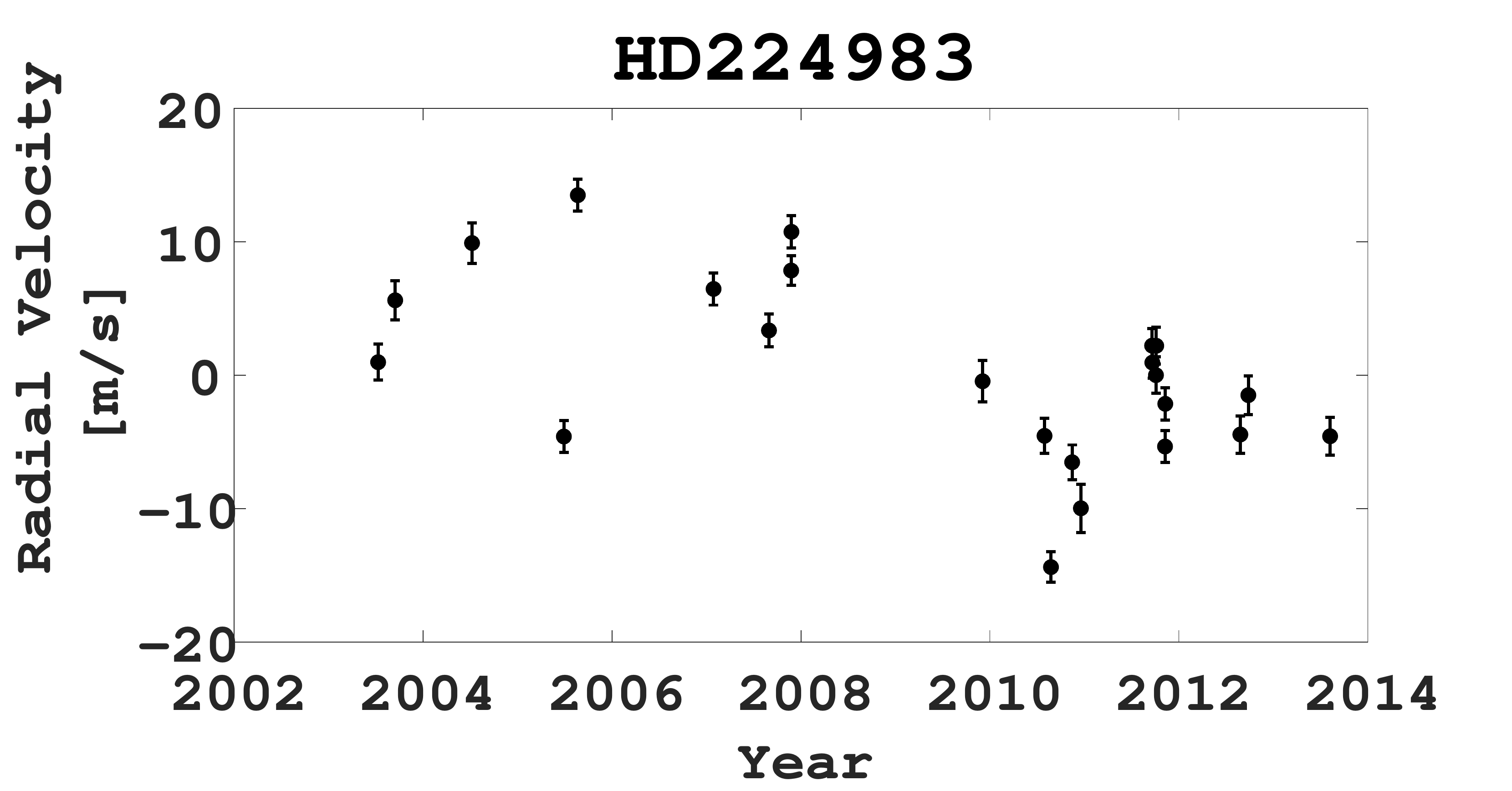}}
\subfloat{\includegraphics[width=.5\textwidth]{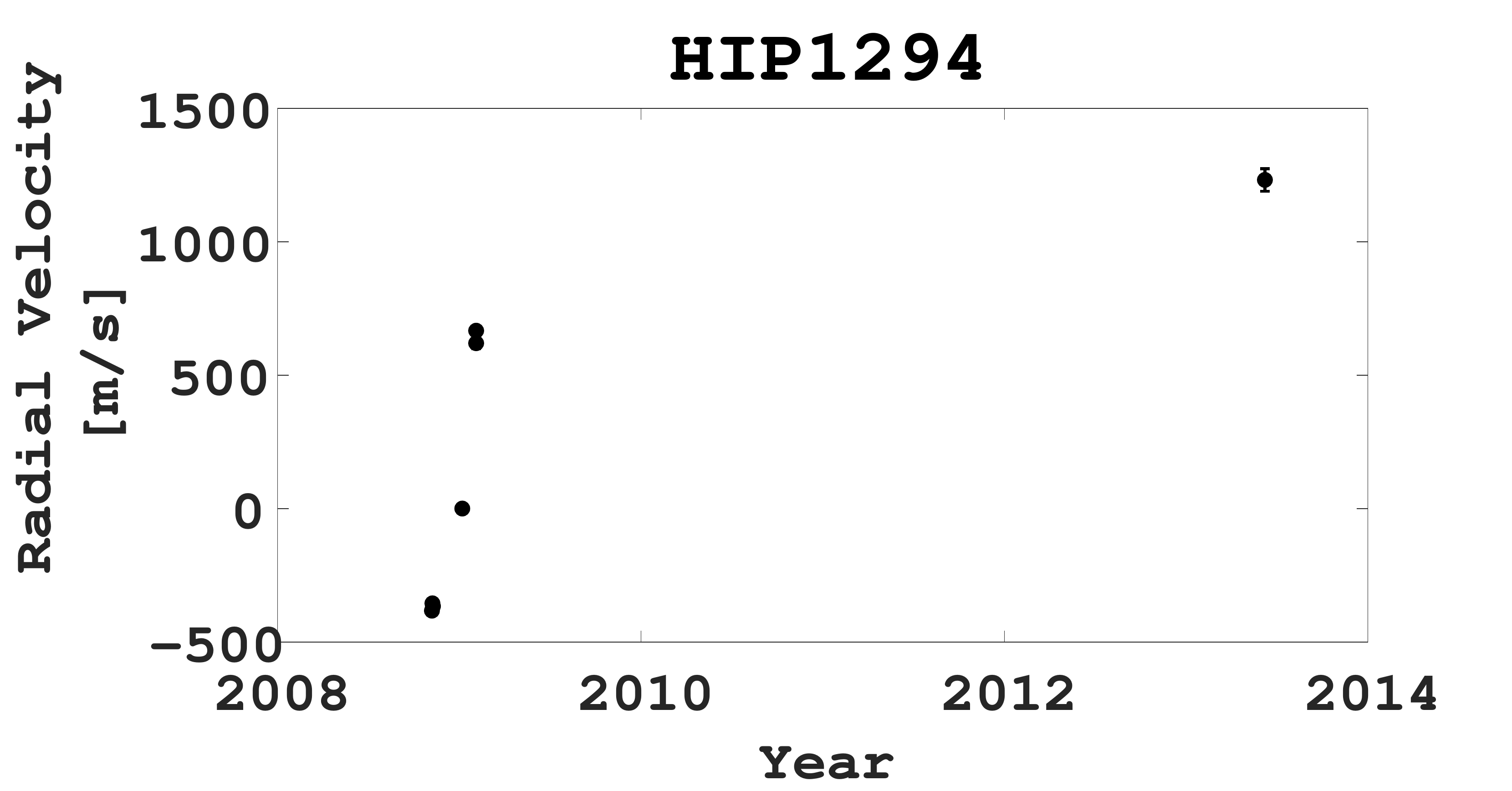}}}\\
\subfloat{\subfloat{\includegraphics[width=.5\textwidth]{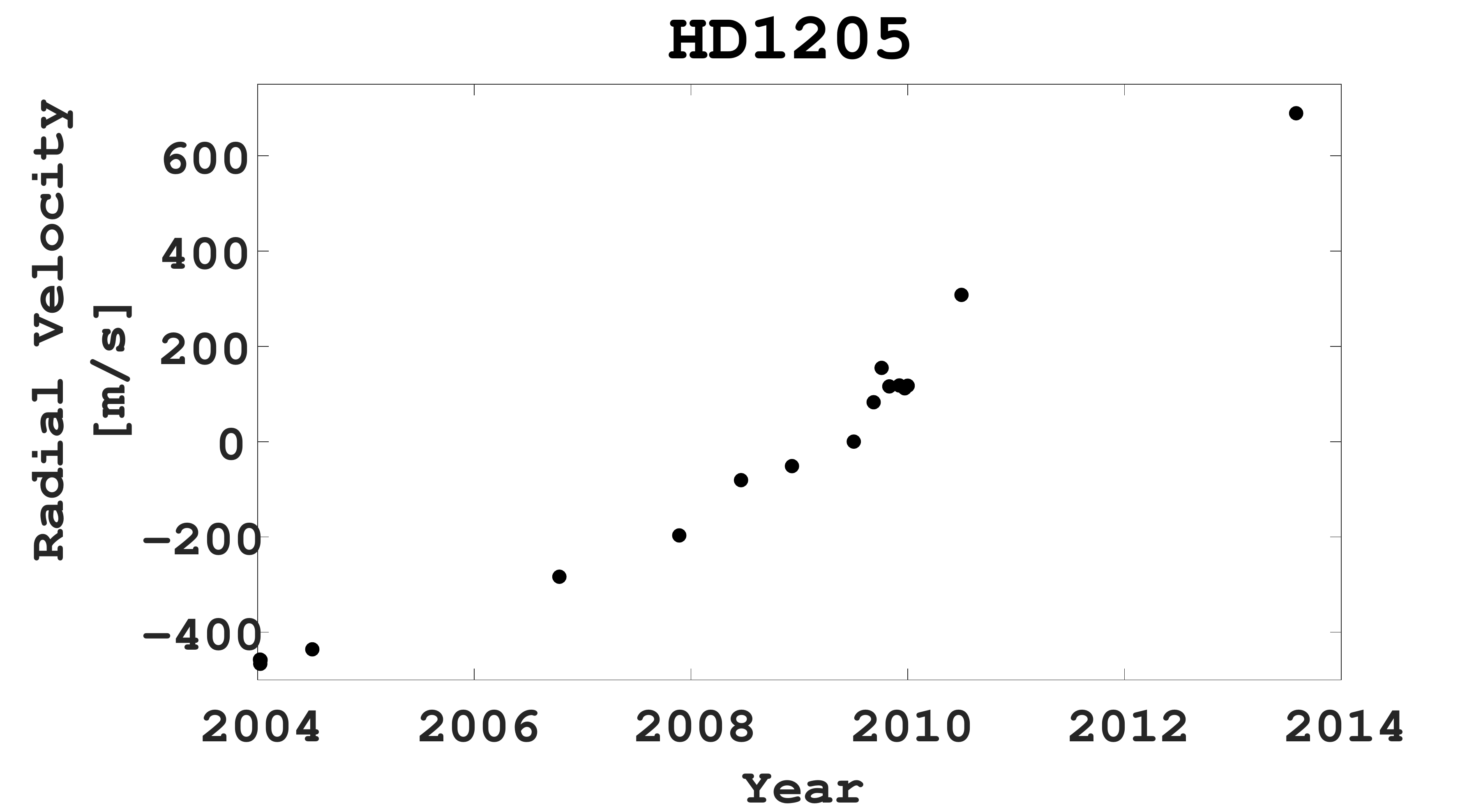}}
\subfloat{\includegraphics[width=.5\textwidth]{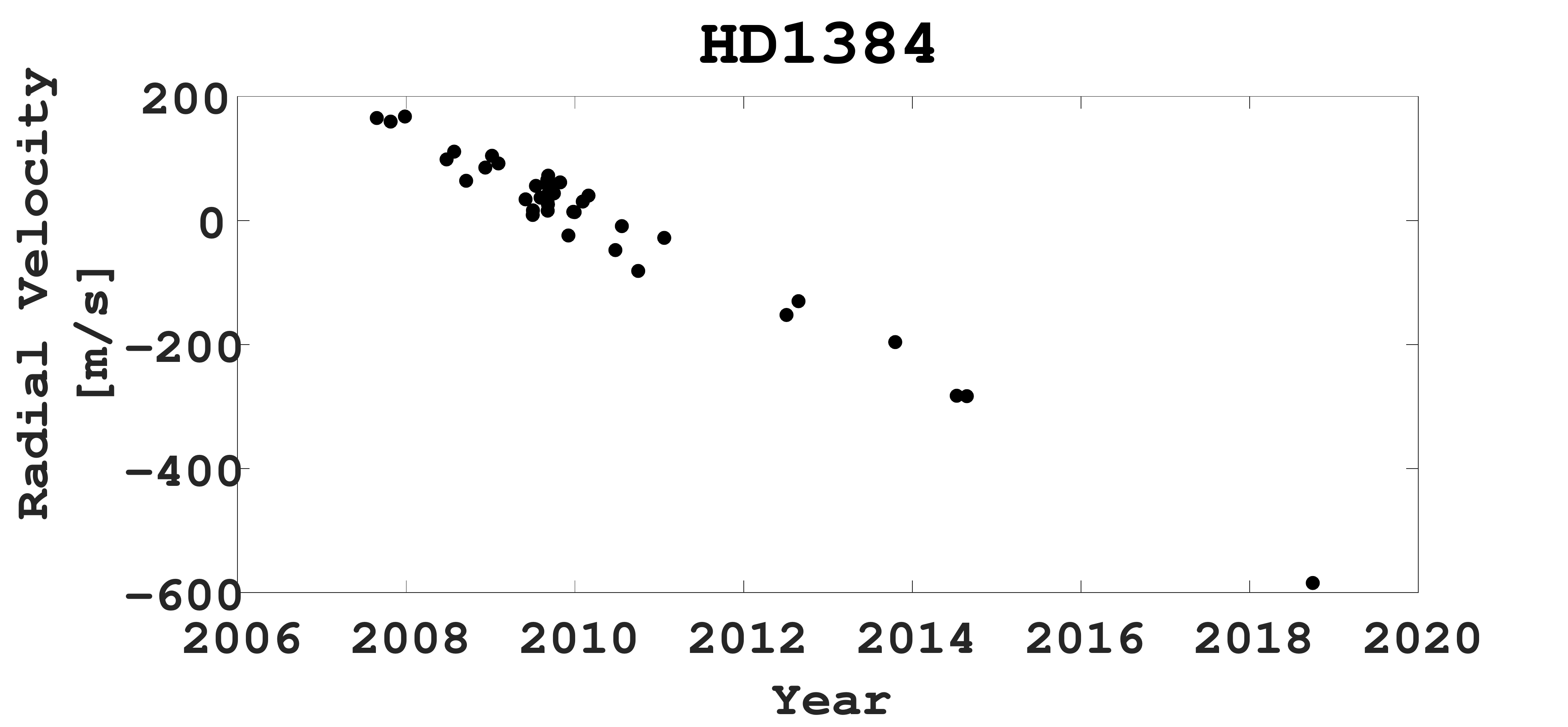}}}\\
\subfloat{\subfloat{\includegraphics[width=.5\textwidth]{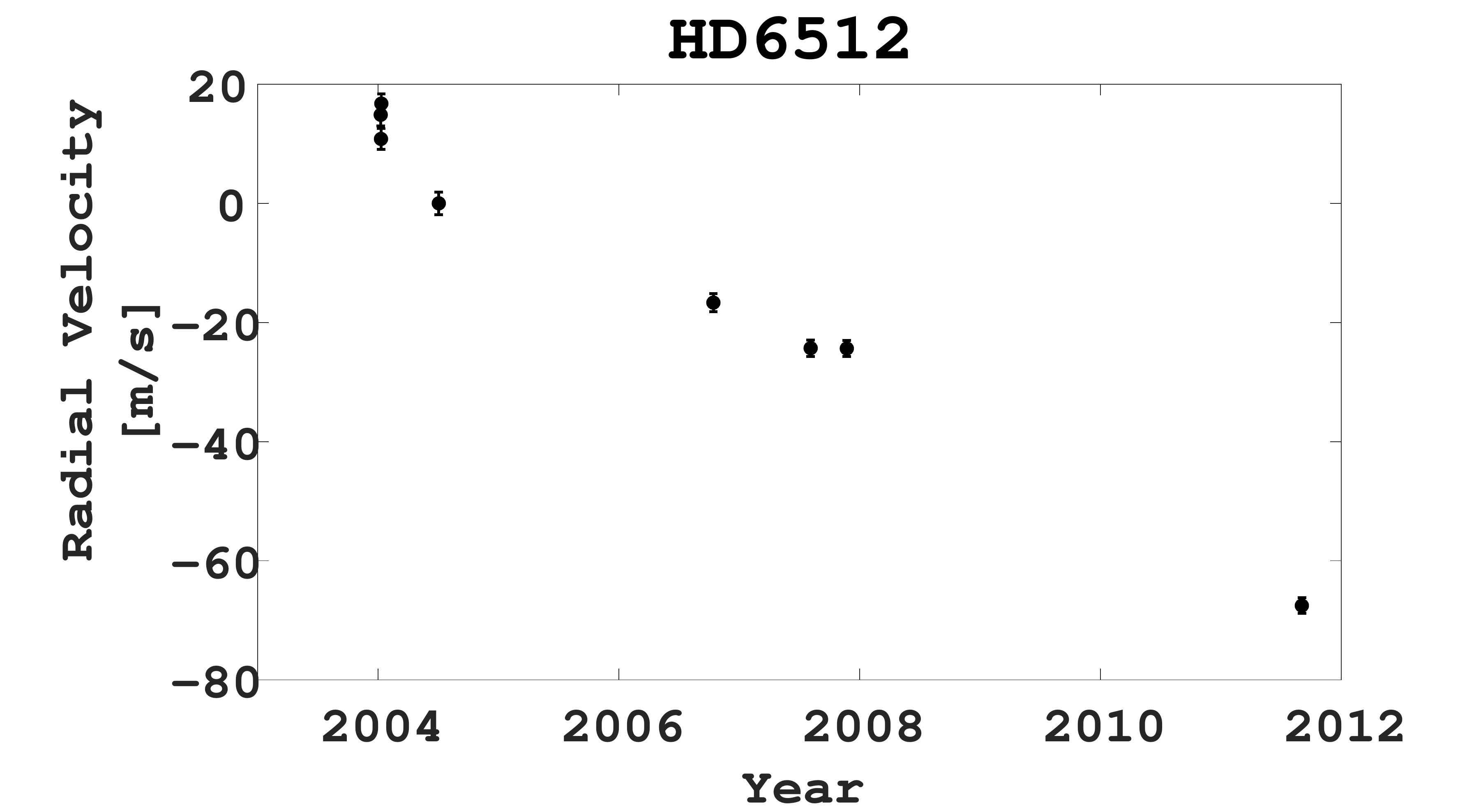}}
\subfloat{\includegraphics[width=.5\textwidth]{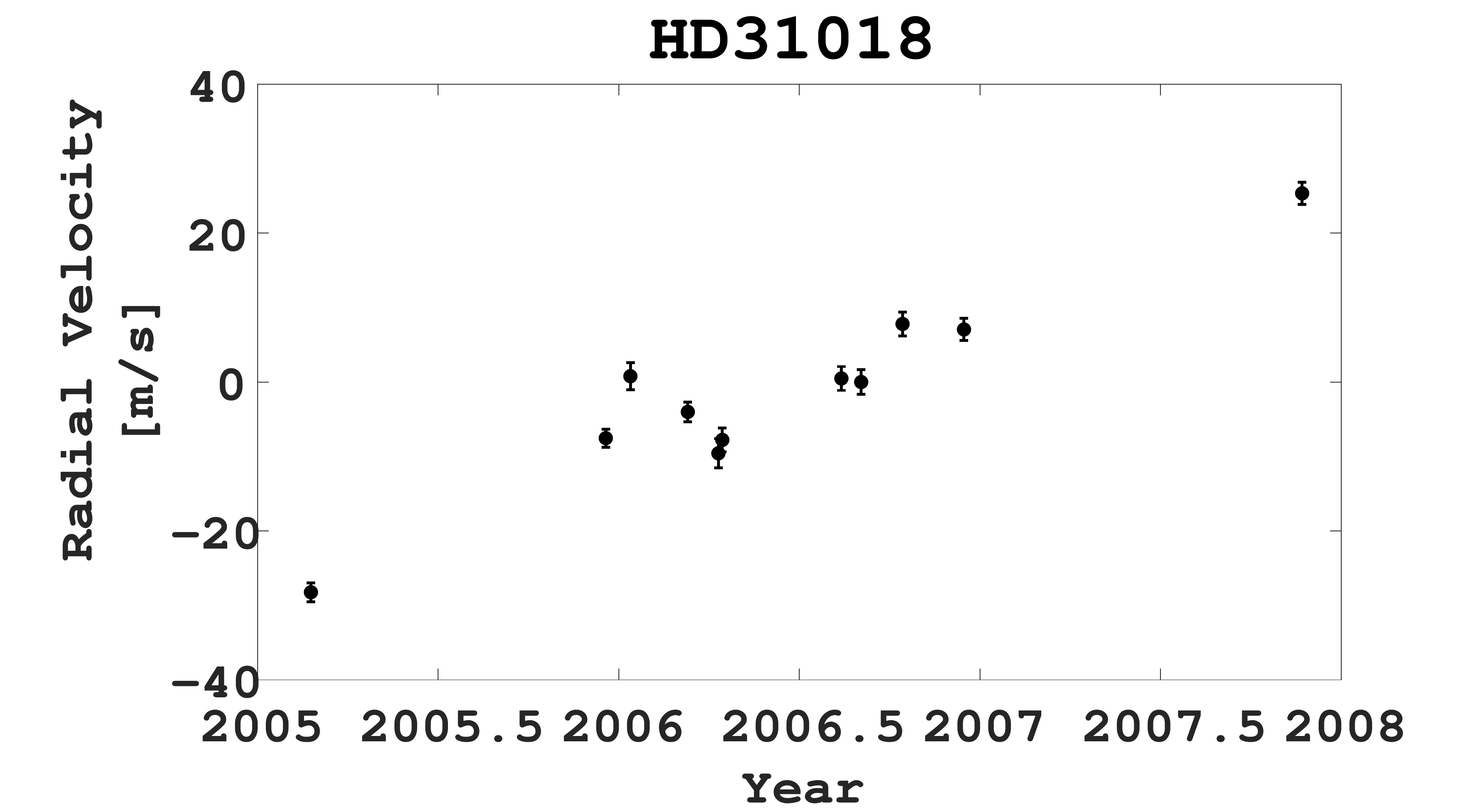}}}\\
\subfloat{\subfloat{\includegraphics[width=.5\textwidth]{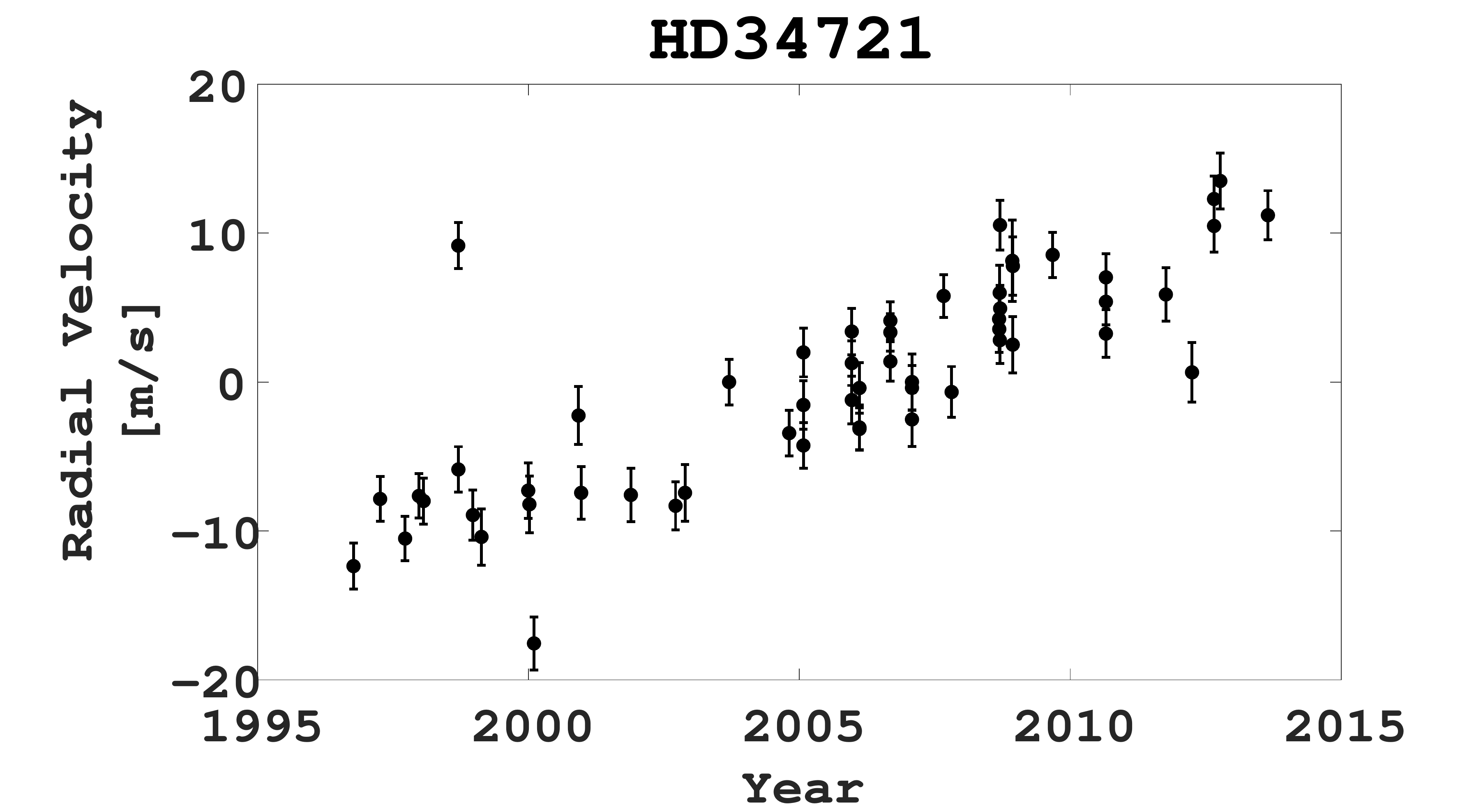}}
\subfloat{\includegraphics[width=.5\textwidth]{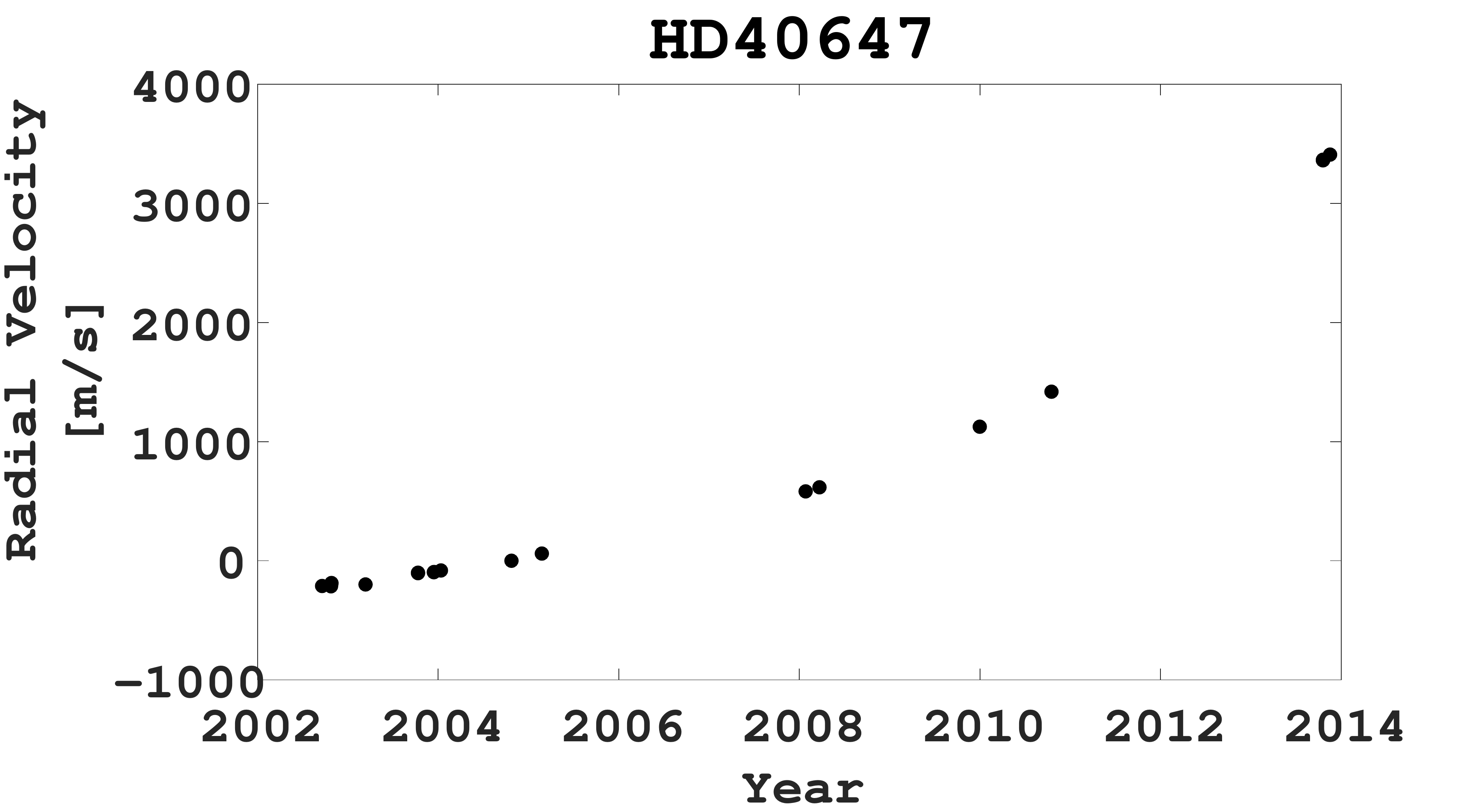}}}\\
\caption{Relative radial velocity plots for stars with confirmed companions.}
\label{fig7}
\end{figure}

\begin{figure}[!htb]
\subfloat{\subfloat{\includegraphics[width=.5\textwidth]{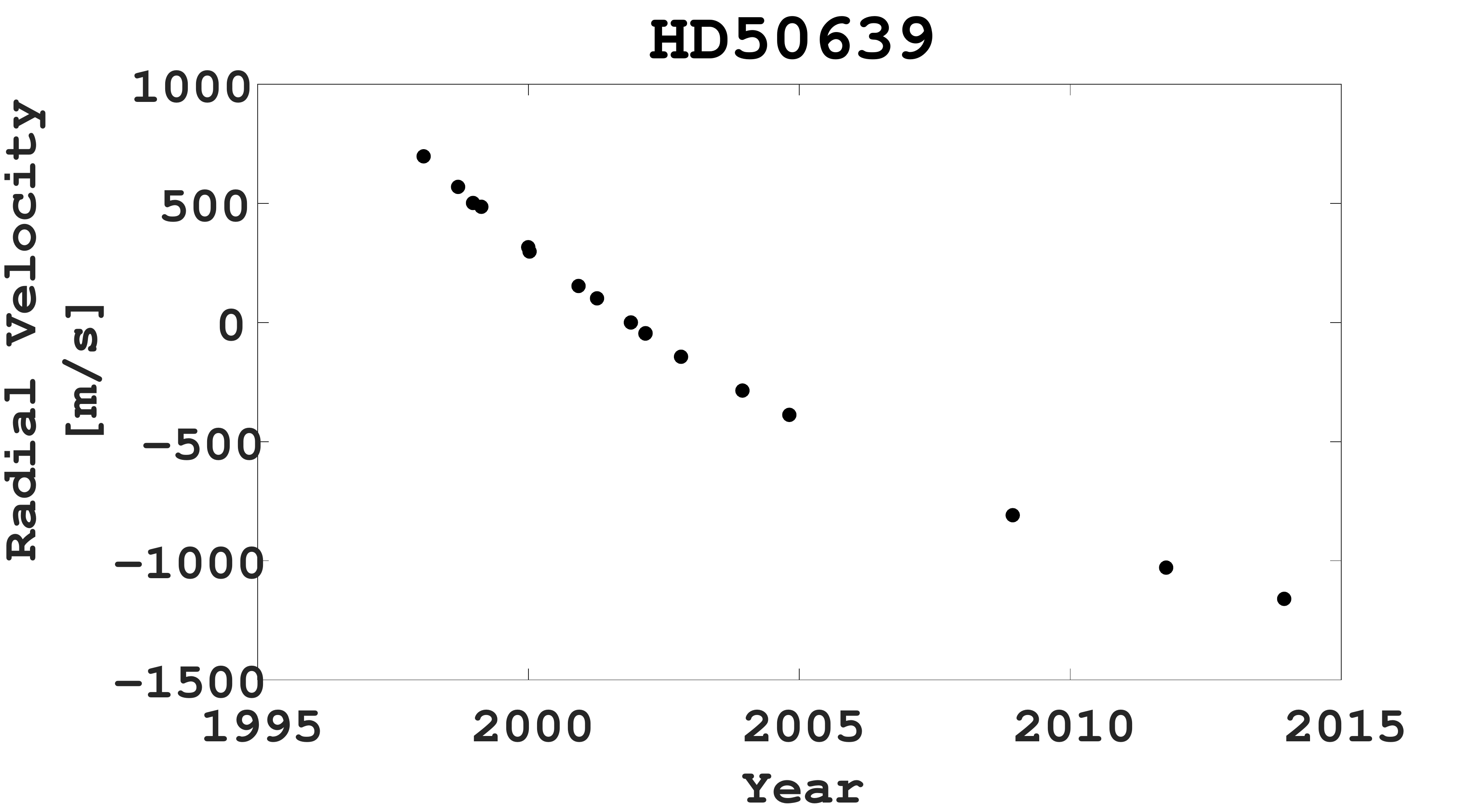}}
\subfloat{\includegraphics[width=.5\textwidth]{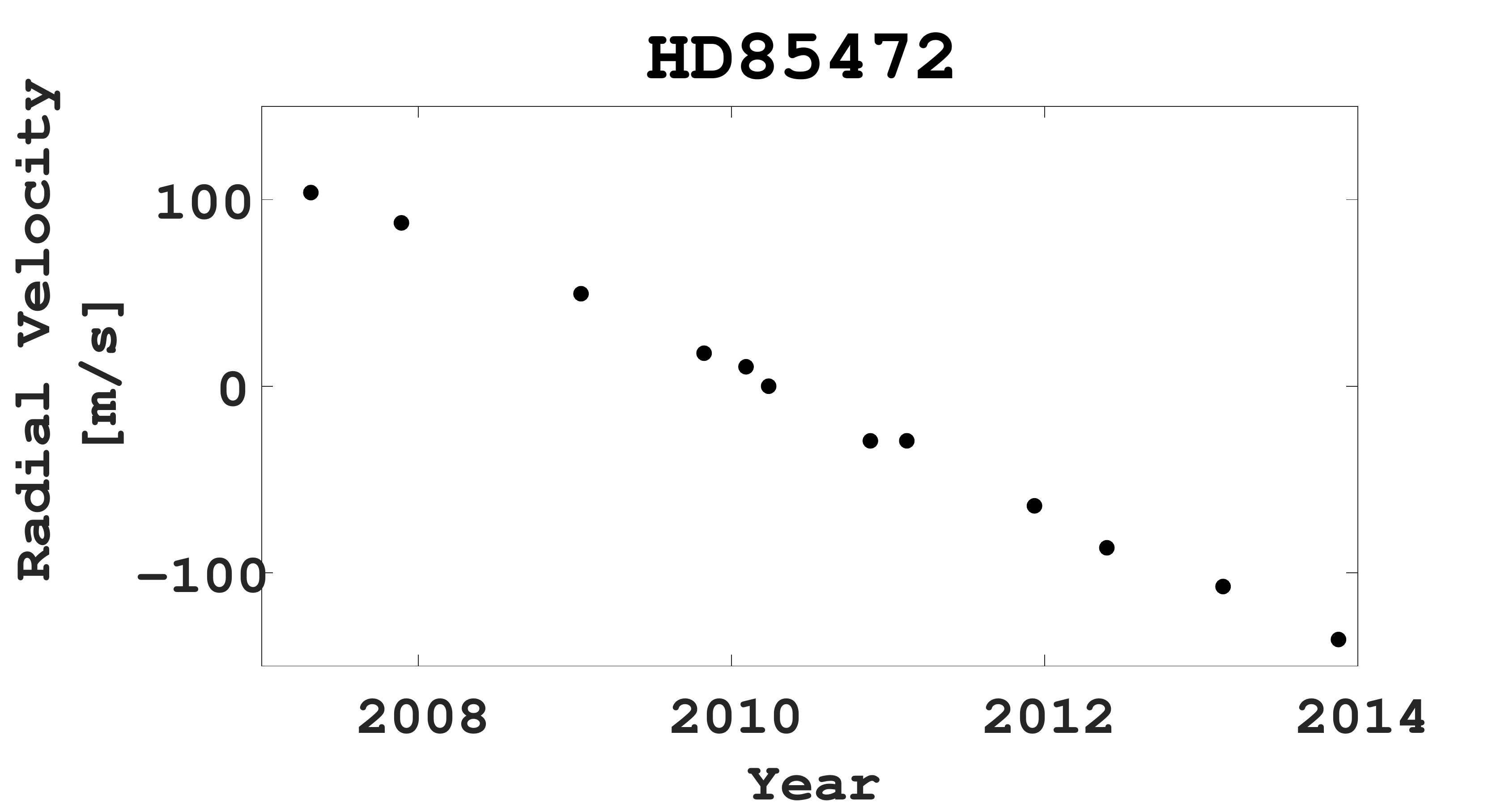}}}\\
\subfloat{\subfloat{\includegraphics[width=.5\textwidth]{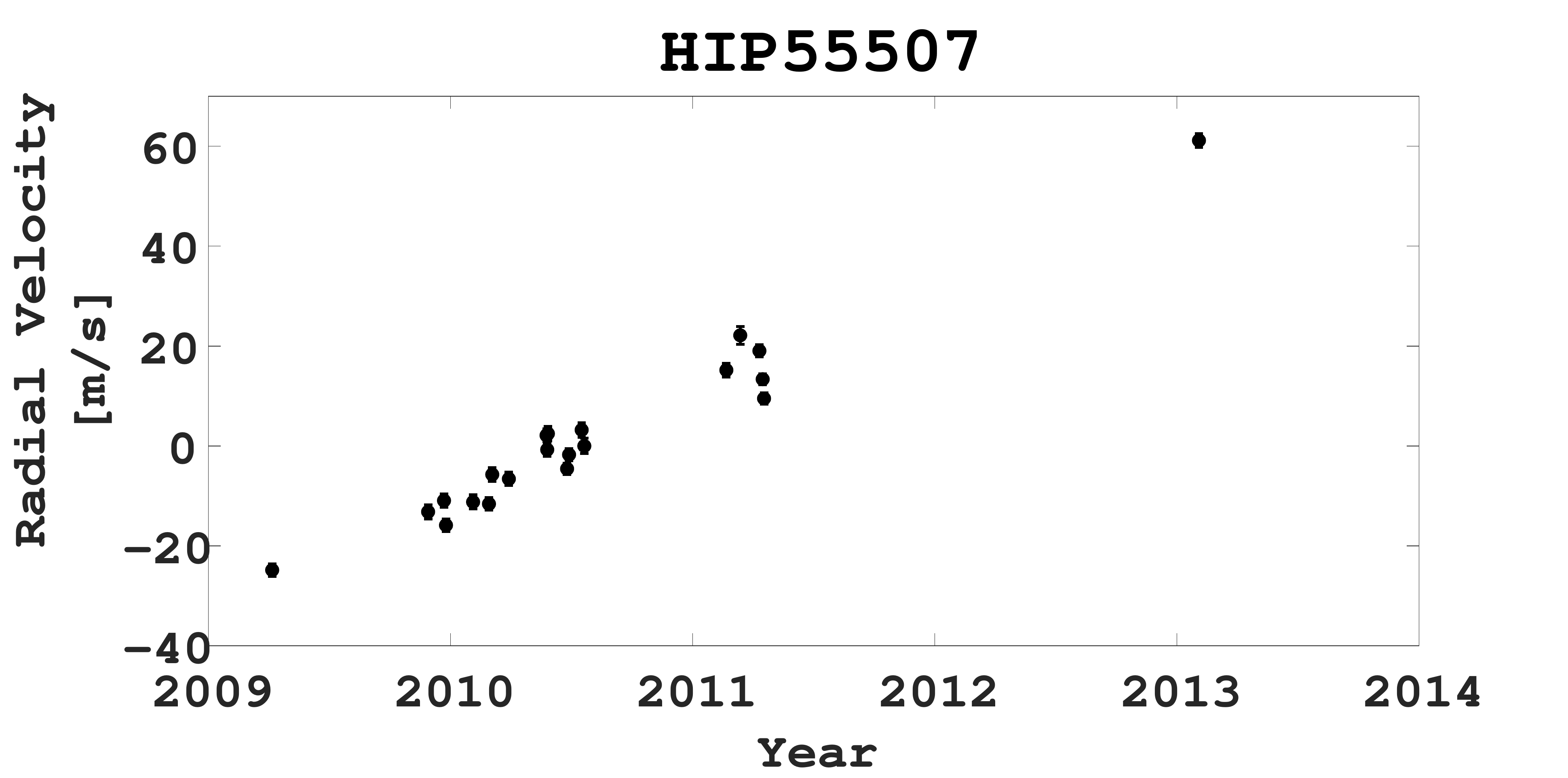}}
\subfloat{\includegraphics[width=.5\textwidth]{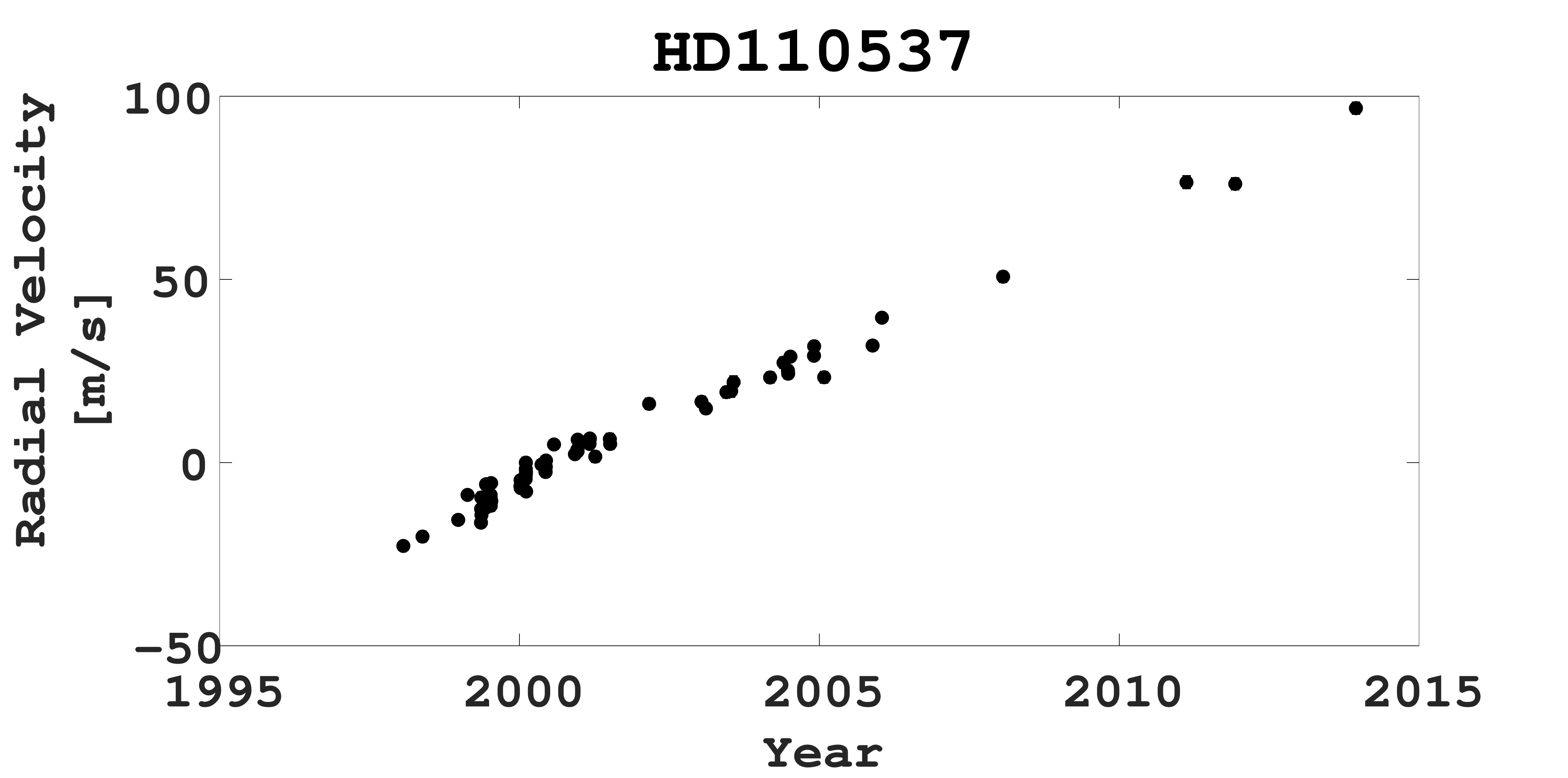}}}\\
\subfloat{\subfloat{\includegraphics[width=.5\textwidth]{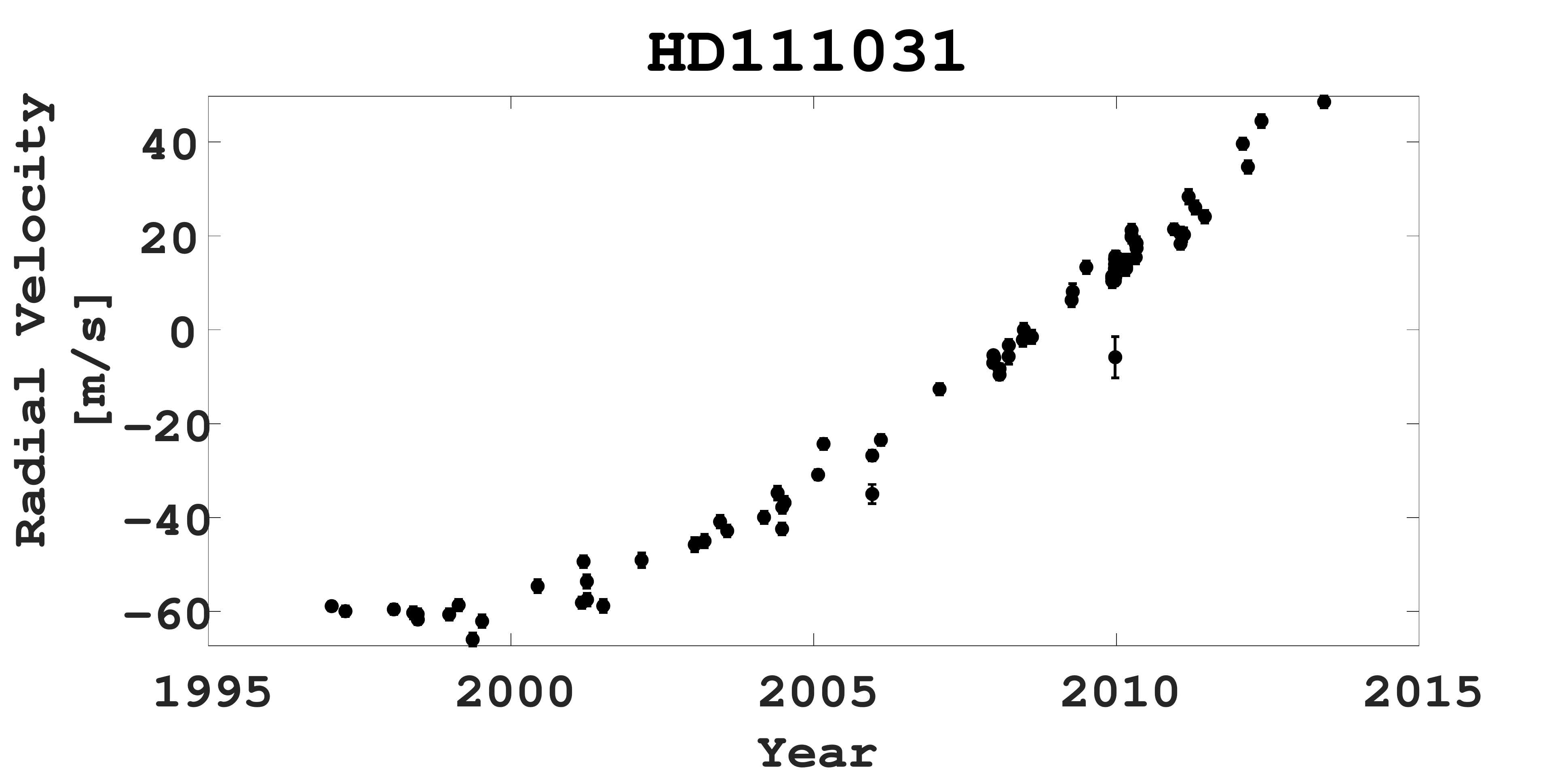}}
\subfloat{\includegraphics[width=.5\textwidth]{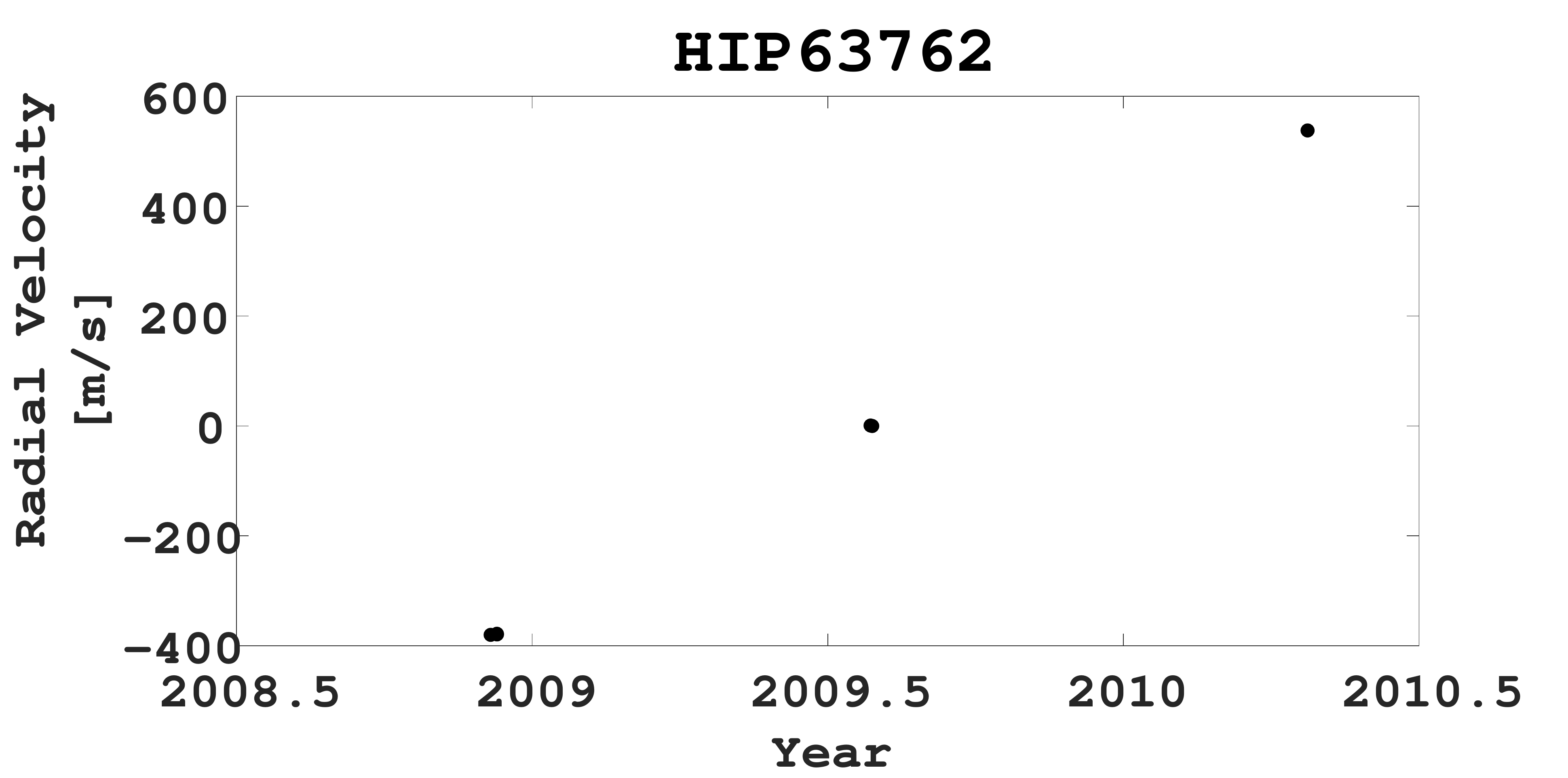}}}\\
\subfloat{\subfloat{\includegraphics[width=.5\textwidth]{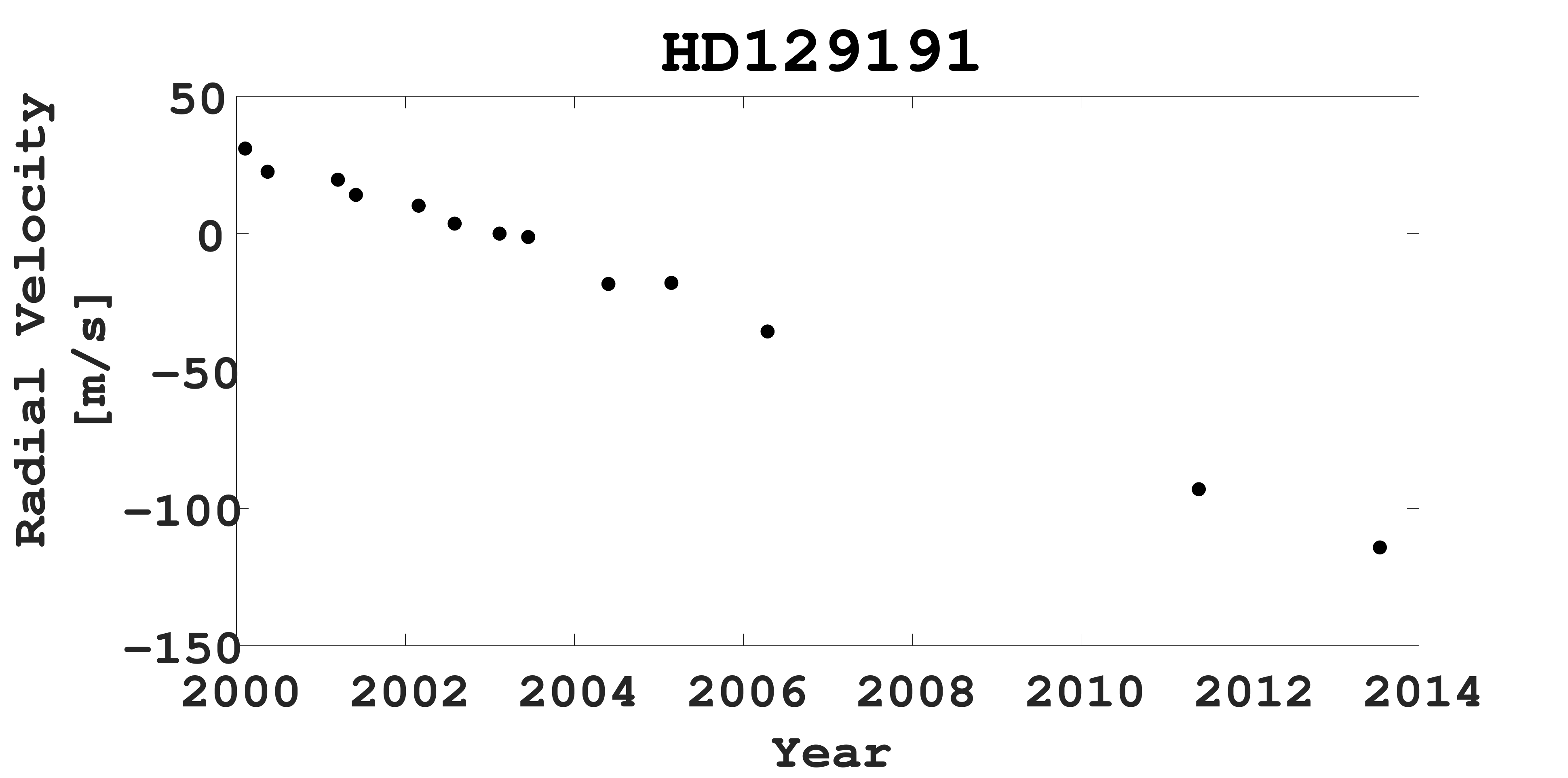}}
\subfloat{\includegraphics[width=.5\textwidth]{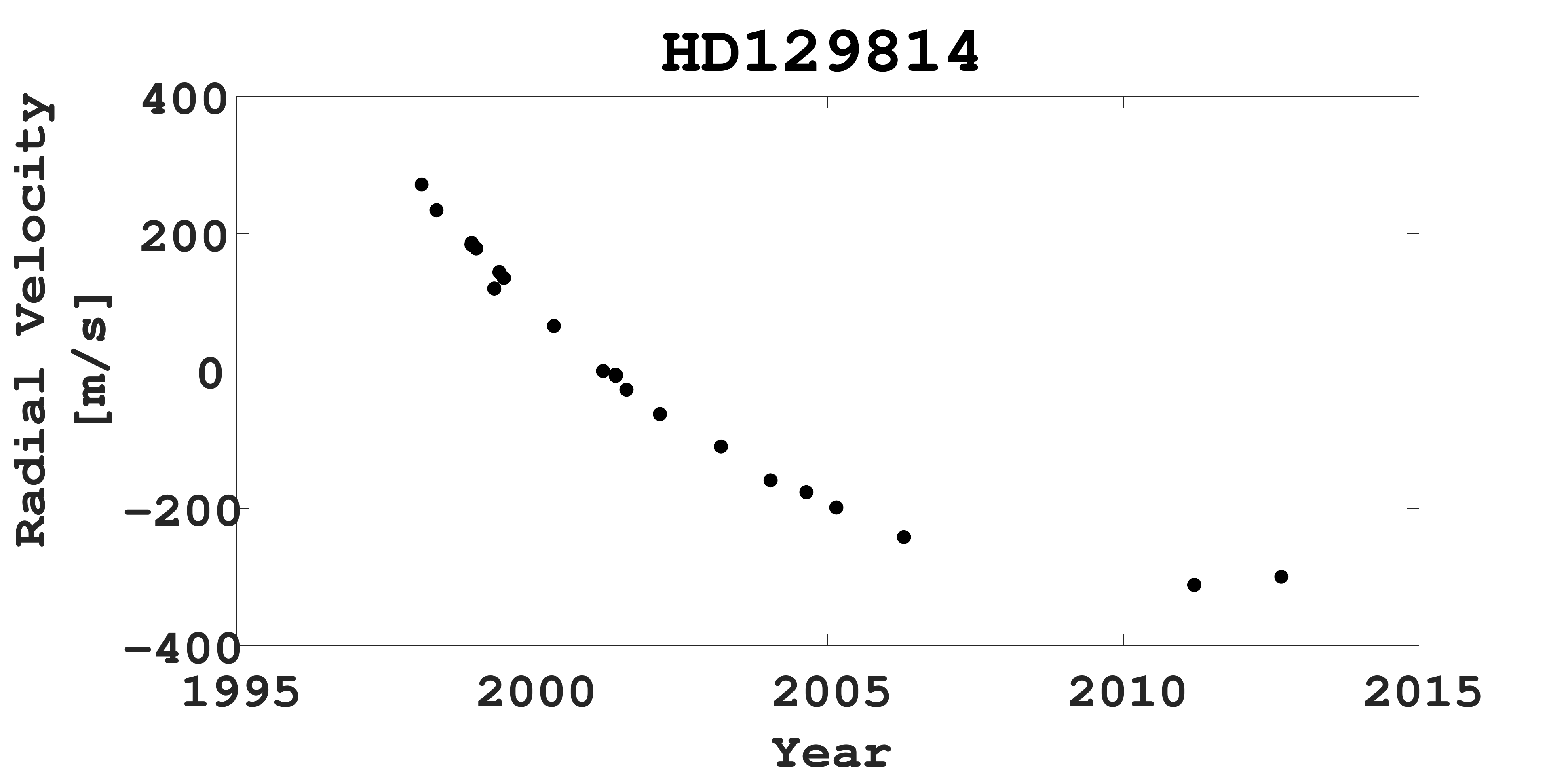}}}\\
\caption{Relative radial velocity plots for stars with confirmed companions. Continuation of Figure \ref{fig7}.}
\label{fig8}
\end{figure}
\clearpage

\begin{figure}[h]
\subfloat{\subfloat{\includegraphics[width=.5\textwidth]{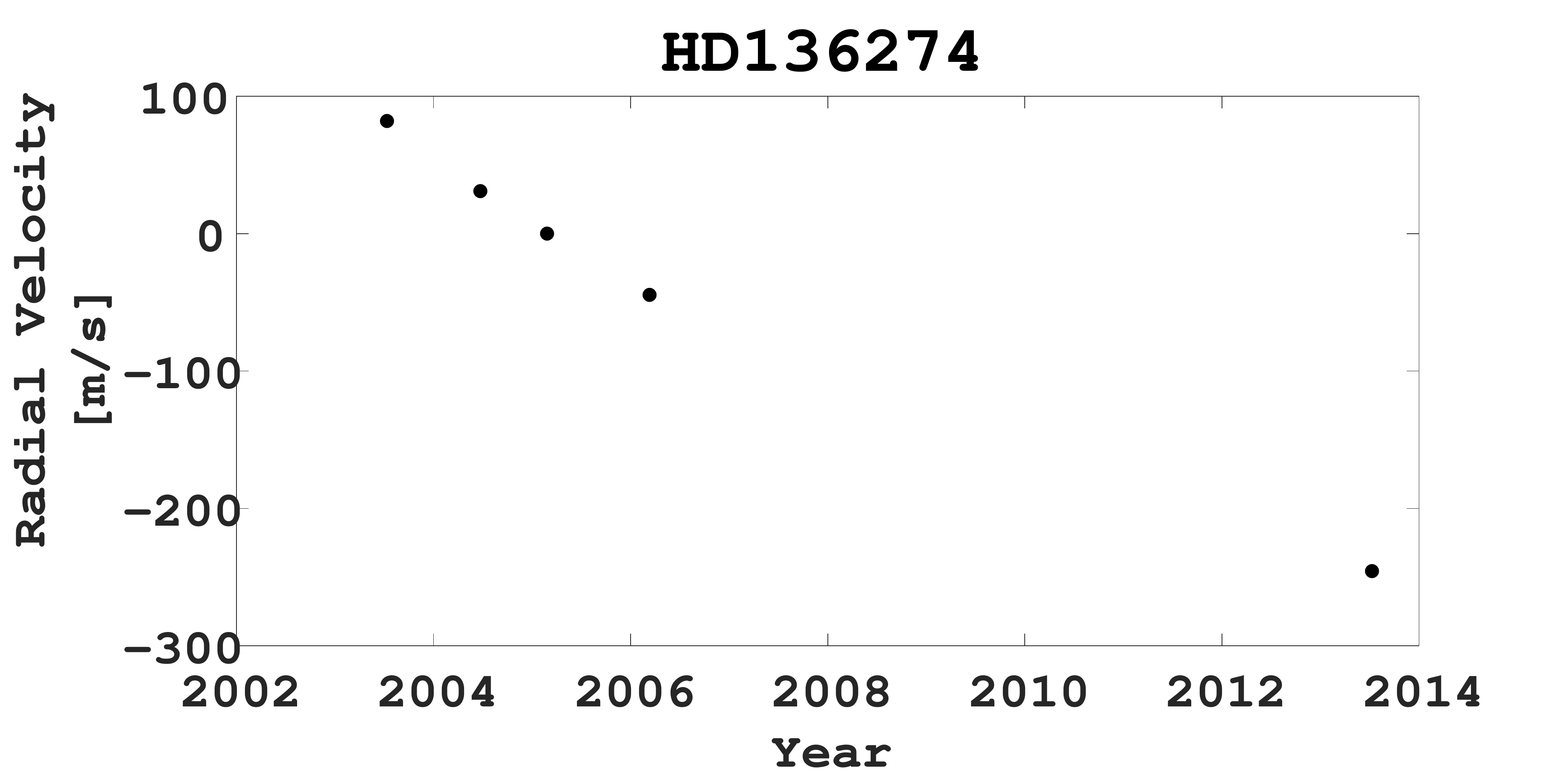}}
\subfloat{\includegraphics[width=.5\textwidth]{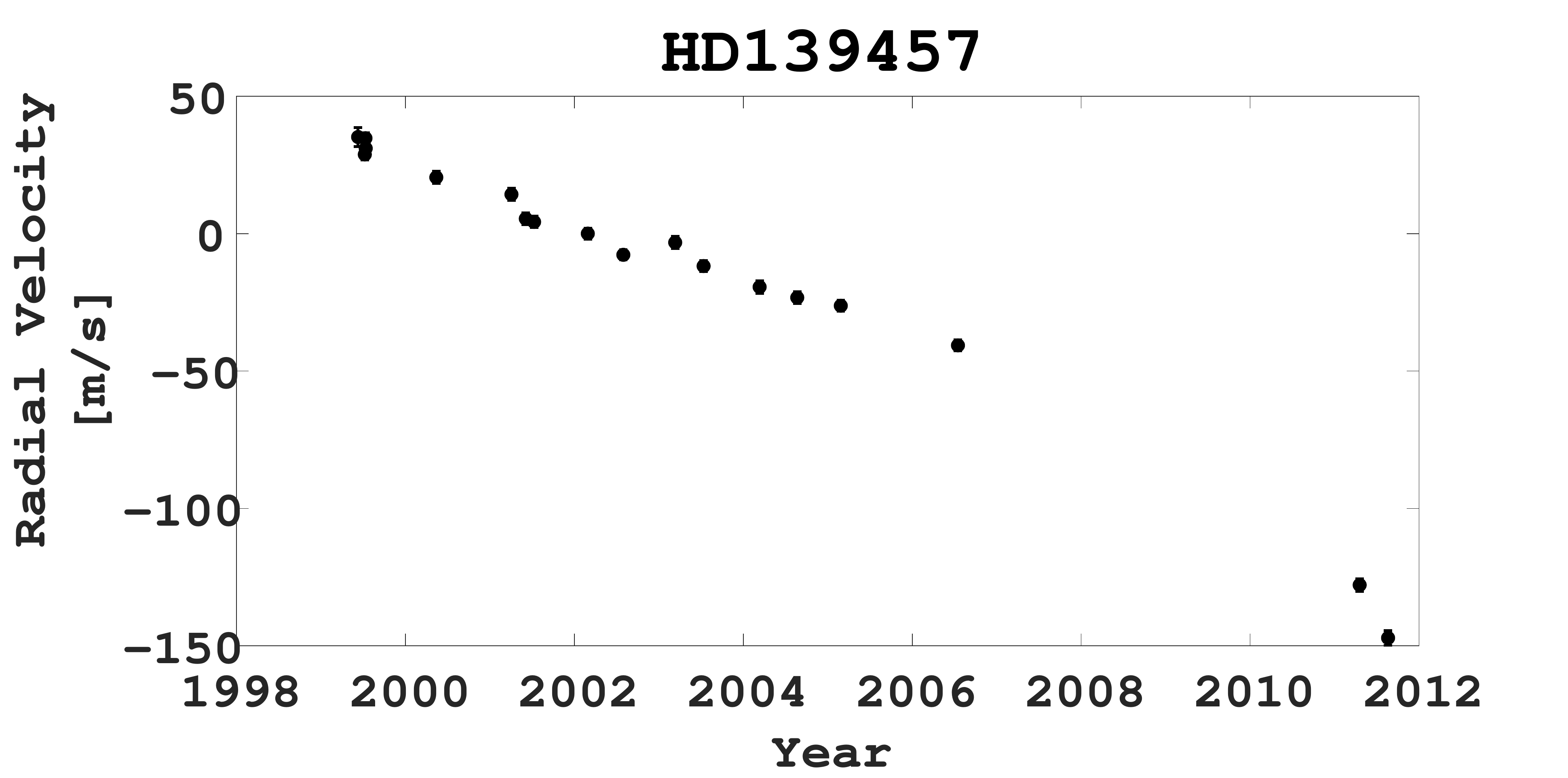}}}\\
\subfloat{\subfloat{\includegraphics[width=.5\textwidth]{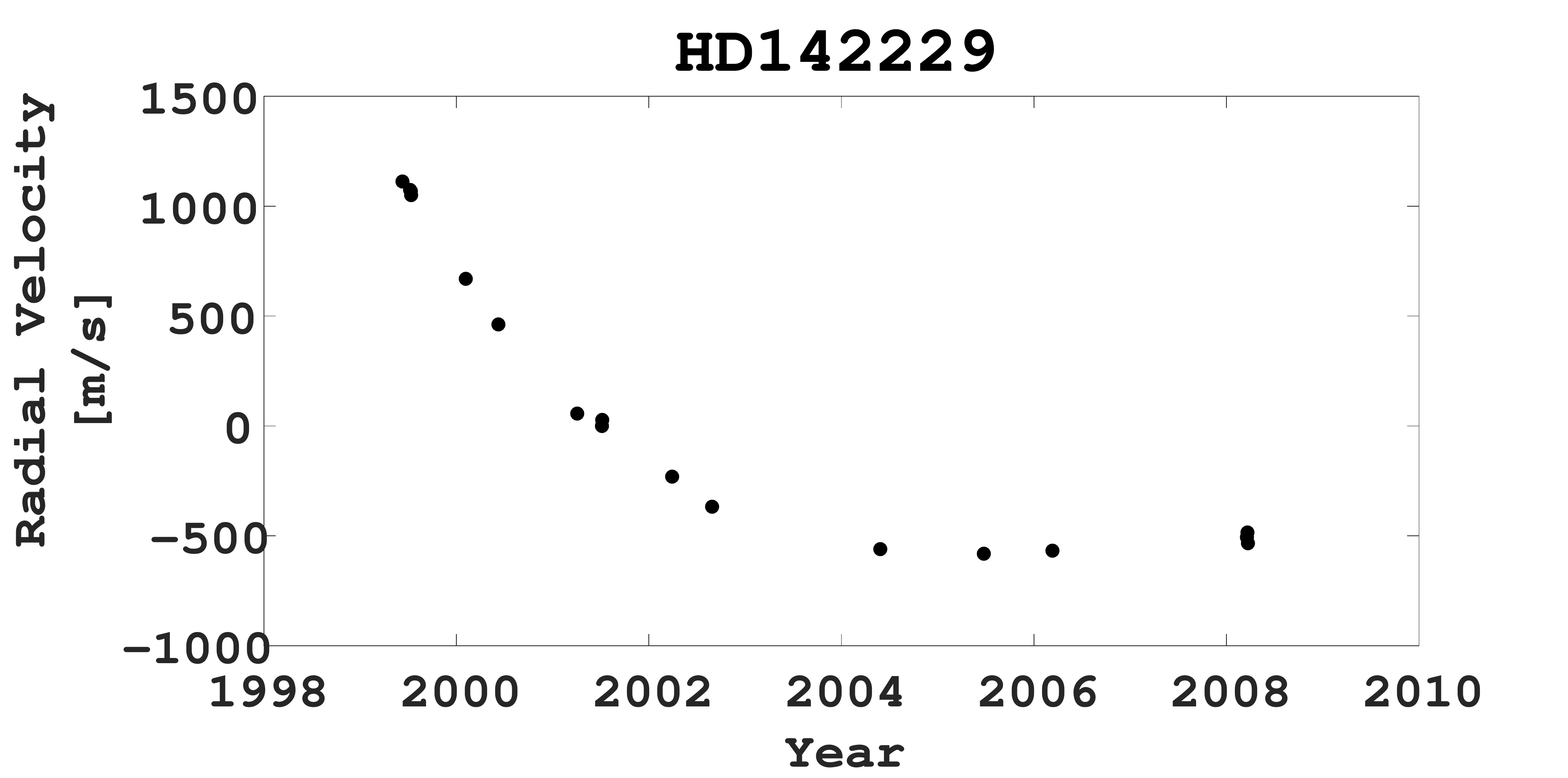}}
\subfloat{\includegraphics[width=.5\textwidth]{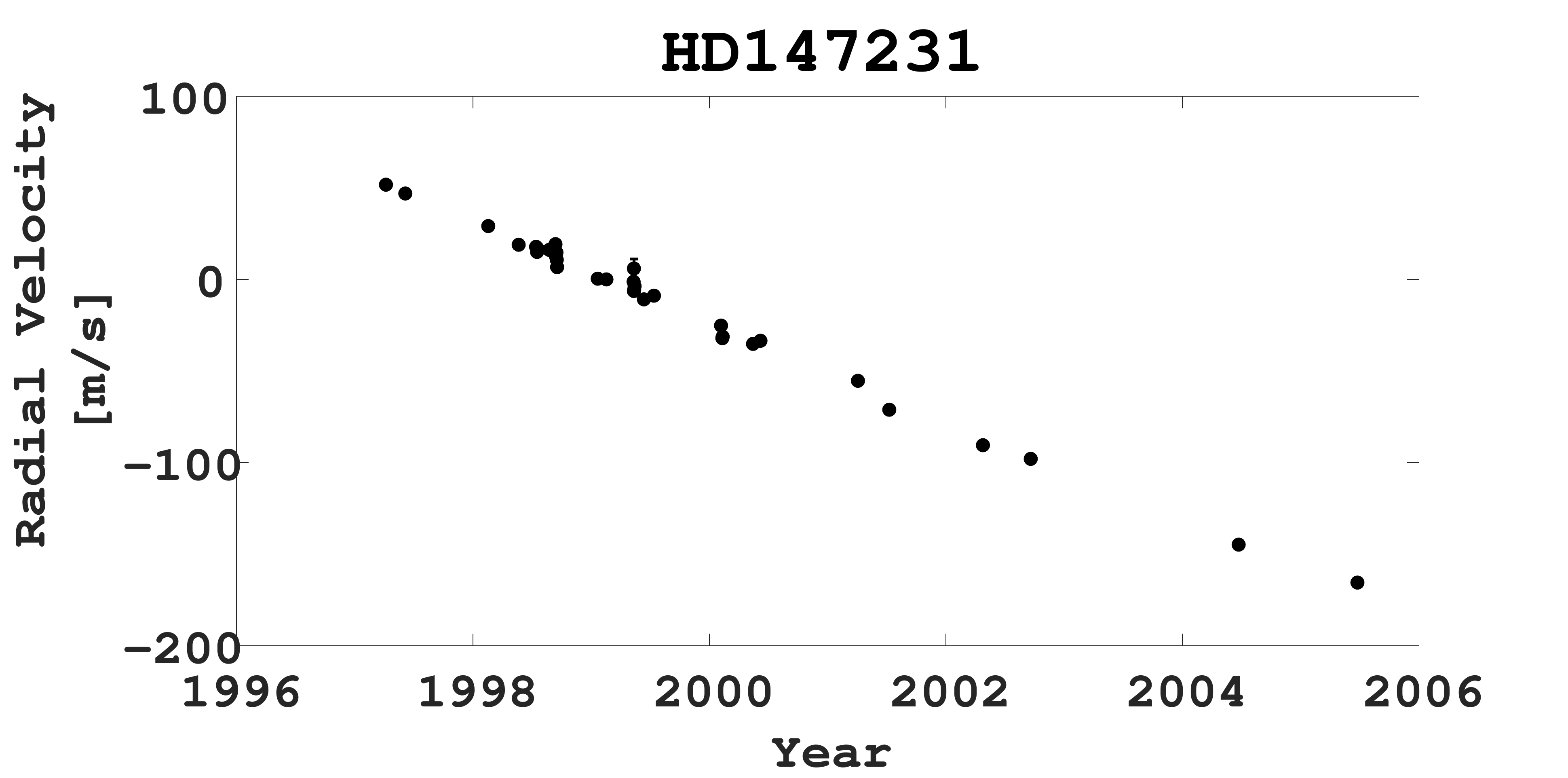}}}\\
\subfloat{\subfloat{\includegraphics[width=.5\textwidth]{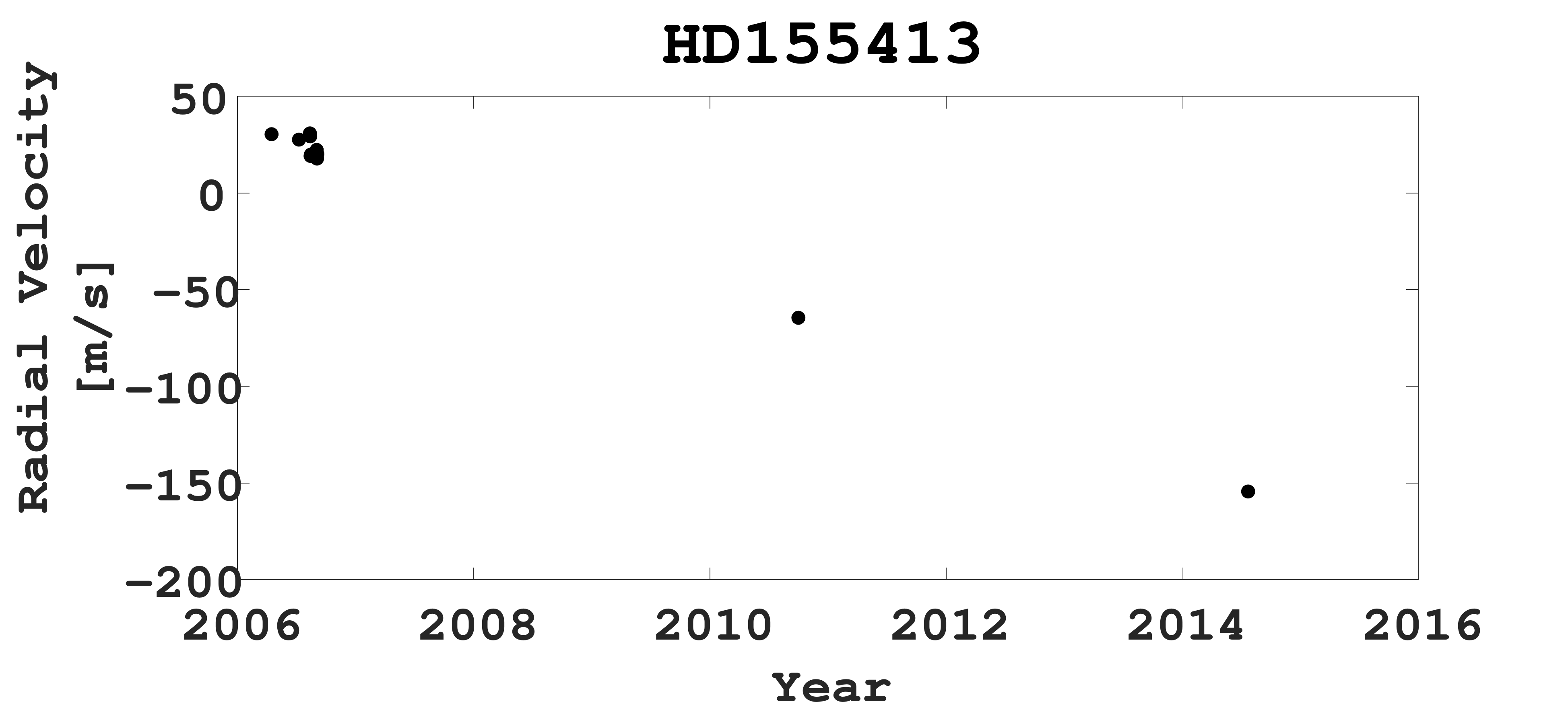}}
\subfloat{\includegraphics[width=.5\textwidth]{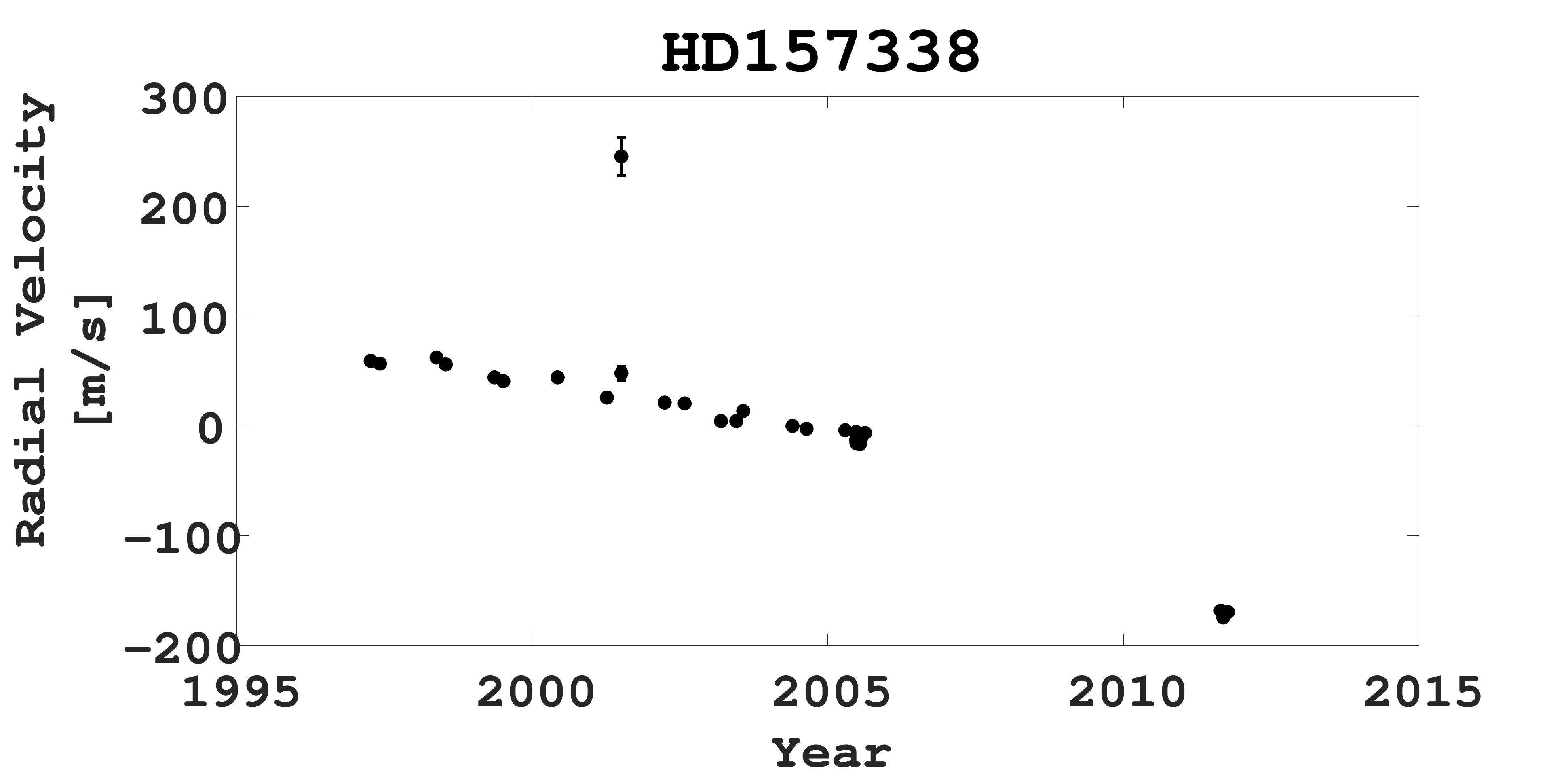}}}\\
\subfloat{\subfloat{\includegraphics[width=.5\textwidth]{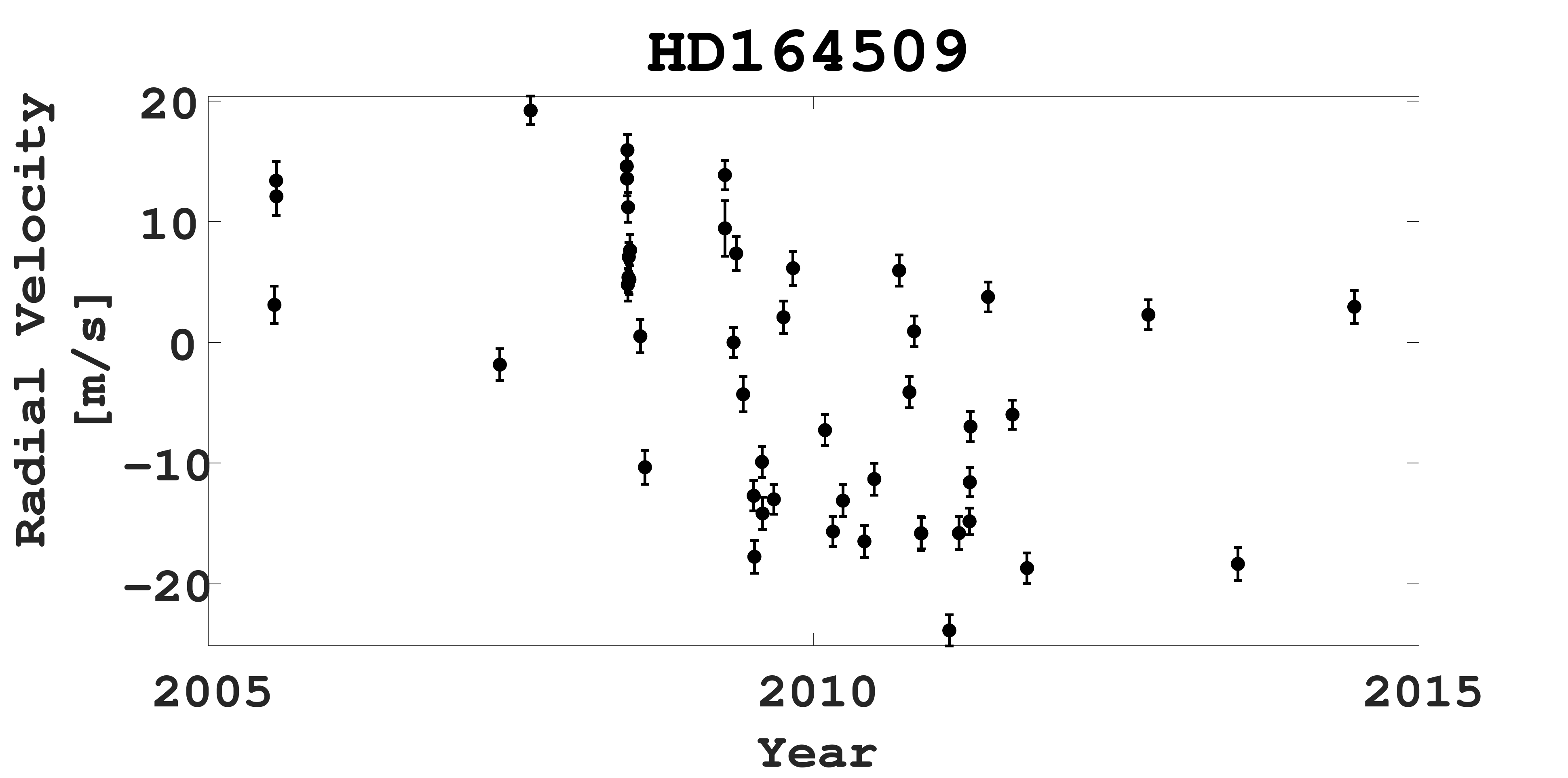}}
\subfloat{\includegraphics[width=.5\textwidth]{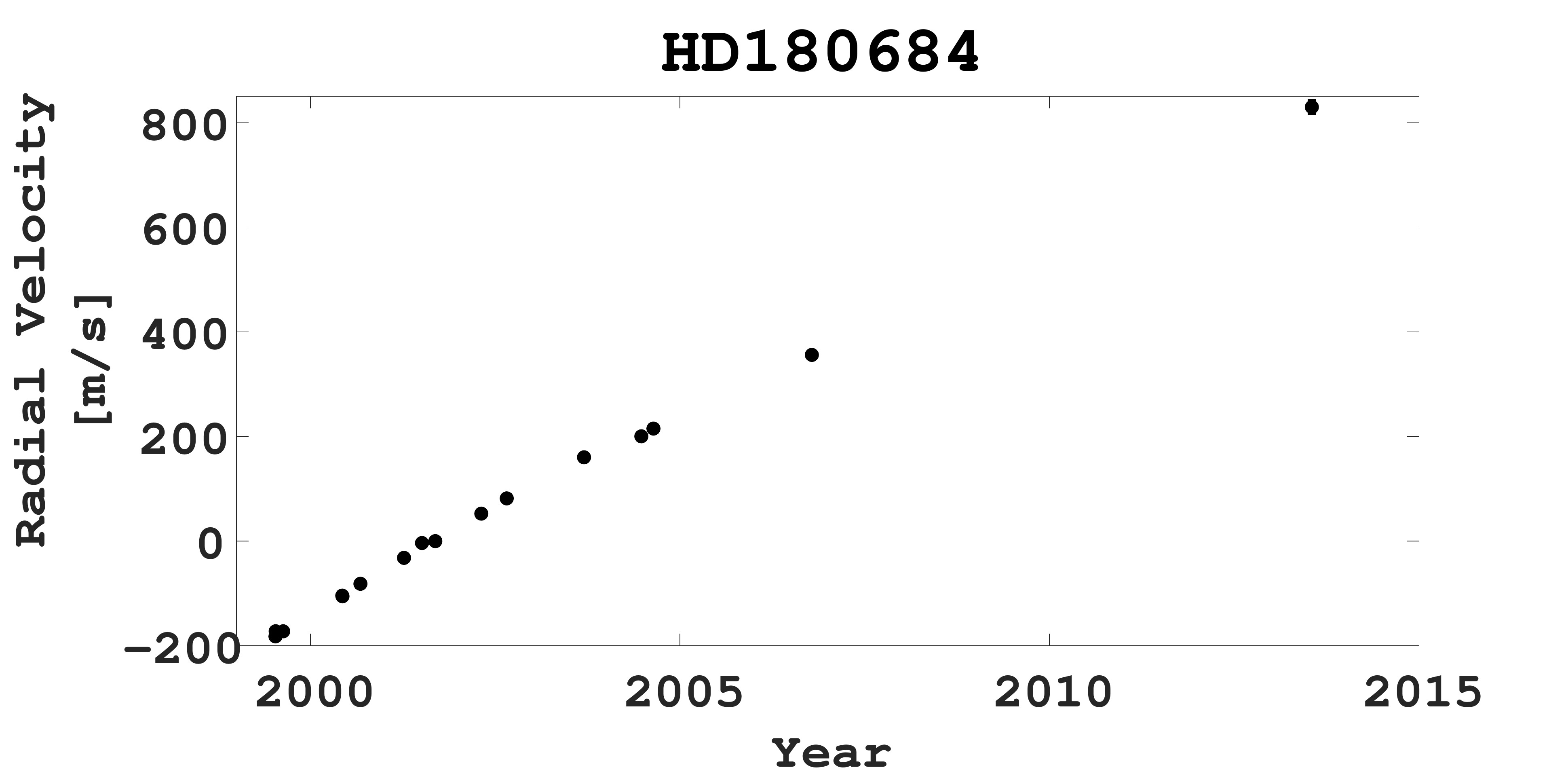}}}\\
\caption{Relative radial velocity plots for stars with confirmed companions.Continuation of Figure \ref{fig8}.}
\label{fig9}
\end{figure}
\clearpage

\begin{figure}[h]
\subfloat{\subfloat{\includegraphics[width=.5\textwidth]{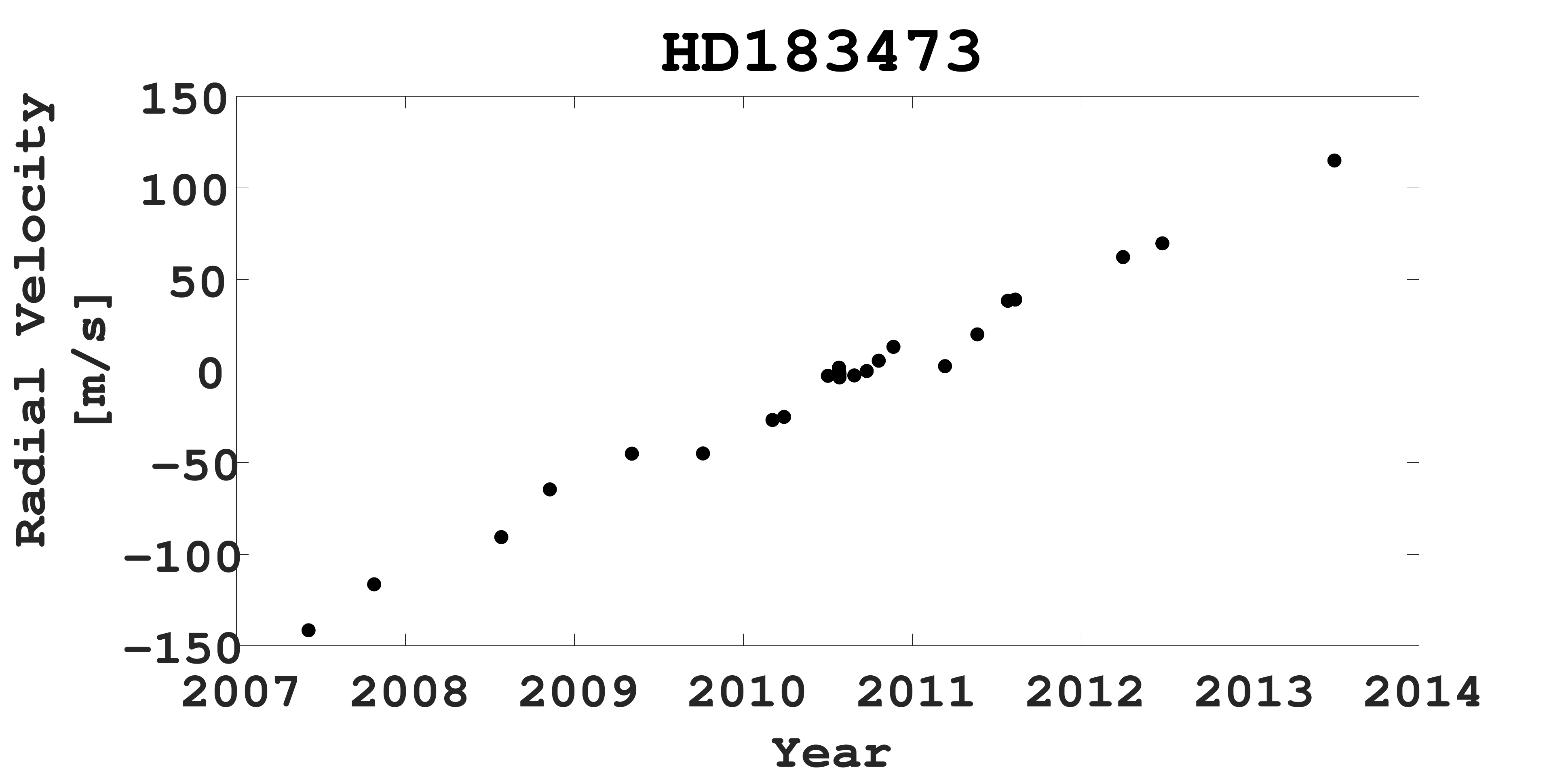}}
\subfloat{\includegraphics[width=.5\textwidth]{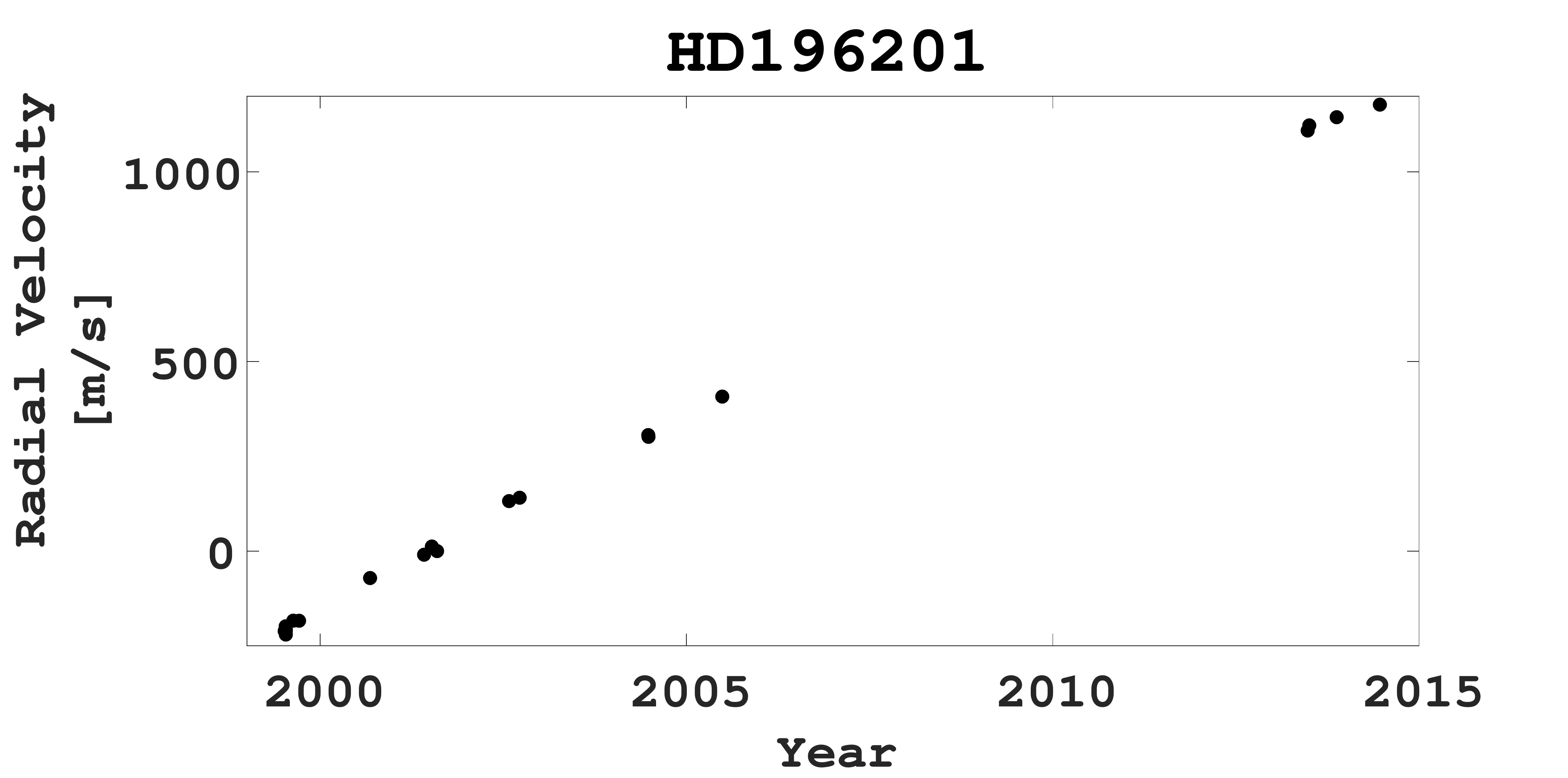}}}\\
\subfloat{\subfloat{\includegraphics[width=.5\textwidth]{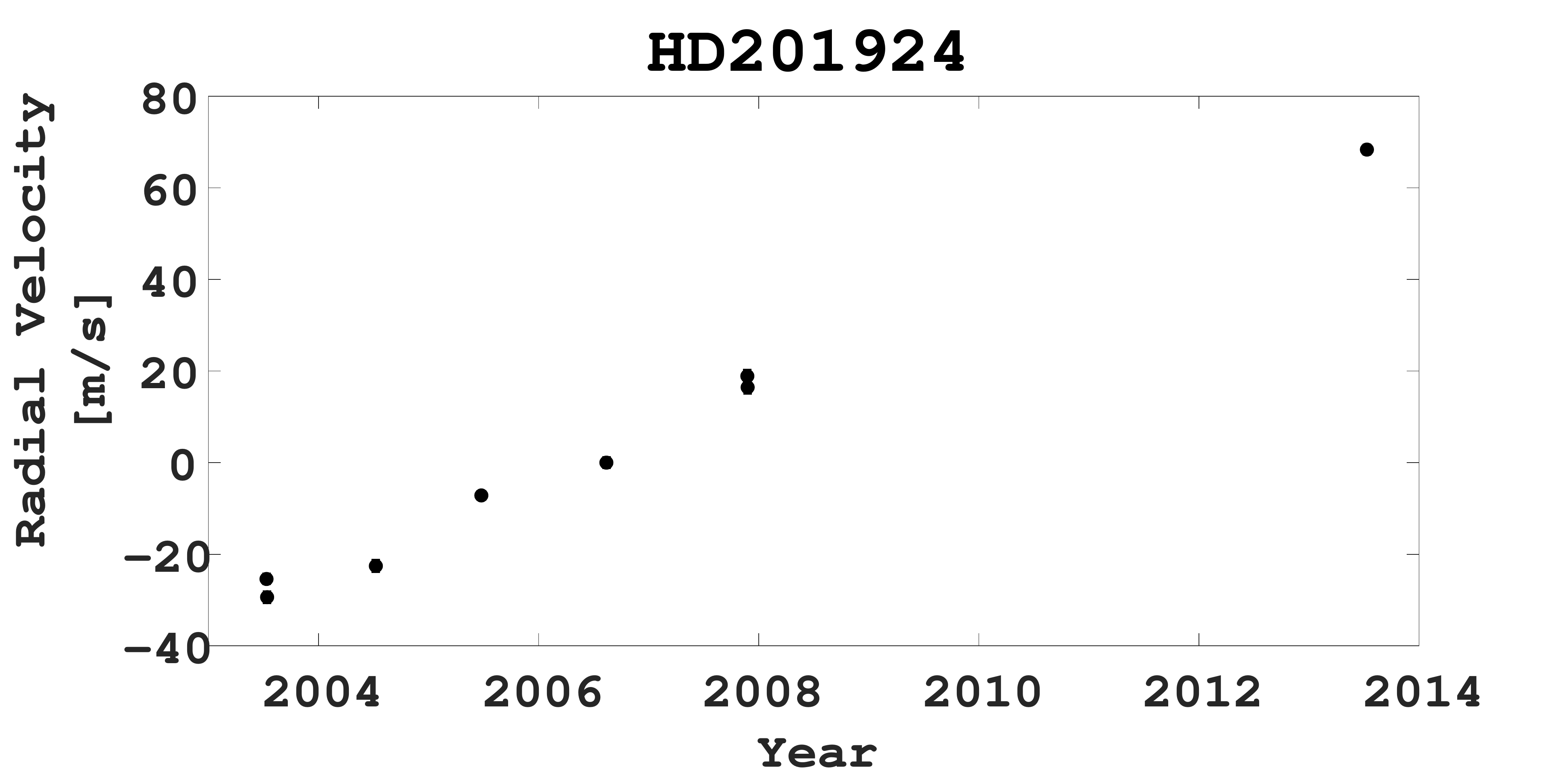}}
\subfloat{\includegraphics[width=.5\textwidth]{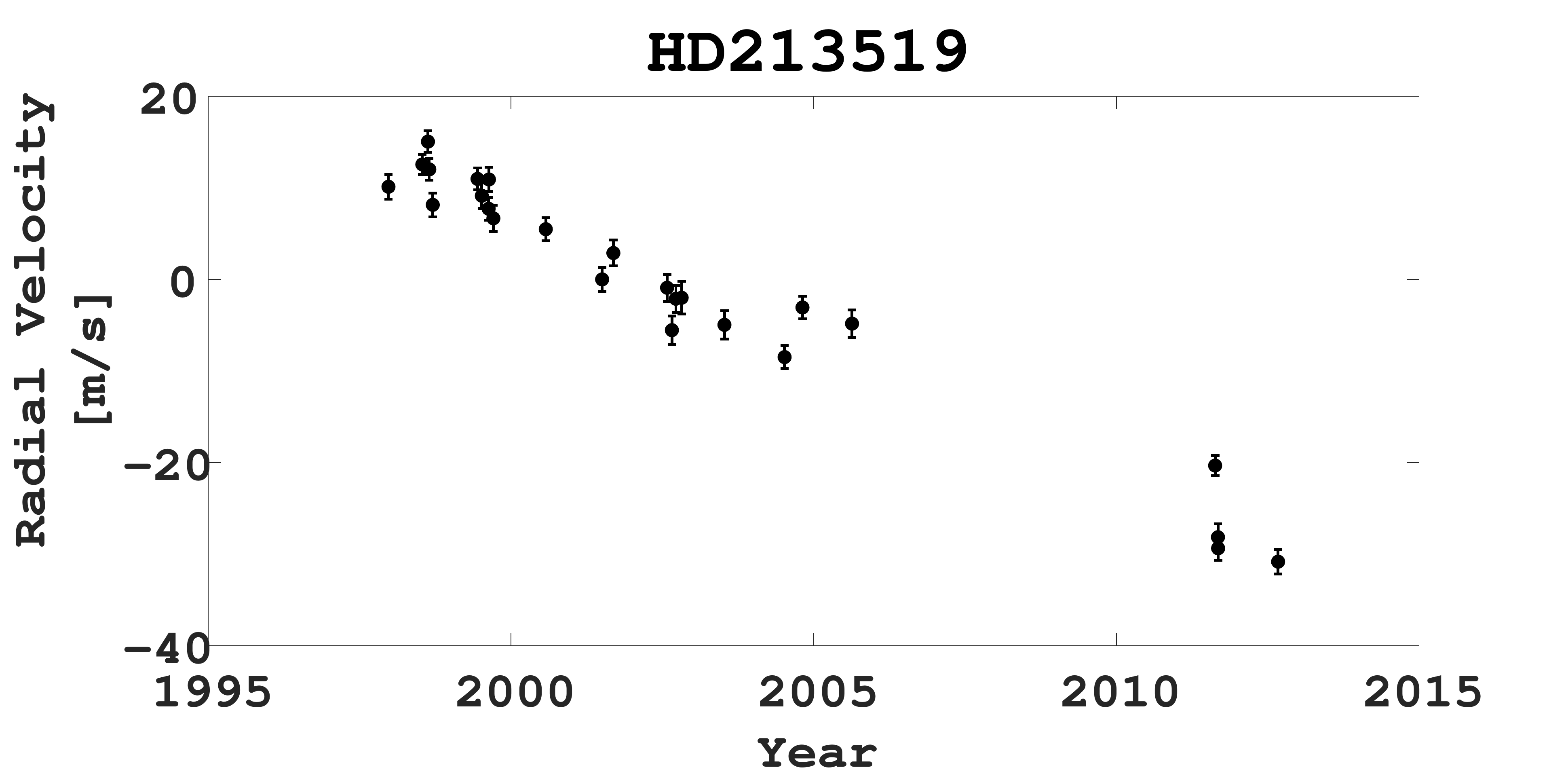}}}\\
\caption{Relative radial velocity plots for stars with confirmed companions.Continuation of Figure \ref{fig9}.}
\label{fig10}
\end{figure}

\begin{figure*}[h]
\subfloat{\subfloat{\includegraphics[width=.5\textwidth]{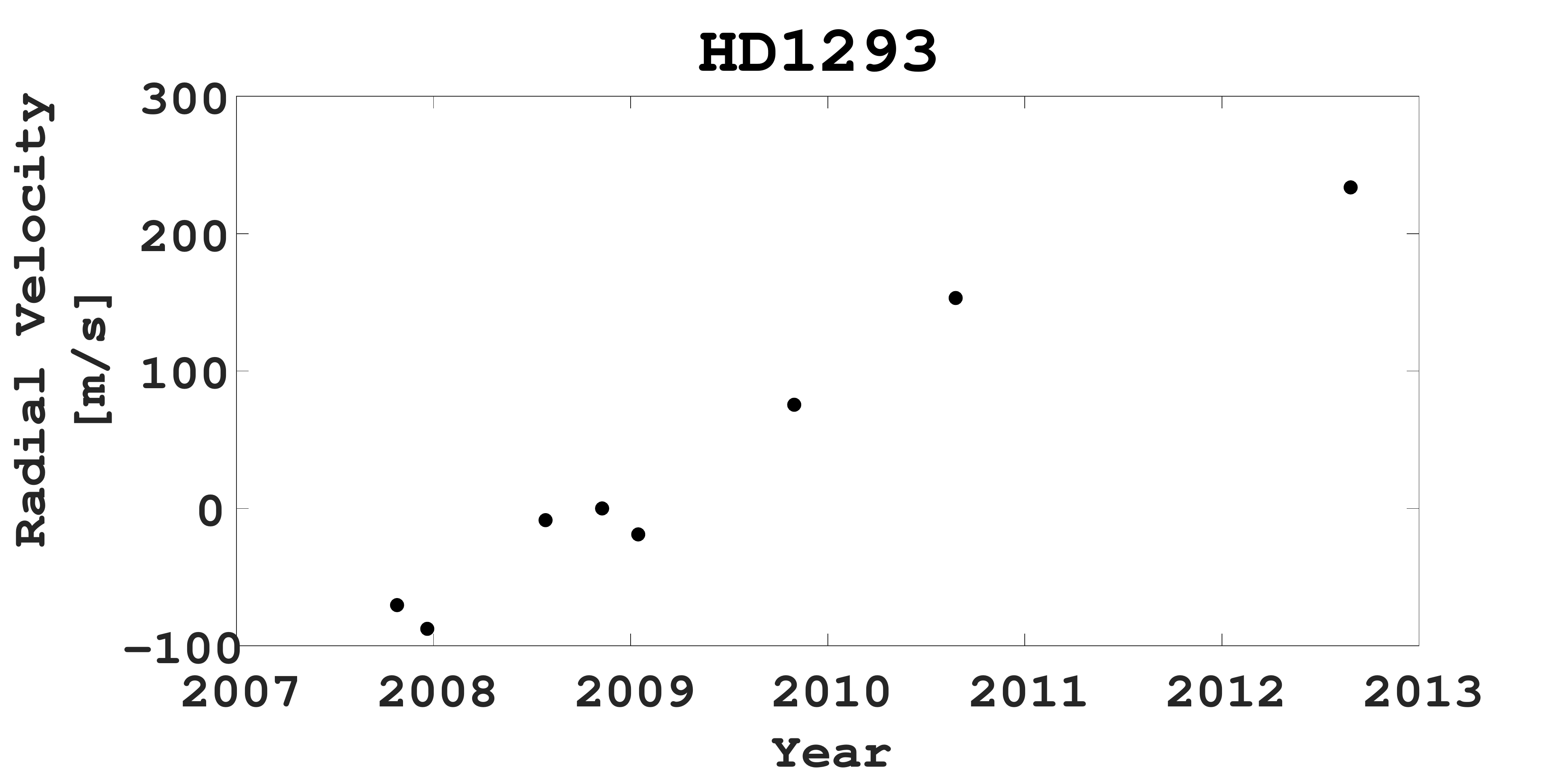}}
\subfloat{\includegraphics[width=.5\textwidth]{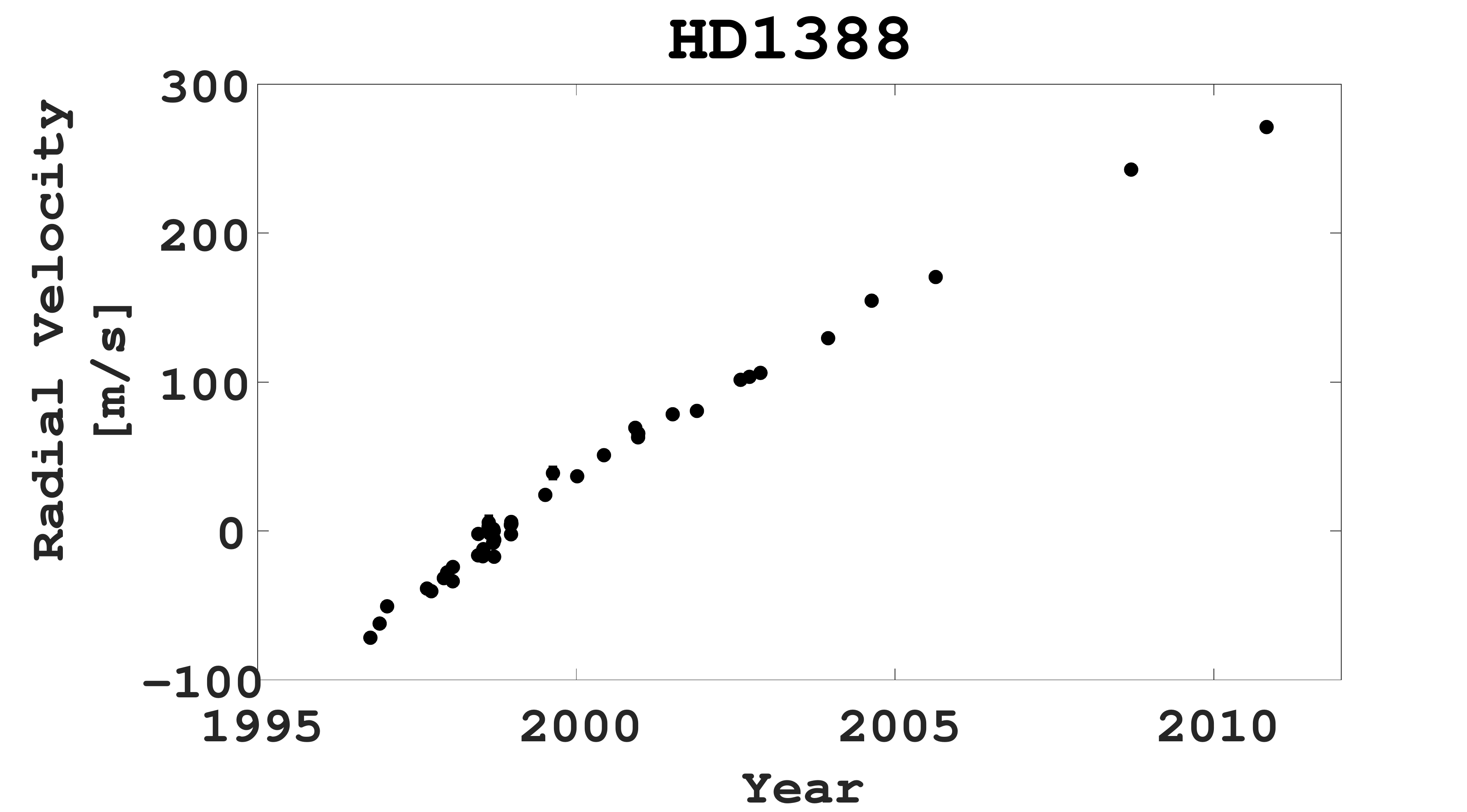}}}\\
\subfloat{\subfloat{\includegraphics[width=.5\textwidth]{4406_RV_plot.pdf}}
\subfloat{\includegraphics[width=.5\textwidth]{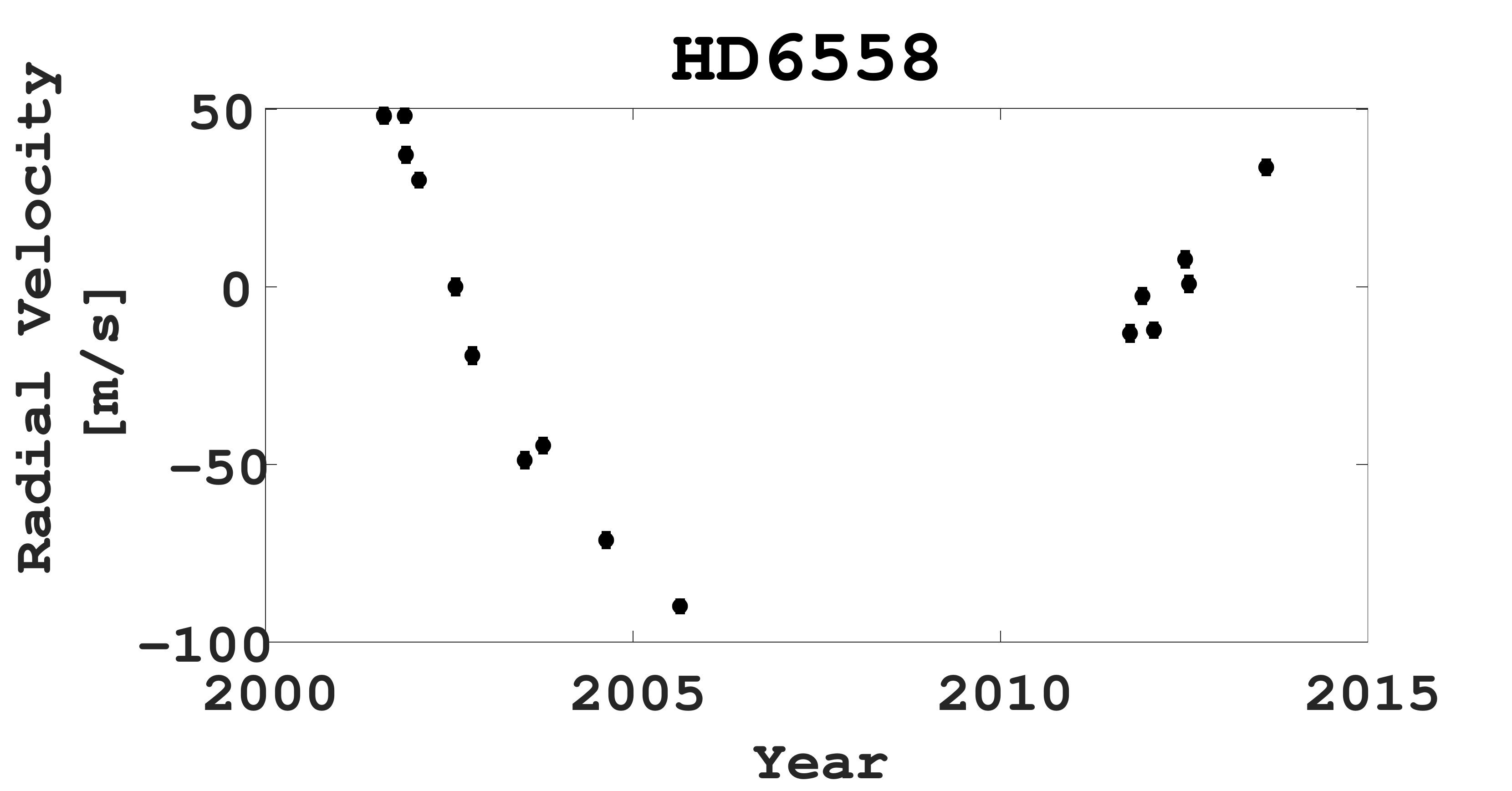}}}\\
\subfloat{\subfloat{\includegraphics[width=.5\textwidth]{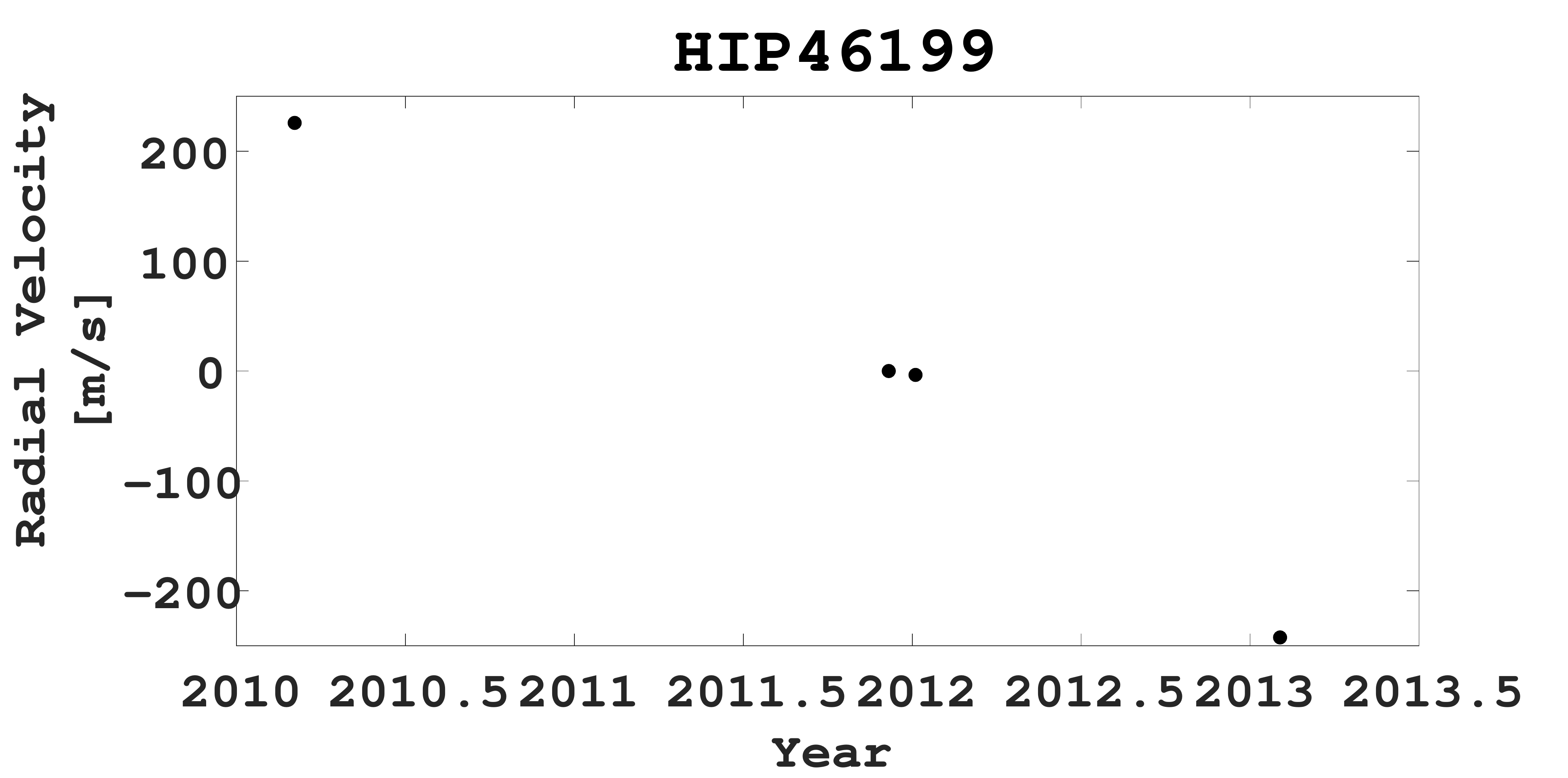}}
\subfloat{\includegraphics[width=.5\textwidth]{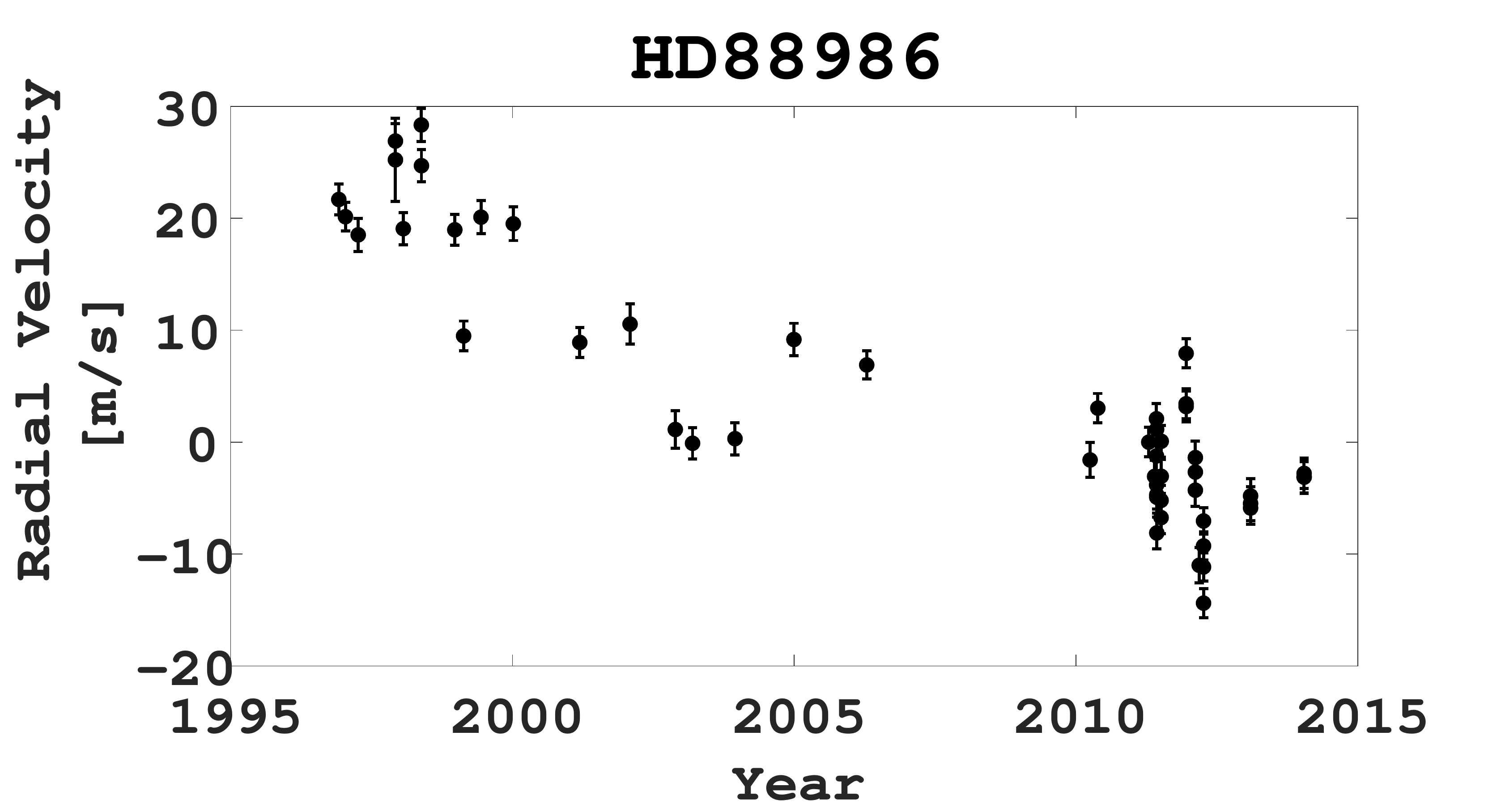}}}\\
\subfloat{\subfloat{\includegraphics[width=.5\textwidth]{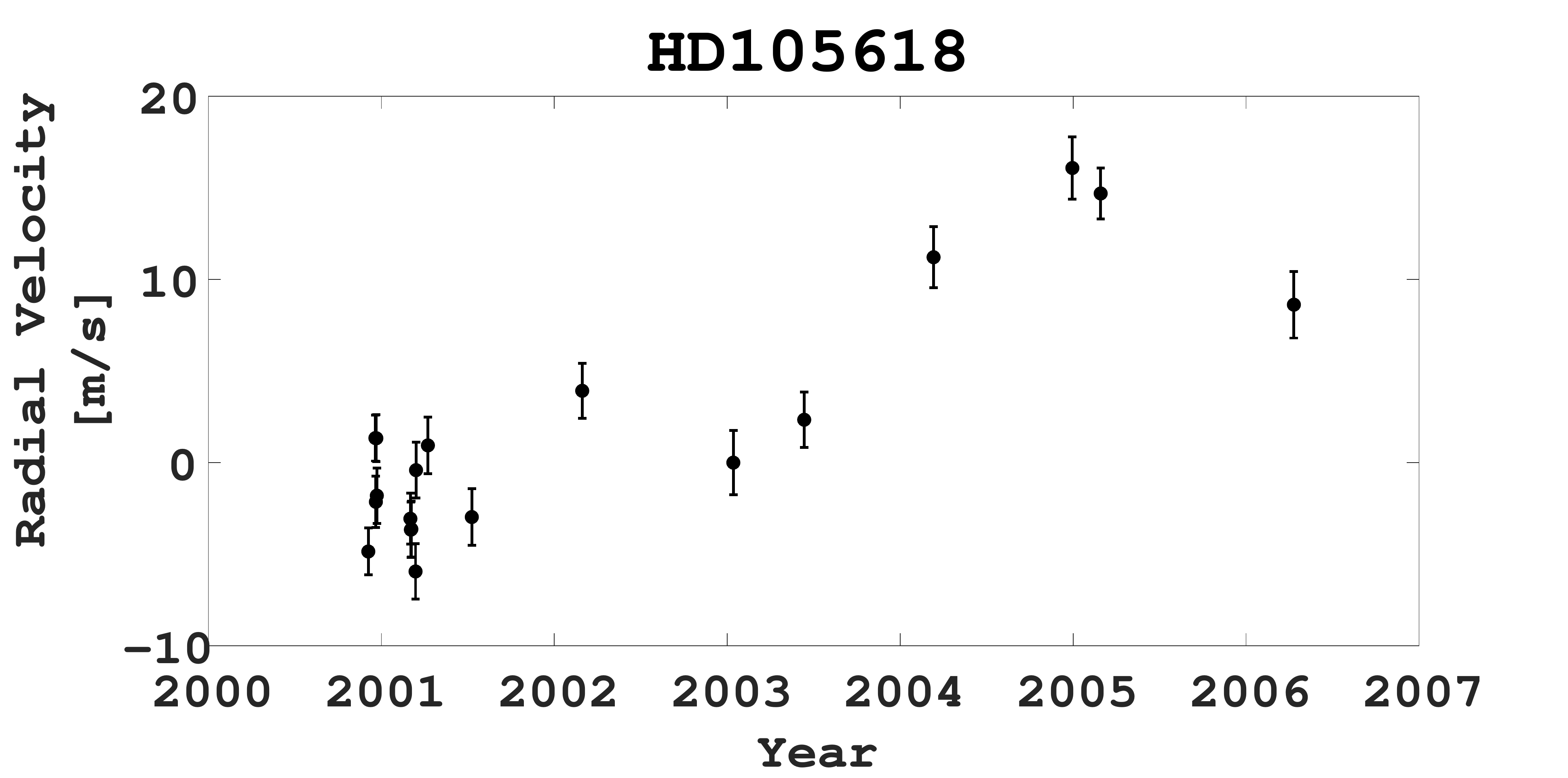}}
\subfloat{\includegraphics[width=.5\textwidth]{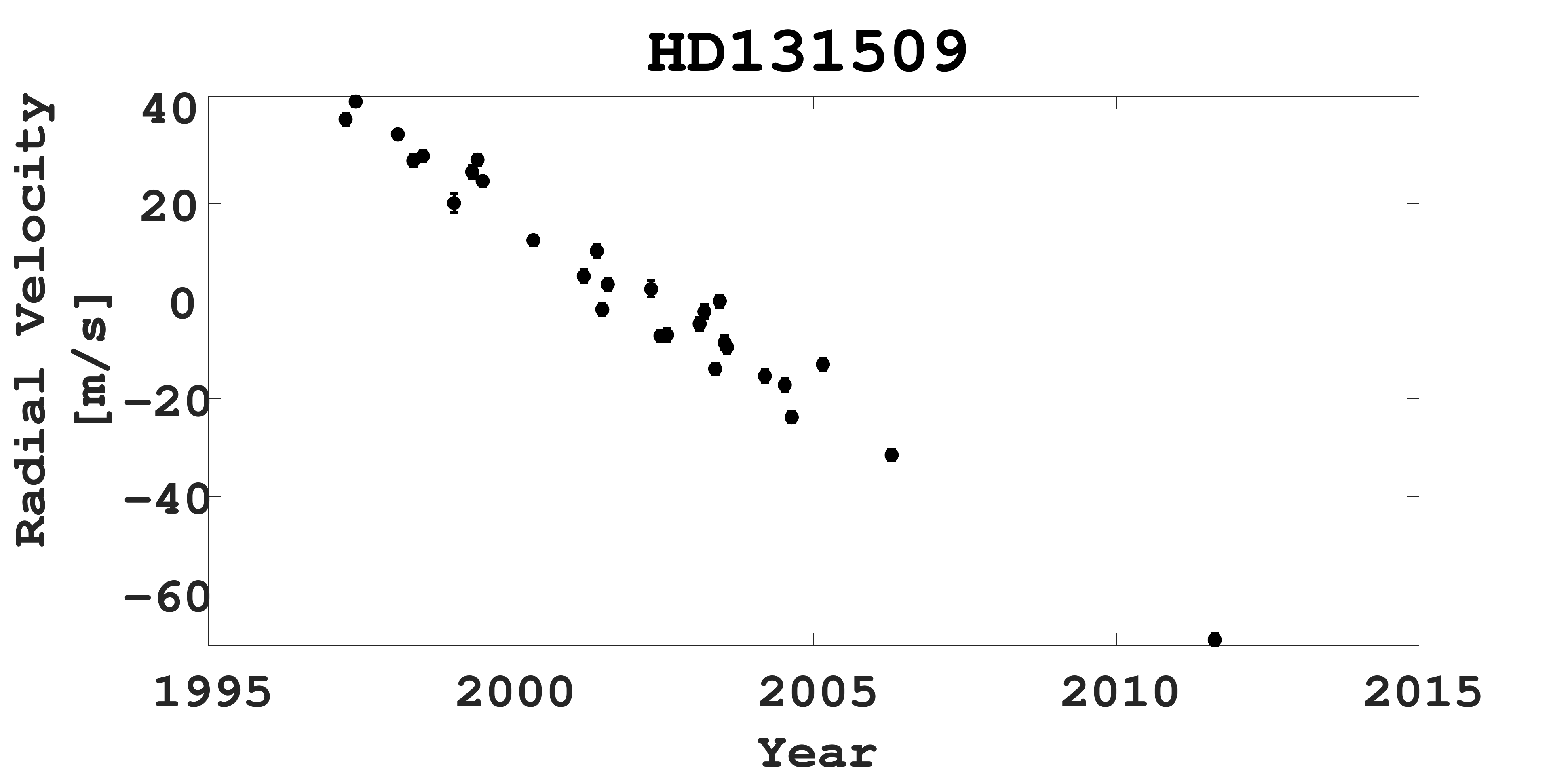}}}\\
\caption{Relative radial velocity plots for stars with candidate companions.}
\label{fig11}
\end{figure*}

\begin{figure}
\subfloat{\subfloat{\includegraphics[width=.5\textwidth]{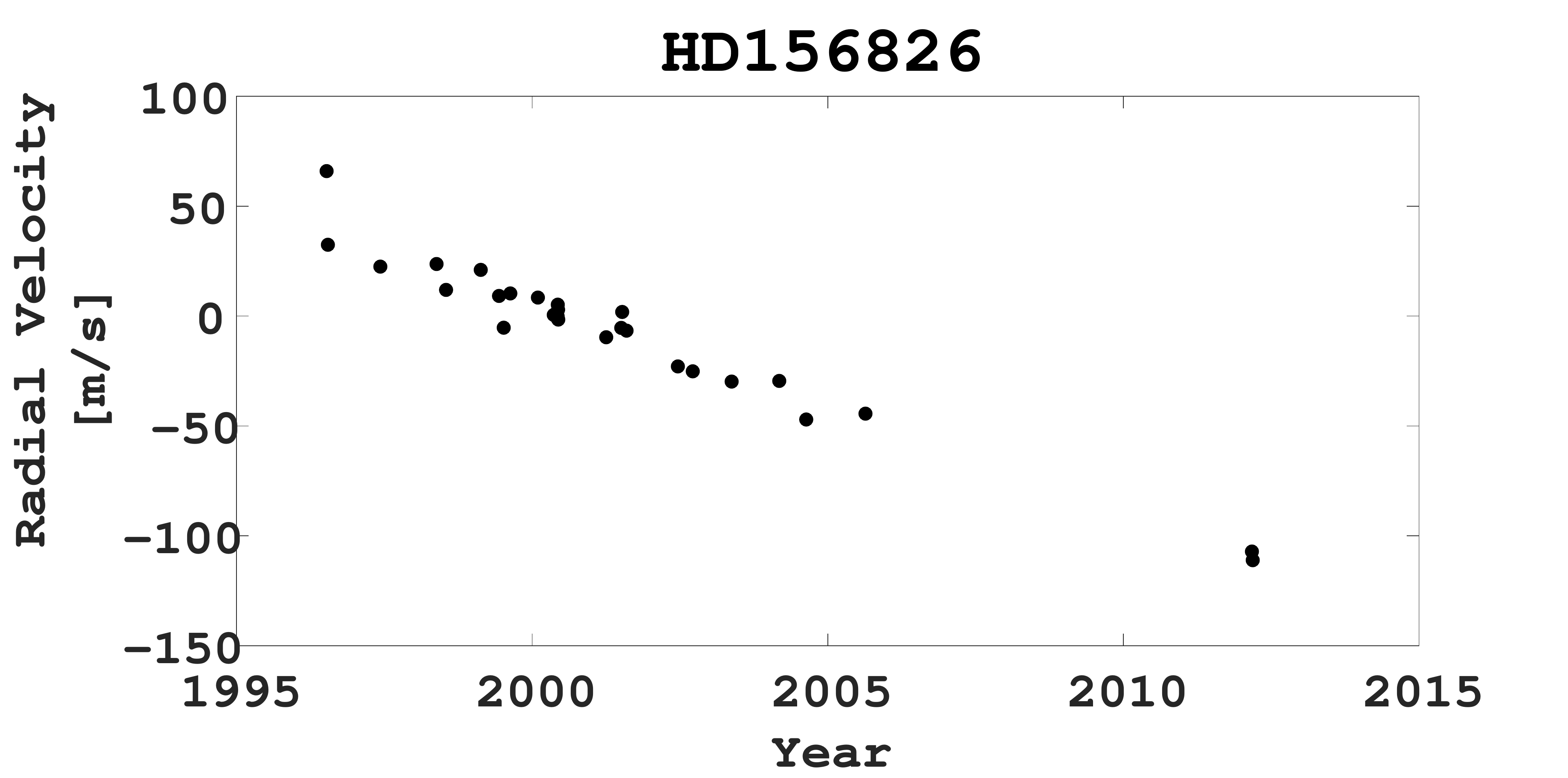}}
\subfloat{\includegraphics[width=.5\textwidth]{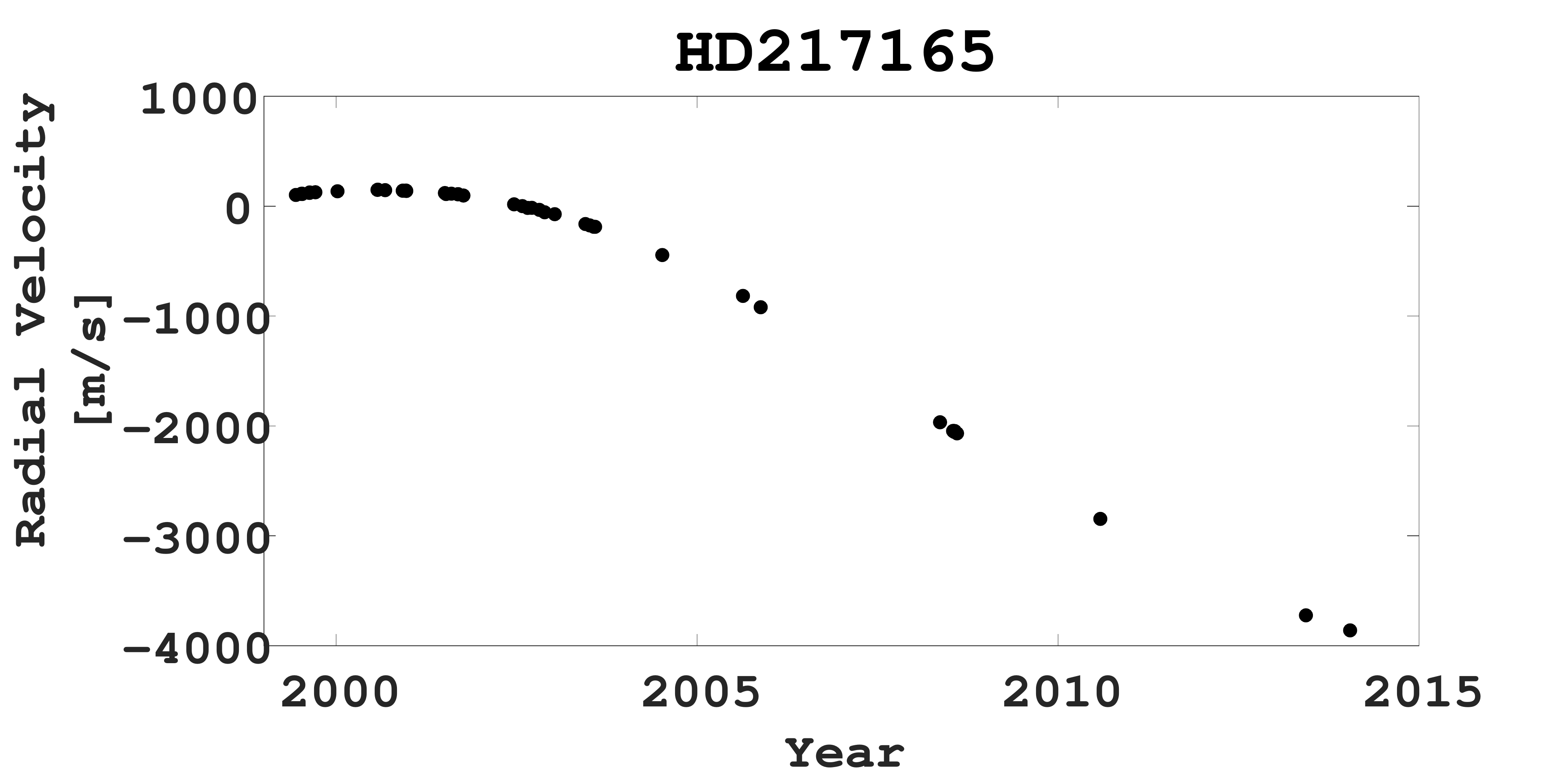}}}\\
\caption{Relative radial velocity plots for stars with confirmed companions.Continuation of Figure \ref{fig11}.}
\label{fig12}
\end{figure}
\clearpage

\subsection{High Contrast Images of Compendium Objects.}

\begin{figure}[ht]
 \subfloat{\subfloat{\includegraphics[width = .5\textwidth]{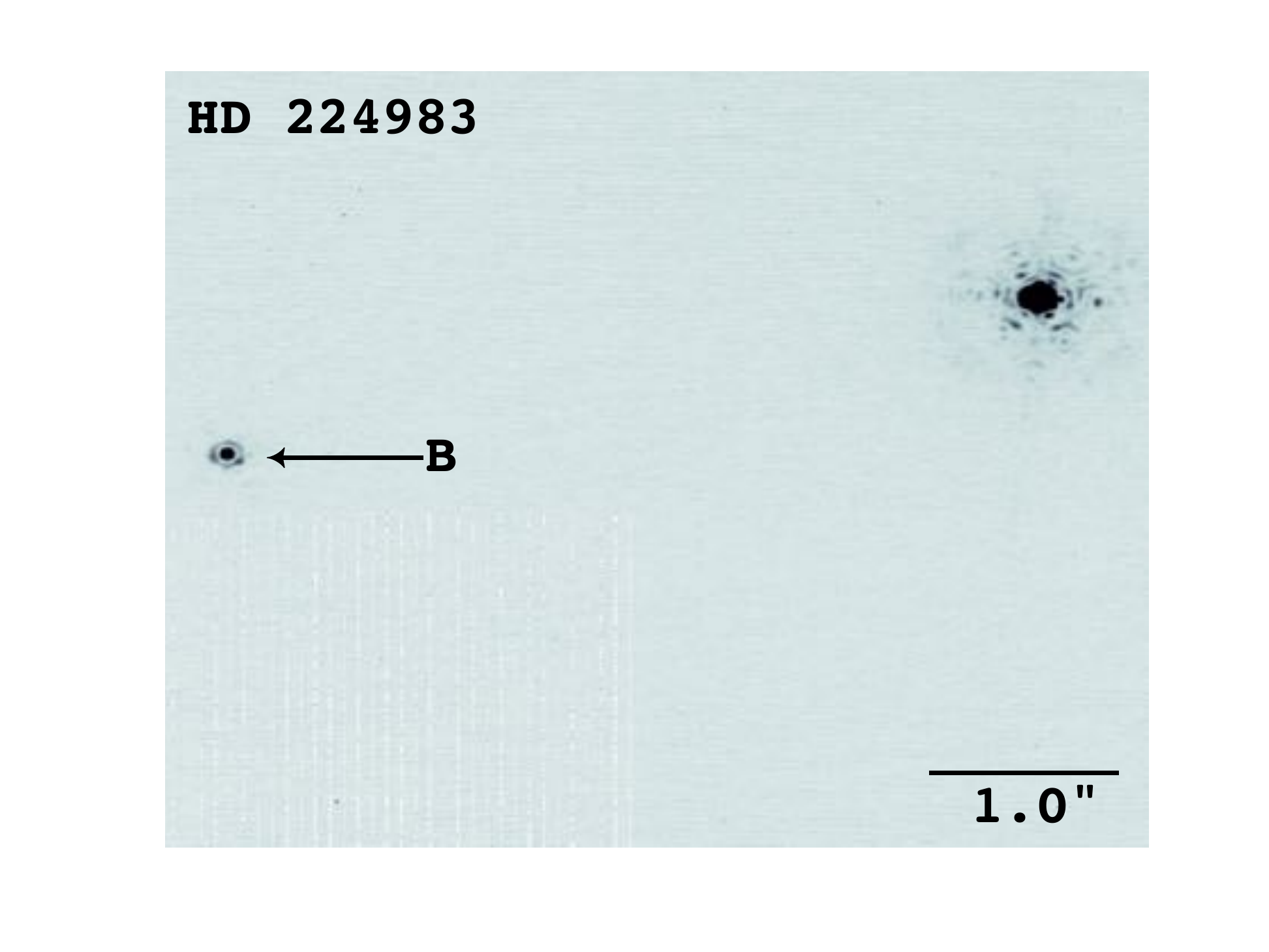}}
 \subfloat{\includegraphics[width= .5\textwidth]{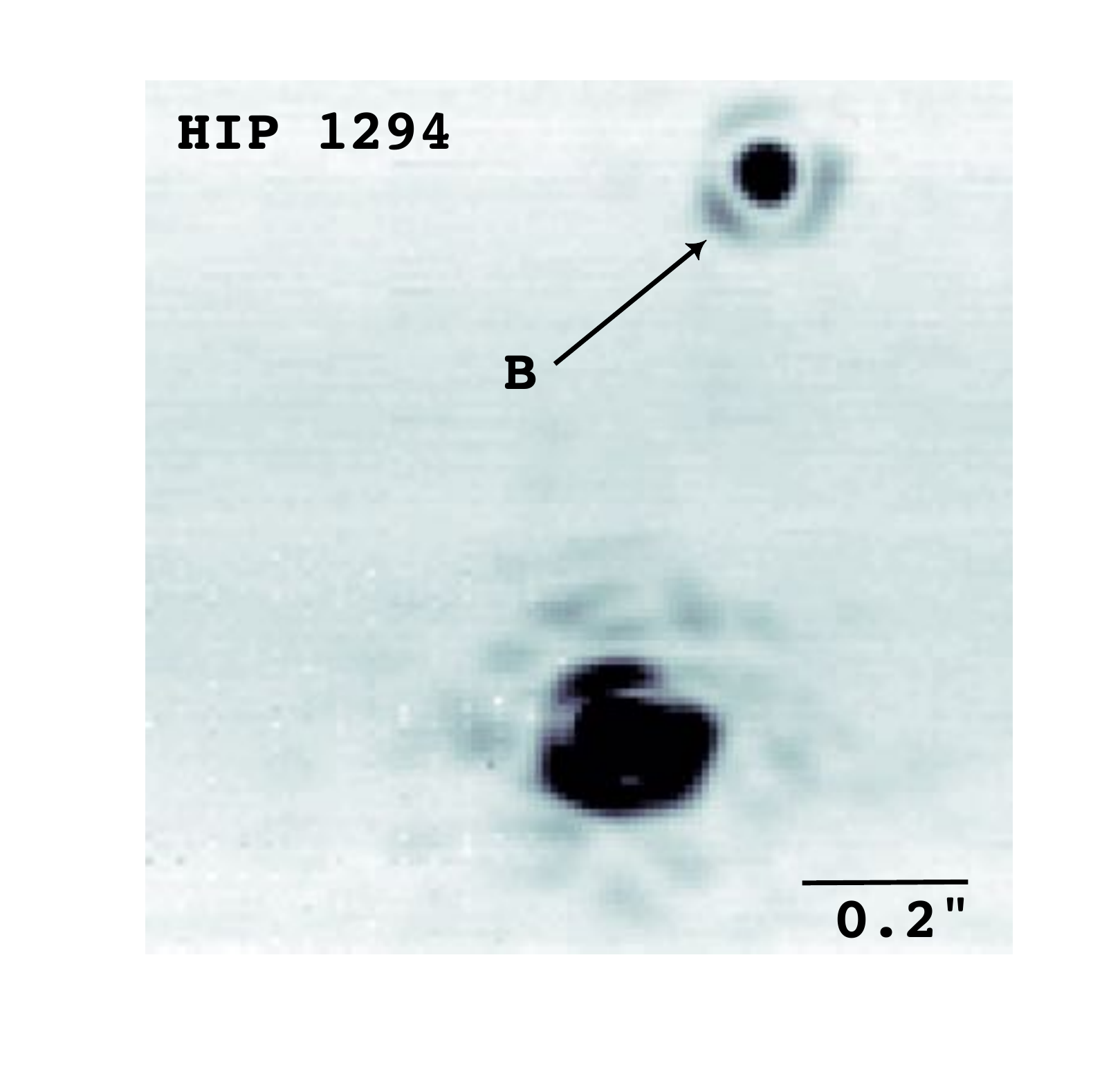}}}\\ \subfloat{\subfloat{\includegraphics[width= .5\textwidth]{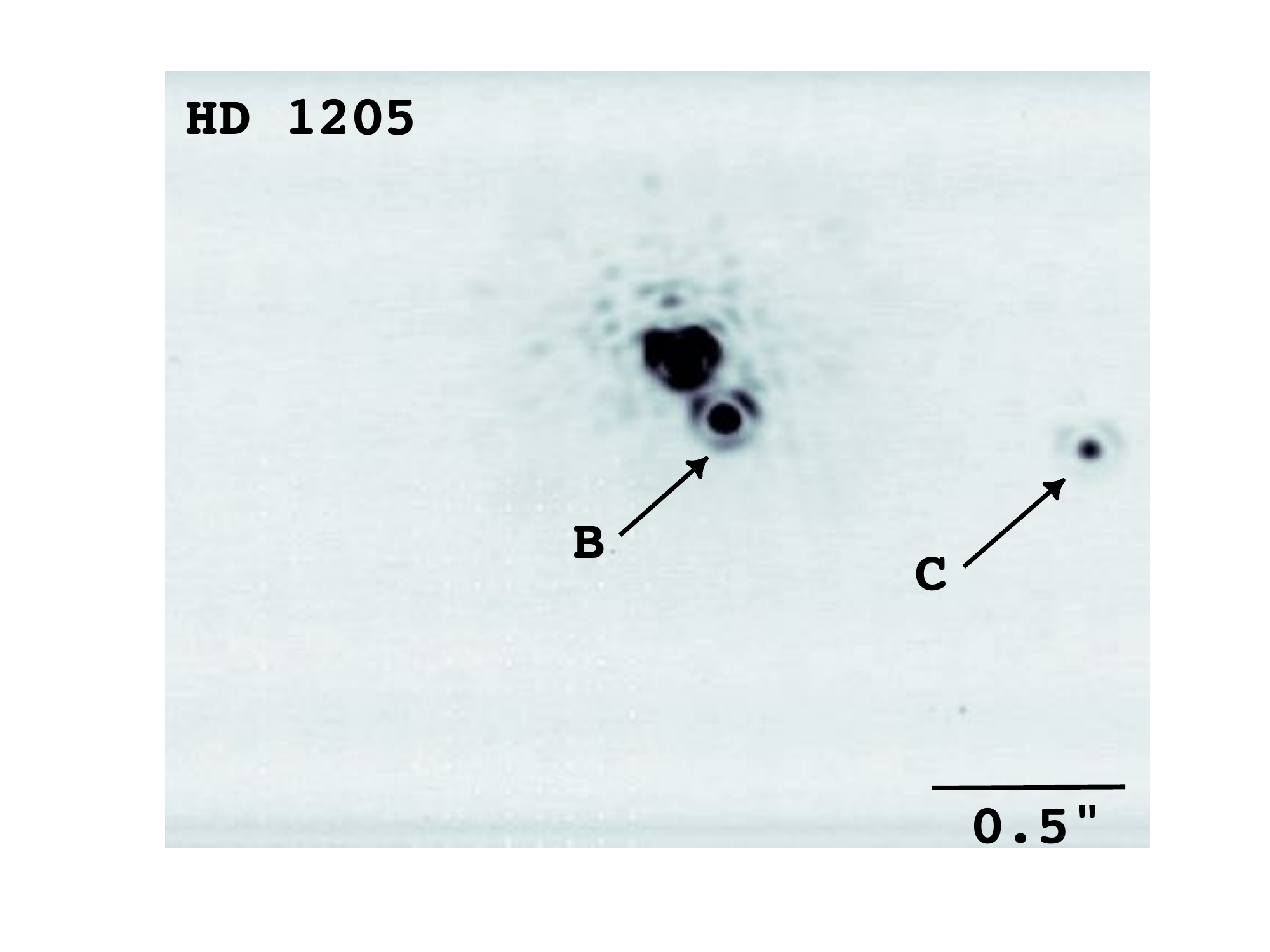}}\subfloat{\includegraphics[width= .5\textwidth]{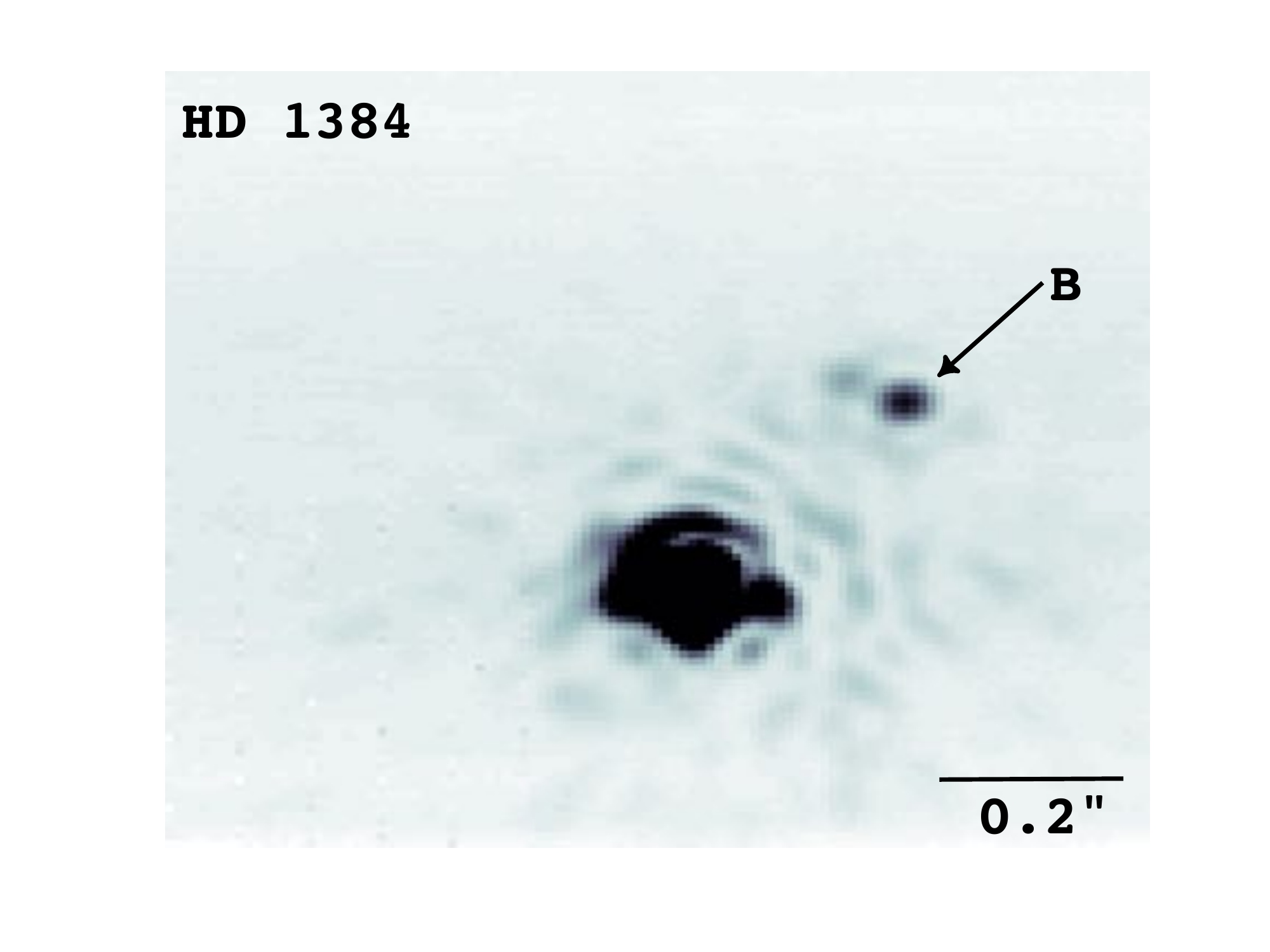}}}\\
  \subfloat{\subfloat{\includegraphics[width = .5\textwidth]{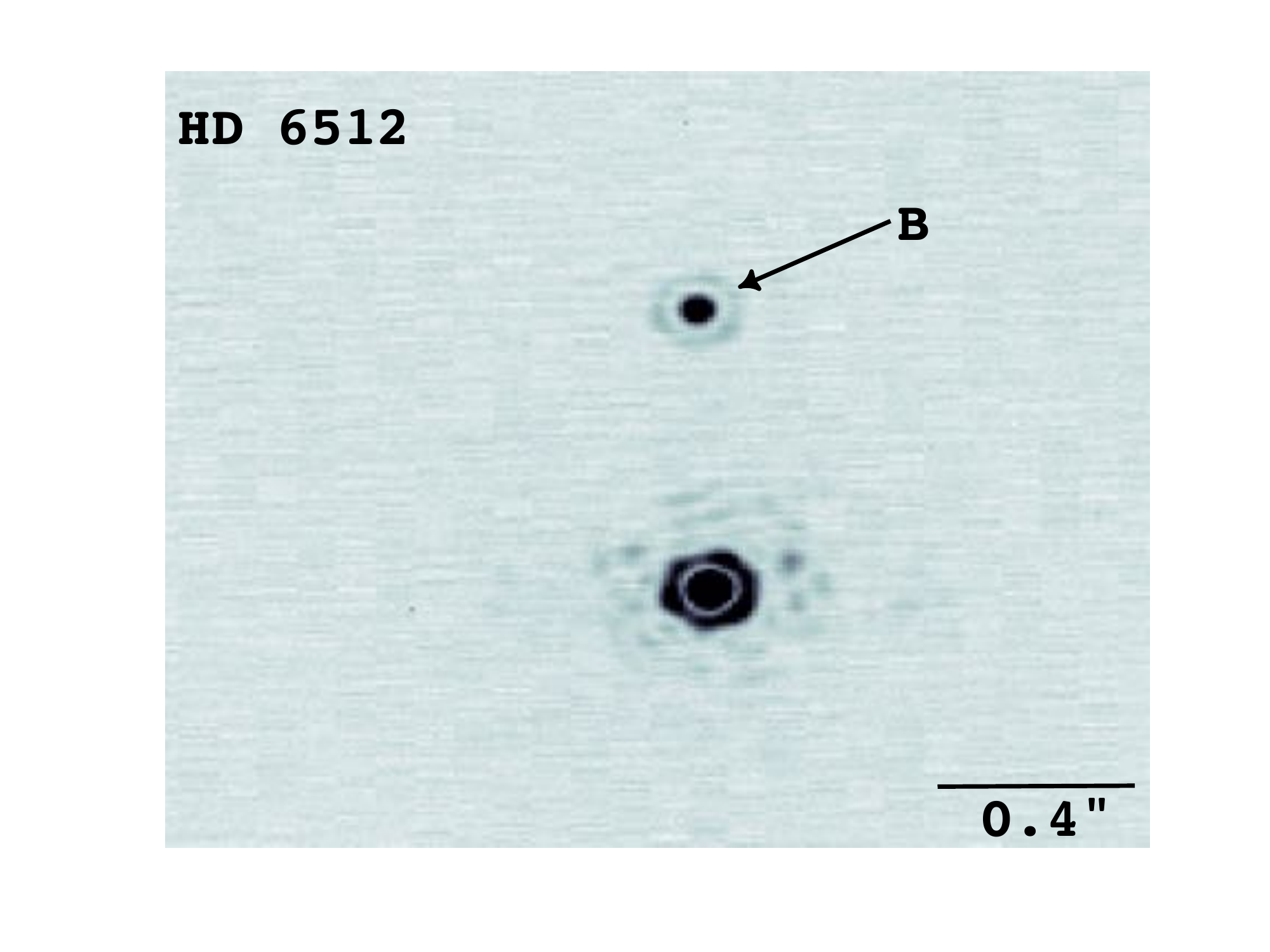}}
 \subfloat{\includegraphics[width= .5\textwidth]{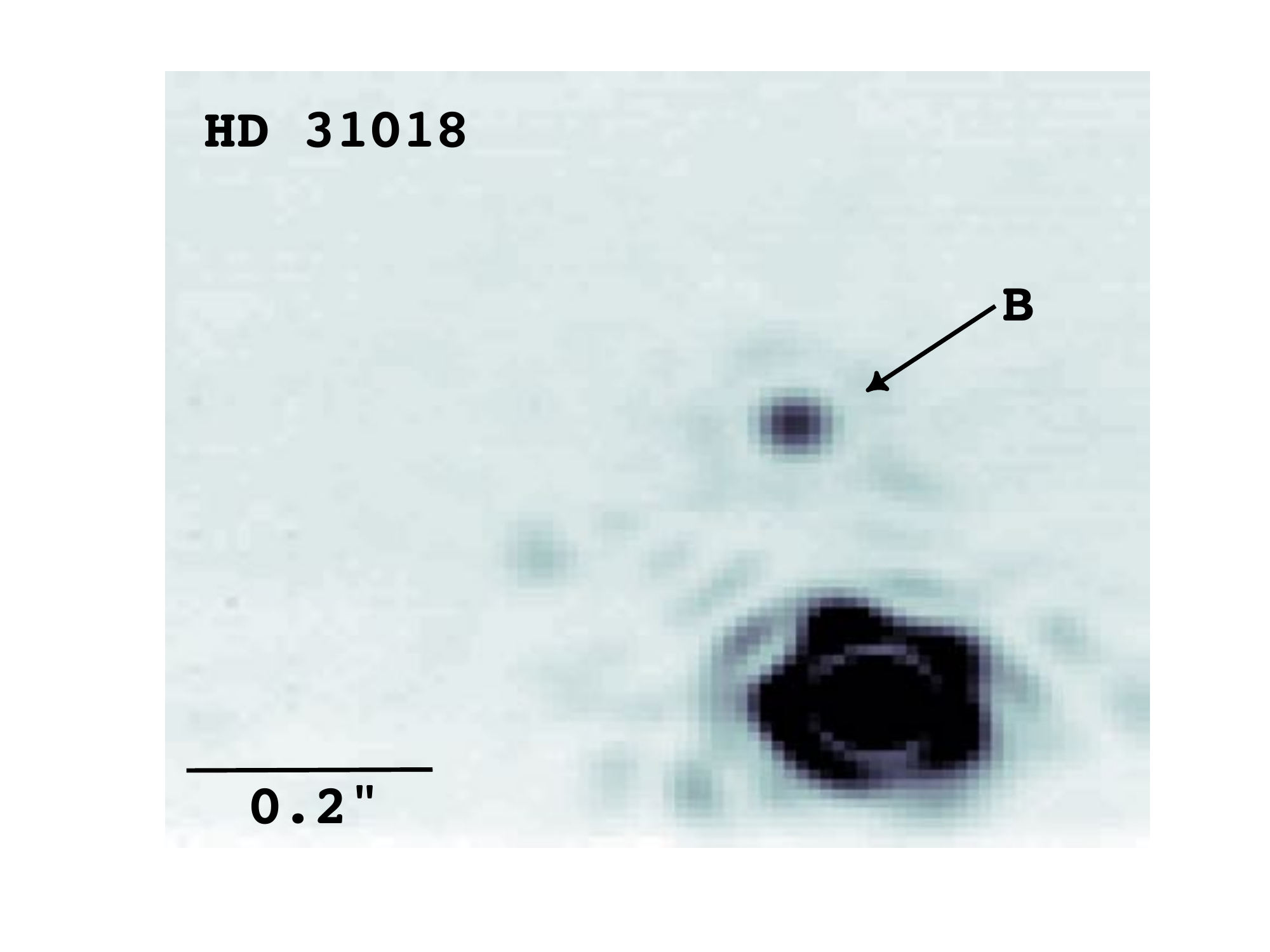}}}\\ 
 \caption{High-resolution adaptive optics images of confirmed companions.}
  \label{fig13}
\end{figure}

\begin{figure}[htp]
 \subfloat{\subfloat{\includegraphics[width = .5\textwidth]{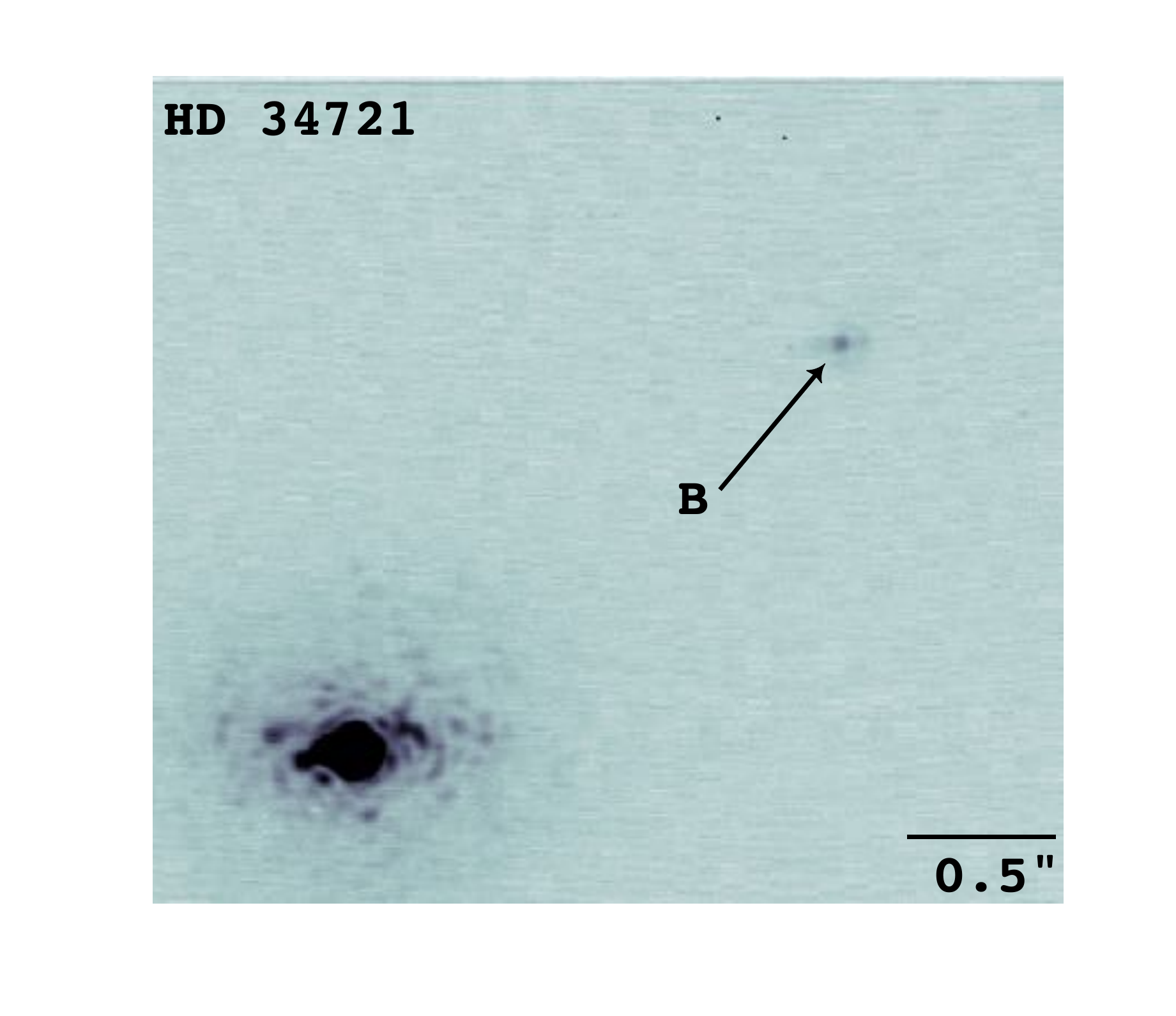}}
 \subfloat{\includegraphics[width= .5\textwidth]{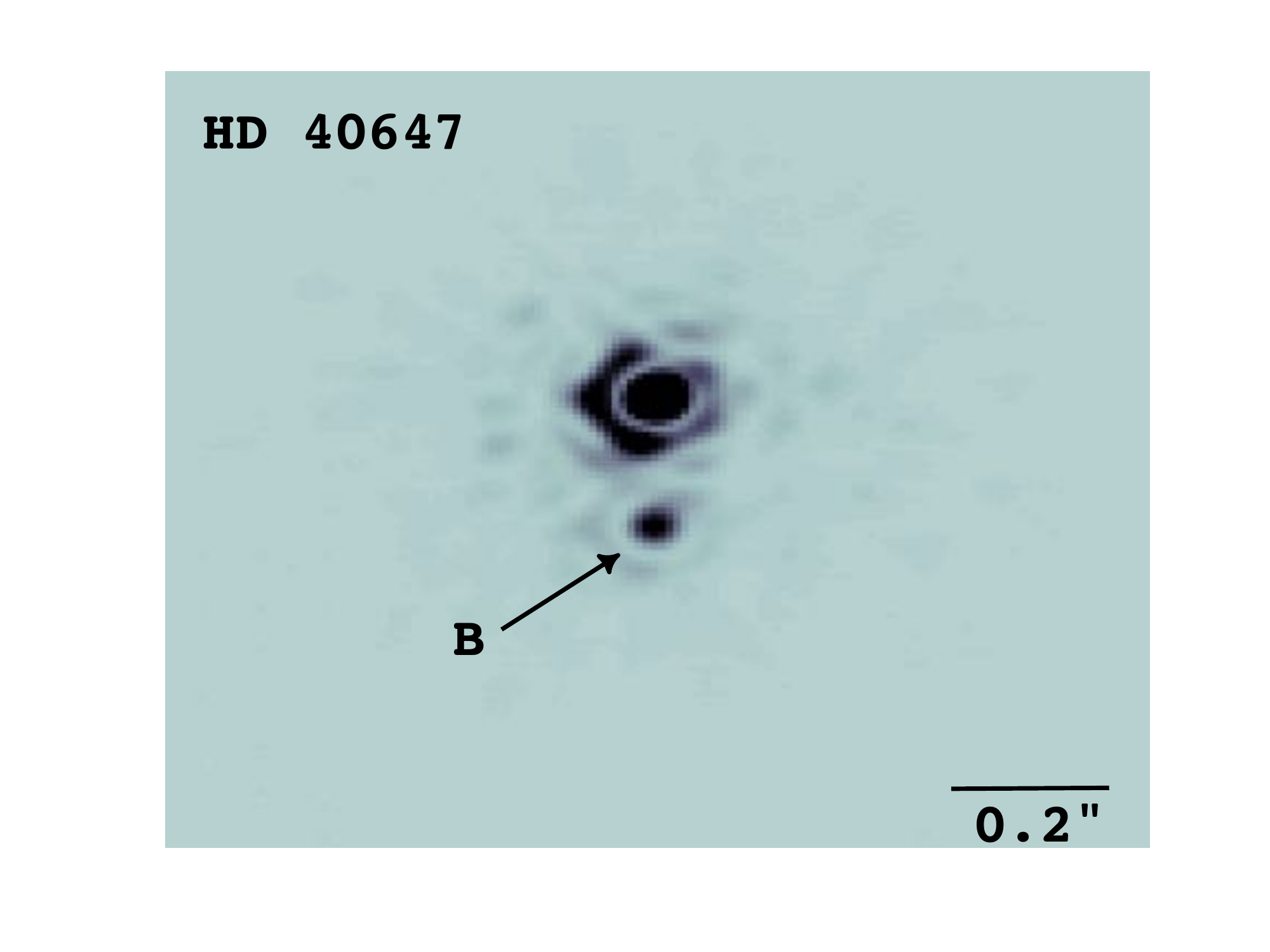}}}\\ \subfloat{\subfloat{\includegraphics[width= .5\textwidth]{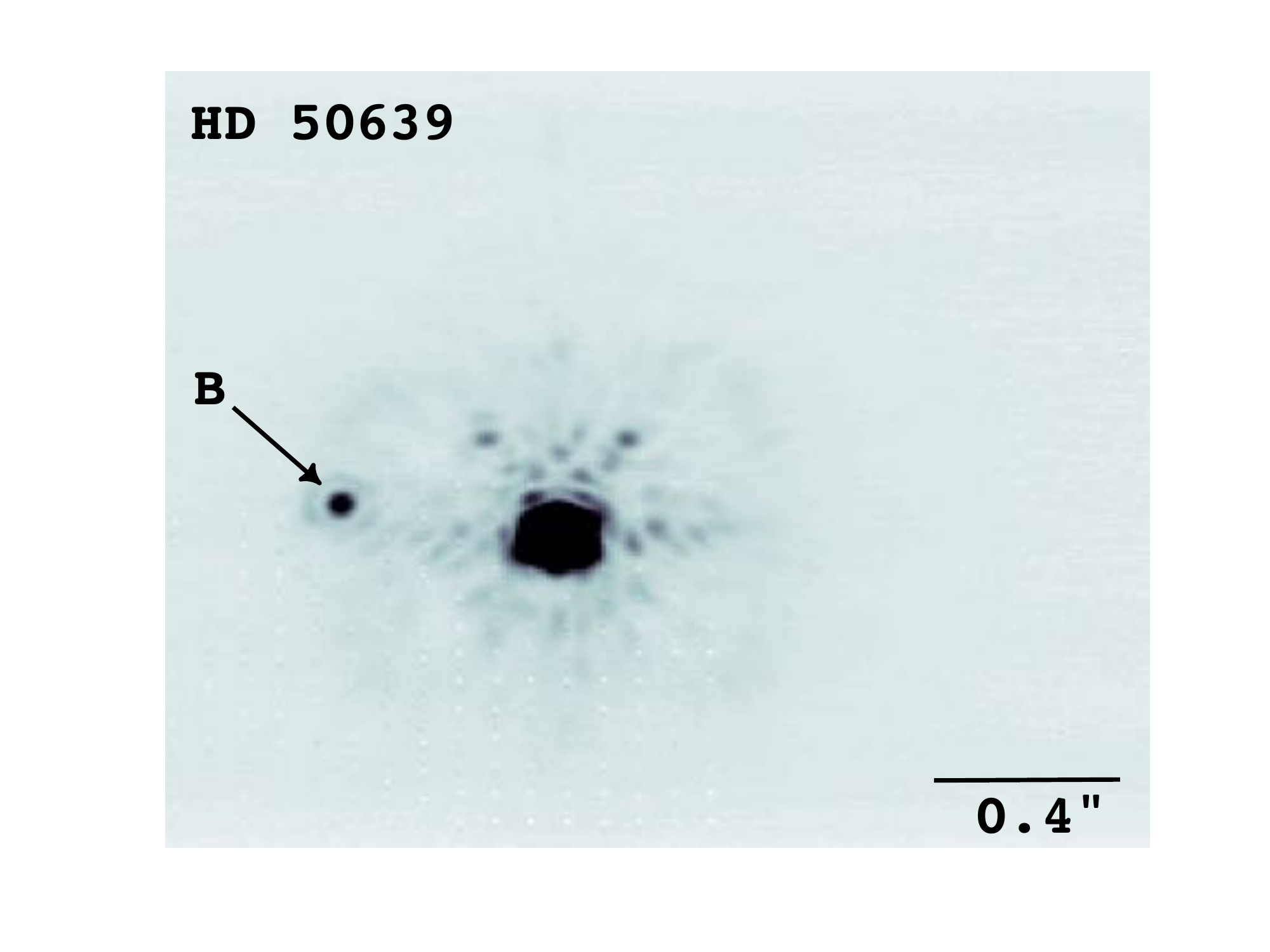}}\subfloat{\includegraphics[width= .5\textwidth]{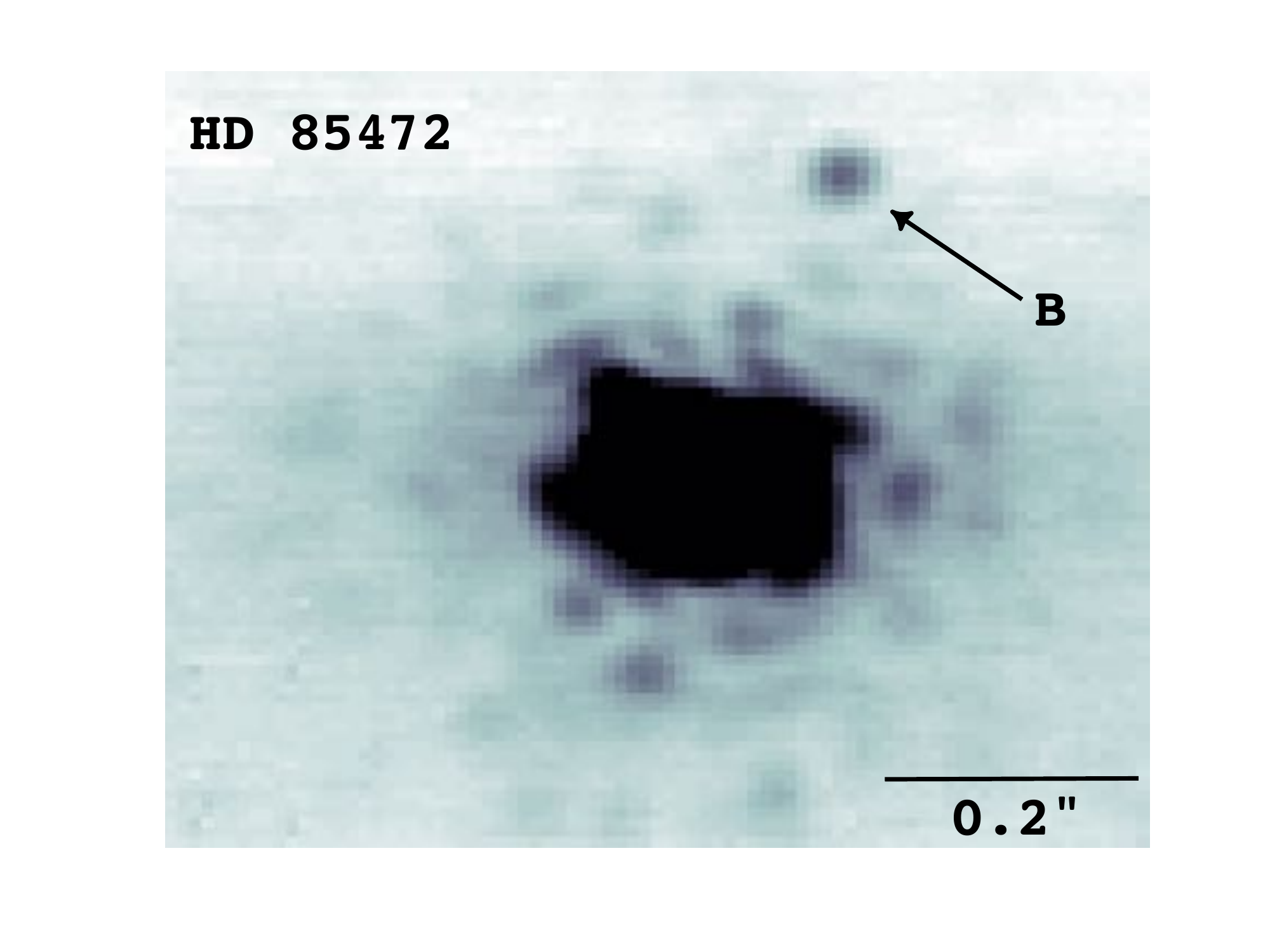}}}\\
  \subfloat{\subfloat{\includegraphics[width = .5\textwidth]{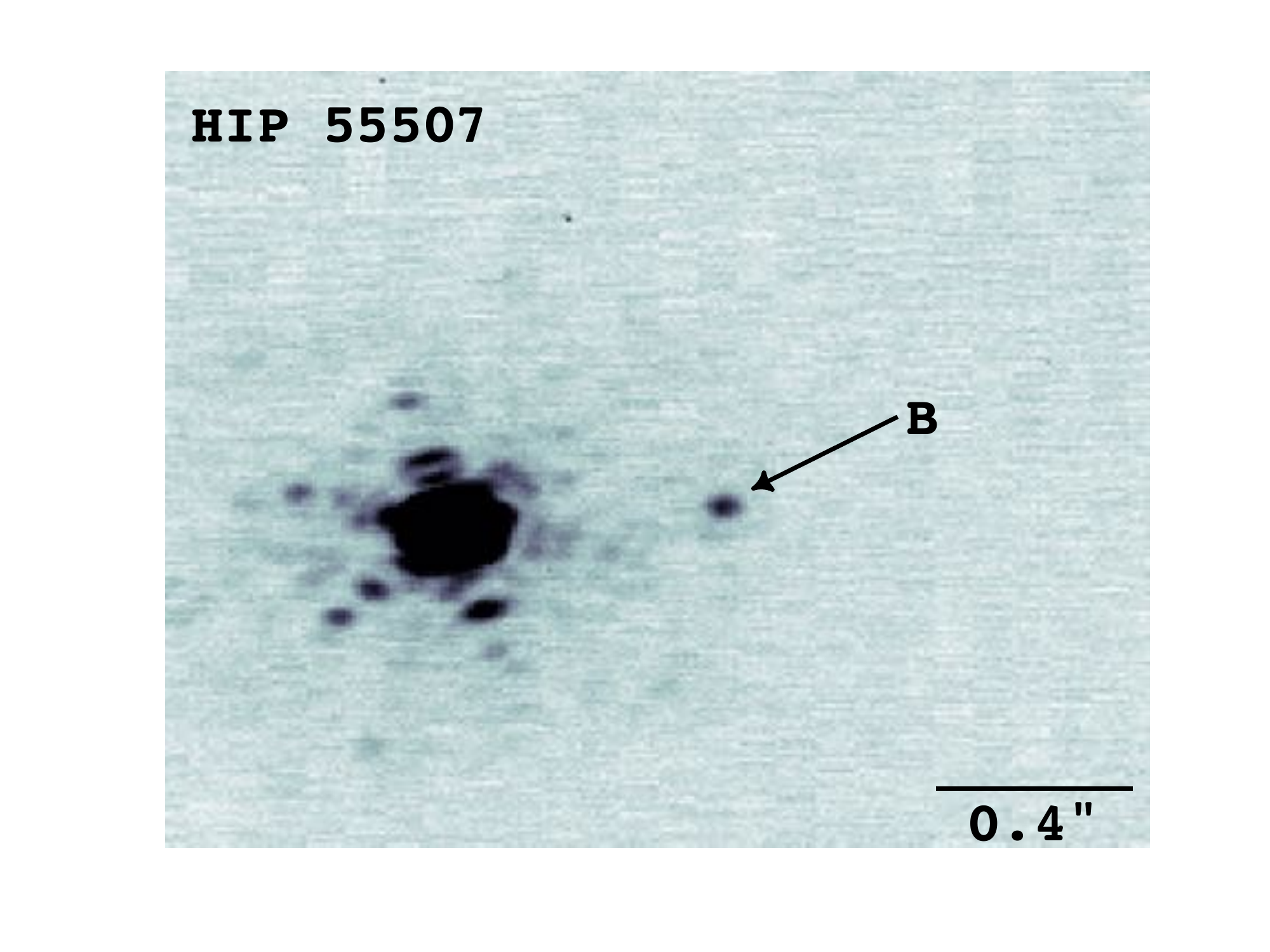}}
 \subfloat{\includegraphics[width= .5\textwidth]{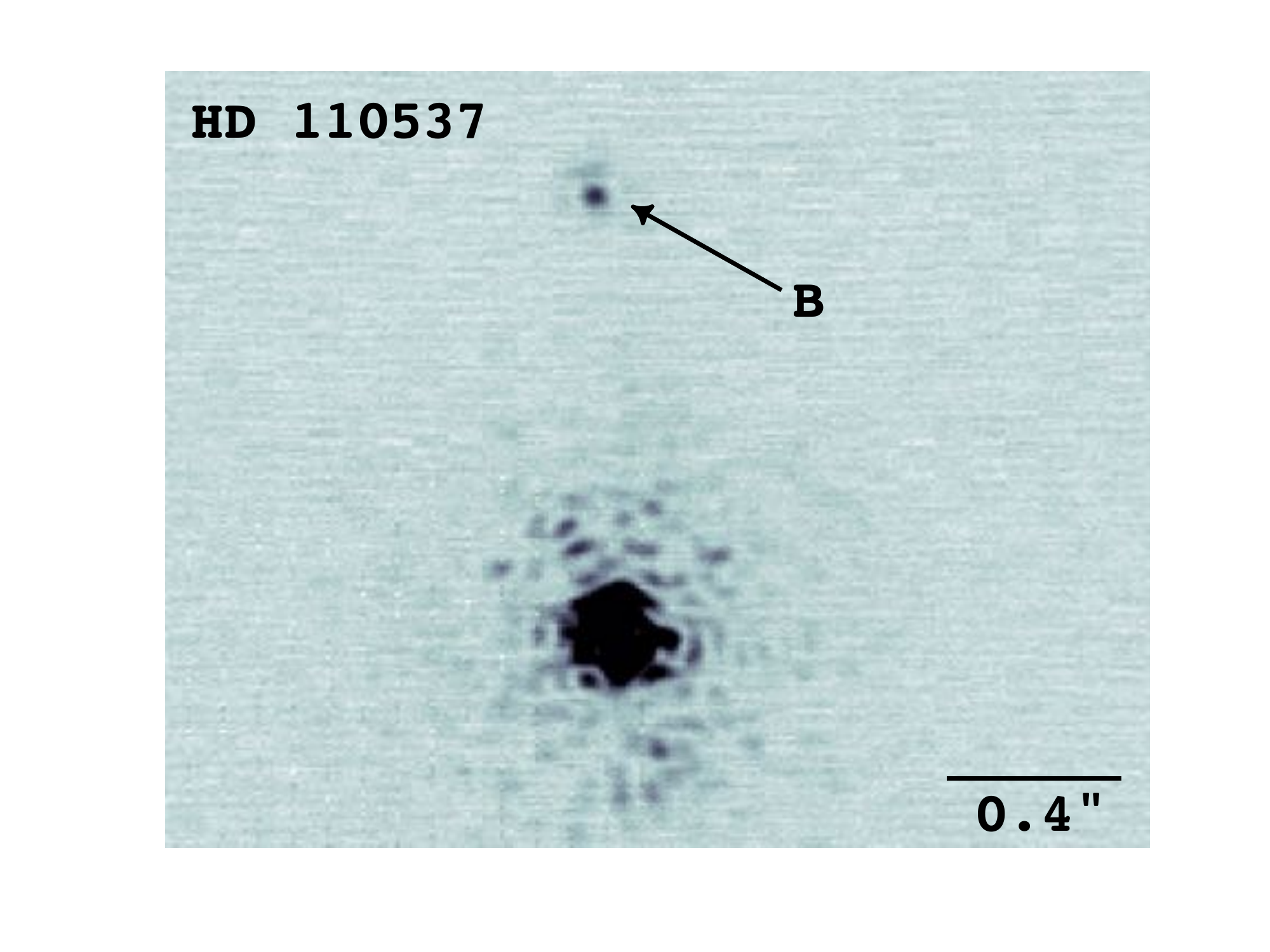}}}\\ 
 \caption{High-resolution adaptive optics images of confirmed companions. Continuation of \ref{fig13}.}
  \label{fig14}
\end{figure}

\begin{figure}[htp]
 \subfloat{\subfloat{\includegraphics[width = .5\textwidth]{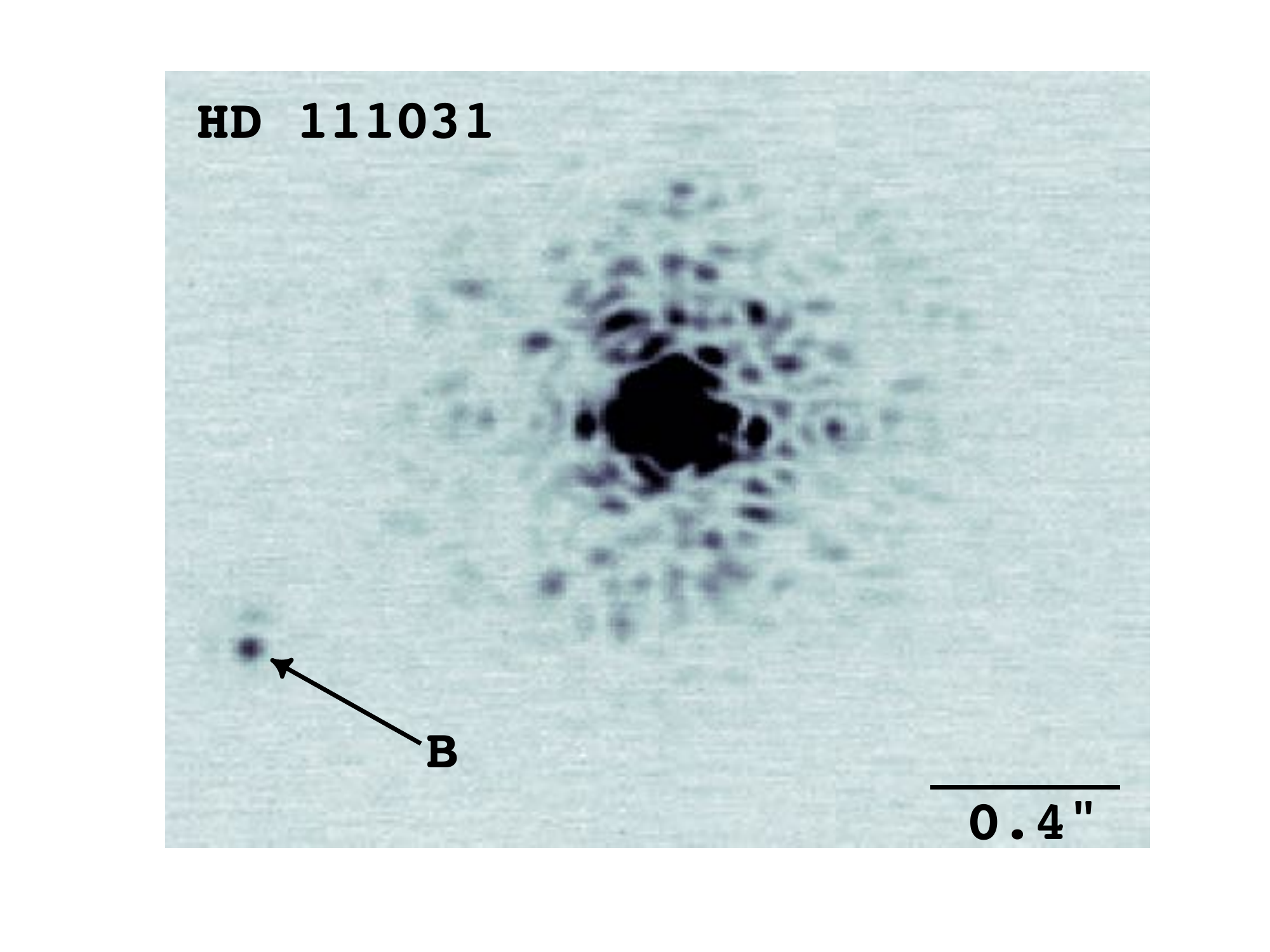}}
 \subfloat{\includegraphics[width= .5\textwidth]{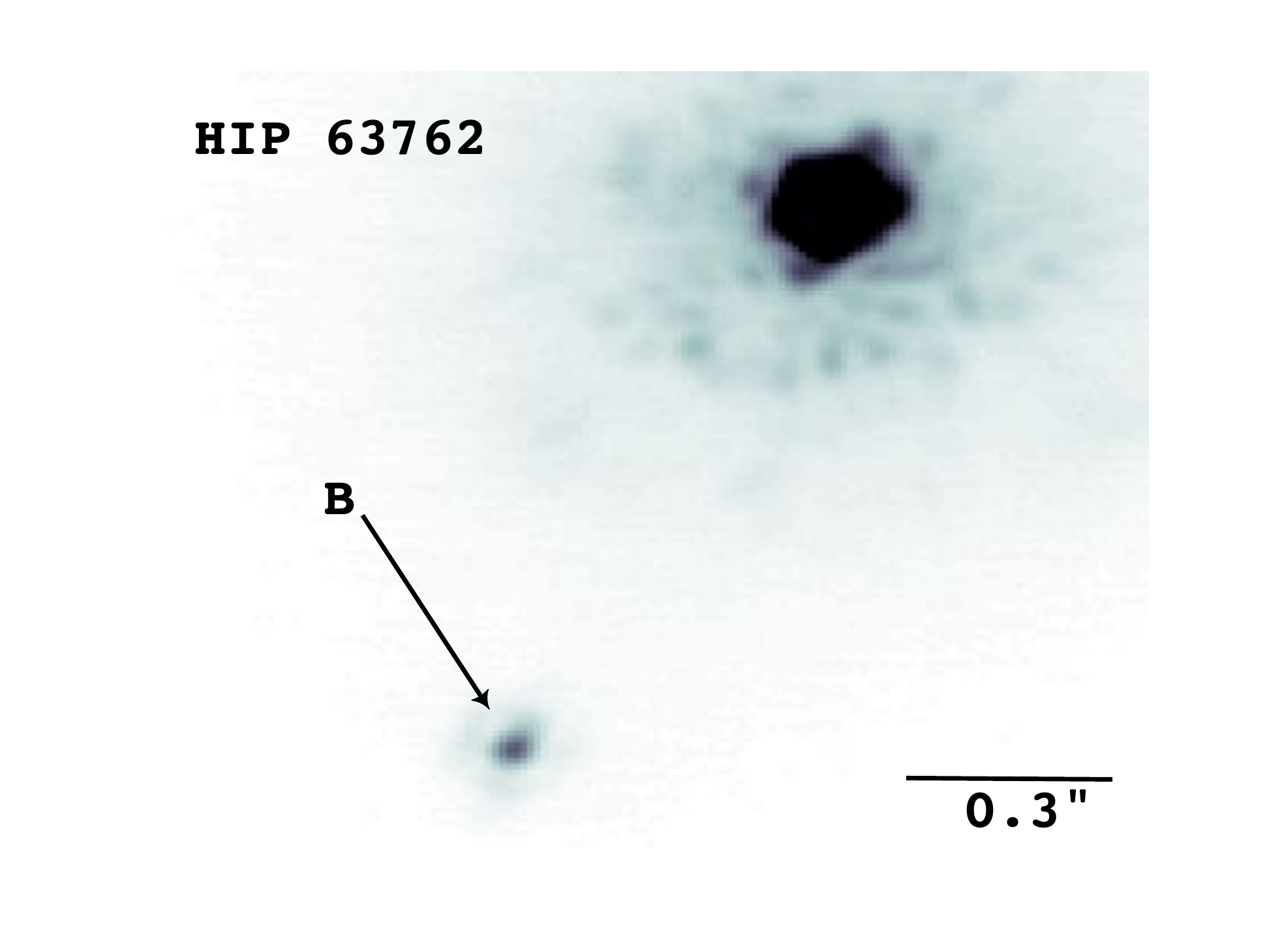}}}\\ \subfloat{\subfloat{\includegraphics[width= .5\textwidth]{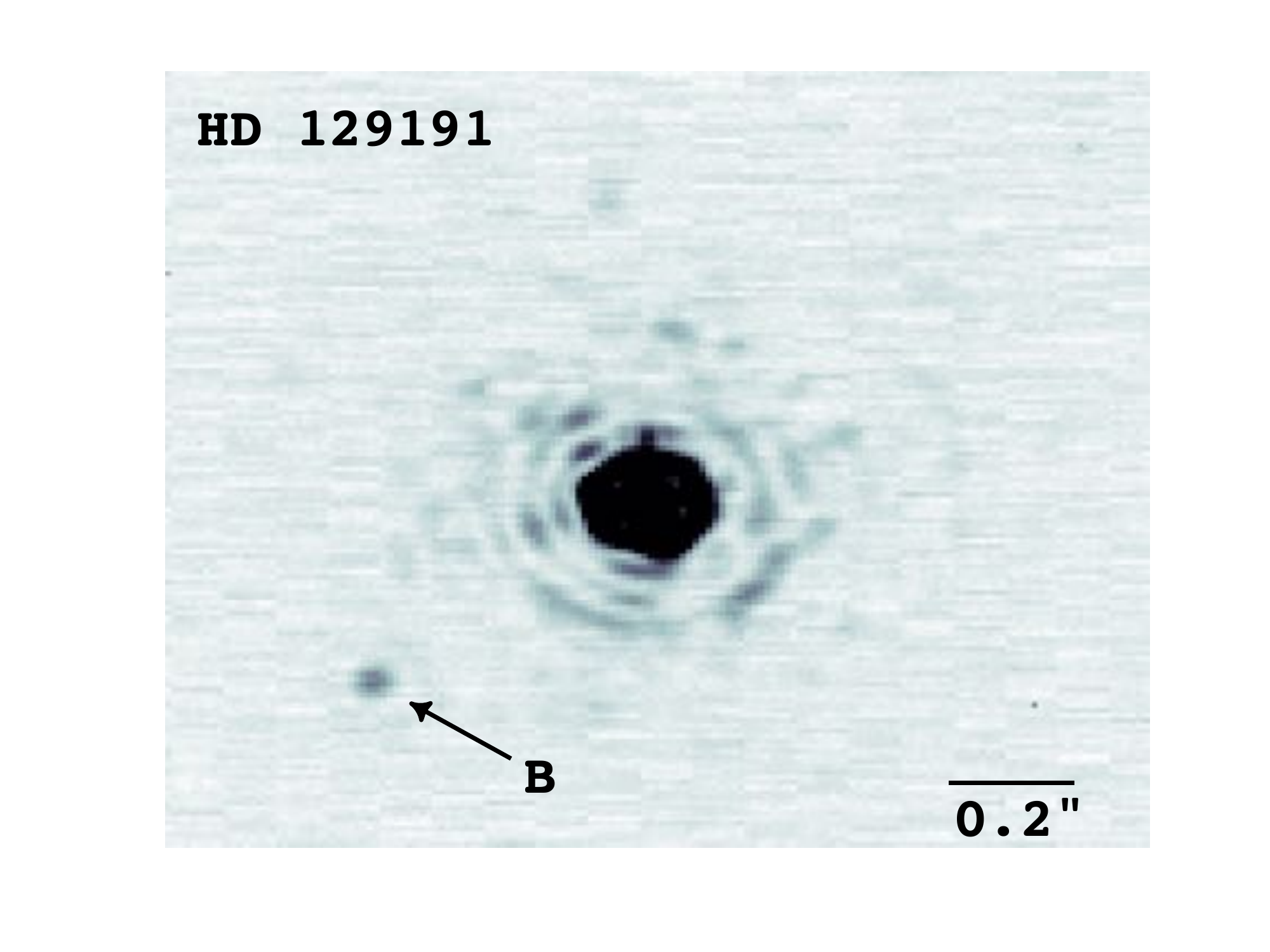}}\subfloat{\includegraphics[width= .5\textwidth]{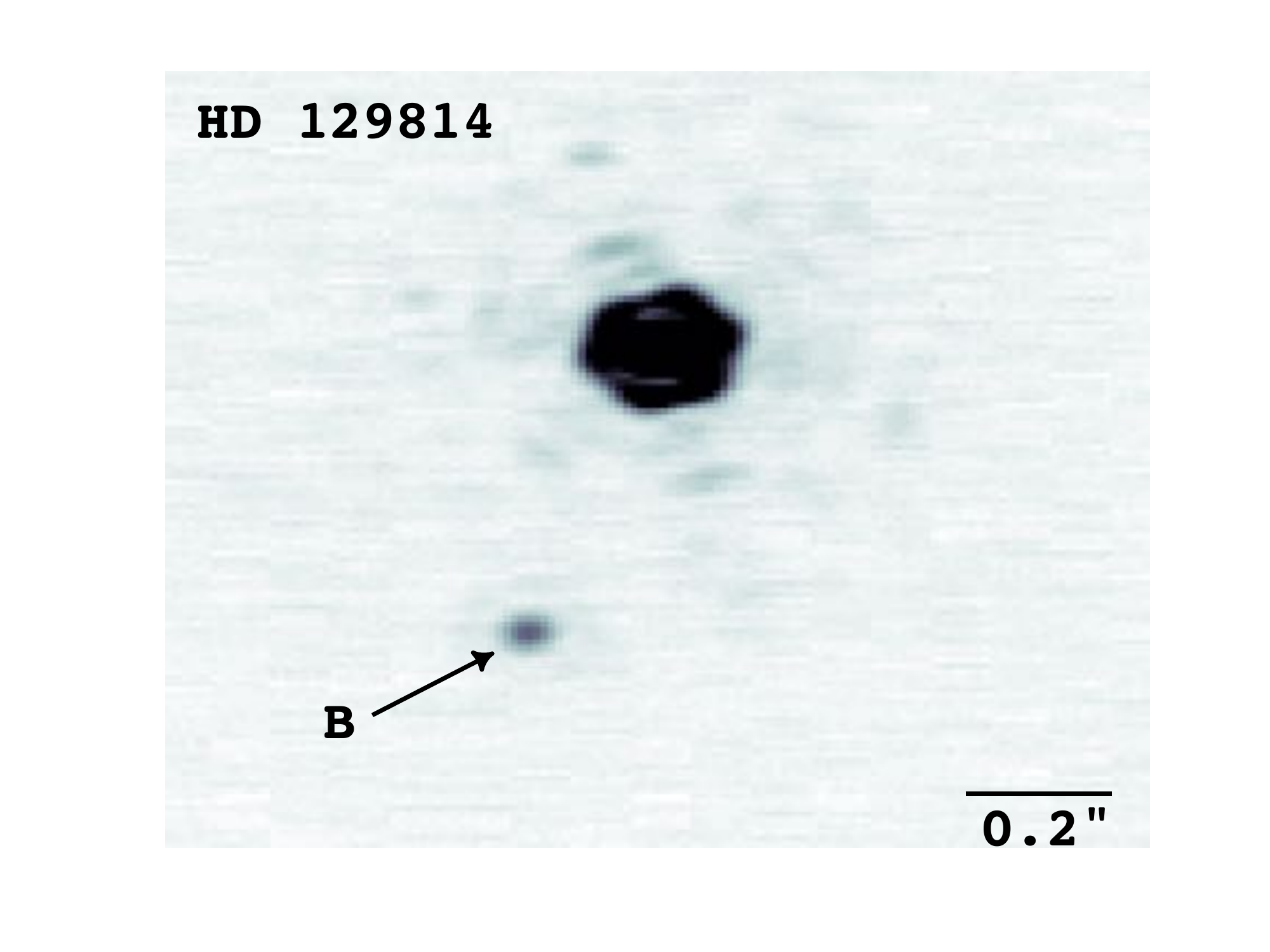}}}\\
  \subfloat{\subfloat{\includegraphics[width = .5\textwidth]{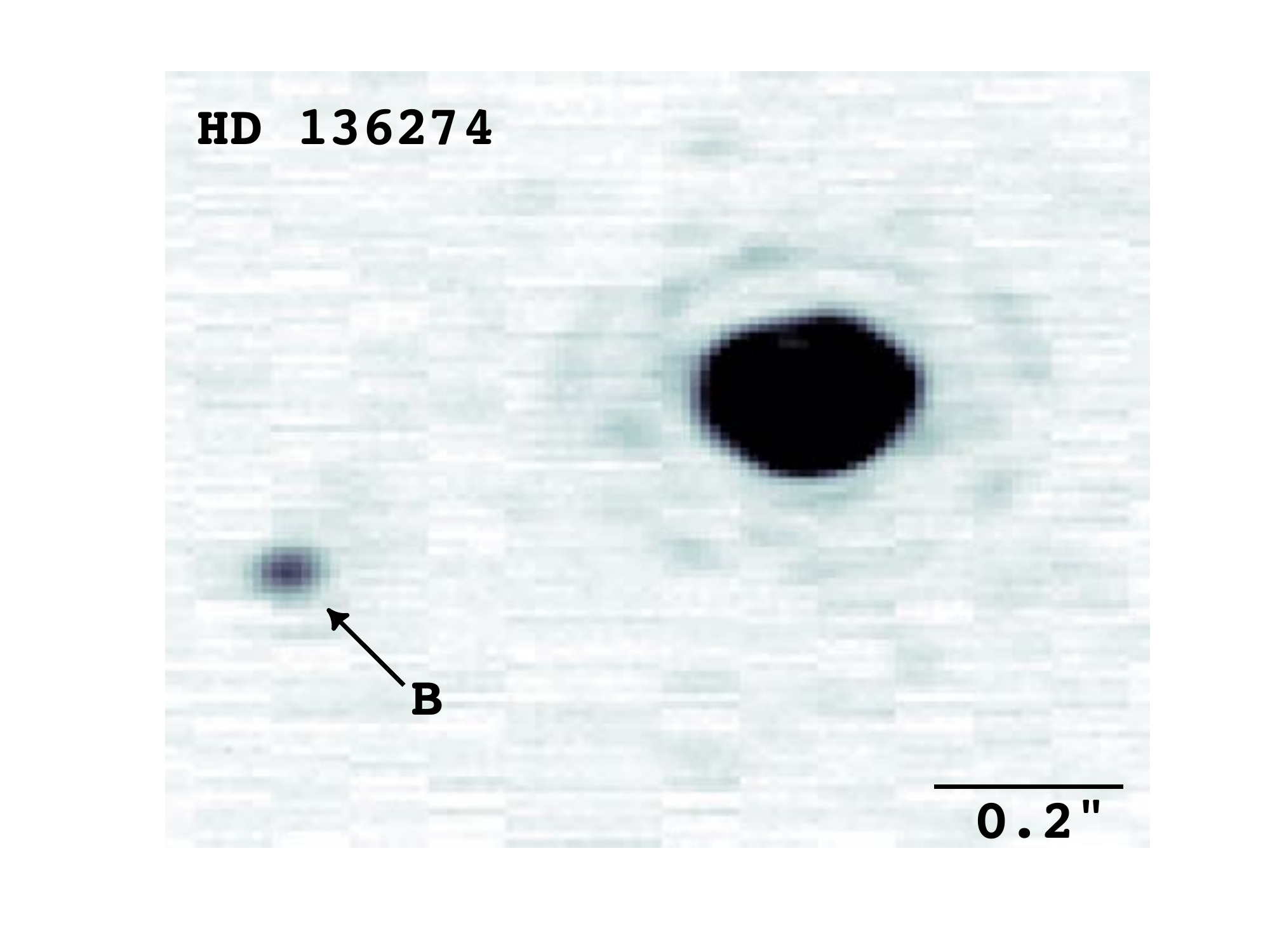}}
 \subfloat{\includegraphics[width= .5\textwidth]{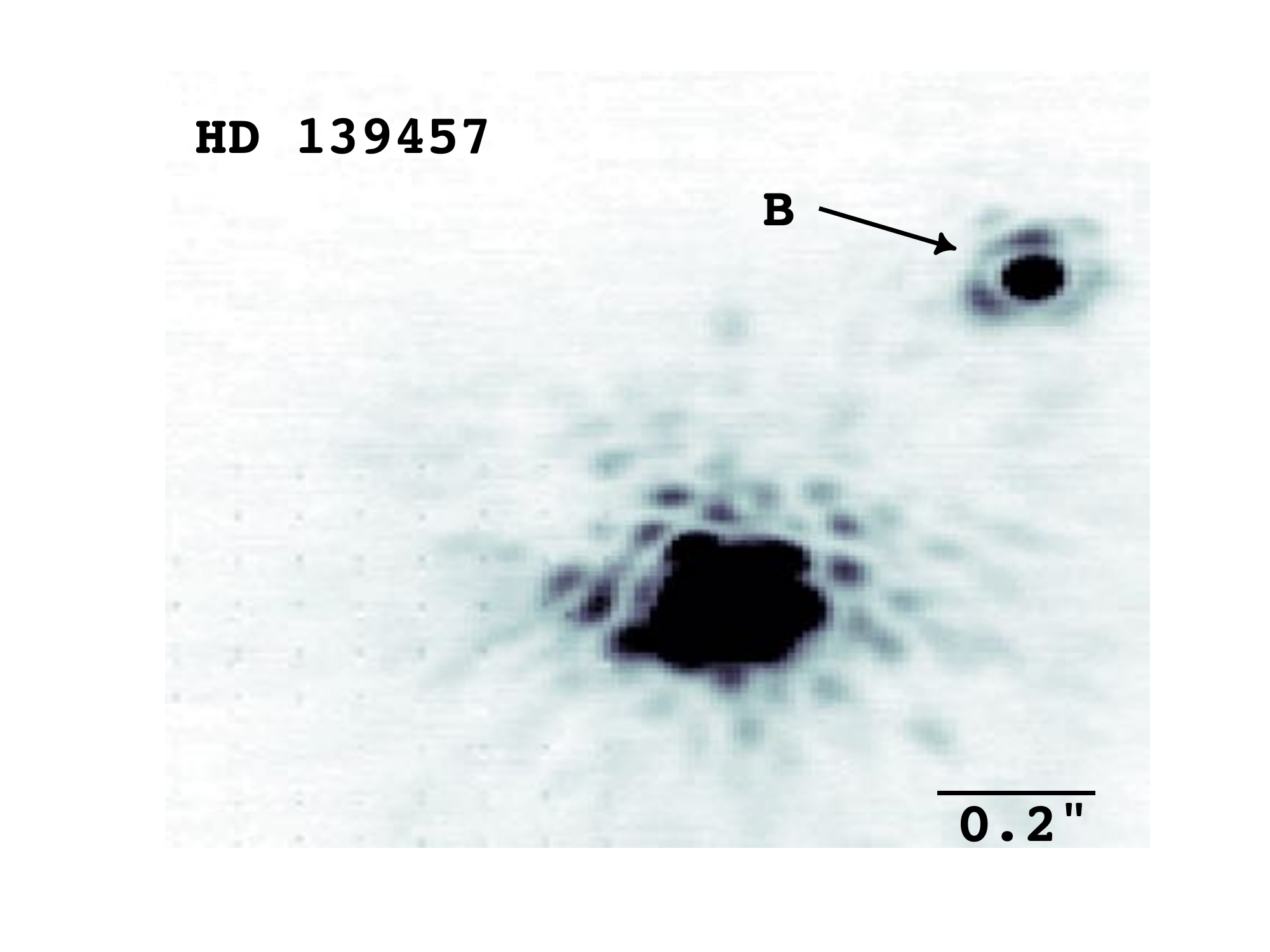}}}\\ 
 \caption{High-resolution adaptive optics images of confirmed companions. Continuation of \ref{fig14}.}
  \label{fig15}
\end{figure}

\begin{figure}[htp]
 \subfloat{\subfloat{\includegraphics[width = .5\textwidth]{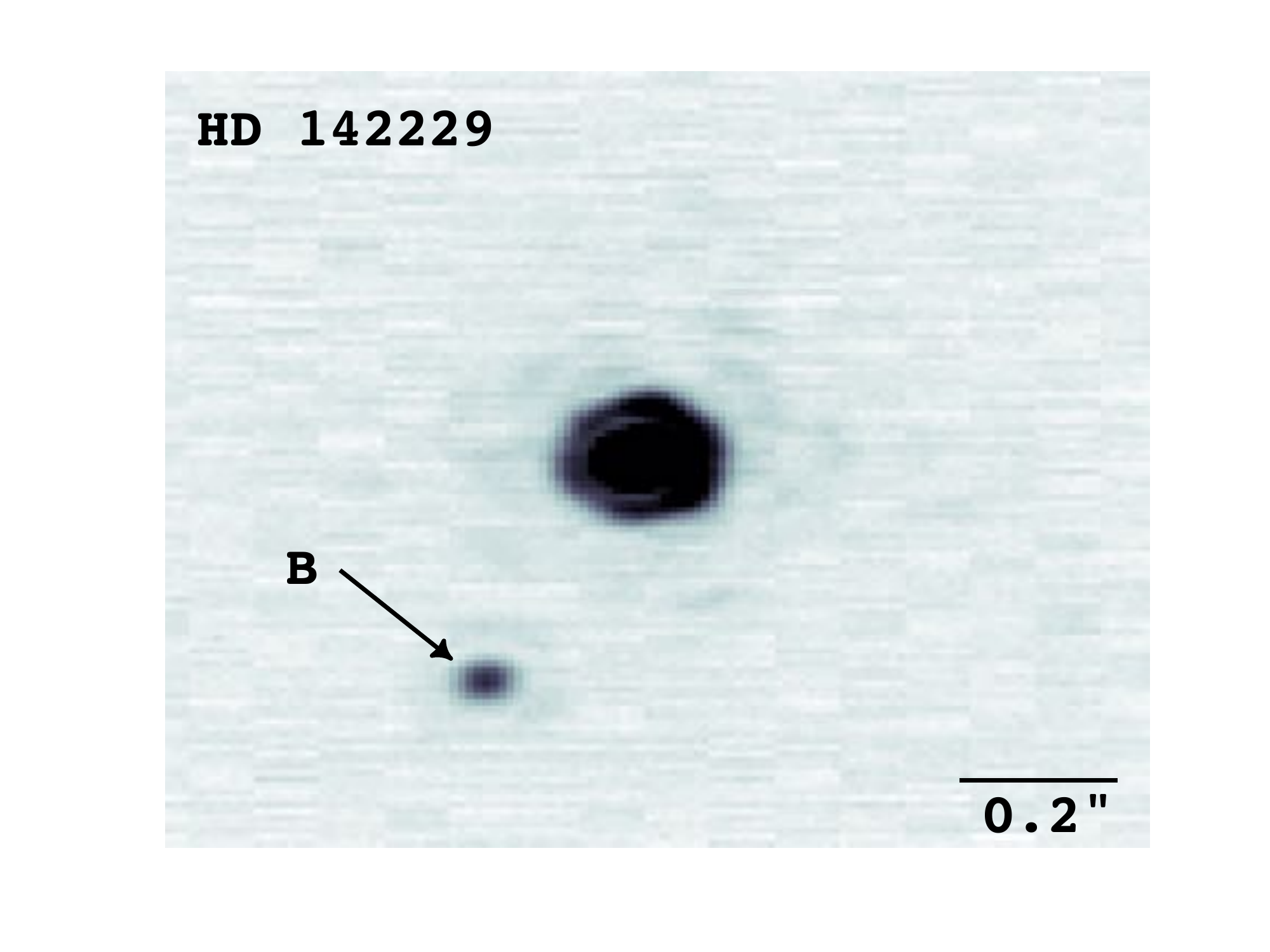}}
 \subfloat{\includegraphics[width= .5\textwidth]{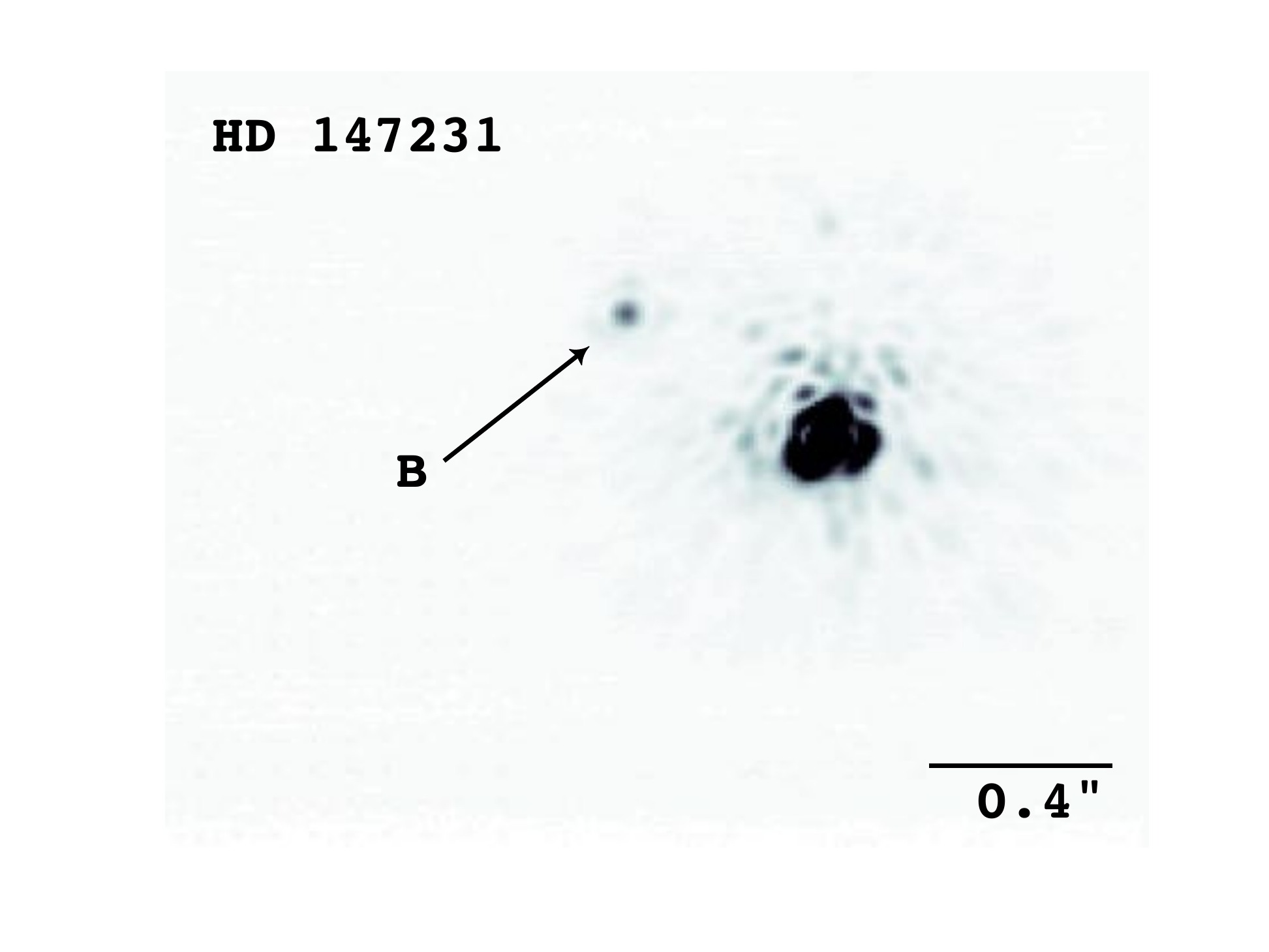}}}\\ \subfloat{\subfloat{\includegraphics[width= .5\textwidth]{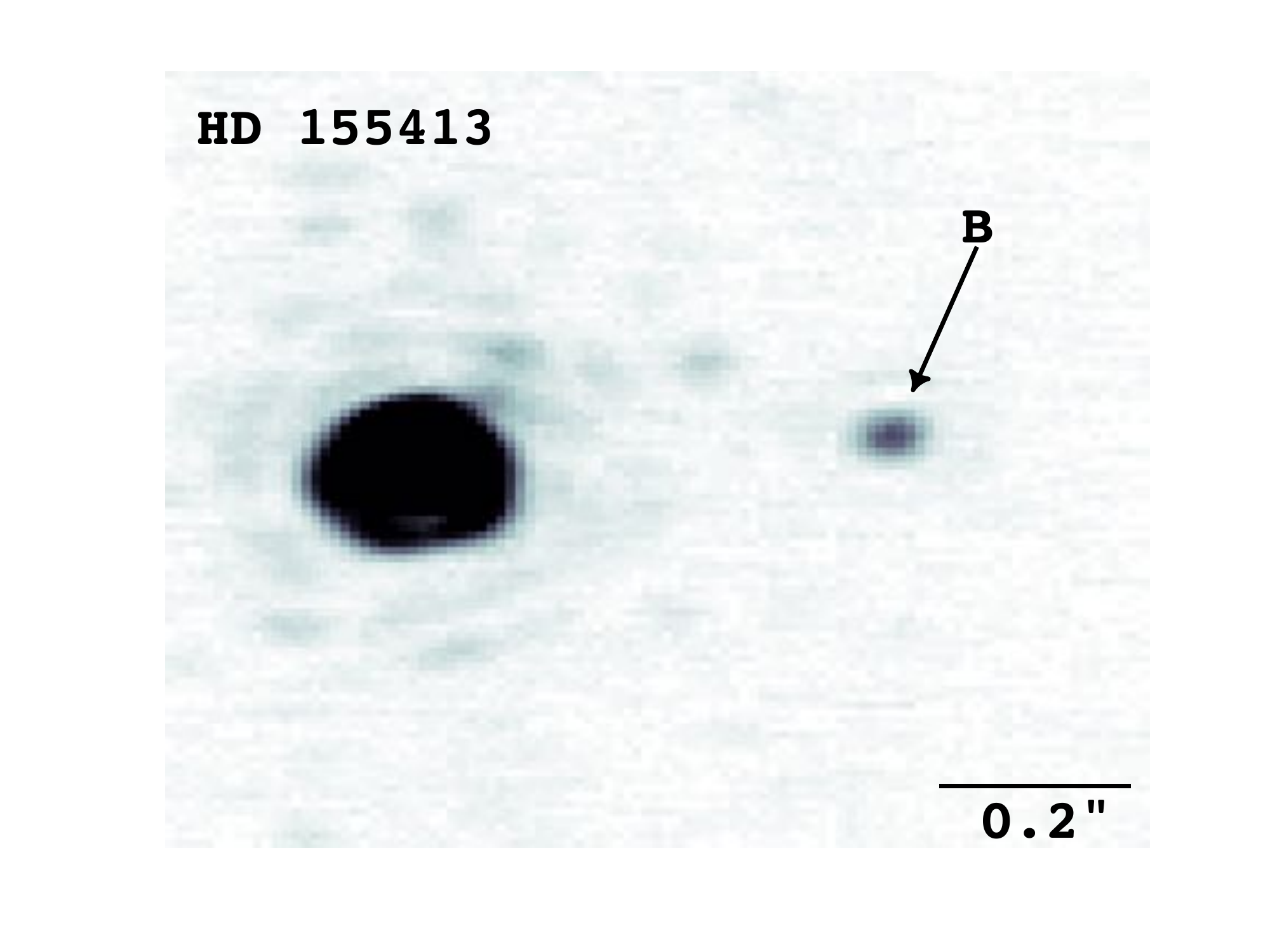}}\subfloat{\includegraphics[width= .5\textwidth]{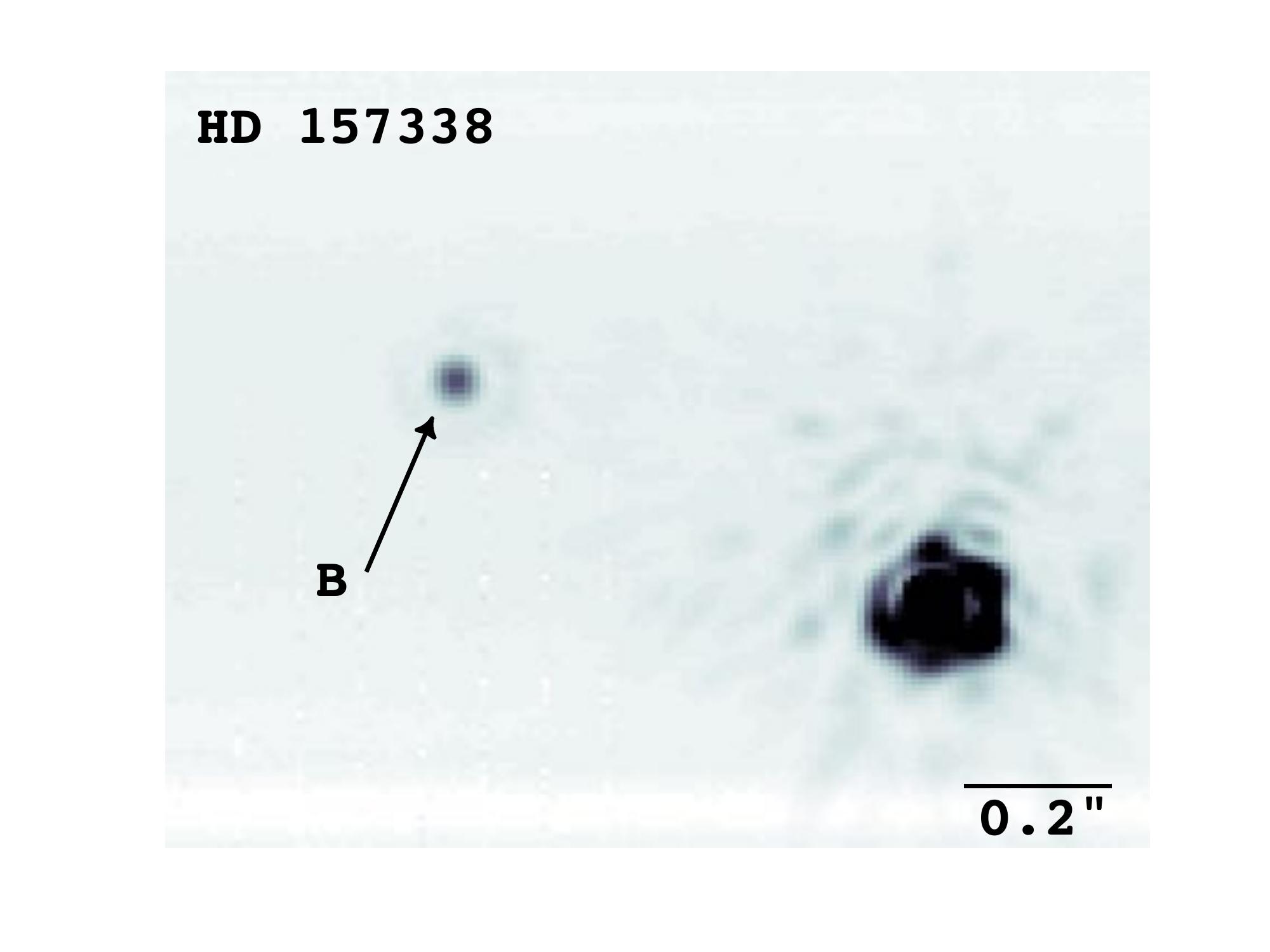}}}\\
  \subfloat{\subfloat{\includegraphics[width = .5\textwidth]{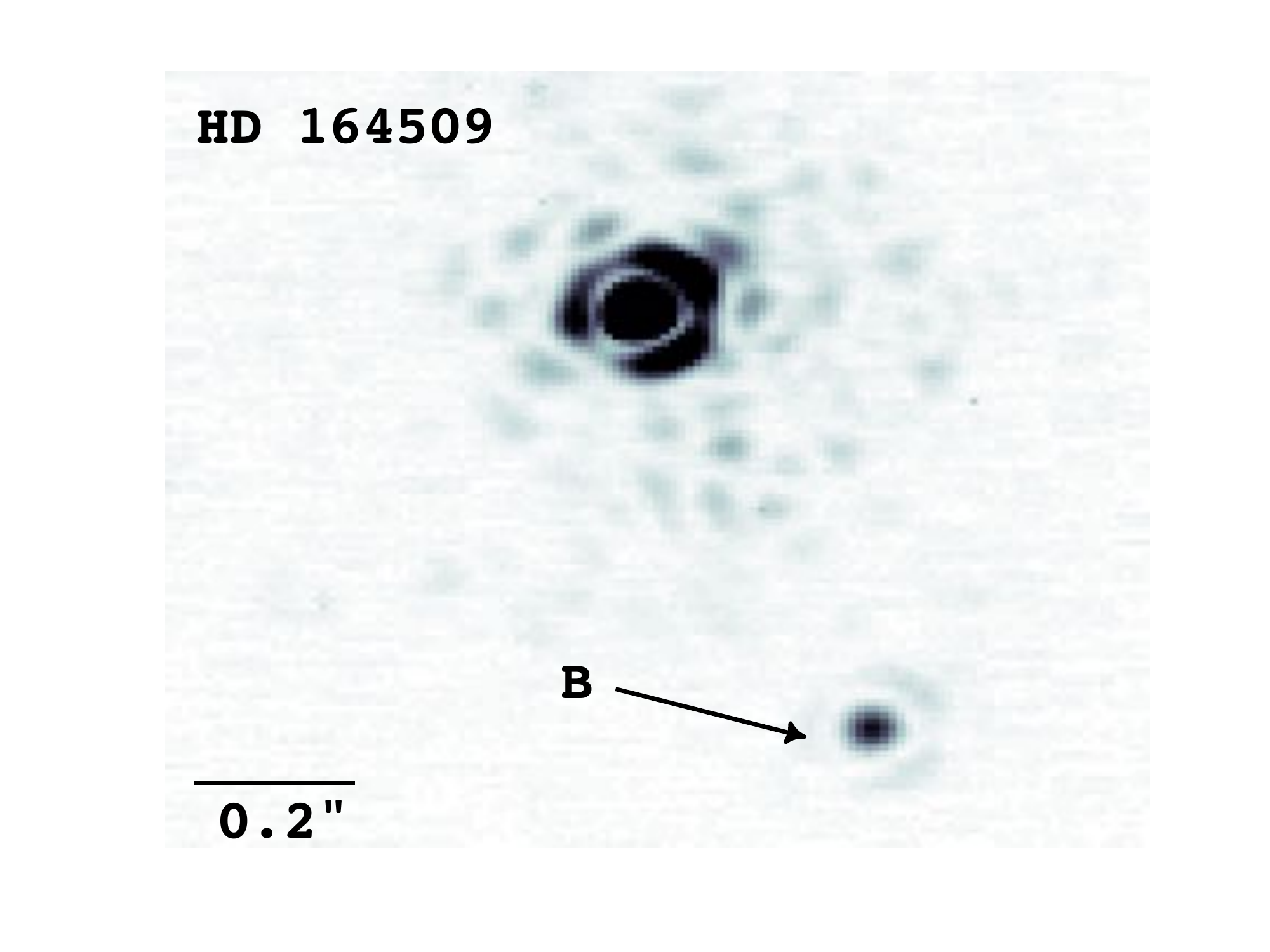}}
 \subfloat{\includegraphics[width= .5\textwidth]{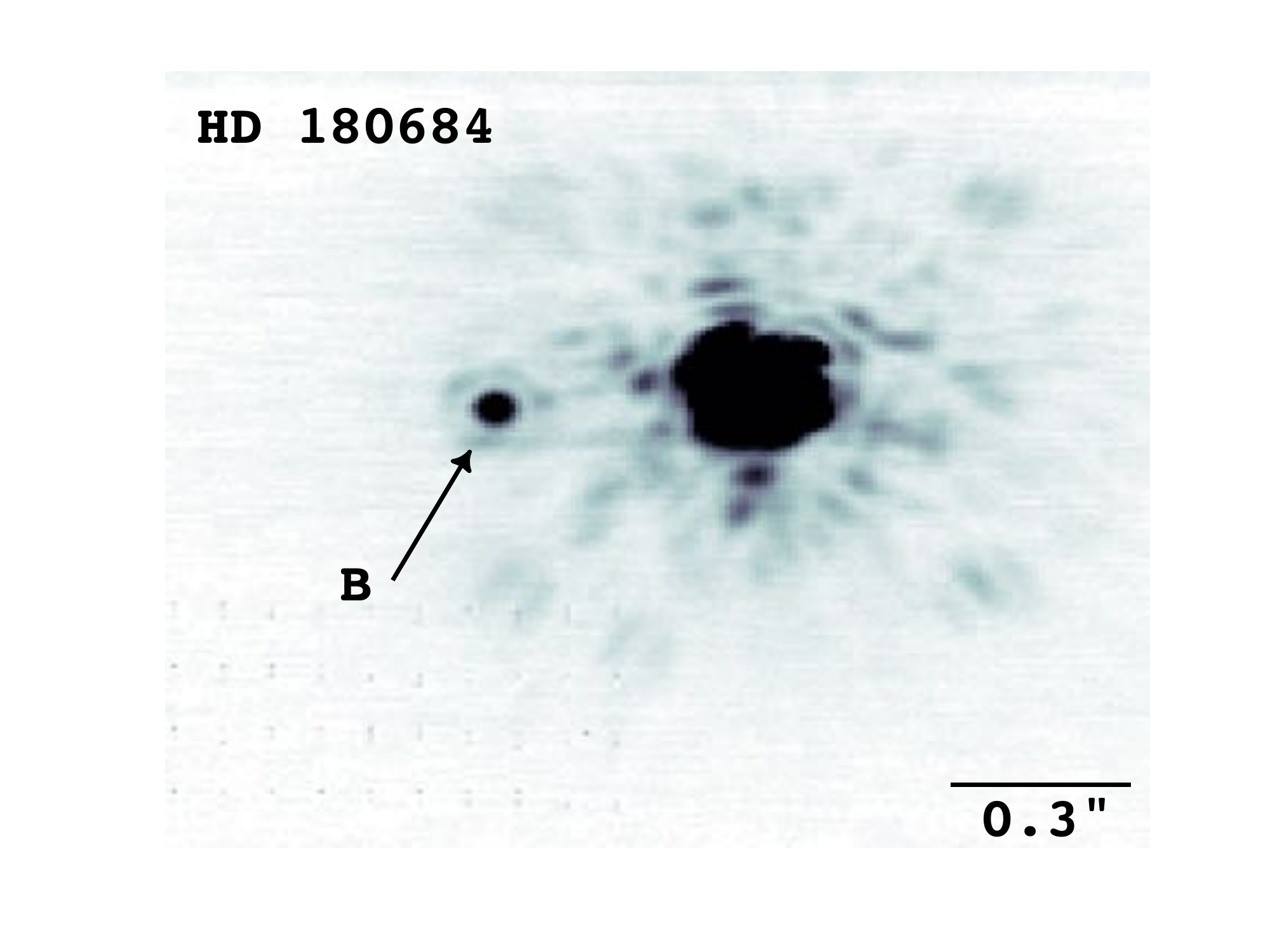}}}\\ 
 \caption{High-resolution adaptive optics images of confirmed companions. Continuation of \ref{fig15}.}
  \label{fig16}
\end{figure}

\begin{figure}[htp]
 \subfloat{\subfloat{\includegraphics[width = .5\textwidth]{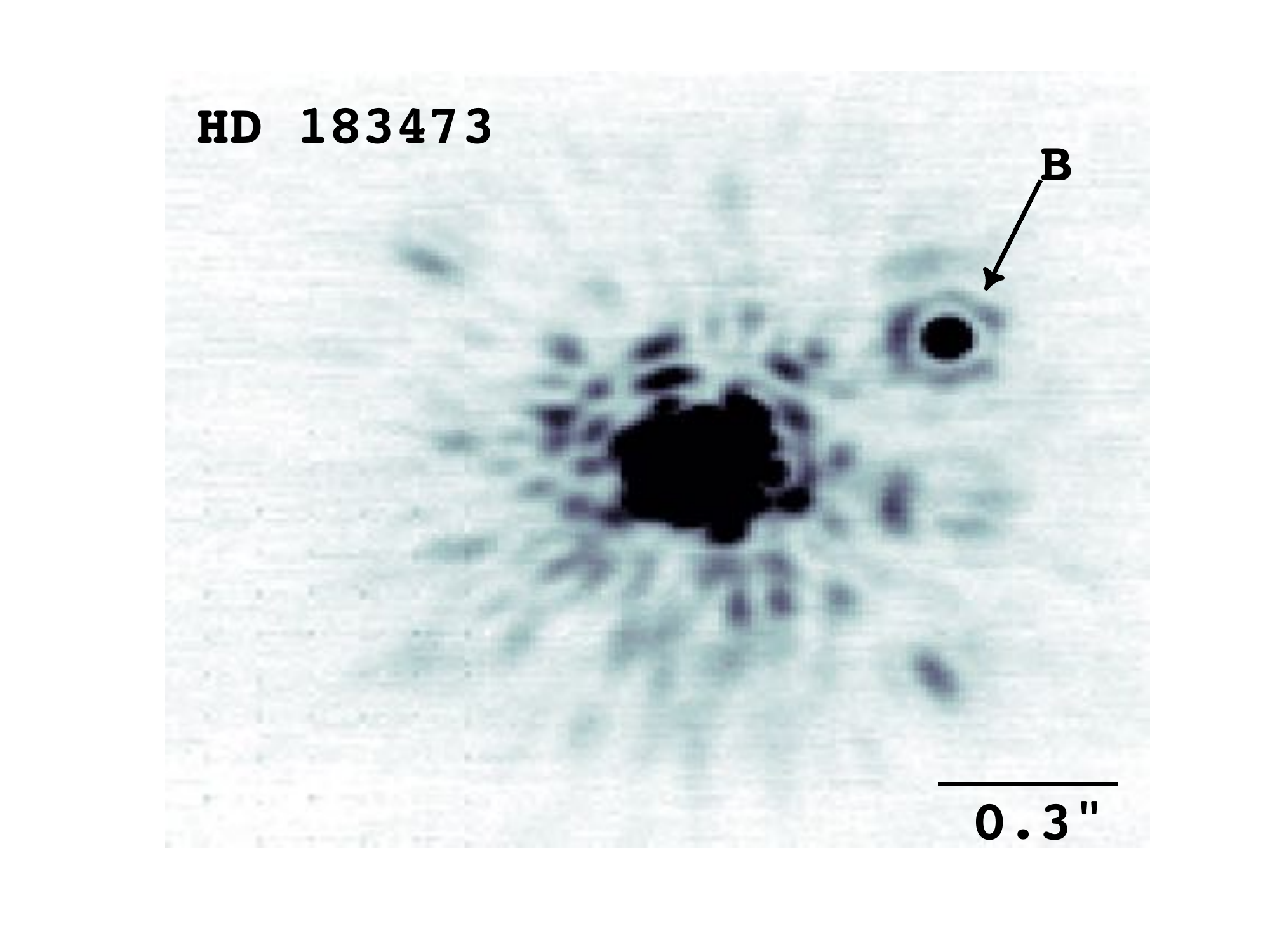}}
 \subfloat{\includegraphics[width= .5\textwidth]{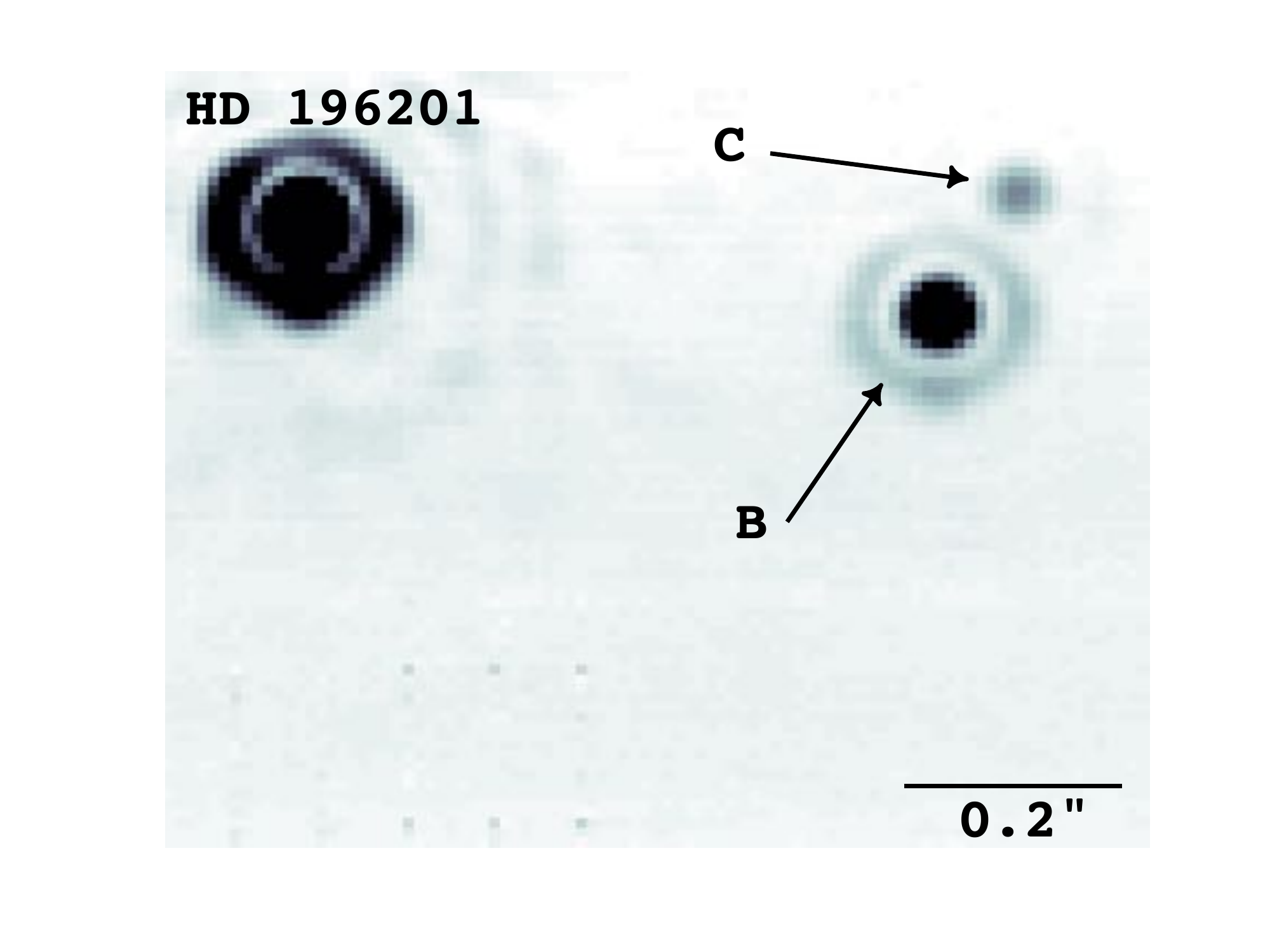}}}\\ \subfloat{\subfloat{\includegraphics[width= .5\textwidth]{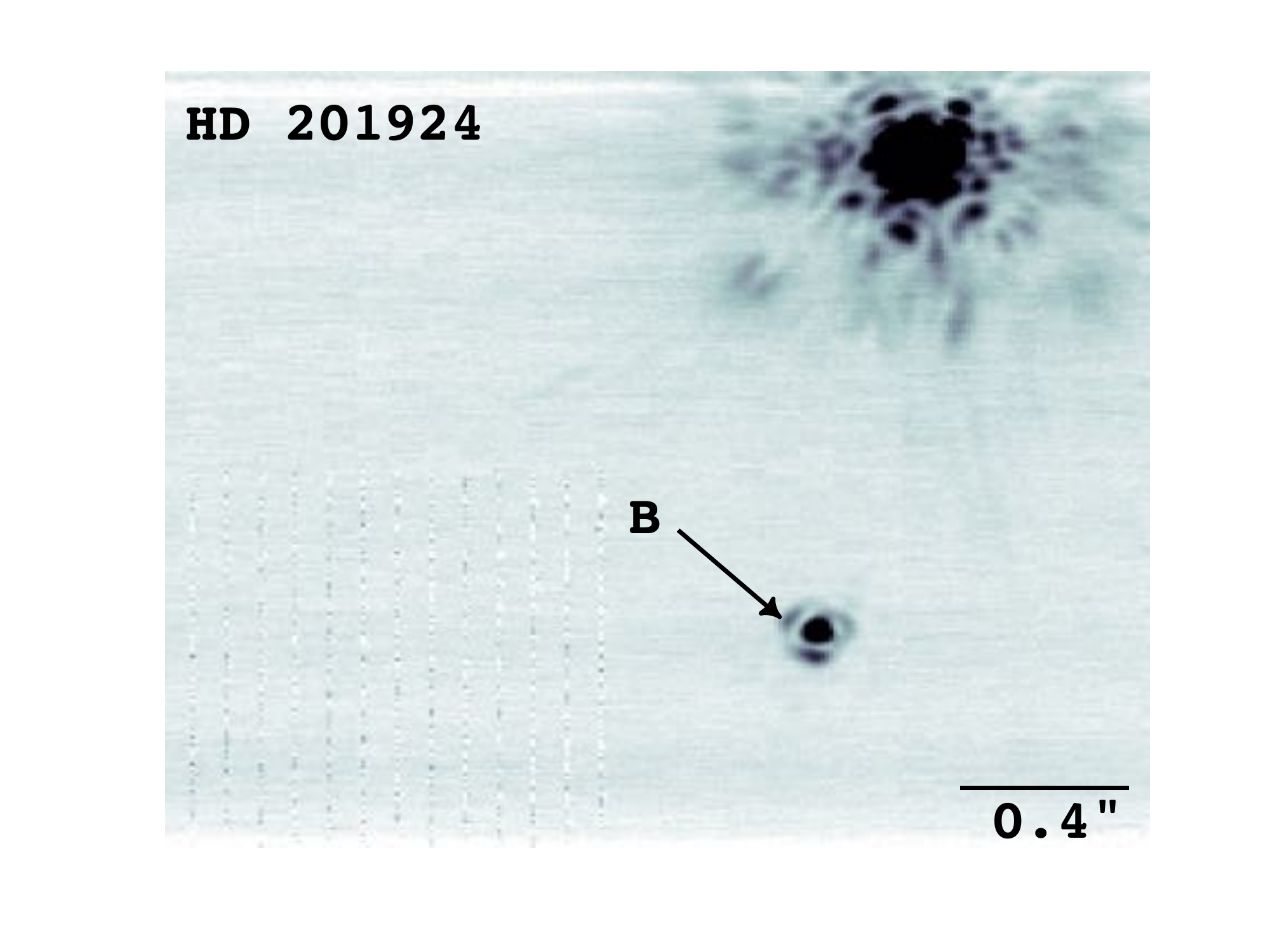}}\subfloat{\includegraphics[width= .5\textwidth]{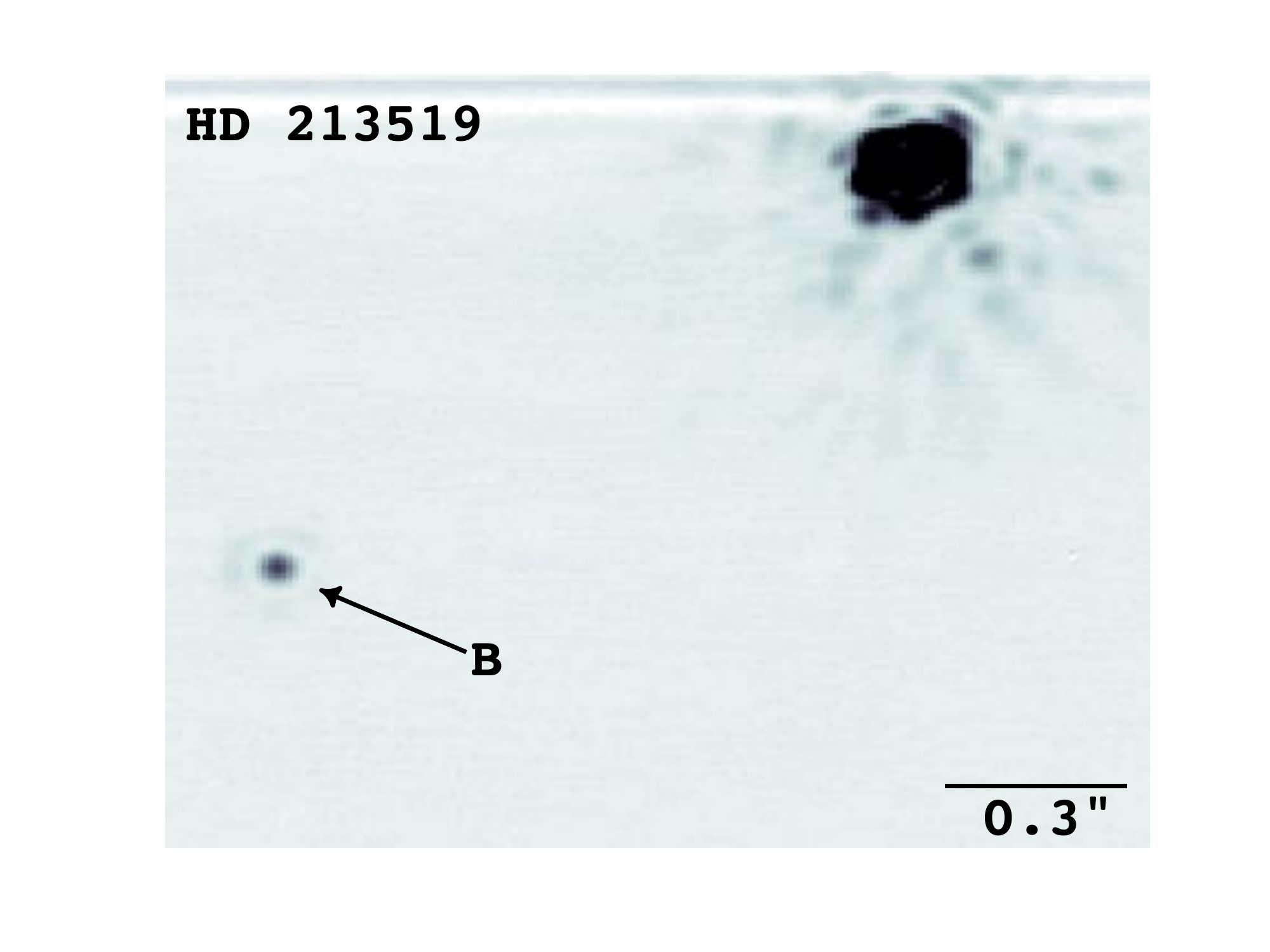}}}\\
 \caption{High-resolution adaptive optics images of confirmed companions. Continuation of \ref{fig16}.}
  \label{fig17}
\end{figure}

\begin{figure}[htp]
 \subfloat{\subfloat{\includegraphics[width = .5\textwidth]{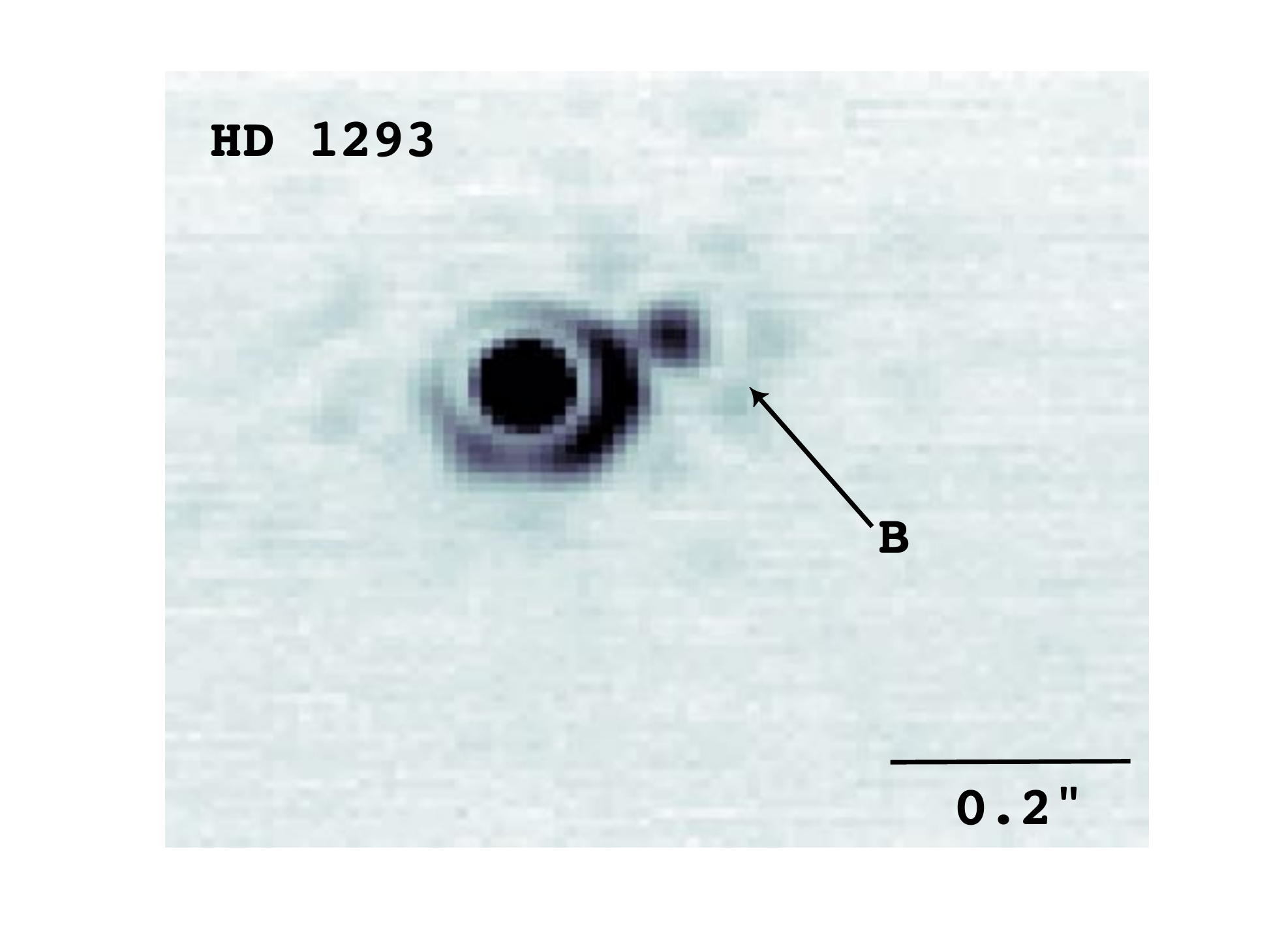}}
 \subfloat{\includegraphics[width= .5\textwidth]{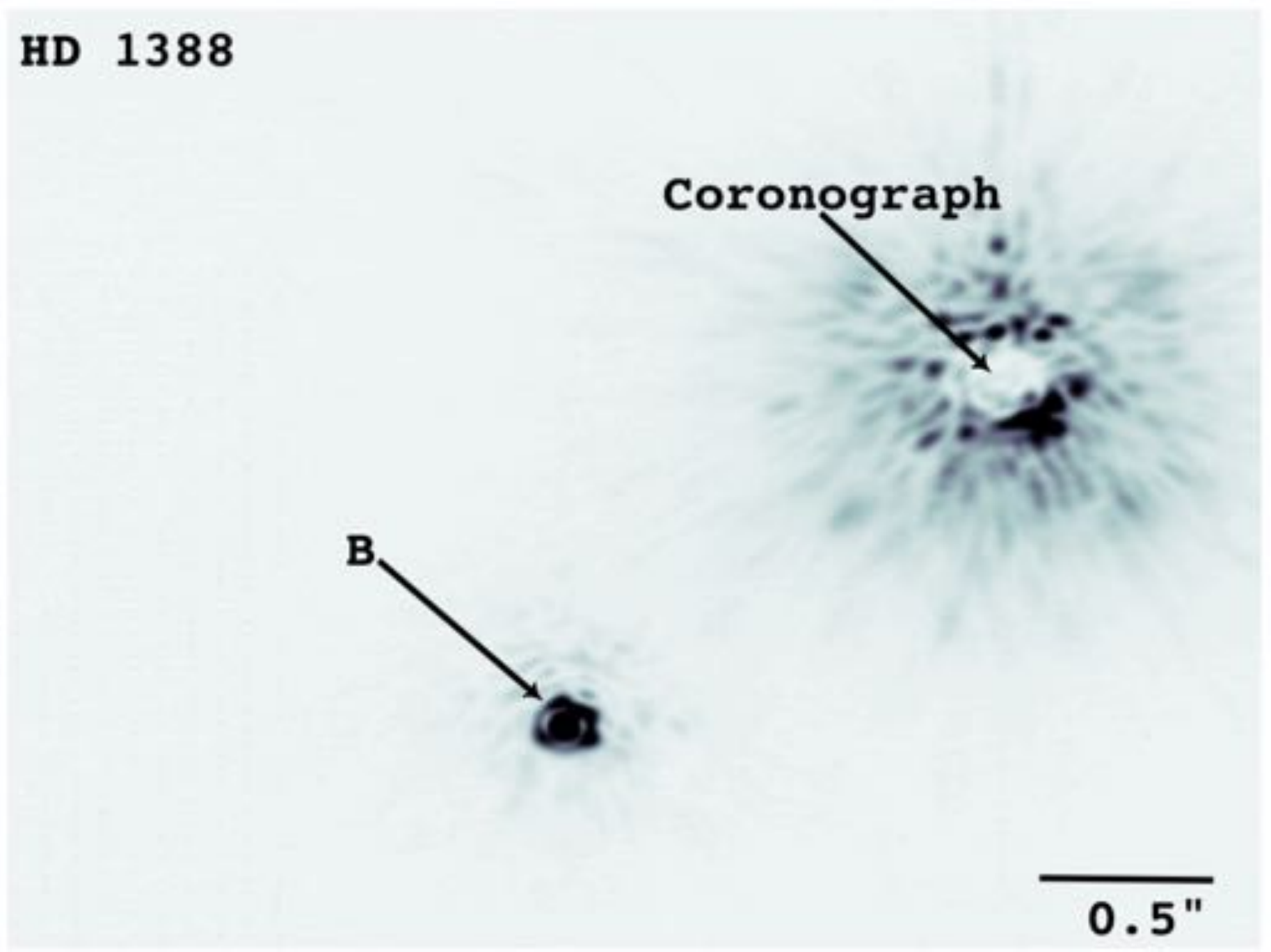}}}\\ \subfloat{\subfloat{\includegraphics[width= .5\textwidth]{4406_v2.pdf}}\subfloat{\includegraphics[width= .5\textwidth]{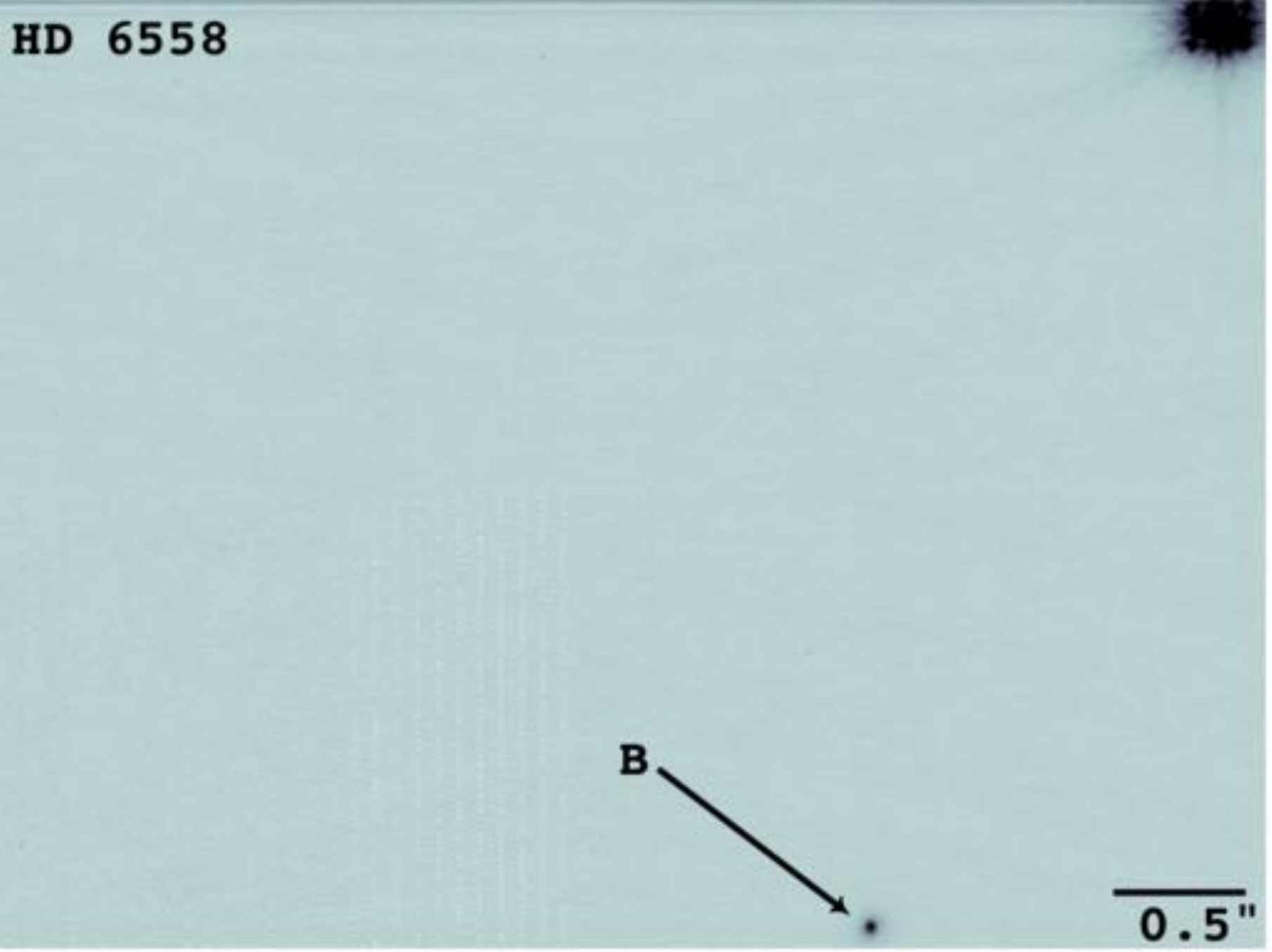}}}\\
  \subfloat{\subfloat{\includegraphics[width = .5\textwidth]{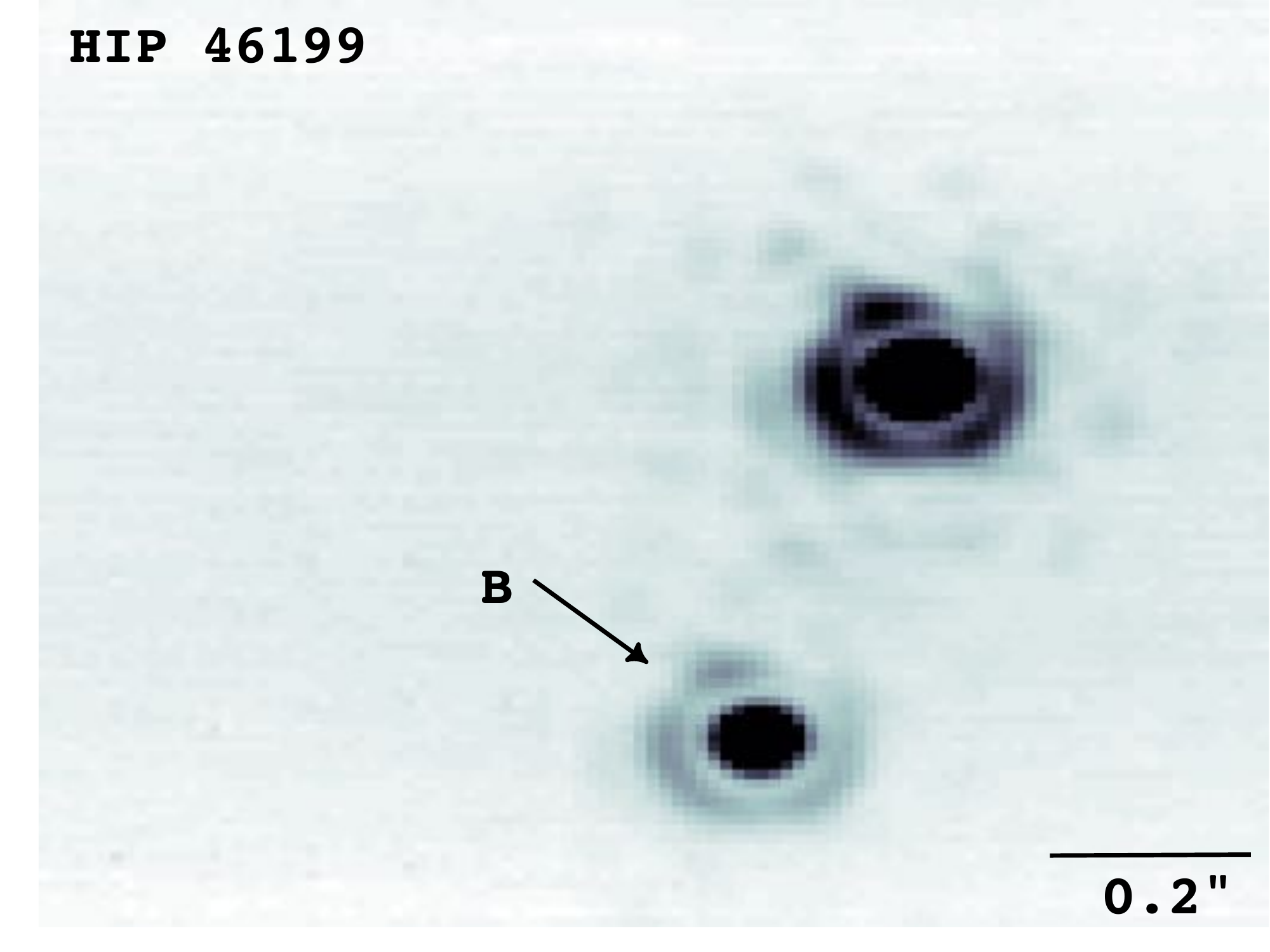}}
 \subfloat{\includegraphics[width= .5\textwidth]{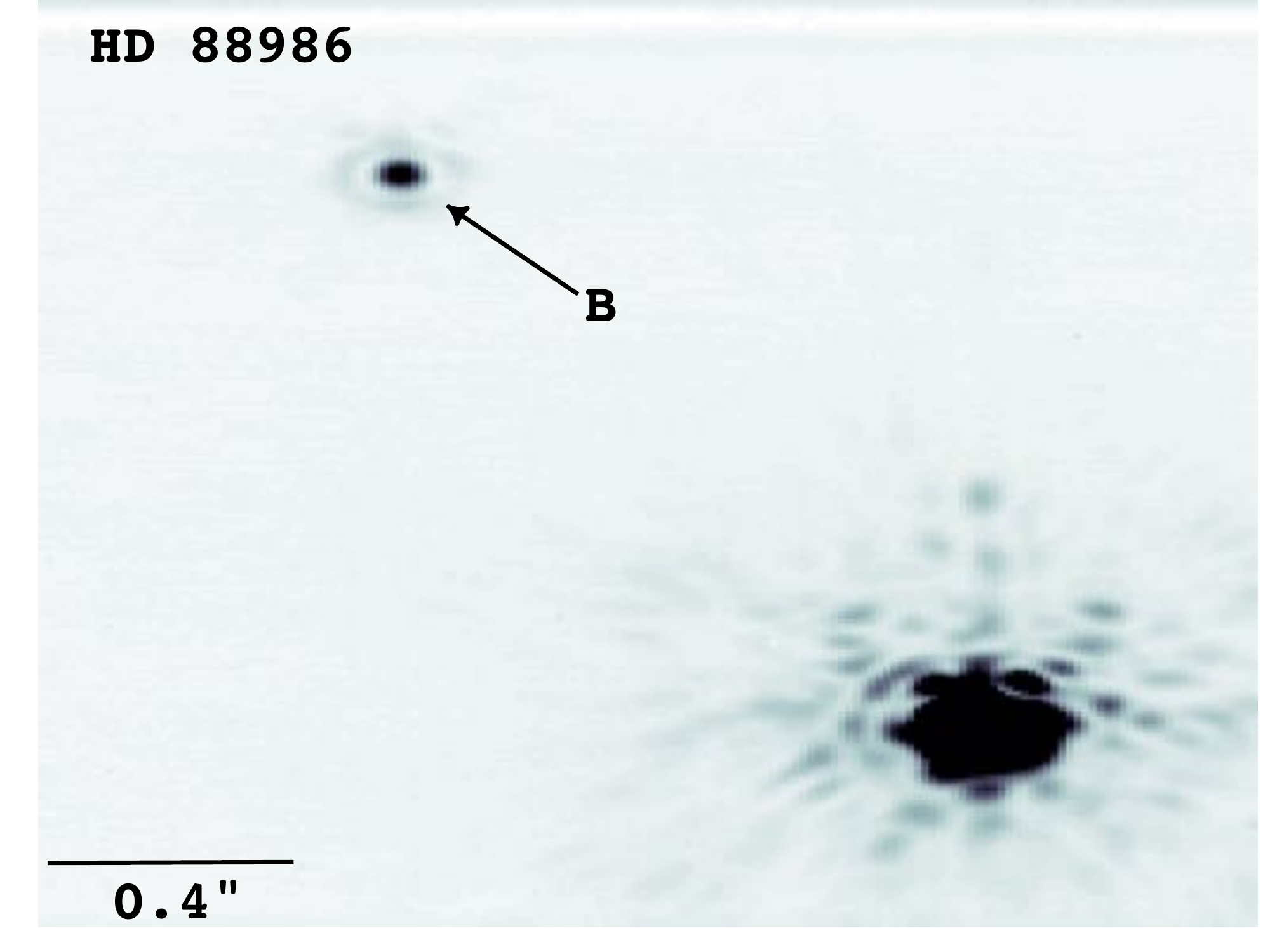}}}\\ 
 \caption{High-resolution adaptive optics images of candidate companions.}
  \label{fig18}
\end{figure}

\begin{figure}[htp]
 \subfloat{\subfloat{\includegraphics[width = .5\textwidth]{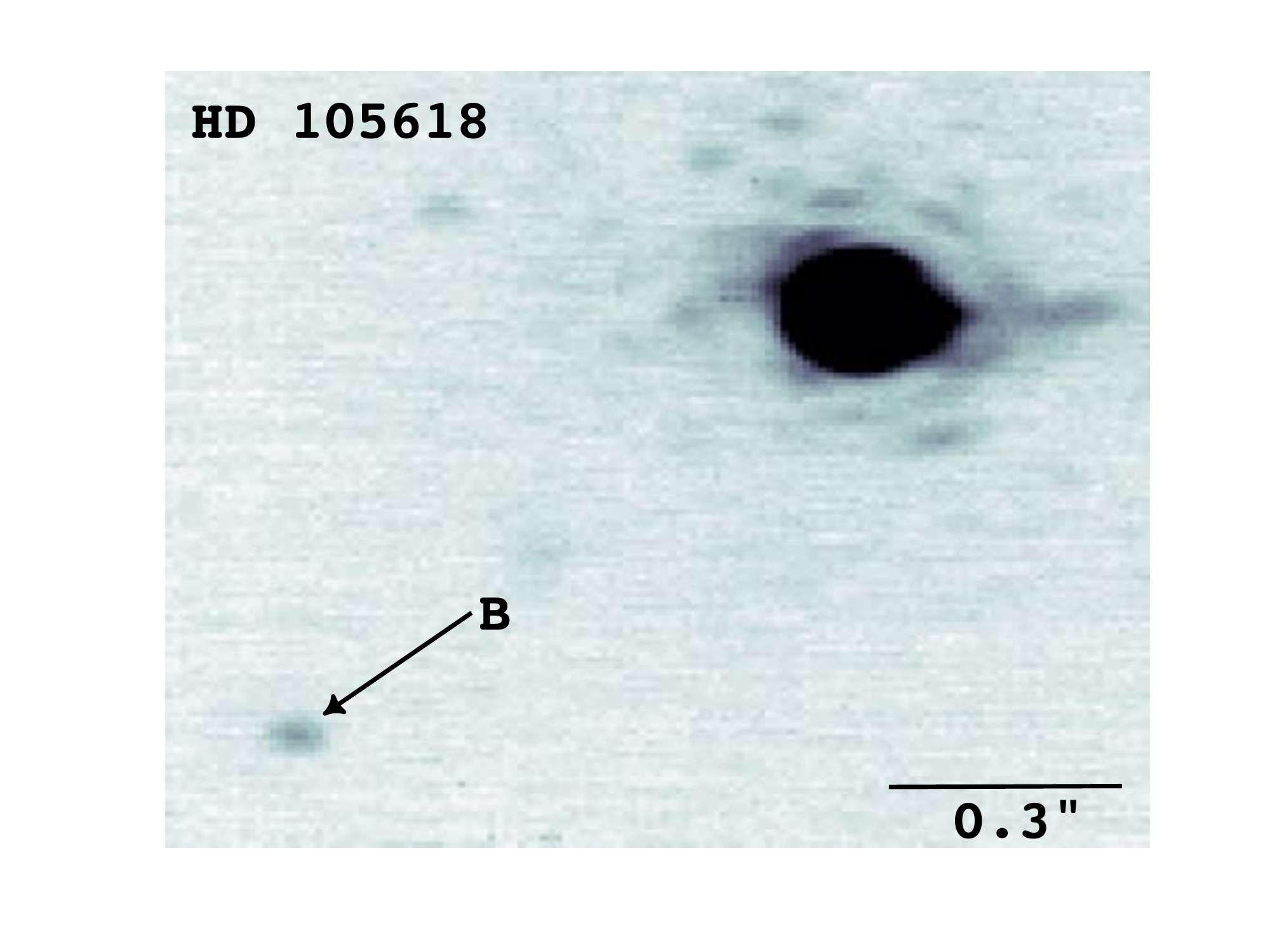}}
 \subfloat{\includegraphics[width= .5\textwidth]{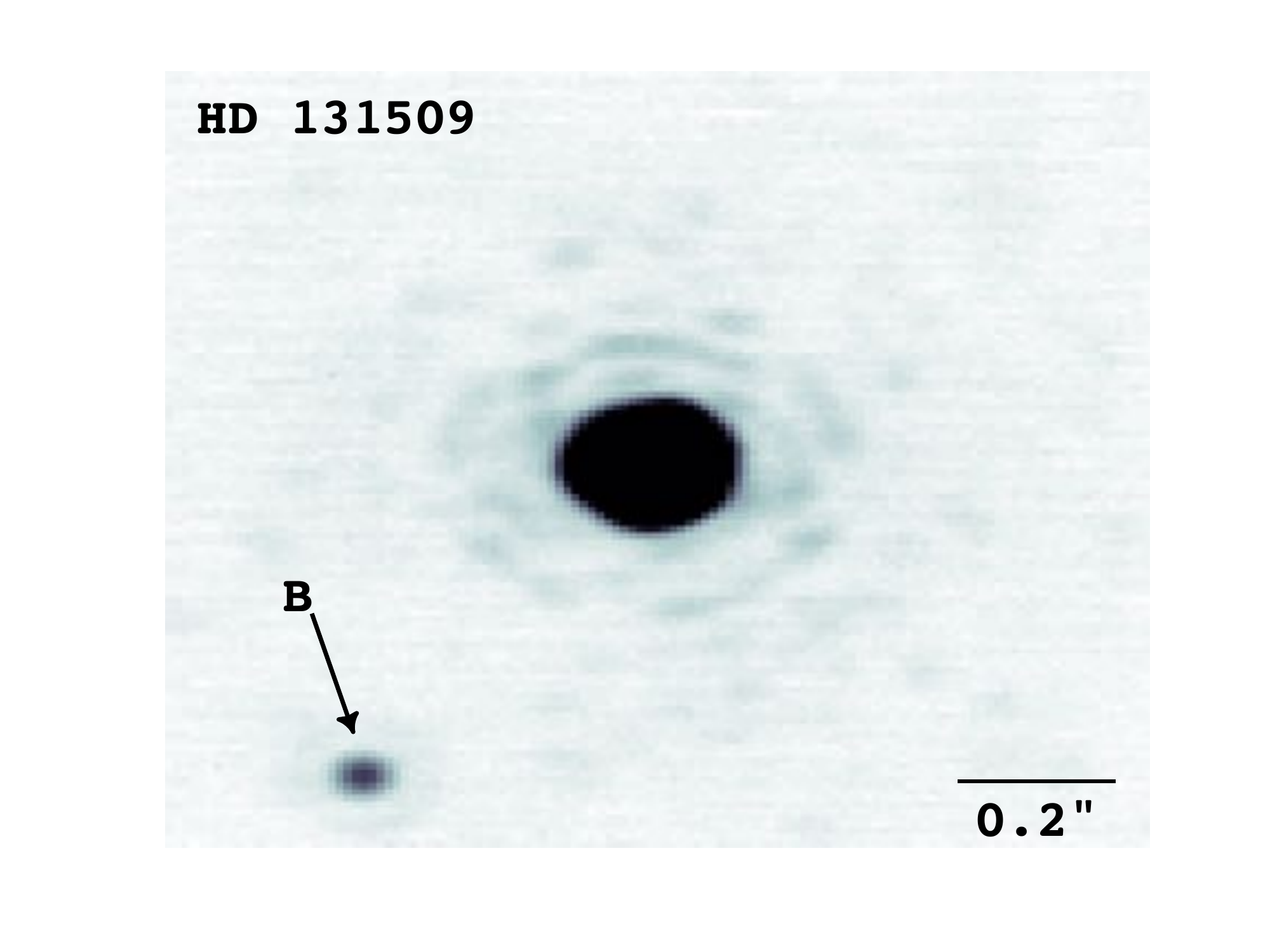}}}\\ \subfloat{\subfloat{\includegraphics[width= .5\textwidth]{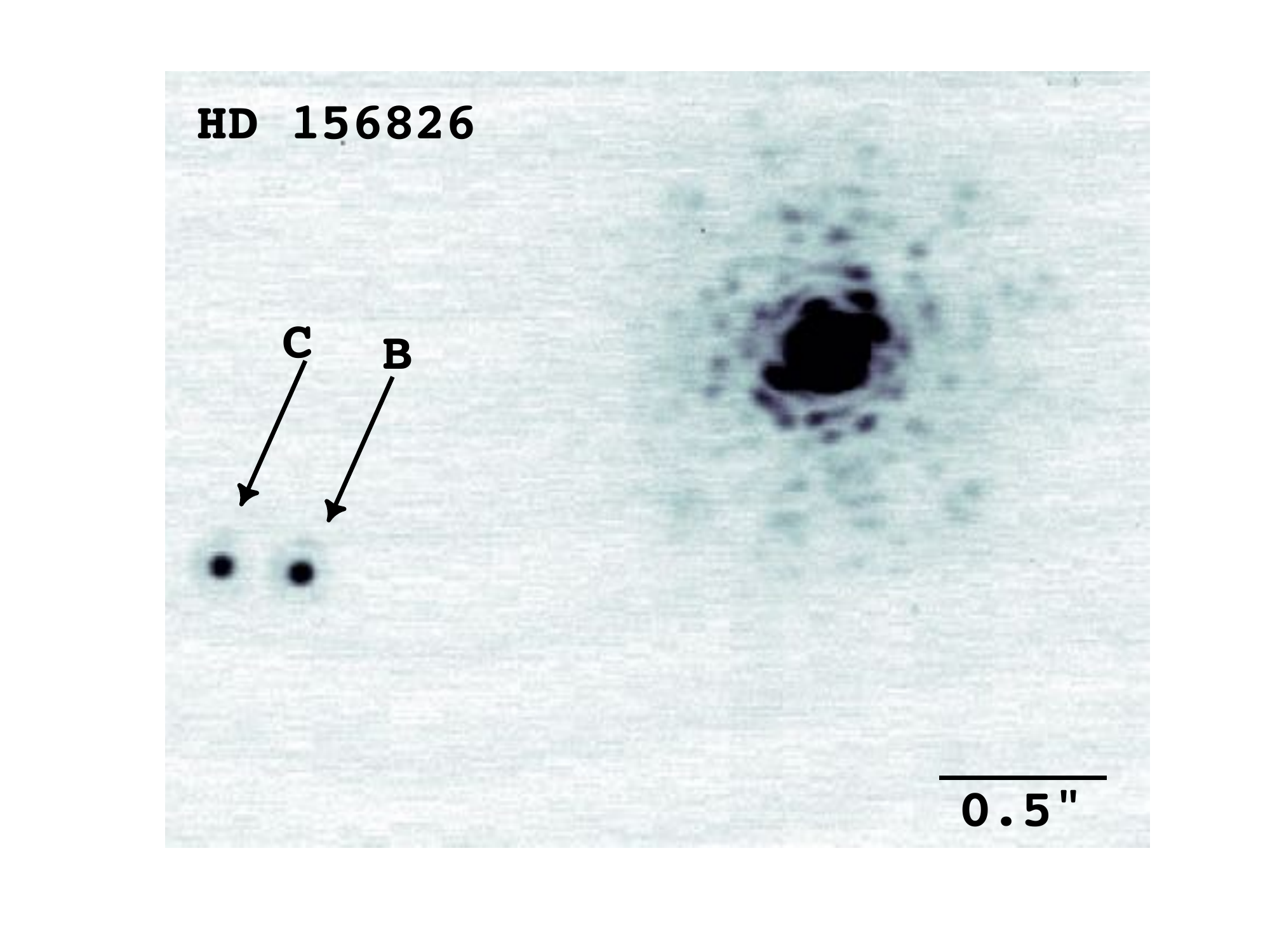}}\subfloat{\includegraphics[width= .5\textwidth]{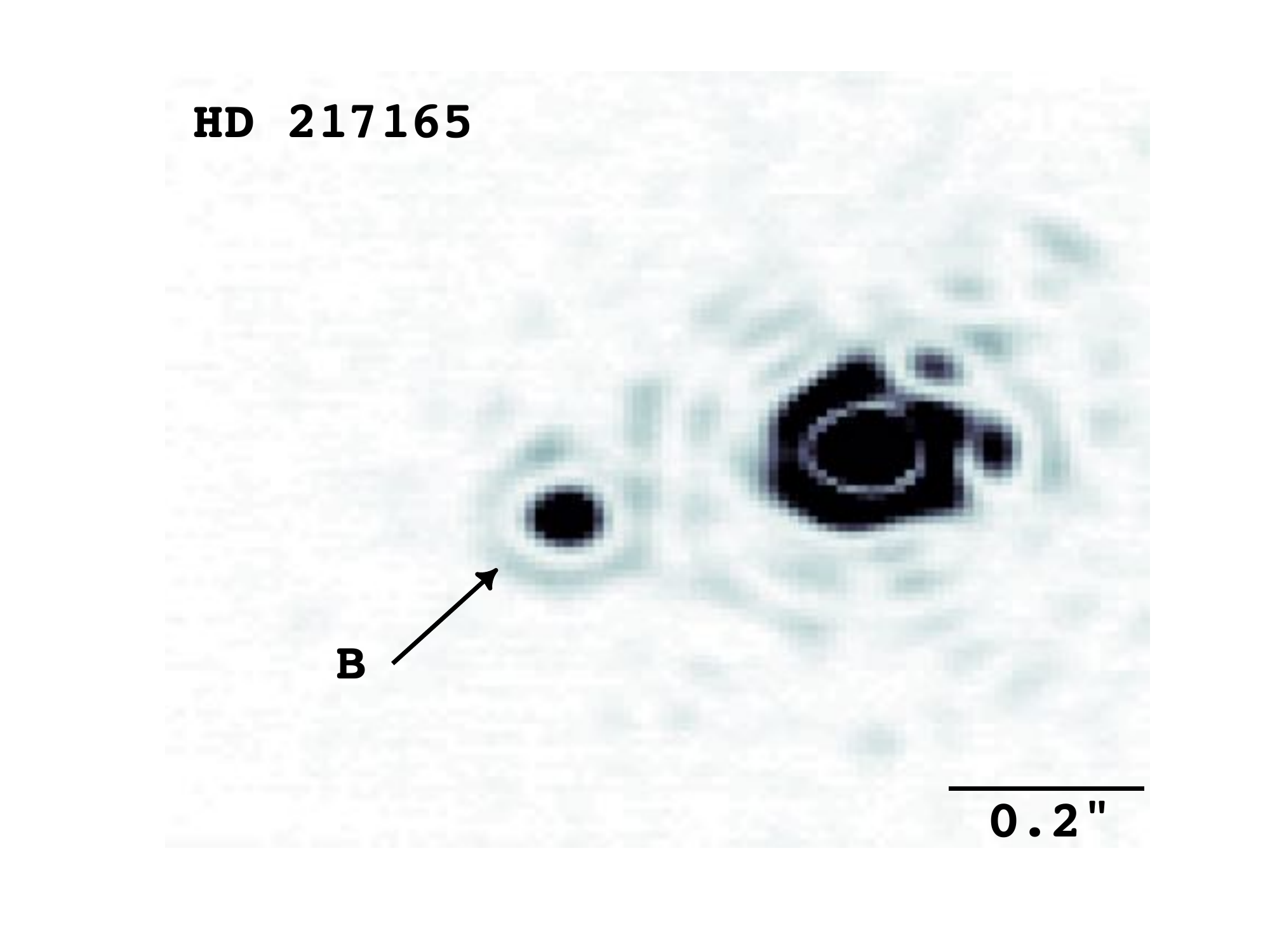}}}\\
 \caption{High-resolution adaptive optics images of candidate companions. Continuation of \ref{fig18}.}
  \label{fig19}
\end{figure}
\clearpage

\subsection{Common Proper Motion Plots of Confirmed Companions.}

\begin{figure}[htp]
\subfloat{\subfloat{\includegraphics[width = .5\textwidth]{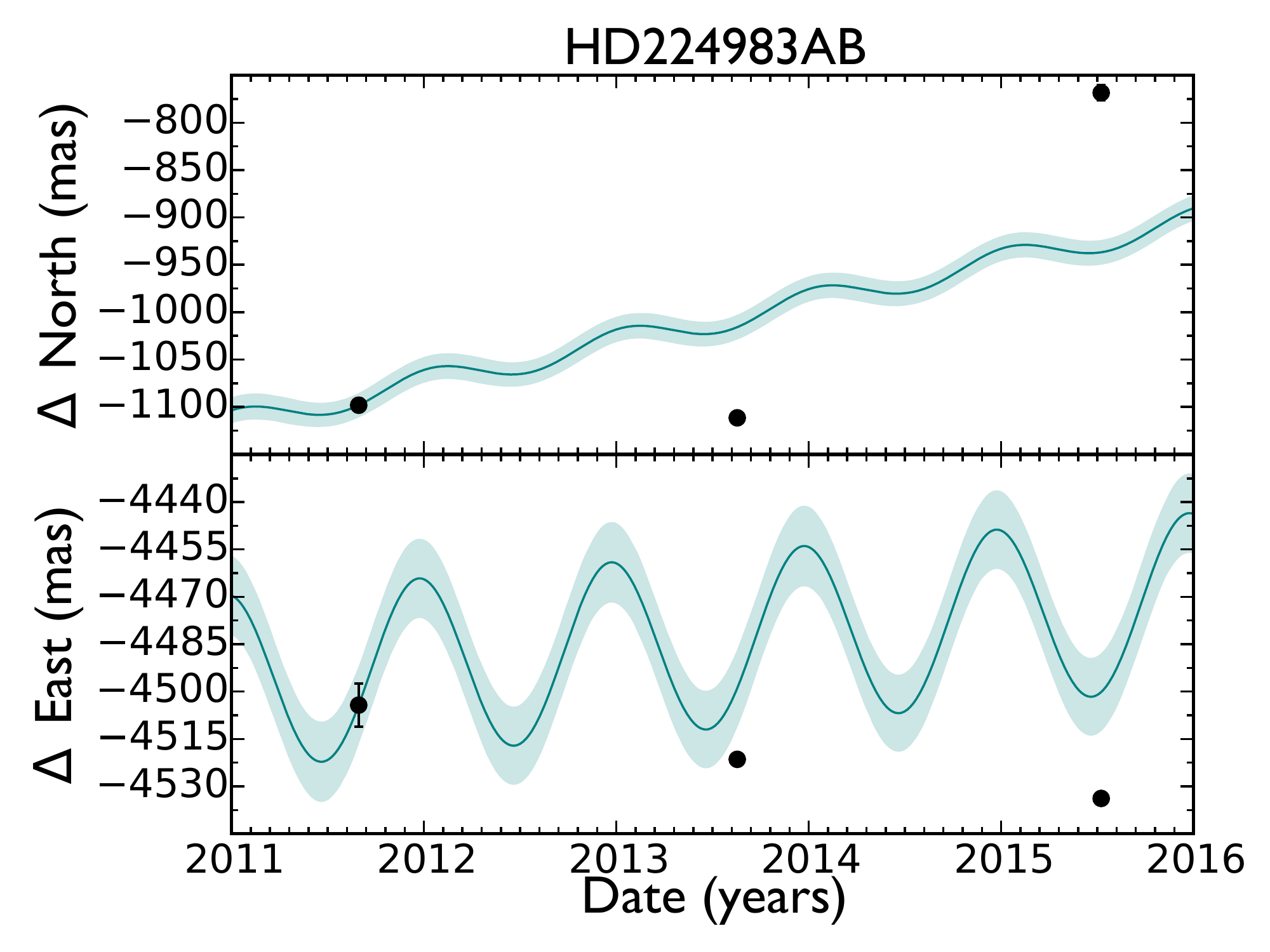}}
\subfloat{\includegraphics[width =.5\textwidth]{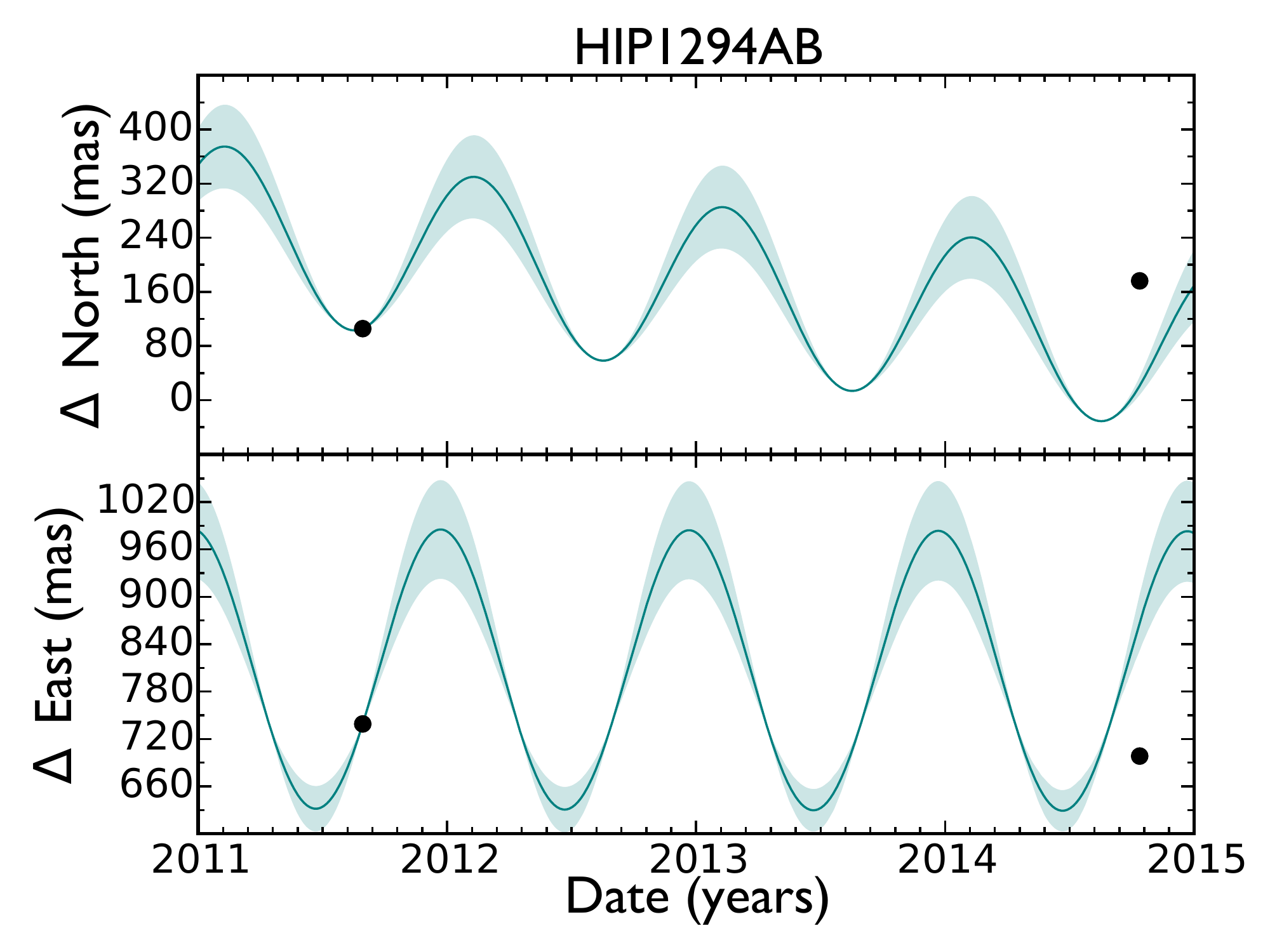}}}\\
\subfloat{\subfloat{\includegraphics[width =.5\textwidth ]{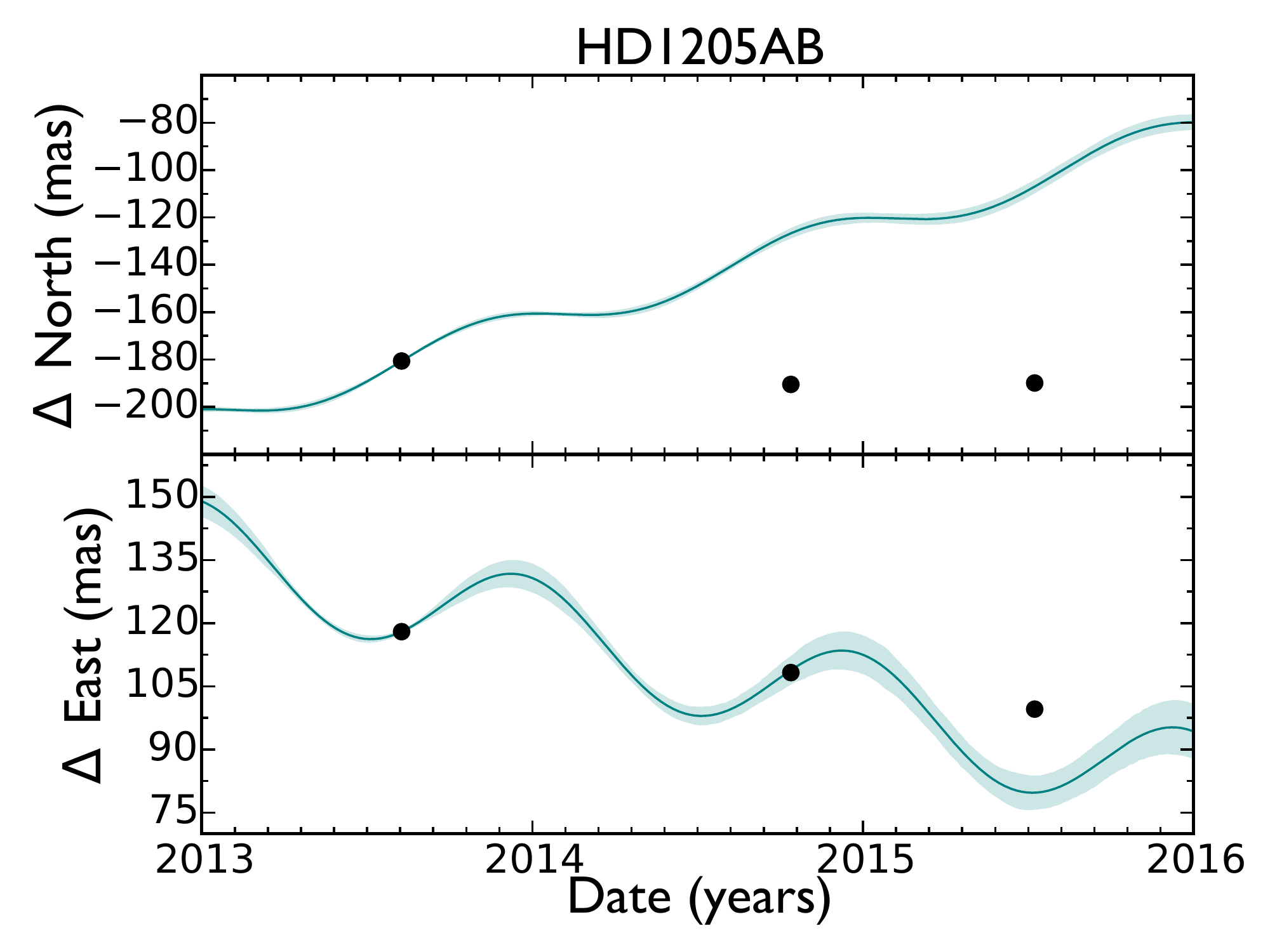}}
\subfloat{\includegraphics[width =.5\textwidth]{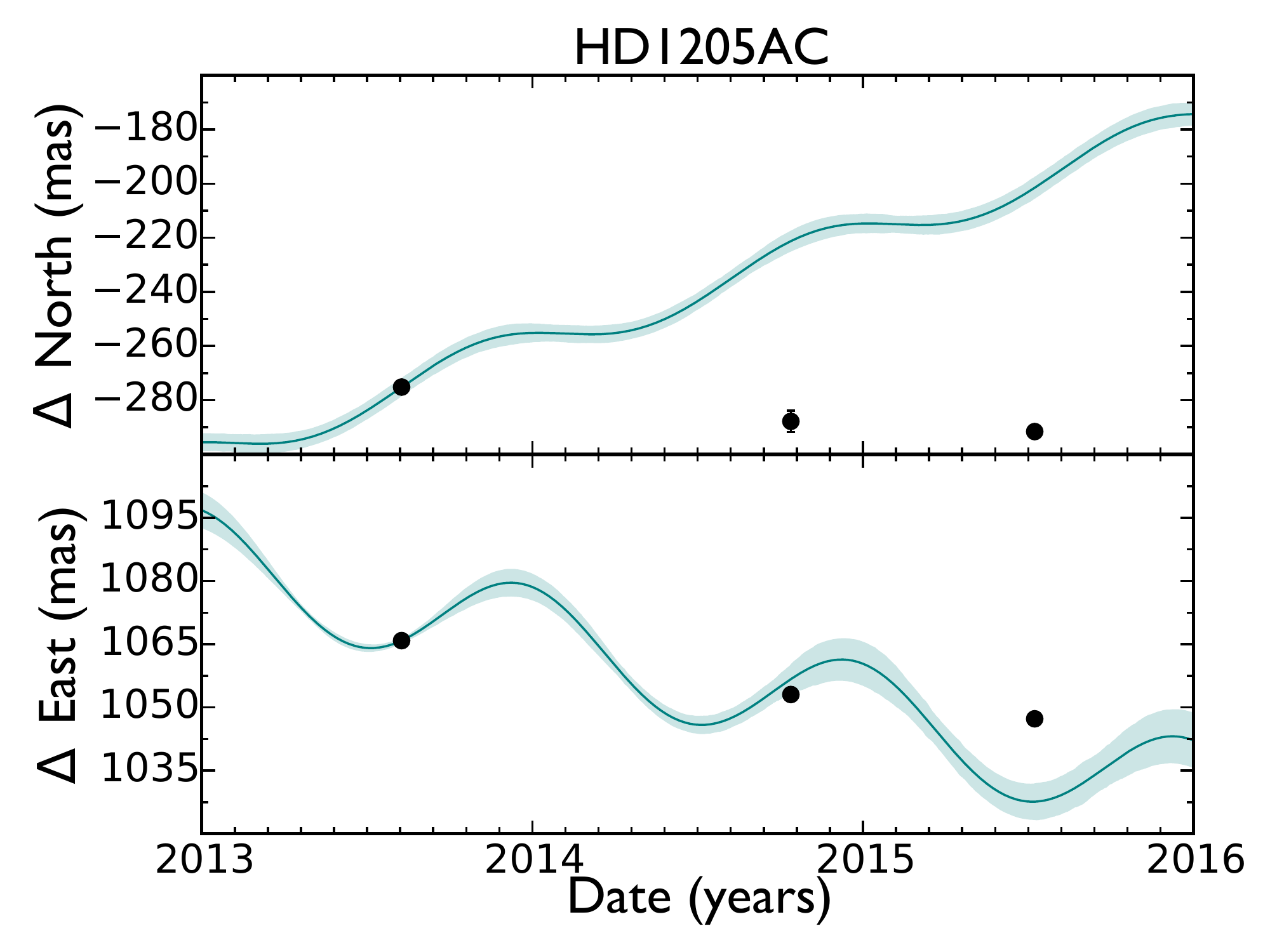}}}\\
\subfloat{\subfloat{\includegraphics[width =.5\textwidth ]{HD1384AB_dNdE.pdf}}
\subfloat{\includegraphics[width =.5\textwidth]{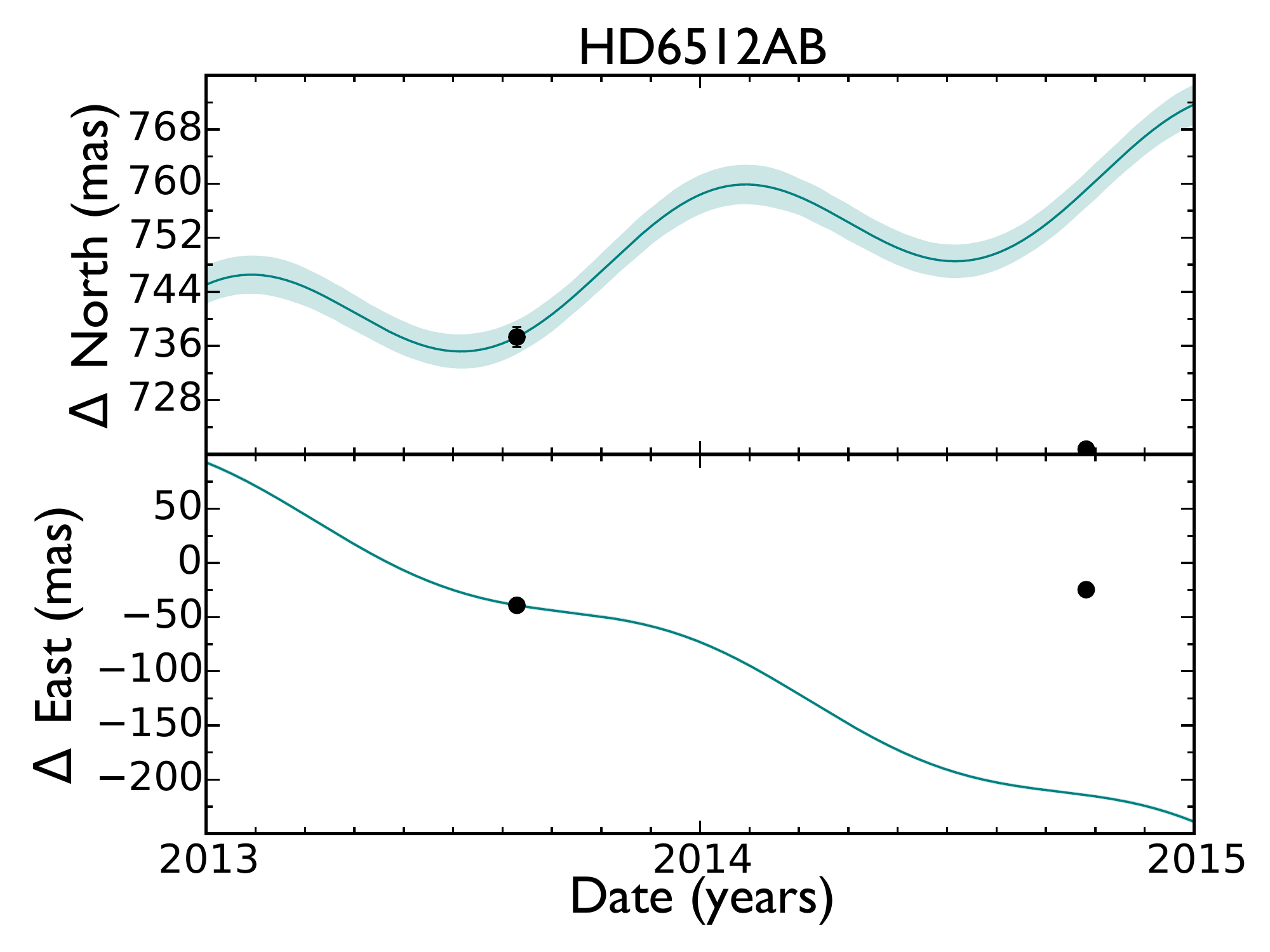}}}\\
\caption{\textbf{Confirmation of Co-movement of Companion:} — The curved lines are the path taken of an infinitely distant background object. Plotted points are the astrometric measurements of the companion. Given that the measurements do not fall along the path of a background object, the companion shares the same space motion and is gravitationally associated.} 
\label{fig:20}
\end{figure}

\begin{figure}[htp]
\subfloat{\subfloat{\includegraphics[width = .5\textwidth]{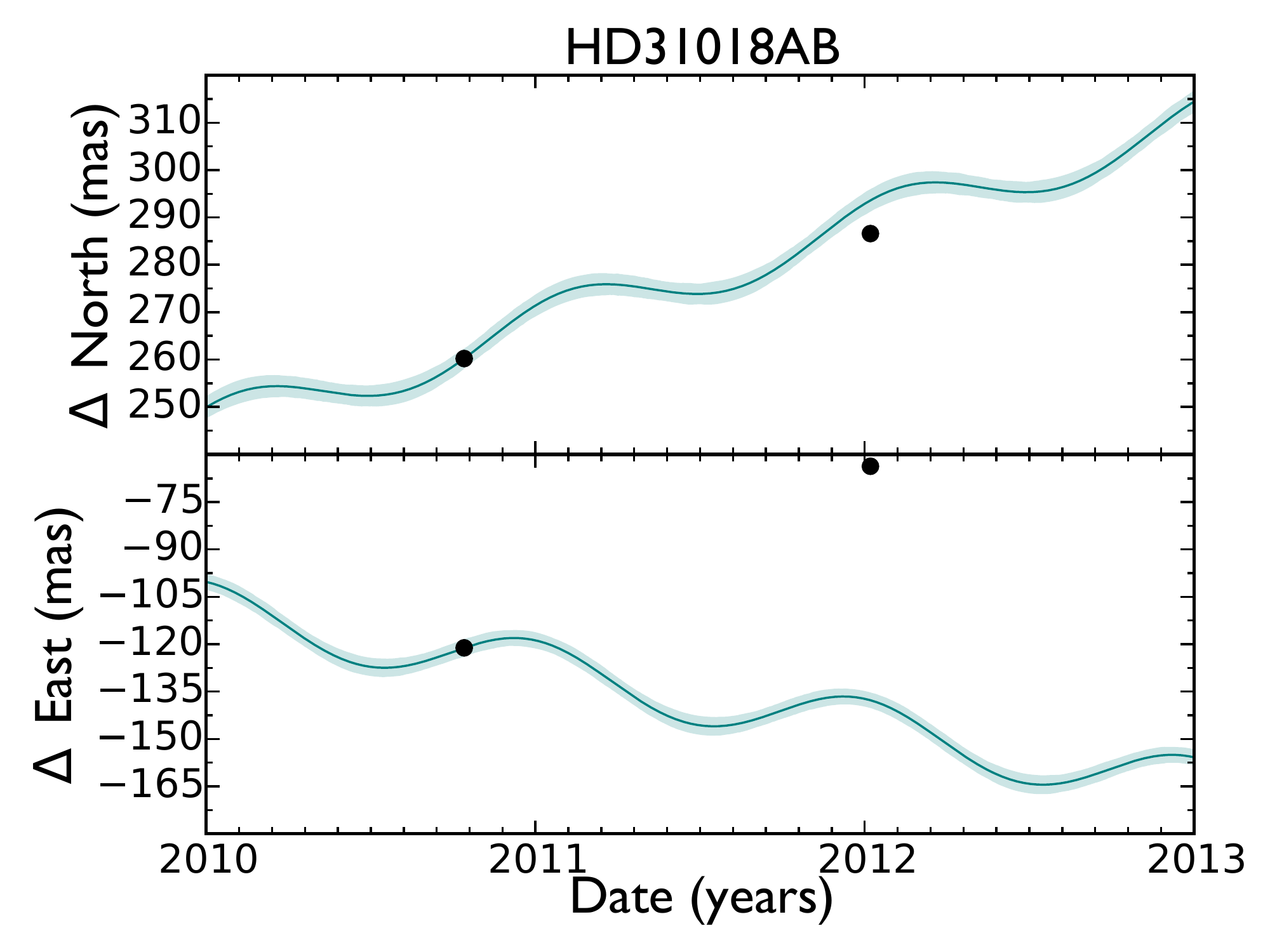}}
\subfloat{\includegraphics[width =.5\textwidth]{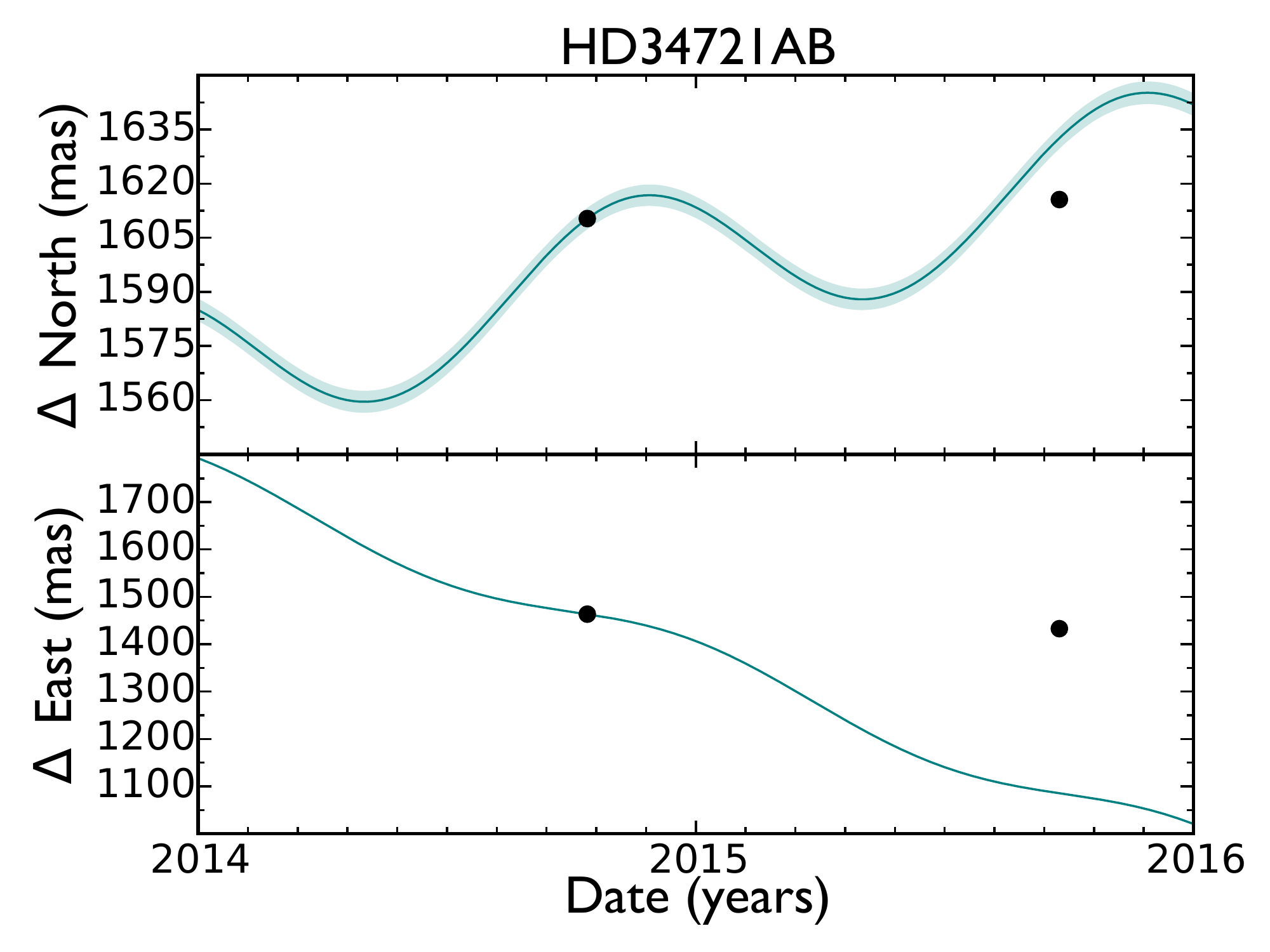}}}\\
\subfloat{\subfloat{\includegraphics[width =.5\textwidth ]{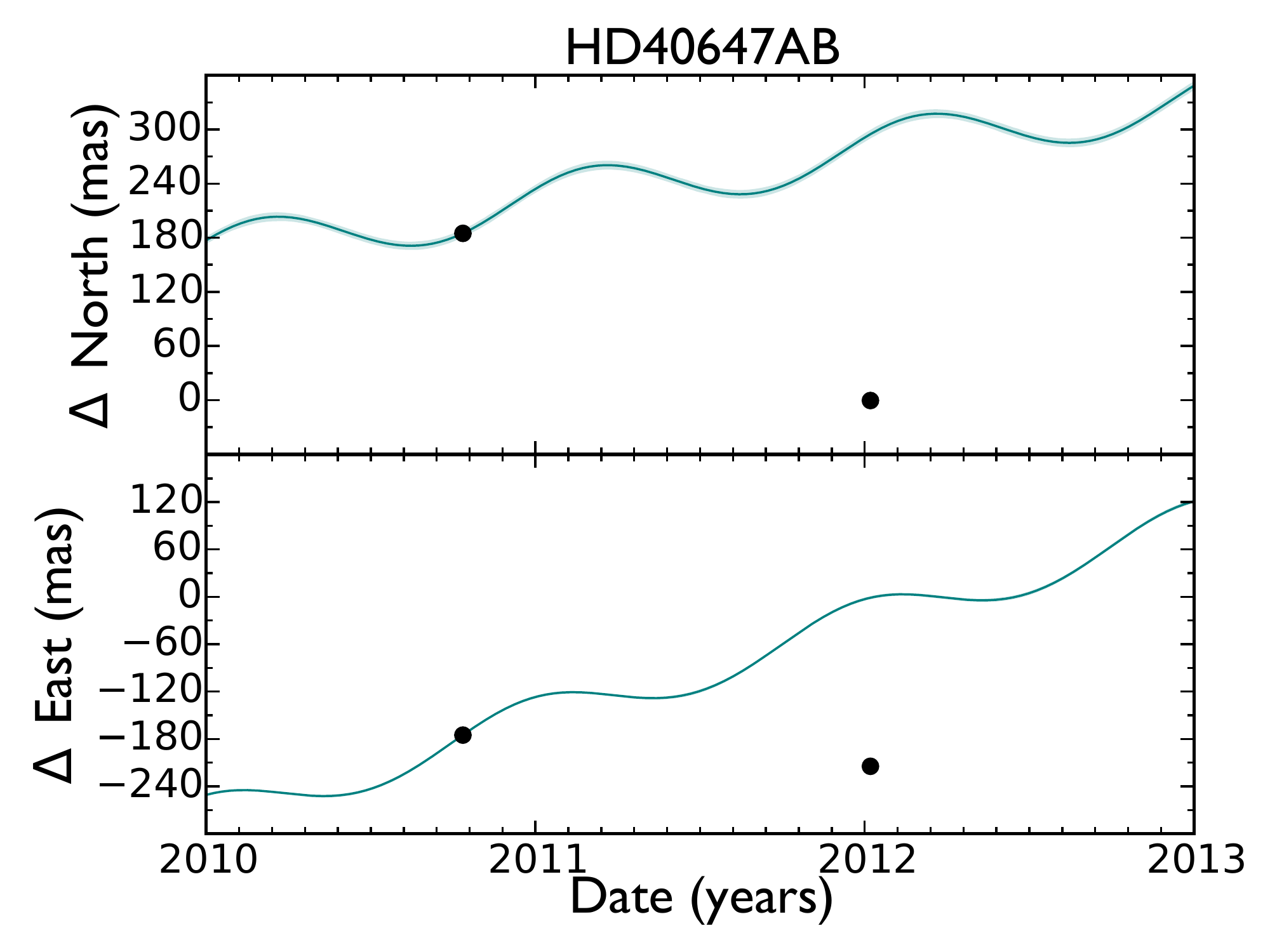}}
\subfloat{\includegraphics[width =.5\textwidth]{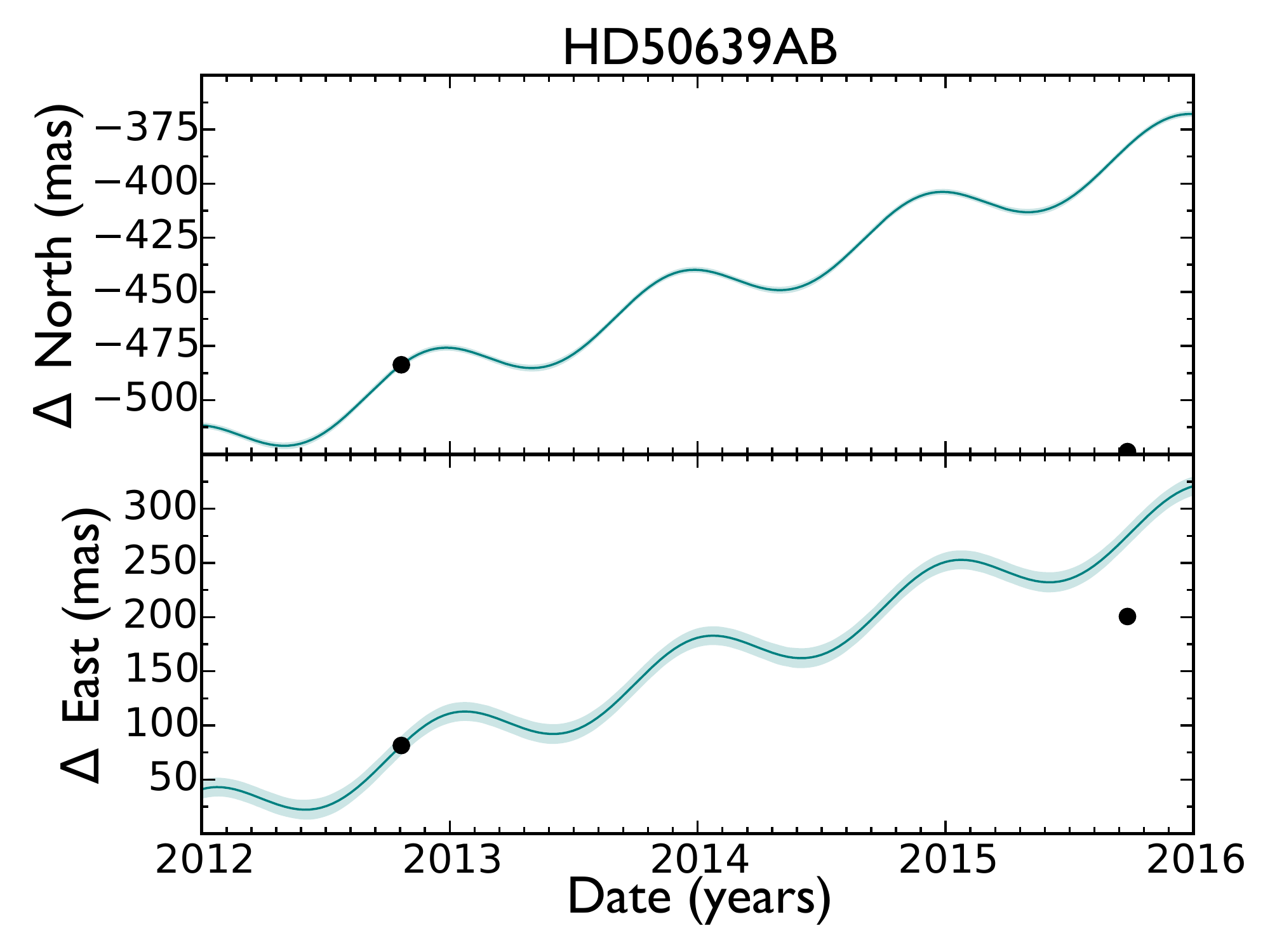}}}\\
\subfloat{\subfloat{\includegraphics[width =.5\textwidth ]{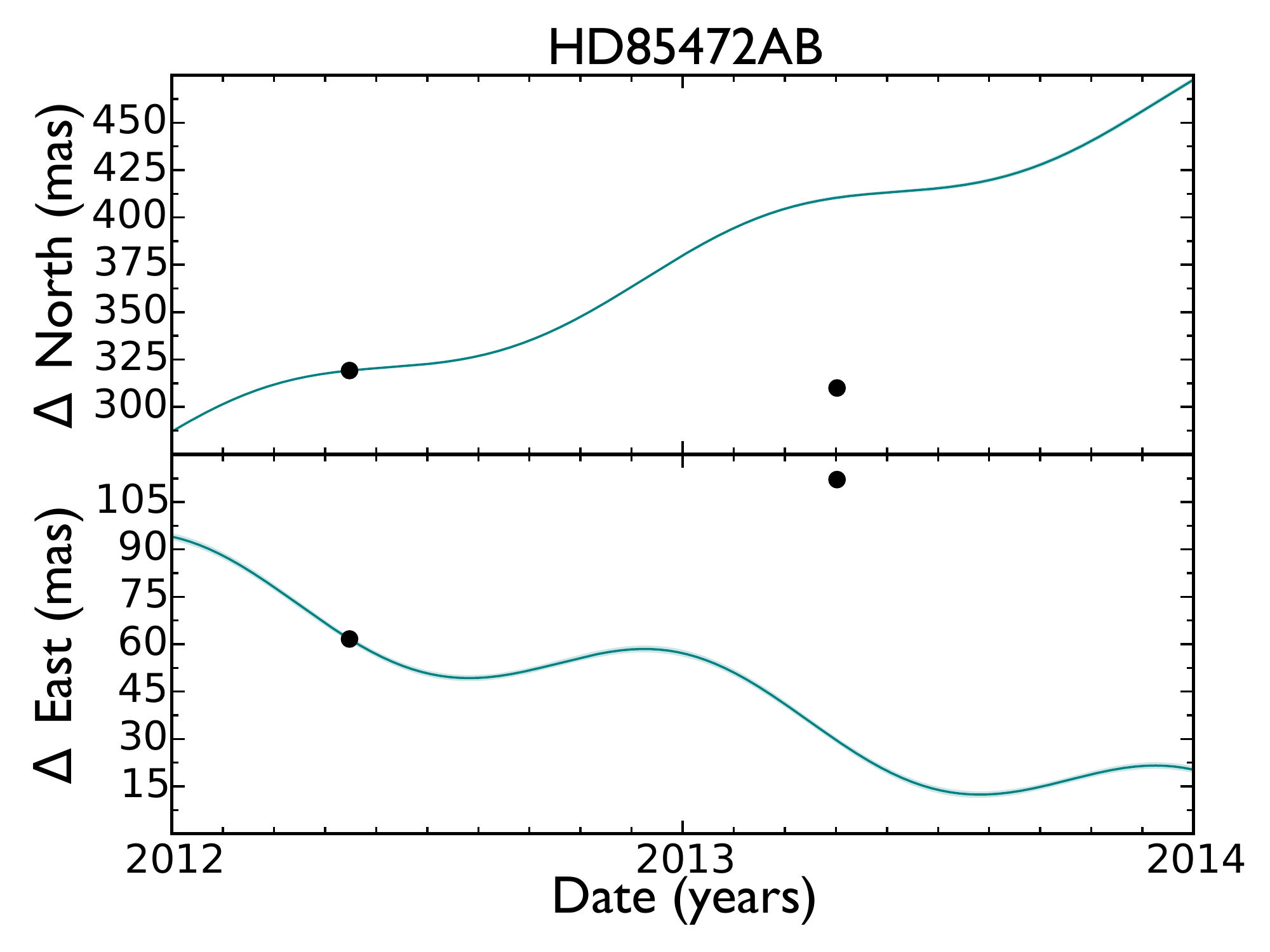}}
\subfloat{\includegraphics[width =.5\textwidth]{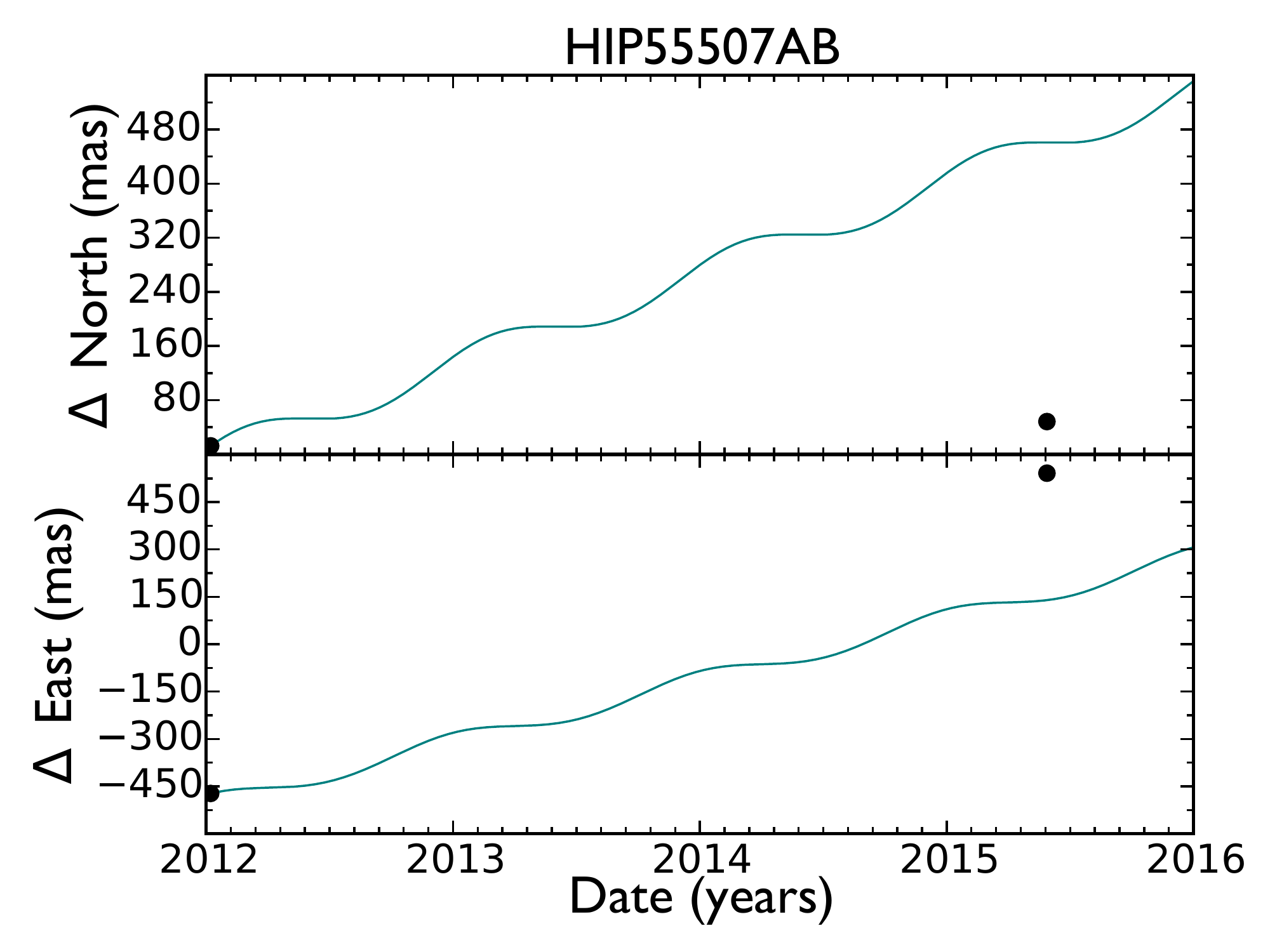}}}\\
\caption{Continuation of Figure \ref{fig:20}.} 
\label{fig:21}
\end{figure}

\begin{figure}[htp]
\subfloat{\subfloat{\includegraphics[width = .5\textwidth]{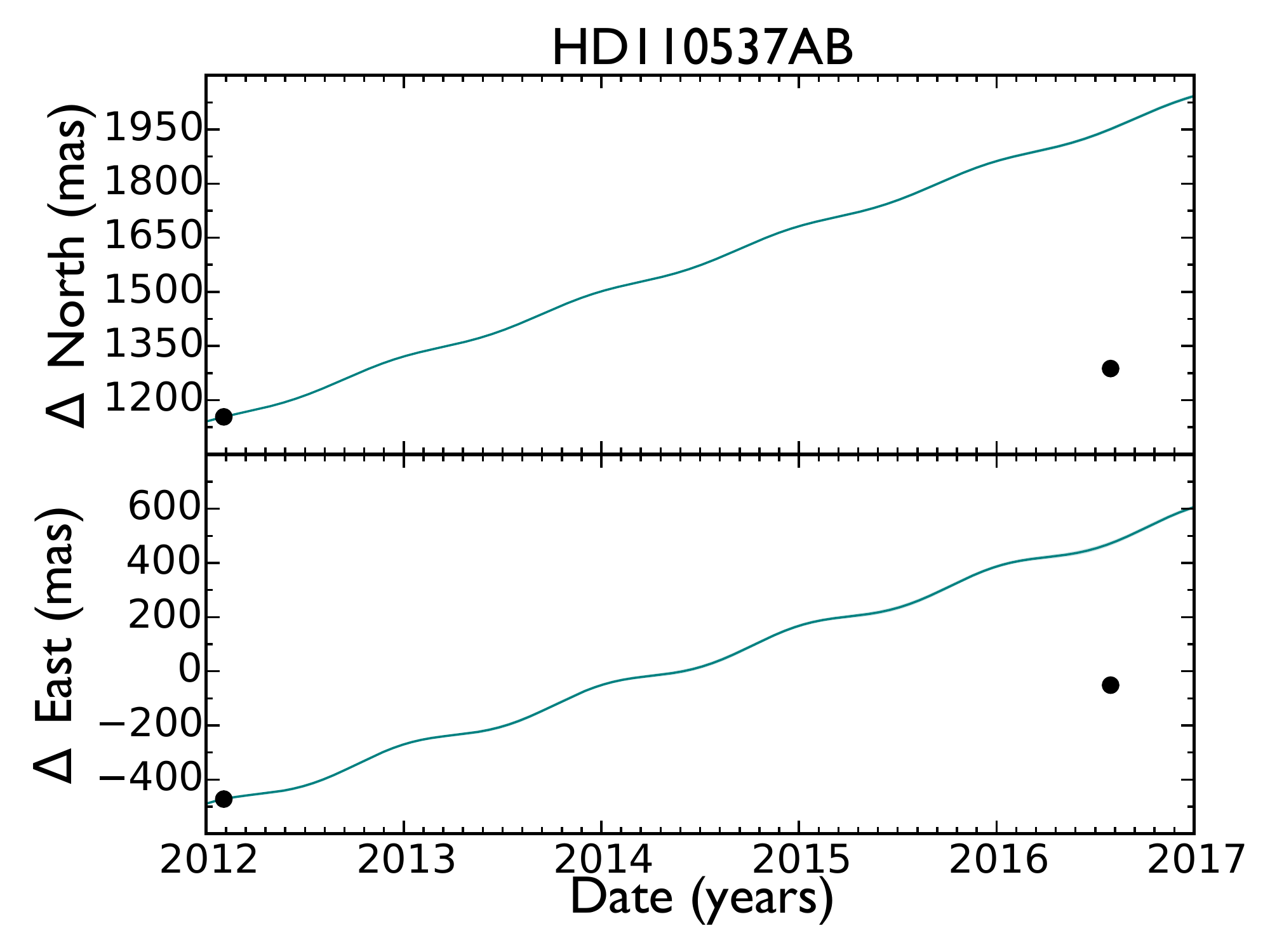}}
\subfloat{\includegraphics[width =.5\textwidth]{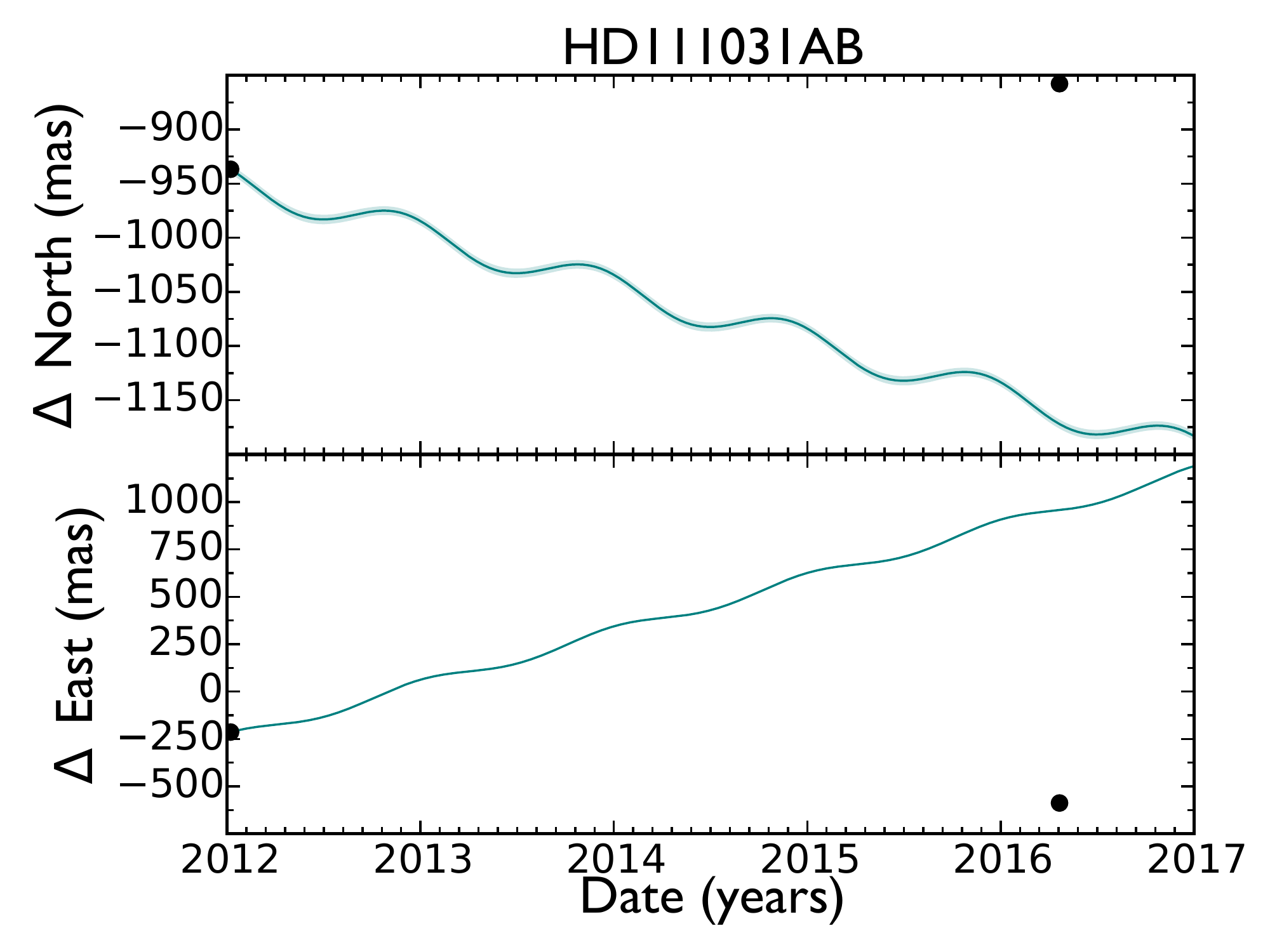}}}\\
\subfloat{\subfloat{\includegraphics[width =.5\textwidth ]{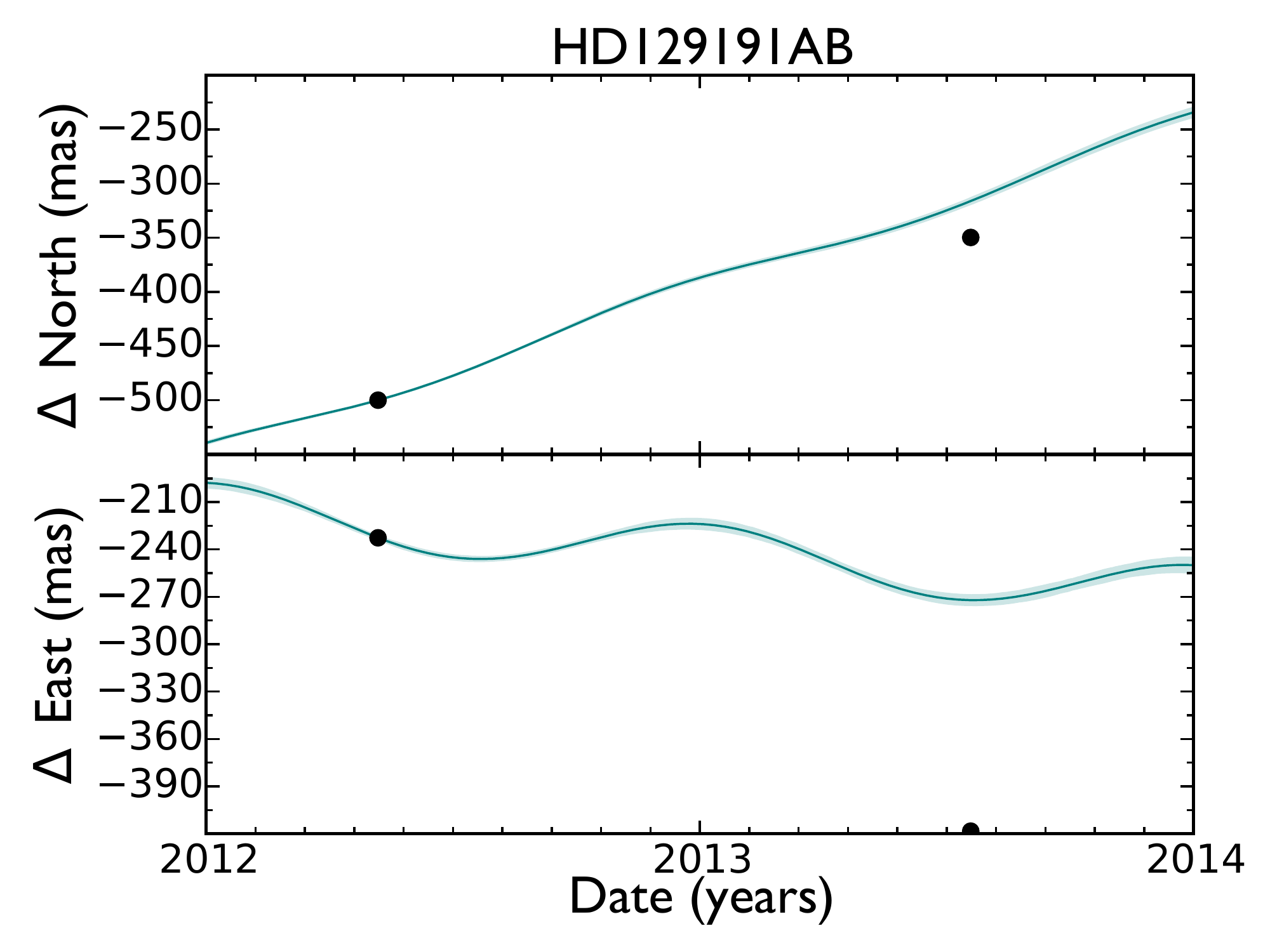}}
\subfloat{\includegraphics[width =.5\textwidth]{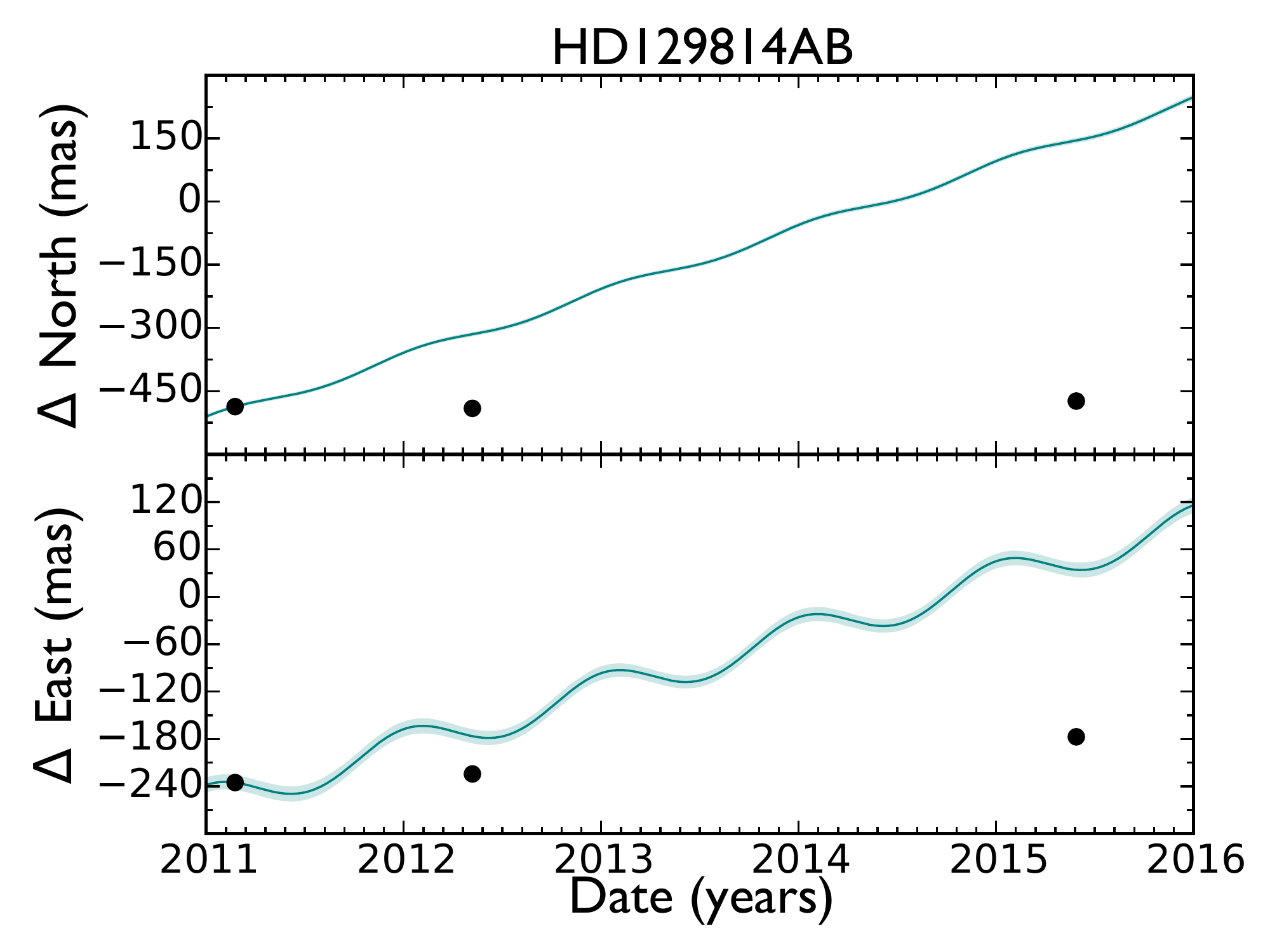}}}\\
\subfloat{\subfloat{\includegraphics[width =.5\textwidth ]{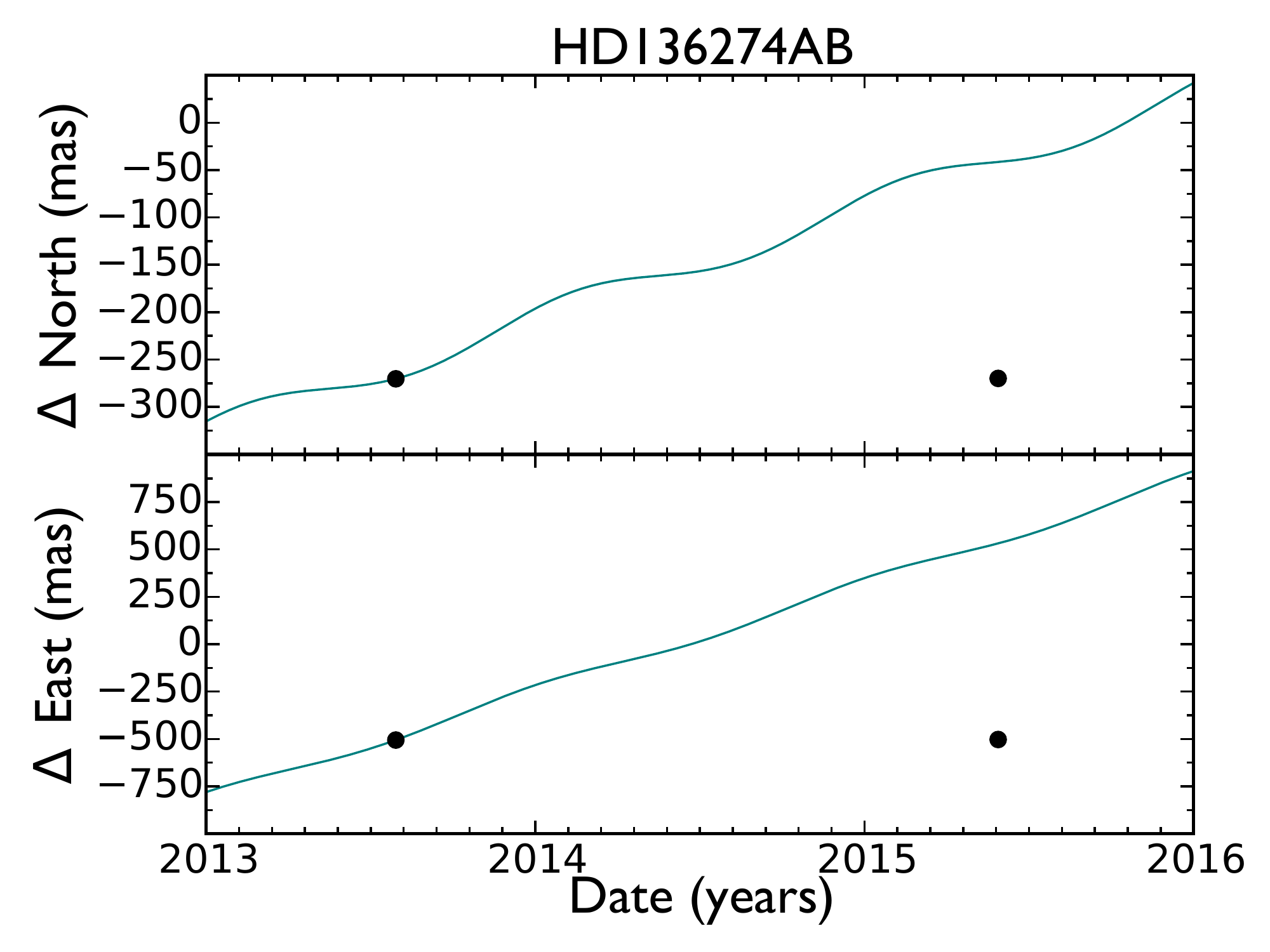}}
\subfloat{\includegraphics[width =.5\textwidth]{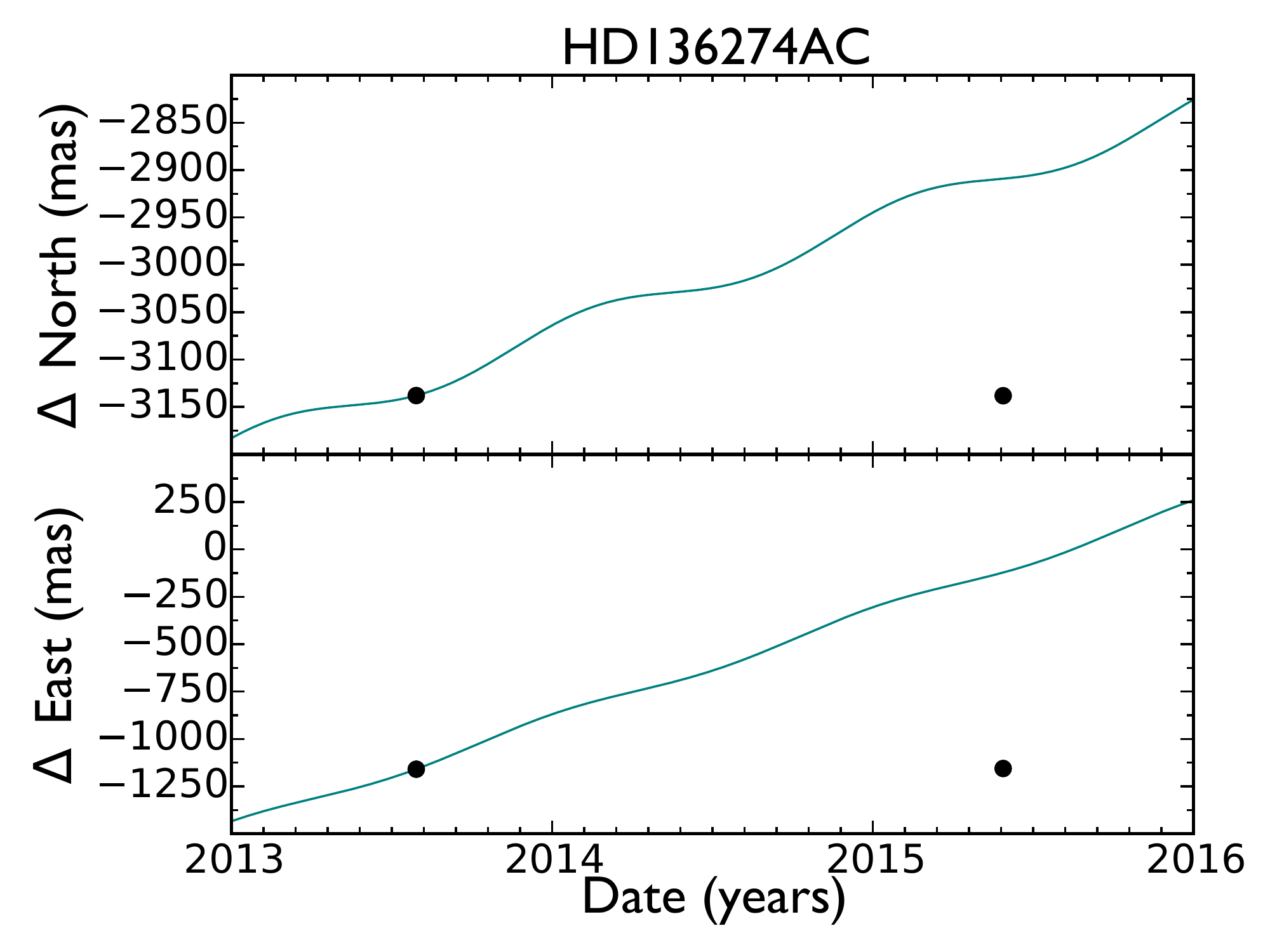}}}\\
\caption{Continuation of Figure \ref{fig:21}.} 
\label{fig:22}
\end{figure}

\begin{figure}[htp]
\subfloat{\subfloat{\includegraphics[width = .5\textwidth]{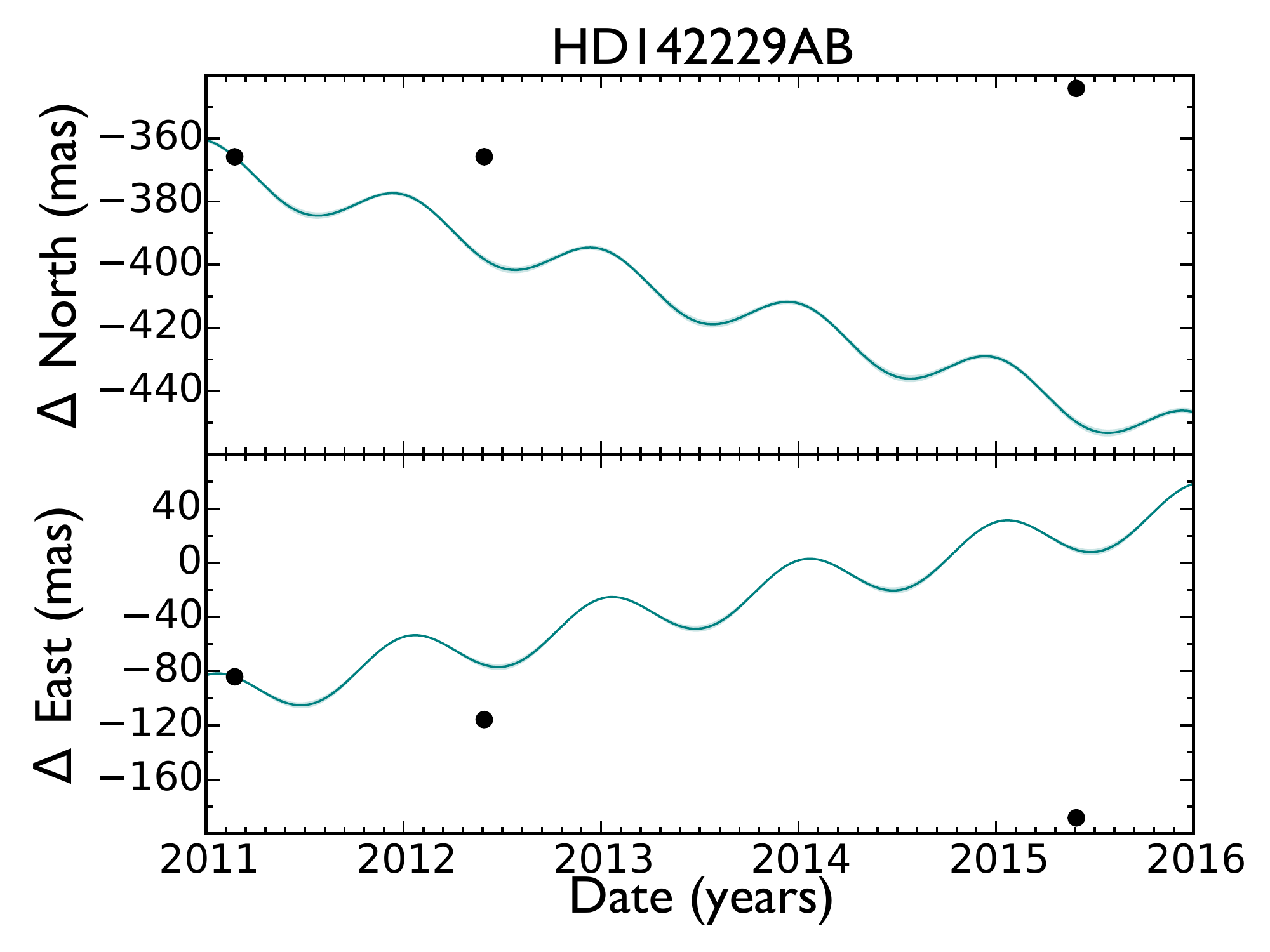}}
\subfloat{\includegraphics[width =.5\textwidth]{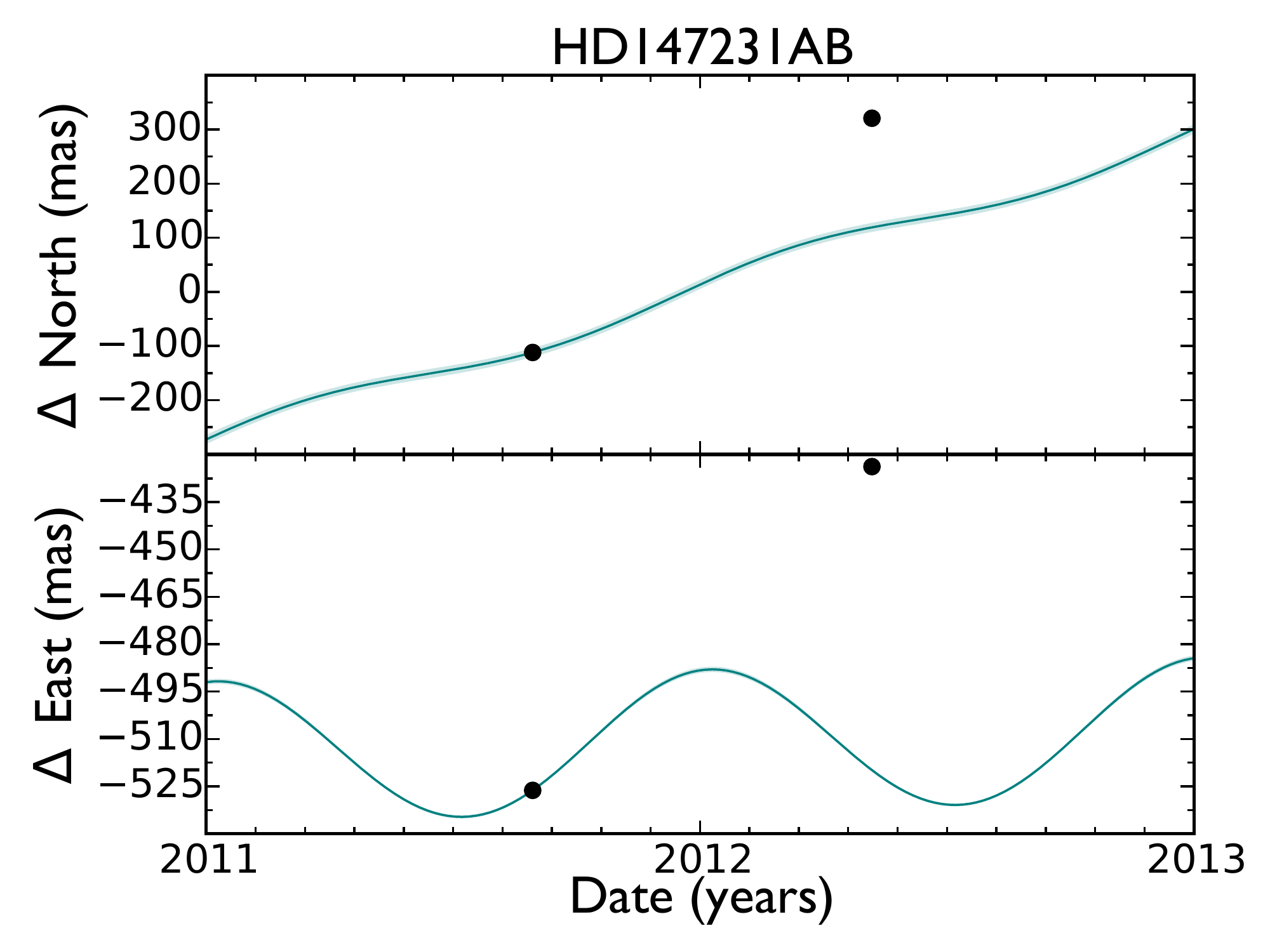}}}\\
\subfloat{\subfloat{\includegraphics[width =.5\textwidth ]{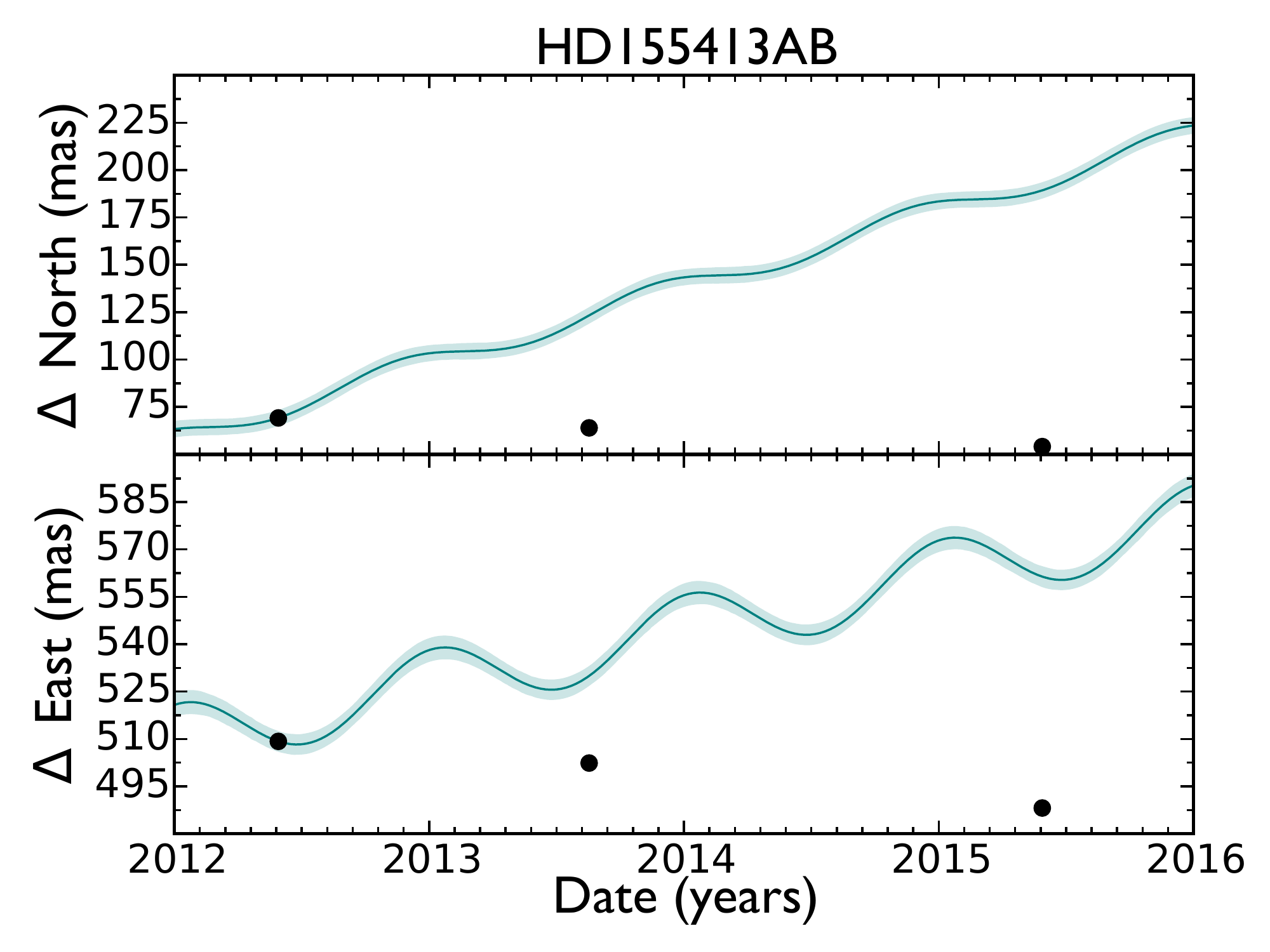}}
\subfloat{\includegraphics[width =.5\textwidth]{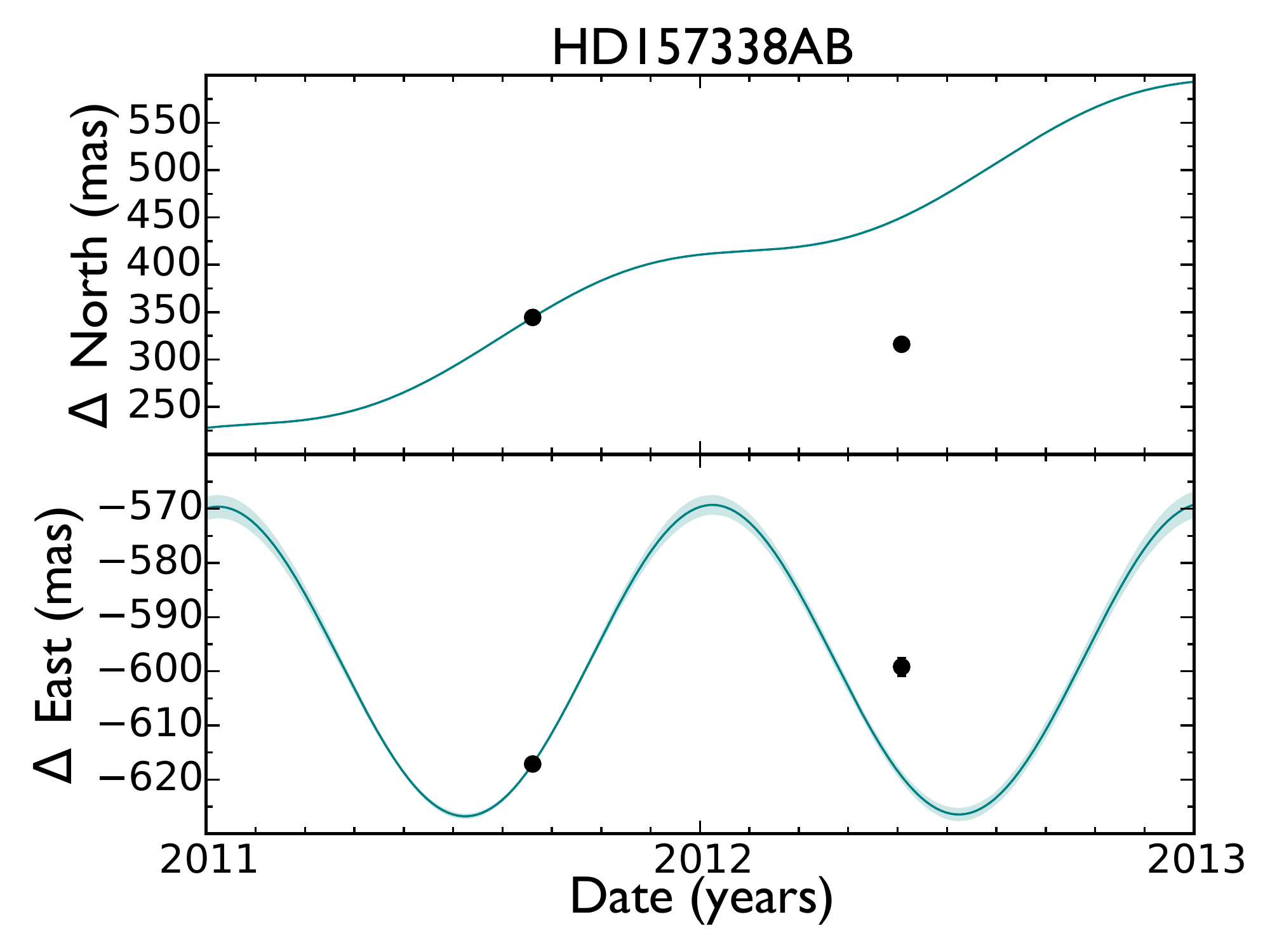}}}\\
\subfloat{\subfloat{\includegraphics[width =.5\textwidth ]{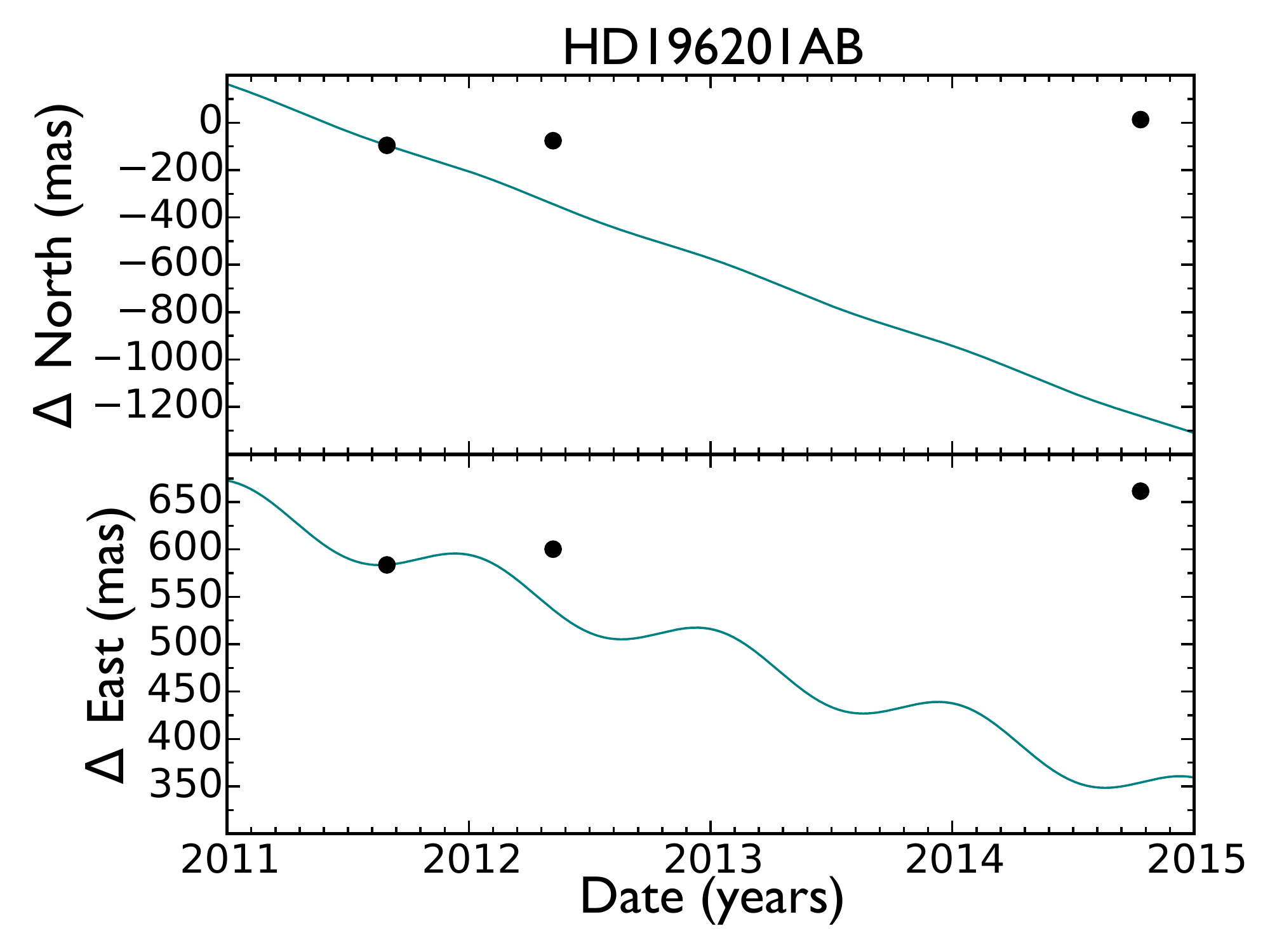}}
\subfloat{\includegraphics[width =.5\textwidth]{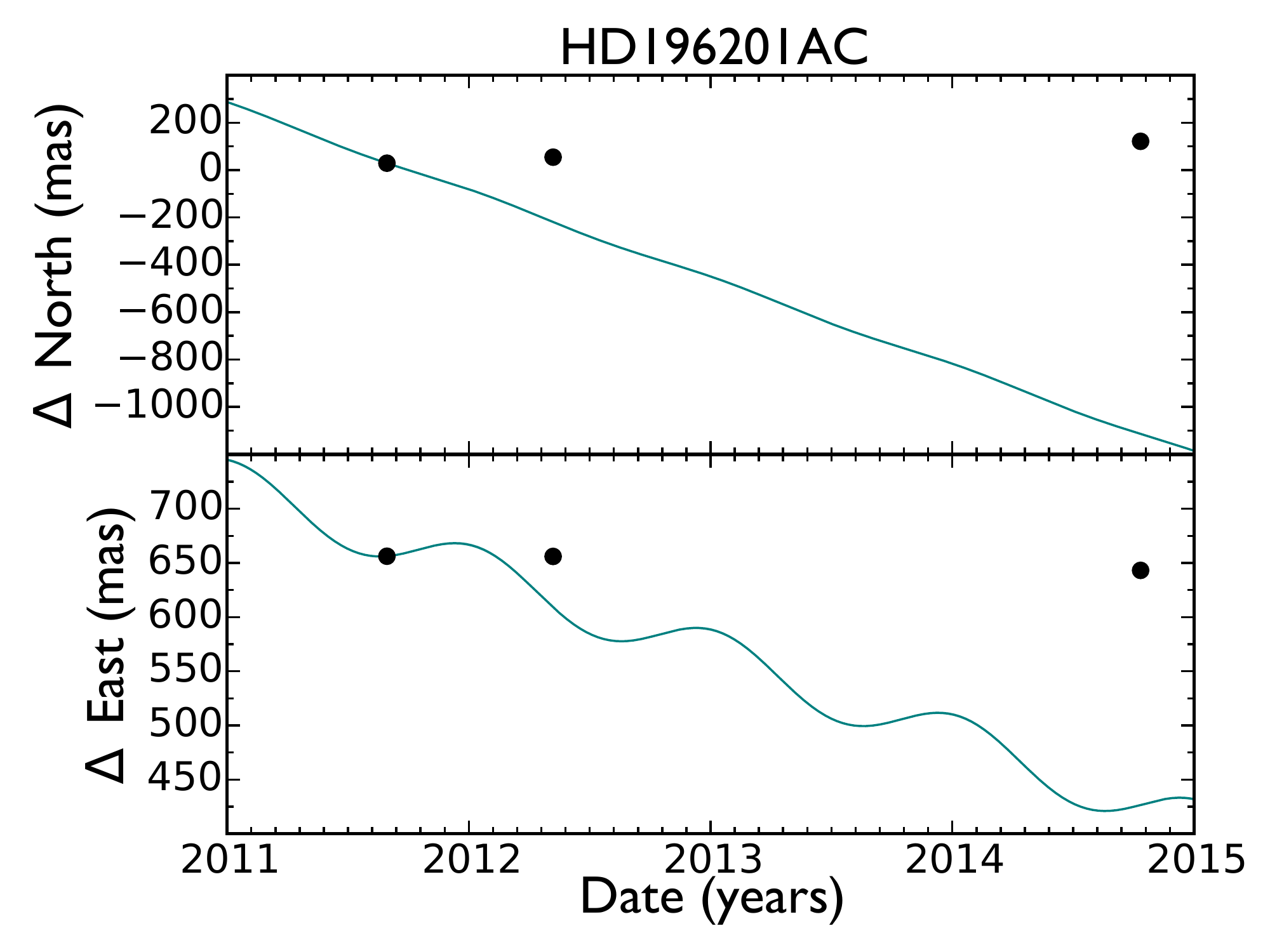}}}\\
\caption{Continuation of Figure \ref{fig:22}.} 
\label{fig:23}
\end{figure}

\begin{figure}[htp]
\subfloat{\subfloat{\includegraphics[width = .5\textwidth]{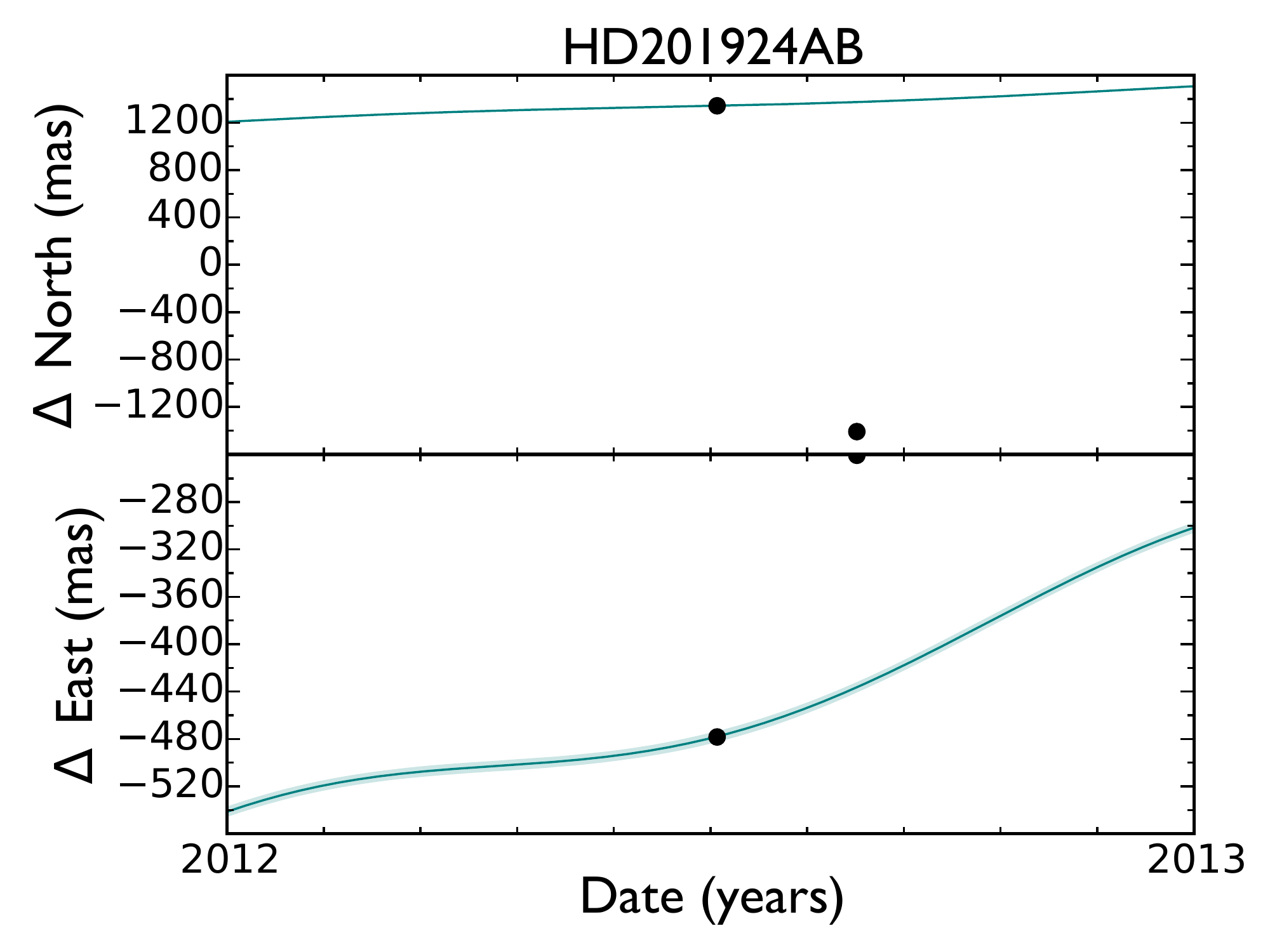}}
\subfloat{\includegraphics[width =.5\textwidth]{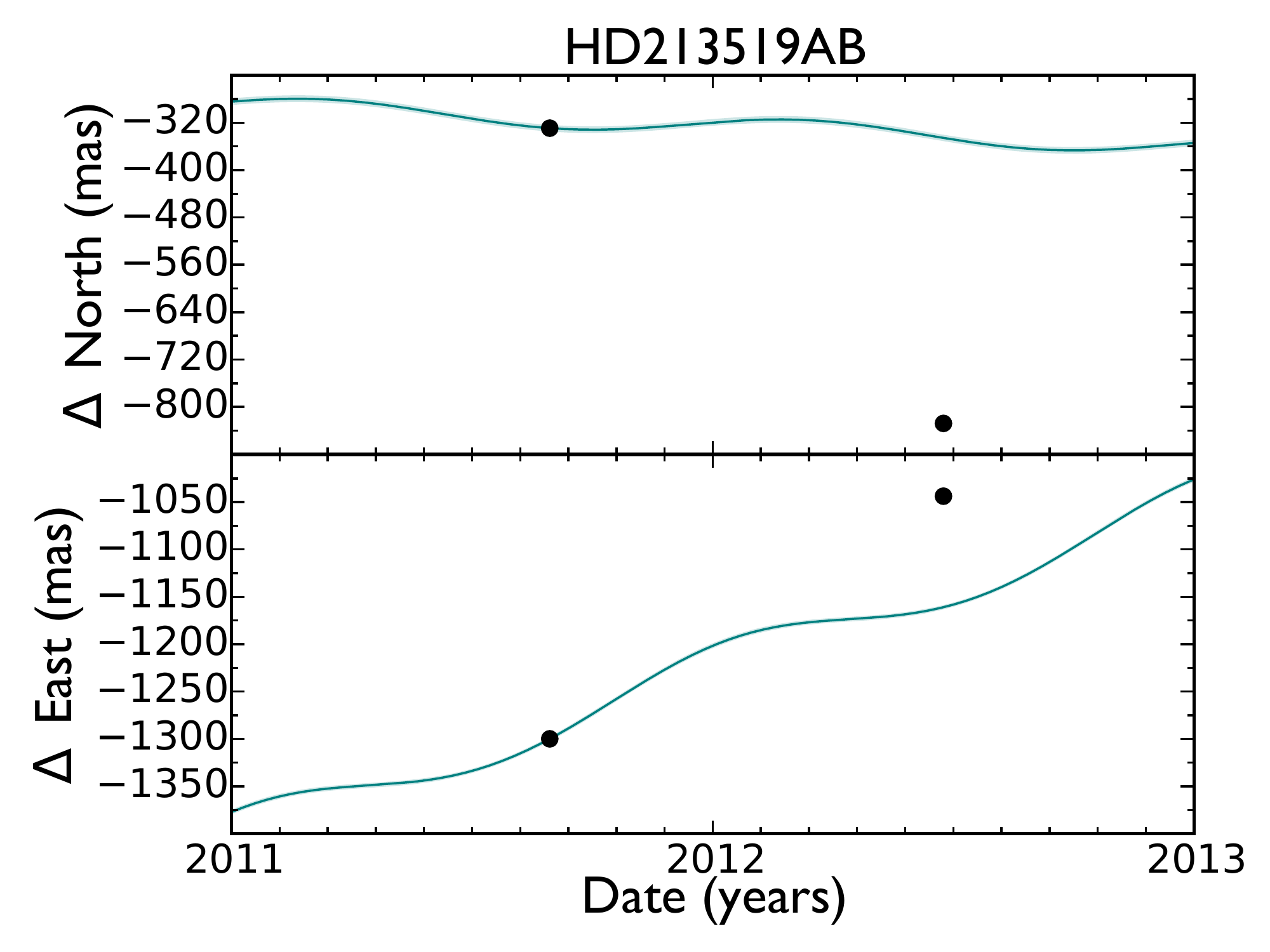}}}\\

\caption{Continuation of Figure \ref{fig:23}.} 
\label{fig:24}
\end{figure}

\end{document}